\documentclass[12pt,a4paper,openright,oldfontcommands,twosided]{memoir}
\usepackage[a4paper,margin=2.7cm]{geometry}
\usepackage{graphicx}
\usepackage[dvips]{color}
\usepackage{pstricks} 
\usepackage{natbib}
\usepackage[latin1]{inputenc} 
\usepackage{epsfig}
\usepackage{bm}
\usepackage{url}
\usepackage{rotating}
\usepackage{aas_macros}
\usepackage{amssymb,amsmath}
\newsubfloat{figure}

\chapterstyle{madsen}

\newcommand{\e}[1]{\cdot10^{#1}}

\psset{xunit=\columnwidth,yunit=0.85\columnwidth}

\newcommand{\be}{ \begin {equation}}
\newcommand{\ee}{ \end {equation}}
\newcommand{\G}{ G}
\newcommand{\degree}{\ensuremath{^\circ}}
\newcommand{\Fig}{figure }
\newcommand{\Figs}{figures }
\newcommand{\Eq}{equation }
\newcommand{\Eqs}{equations }
\newcommand{\Prom}{Prometheus}

\newcommand*{\setasuspacing}[1]{%
\let\AsuSpacing#1
\AsuSpacing} 
\maxsecnumdepth{subsection}

\author{Hanno Rein}
\title{Thesis}

\begin{document}
\frontmatter
\pagestyle{empty}
\begin{titlingpage}
~	\vspace{-0.2cm}
	\begin{flushleft}
	\resizebox{1.5cm}{!}{\includegraphics{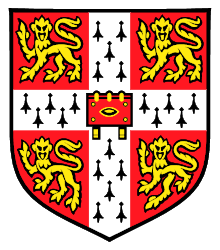}}\\
	\Large University of Cambridge\\
	\normalsize
	Department of Applied Mathematics\\
	and Theoretical Physics

	\vspace{2.2cm} 
	\hrule
	\vspace{2.2cm} 

	\Large \chaptitlefont\flushleft
	 The effects of stochastic forces \\on the evolution of planetary \\ systems and Saturn's rings \\ 
	\vspace{0.5cm} 
	\Large Hanno Rein
	\normalfont 
	\vspace{1.9cm} 

	\normalsize This dissertation is submitted for the degree of Doctor of Philosophy.\\
	\vspace{0.5cm} 
	\normalsize June 14, 2010
	
	\vspace{2.2cm} 
	\hrule 
	\vspace{2.2cm} 

	\resizebox{1.5cm}{!}{\includegraphics{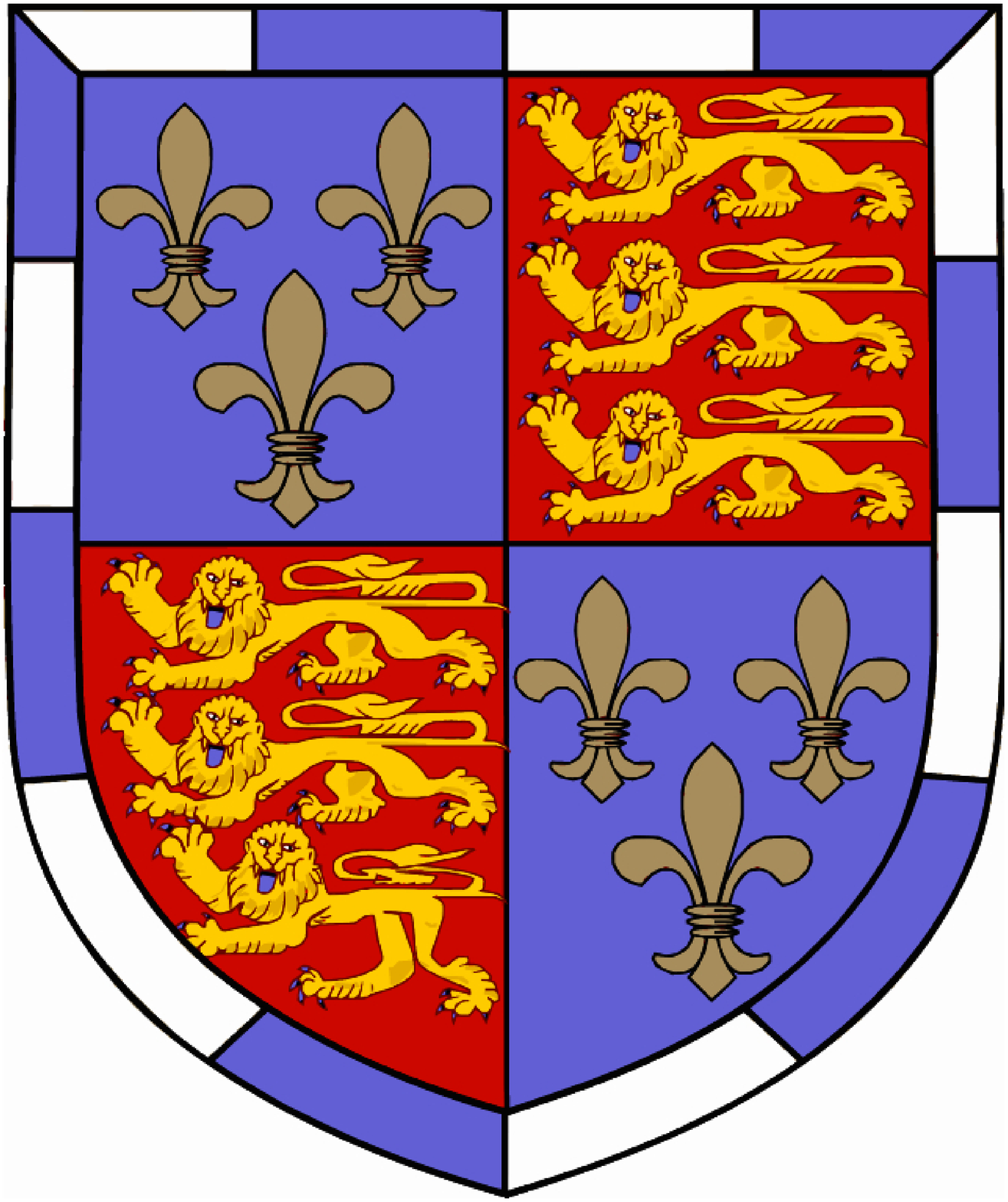}}\\
	\Large
	St John's College\\ 
	\normalsize
	Cambridge
	\end{flushleft}
\end{titlingpage}

\begin{center}
This version has been modified for the arXiv. The original version with higher quality figures can be found at \url{http://sns.ias.edu/~rein/} .
\end{center}

\mbox{}\vfill
{\centering \par { \begin{minipage}{0.94\textwidth}\includegraphics[width=\textwidth]{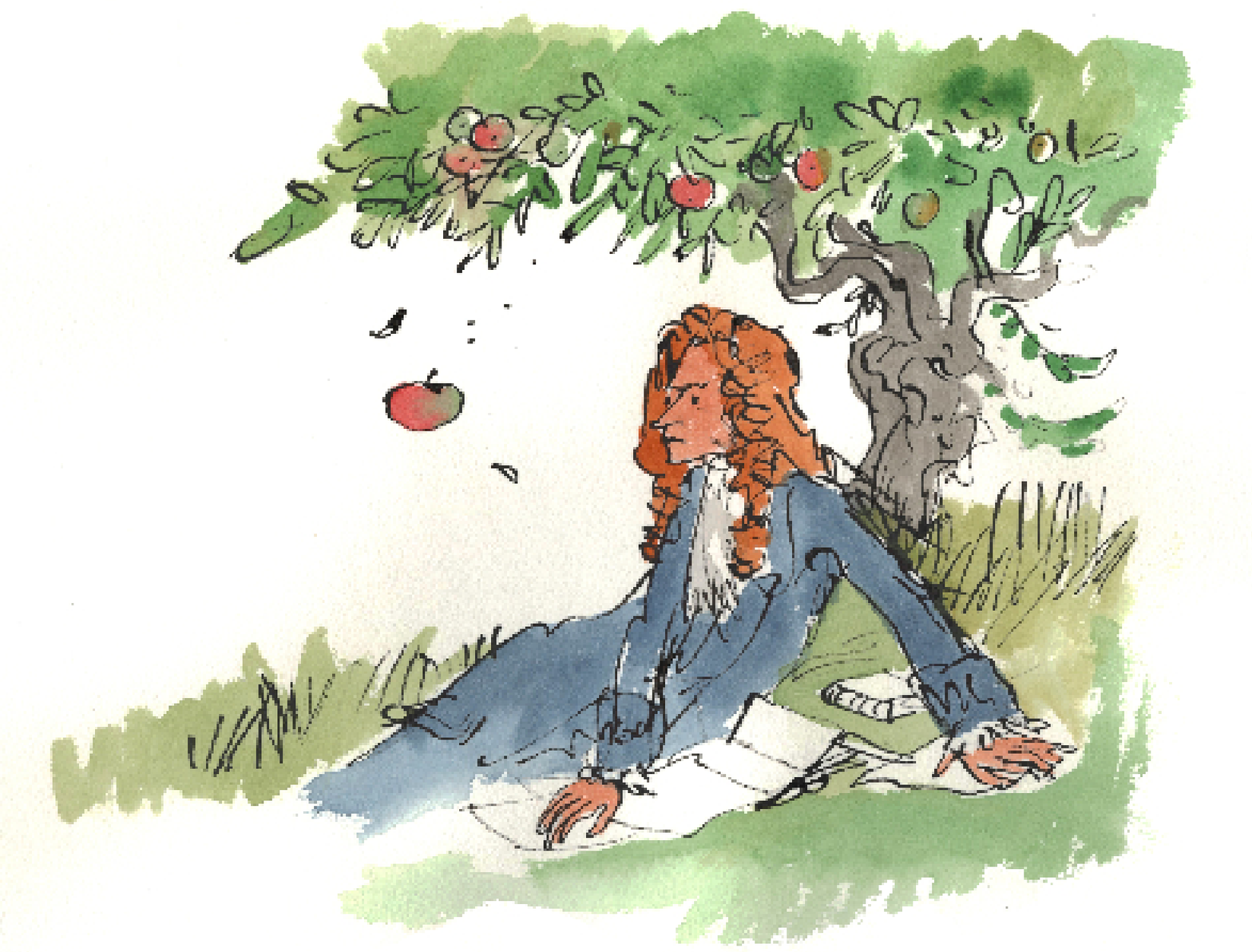} 
\end{minipage} \par } }
\vfill

\chapter{Abstract}
The increasing number of discovered extra-solar planets opens a new opportunity for studies of the formation of planetary systems. Their diversity keeps challenging the long-standing theories which were based on data primarily from our own solar system.
Resonant planetary systems are of particular interest because their dynamical configuration provides constraints on the otherwise unobservable formation and migration phase. 

In this thesis, formation scenarios for the planetary systems HD128311 and HD45364 are presented. 
N-body simulations of two planets and two dimensional hydrodynamical simulations of proto-planetary discs are used to realistically model the convergent migration phase and the capture into resonance. 
The results indicate that the proto-planetary disc initially has a larger surface density than previously thought. 

Proto-planets are exposed to stochastic forces, generated by density fluctuations in a turbulent disc. 
A generic model of both a single planet, and two planets in mean motion resonance, being stochastically forced is presented and applied to the system GJ876. 
It turns out that GJ876 is stable for reasonable strengths of the stochastic forces, but systems with lighter planets can get disrupted. 
Even if a resonance is not completely disrupted, stochastic forces create characteristic, observable libration patterns.

As a further application, the stochastic migration of small bodies in Saturn's rings is studied. 
Analytic predictions of collisional and gravitational interactions of a moonlet with ring particles are compared to
realistic three dimensional collisional N-body simulations with up to a million particles.
Estimates of both the migration rate and the eccentricity evolution of embedded moonlets are confirmed. 
The random walk of the moonlet is fast enough to be directly observable by the Cassini spacecraft. 

Turbulence in the proto-stellar disc also plays an important role during the early phases of the planet formation process. 
In the core accretion model, small, metre-sized particles are getting concentrated in pressure maxima and will eventually undergo a rapid gravitational collapse to form a gravitationally bound planetesimal. 
Due to the large separation of scales, this process is very hard to model numerically. 
A scaled method is presented, that allows for the correct treatment of self-gravity for a marginally collisional system by taking into account the relevant small scale processes.
Interestingly, this system is dynamically very similar to Saturn's rings. 

\chapter{Preface and Acknowledgements}
This thesis is the result of my own work and includes nothing which is the outcome of work done in collaboration except where specifically indicated.

The formation scenario presented in chapter \ref{ch:threetwo} was done in collaboration with Professor Wilhelm Kley and has been published as a highlighted article in Astronomy and Astrophysics \citep{ReinPapaloizouKley2010}. 
The analysis of chapter \ref{ch:randwalk} is based on \cite{ReinPapaloizou2009}. 
The results of chapter \ref{ch:planetesimals} are the outcome of work done in collaboration with Geoffroy Lesur and Zo\"{e} Leinhardt. 
I have contributed more than two thirds to all of these studies. In particular, all simulations and their analysis have been performed by myself. 
The analytic calculation in appendix \ref{app:response} has been made by Professor John Papaloizou.

I would like to thank my supervisor, Professor John Papaloizou, for his time and support during the last three years. I would like to thank Zo\"{e} Leinhardt, Geoffroy Lesur, Sijme-Jan Paardekooper, Gordon Ogilvie, Aur\'{e}lien Crida, James Stone and Adrian Barker for stimulating discussions.
For proofreading a draft of this thesis, I would like to thank Min-Kai Lin, Adrian Barker, Katy Richardson, Zo\"{e} Leinhardt, Aur\'{e}lien Crida and Chris Donnelly.
I appreciate the help of Professor Wilhelm Kley who provided me with data for various code comparisons.

This work was supported by an Isaac Newton studentship and an STFC studentship (ST/F003803/1). 
I am grateful for financial support for travel from St John's College Cambridge, the Department of Applied Mathematics and Theoretical Physics, the Institute for Advanced Study in Princeton, AstroSim, the Cambridge Philosophical Society and the Tokyo Institute of Technology. I would also like to thank the Isaac Newton Institute in Cambridge for their support and hospitality during the DDP workshop and associated conferences. 

I would like to thank the 800th Anniversary Team for the permission to reprint the drawing of Isaac Newton, copyright Quentin Blake 2009.

\cleartoverso 
\tableofcontents

\cleartorecto 
\mbox{}\vfill
{\centering\par F\"{u}r meine Mutter. \par}
\vfill\clearpage

\mainmatter
\pagestyle{plain}
\chapter{Introduction}\label{ch:introduction}
\setlength{\epigraphwidth}{9cm}
\epigraph{
{
And if the fixed Stars are the Centers of other like systems, these, being form'd by the like wise counsel, must be all subject to the dominion of One, [...].
}}{\tiny Isaac Newton, General Scholium, translated by Motte, 1729}

\noindent In 1713 Isaac Newton wrote this sentence in an essay attached to the third edition of his famous Principia Mathematica.
In other words, he expected to see planets around other stars. 
Almost three centuries later, astronomers discovered the first planet beyond our own solar system, a so-called exo-planet. 
The number of known planets has increased rapidly ever since. To date, 461 extra-solar planets have been discovered \citep{exoplanet}. 
At least 10\% of all nearby solar type stars host planets \citep{Cumming2008}.
With this tremendous observational success, it is now the theoreticians' turn to explain the discovered systems. 
One important aspect is to find out if, and if so why, these systems formed differently compared to the solar system.

In this chapter, first the discoveries of exo-planets in recent years are presented. Then, suggested formation scenarios of planets, planetary systems and their evolution are reviewed.

This thesis discusses stochastic phenomena in a range of astrophysical systems. 
The analytic model presented in chapter \ref{ch:randwalk} is the key in understanding the effects of stochastic forces and turbulence in those systems. 
It forms the basis of the physical understanding in many celestial systems in which stochastic forces are present.

A detailed formation scenario of the planetary system HD45364 is presented in chapter \ref{ch:threetwo}. 
In chapter \ref{ch:randwalk} another formation scenario is presented, this time for the planetary system HD128311. 
Both systems are resonant systems and their dynamical states provide important constraints on their formation history. 

HD45364 formed most likely in a massive disc and had a phase of rapid convergent migration. 
On the other hand, the current observed orbital parameters of the system HD128311 are consistent with the formation in a strongly turbulent disc. 

In chapter \ref{ch:moonlet}, these formalism developed in chapter \ref{ch:randwalk} is applied to Saturn's rings and a moonlet. 
Saturn's rings also exert stochastic forces. The rings, together with embedded moons, resemble a small scale version of the proto-planetary disc.

Finally, the issue of numerical convergence in simulations of planetesimal formation is discussed in section \ref{ch:planetesimals}. 
Planetesimals are likely to form in turbulent proto-stellar discs via gravitational instability. 
It turns out that the system can be simulated consistently only if the relevant small scale processes are included. 
The dynamical evolution is then very similar to Saturn's rings, except that the final clump is gravitationally bound.

In chapter \ref{ch:summary} we summarise the results. 
The main numerical codes that have been used in chapters \ref{ch:threetwo}, \ref{ch:randwalk}, \ref{ch:moonlet} and \ref{ch:planetesimals} are described in the appendices \ref{app:dpmhd3d} and \ref{app:gravtree}.

\section{Methods of detecting extra-solar planets}\label{sec:introduction:methods}
For thousands of years, the observation of planets was limited to the major planets of our own solar system. Since then, our knowledge of the solar system has been growing constantly and has reached an overwhelming magnitude. 
However, the fundamental question of whether we inhabit a special place in the universe or not hasn't been answered yet. This is deeply linked to our spiritual desire to know if there are lifeforms on other planets. Luckily, we are living in a time of great technological and scientific progress and it is not unreasonable to assume that those questions can be answered within the next 50 years. 

The first steps have already been taken, namely the discovery of planets beyond our own solar system. Various groups around the world have successfully detected exo-planets using different techniques. Each method has both advantages and disadvantages which will be summarised in this section. This is important because the sparse information that we get from these observations determines the predictability of theoretical studies. 

\begin{description}
\item[Radial velocity measurements.]
Most planets have been discovered by the radial velocity (RV) method. A periodic Doppler shift in the spectral lines of the host star can be measured with high precision spectroscopy. 
The Doppler effect occurs because the star is moving around the centre of mass (of both the star and the planet). Only the radial part of this movement is measurable. 
The period of the oscillation is simply the planet's orbital period.
The mass of the star can be calculated by stellar evolution models. If we assume that we observe the system in the plane of the planet's orbit, we can then use Newton's third law $M_\odot v_\odot = m_p\,v_p$, where $M_\odot$, $m_p$, $v_\odot$ and $v_p$ are the masses and velocities of the host star and the planet respectively, to calculate the planet's mass. 

This method biases the detection of massive planets on short orbits (hot Jupiters) as the gravitational influence of the planet on the star (and therefore the measured $v_\odot$) is bigger in those cases.
Note that there is a degeneracy in the inclination $i$ because it is not possible to measure the non-radial part of the velocity. 
Thus, we can only get a lower limit on the mass of the planet and the mass might actually be larger:
\begin{eqnarray}
m_{p,\text{real}} \geq m_{p,\text{real}}\cdot \sin i = m_p.
\end{eqnarray}

Because of this degeneracy, only a limited number of parameters can be obtained by the RV method.
Finding orbital parameters of multiplanetary systems beyond the planet masses and the orbital periods is even more challenging. Each new planet adds 7 degrees of freedom, such that solutions are highly degenerate. Especially the fitting of the eccentricities is very unreliable (see discussion in chapter \ref{ch:threetwo}).
Furthermore, it is difficult to find precise orbital parameters for very long period planets as it can take many years to sample a full period of the light curve. 

The first extra-solar planet discovered by the radial velocity method is 51 Pegasi b \citep{Mayor1995}.

\item[Transit light curves.]
Another method that has been very successful is the transit method. 
This is mainly due to space based missions such as CoRoT and Kepler.
The idea is to find planets by observing variations in the star light caused by transits of planets in front of the host stars.
This effect also occurs in our own solar system, known as Mercury and Venus transits. 
However, transits of extra-solar planets are rare because the host star, the exo-planet and the Earth must be aligned exactly in one line for a transit to occur. 
The first exo-planet discovered by the transit method is OGLE-TR-56 b \citep{Konacki2003}.

The transit method is capable of measuring the density $\rho$ of exo-planets, a parameter inaccessible to the RV method.  
It is even possible to do spectroscopy on the atmosphere of the planet during the transit and the secondary transit (when the planet is occulted by the star) to create a temperature map of the exo-planet \citep{tempmapplanet}. 

Furthermore, the Rossiter-McLaughlin effect allows one to measure the sky projected angle between the orbit and the rotation axis of the host star. To do that, one has to combine transit and RV measurements. A recently submitted paper suggests that a large number of planets might be on retrograde orbits (Triaud et al, in preparation). 

Dedicated ground and space based missions promise to detect a large number of planets in the near future. 
Furthermore, by observing tiny variations in the transit timing (TTV) and transit duration (TDV), it might be possible to find other planets in systems in which only one planet is transiting. Even the possibility to discover exo-moons has been discussed \citep{Kipping2009}.

\item[Gravitational microlensing.]
Planets can also be discovered by gravitational microlensing. 
If the star is aligned in one line with the Earth and a bright object in the background (e.g. another star), the light from the background object bends around the star and can be detected by an increase in luminosity. 
A planet in an orbit around the star disturbs the light curve and theoretical models can determine the planet's mass and some orbital parameters. 
This event happens only once per star, so good timing and a global collaboration is needed to perform a continuous measurement of the light curve. 
The orbital parameters cannot be determined with high precision because one only obtains a snapshot of the system without any dynamical evolution.
The first exo-planet was discovered by gravitational microlensing in 2004 by \citeauthor{microlensing}.

\item[Direct imaging.]
In 2004 \citeauthor{directimaging} reported the first detection of a giant planet candidate by direct imaging. Since then, several planetary systems have been imaged. These are all massive planets, far away from the host star (hundreds of AU). Maybe the most interesting of those planets is Fomalhaut b, which seems to be embedded in a debris disc \citep{Kalas2008}. As all those planets have long orbital periods, it is, similar to the gravitational microlensing method, difficult to determine the precise orbital configuration, especially in multi-planetary systems.

\item[Pulsar timing.]
Despite the recent success of the detection methods presented above, the first exo-planet was discovered using the pulsar timing method \citep{Wol1992} by measuring slight variations in the regular timings from a pulsar. This is sensitive down to very small mass planets. However, pulsars are rare compared to normal stars and the focus has shifted away from the pulsar timing method.

\end{description}

\section{Observed objects}
Extrasolar planets have been observed around a variety of parent stars from pulsars to solar-type stars to M-dwarfs \cite[see e.g.][]{directimaging,Wol1992} indicating that planet formation is common and successful in a broad range of environments. 
Almost all extra-solar planetary systems are distinct from the solar system. Many of the detected objects are so-called hot Jupiters. Their size and mass is comparable to Jupiter, but their orbits are very close to their host star ($\approx 0.05$~AU). Most methods described above bias the detection in favour of these objects.

The existence of hot Jupiters was very surprising because in the solar system all gas giants are located beyond several AU. This still results in difficulties for planet formation theories, although planetary migration is one solution, as described below.

Other observed objects are very heavy and are more likely to be brown dwarfs rather than planets. The International Astronomical Union (IAU) defines an extra-solar planet as follows:
\begin{quote}
Objects with true masses below the limiting mass for thermonuclear fusion of deuterium (currently calculated to be 13 Jupiter masses for objects of solar metallicity) that orbit stars or stellar remnants are \textit{planets} (no matter how they formed). The minimum mass/size required for an extra-solar object to be considered a planet should be the same as that used in our Solar System. \citep{planetdef} 
\end{quote}
Thus, many of the massive objects, at separations of hundreds of AU, detected by direct imaging, could actually be brown dwarfs.

In cases where the planet's mean density can be observed, evolutionary models seem to be in broad agreement with observations, although many hot Jupiters have a large surface temperature and are subject to tidal heating. 
Their atmospheres are not well understood and first studies show a broad range of possible configurations. 
Observations furthermore suggest that their densities might be rather low \citep[see e.g.][]{Anderson2010}. 

The Holy Grail for exo-planet hunters is another Earth-like planet that orbits the host star within the habitable zone. At the present day, the planet that is closest to one Earth mass is Gliese 581 e which has a mass of approximately 2~Earth masses \citep{Mayor2009}.

\section{Formation of planets}

\subsection{Accretion disc}\label{sec:introduction:mri}
Stars form out of giant gas clouds that become gravitationally unstable.
Because of angular momentum conservation, an accretion disc forms around every new star. 
As the name suggests, an accretion disc will eventually accrete most of its material onto the star, leaving only a small fraction of it as planets or a debris disc in orbit around the star.
Observations of accretion rates in proto-planetary discs suggest that accretion timescales are of the order of a few $10^5$ years \citep{Enoch2008}. 

Usually one assumes that the Navier-Stokes equations are a good approximation to describe the fluid motion within the disc. However, to reproduce the measured accretion rates, a purely molecular viscosity is not efficient enough.
It is generally believed that this large viscosity originates from turbulence \citep{Pringle1981}.
The mass flux in a steady state Keplerian disc is then given by 
\begin{eqnarray}
\dot M = 3\pi \nu \Sigma,
\end{eqnarray}
where $\nu$ is the effective kinematic viscosity and $\Sigma$ the surface density. From observations of proto-stars, a value of $\nu\sim 10^{-5} \mathrm{AU^2/(yr/}2\pi\mathrm{)}$ has been inferred, but with considerable uncertainty.

The magneto rotational instability \citep[MRI,][]{BalbusHawley91} is the most likely candidate to be responsible for the anomalous value of $\nu$ which is often characterised using the \cite{shakurasyunyaev73} $\alpha$ parameter, defined as $\alpha =\nu/(c_s^2/ \Omega)$, with $c_s$ being the local sound speed and $2\pi/\Omega$ being
the orbital period. For the most likely situation of small or zero net magnetic flux,
recent MHD simulations have indicated $\alpha\approx 10^{-2} - 10^{-3}$. 
This value is very uncertain when realistic values of the actual transport coefficients are employed
due in no small part to numerical resolution issues \citep[see][]{FromangPapaloizou07, Fromangetal07}. 
Which parts of proto-planetary discs are adequately ionised, or constitute a dead zone, is also an issue \citep{Gammie96,Sano2000}.

Putting aside all those issues for a moment, the MRI always creates density fluctuations in the disc, resulting in a stochastic force that is exerted on embedded planets which are otherwise decoupled from the gas motion (ignoring migration, which is a laminar effect and acts on timescales much longer than one orbit).
In this thesis, we do not attempt to simulate the MRI directly and rather describe it in an empirical way. We therefore avoid all problems mentioned above and can understand the physical scaling of our results. 
Only two quantities are needed for our analytic model, the root mean square value of the stochastic gravitational force and the corresponding auto correlation time. 
\cite{LaughlinSteinackerAdams04} propose a more complicated model, in which random modes with random decay times create a gravitational potential that is supposed to mimic the MRI. 
Our, much simpler and more intuitive description, allows us to survey a large parameter space and understand the resulting physical processes, as discussed in chapter \ref{ch:randwalk}.

\subsection{Minimum mass solar nebula and snowline}\label{sec:introduction:snowline}
A standard model of a proto-planetary disc assumes a steady state and a vertical equilibrium (see also appendix \ref{sec:dpmhd:twodim}). 
The gas and dust components are furthermore assumed to be fully mixed. The resulting disc is flared, although still geometrically thin. 
For sufficiently low mass accretion rates ($\leq 10^{-8}M_\odot/\text{yr}$) the dominant heating source is stellar irradiation \citep{ChiangGoldreich1997}.

\cite{Hayashi1981} prescribes the so-called minimum mass solar nebula (MMSN) which is comprised of just enough mass to make all planets of the solar system. 
This model became the standard disc model and has been used excessively in recent years with different normalisations. 
One can describe the surface density and temperature profiles as
\begin{eqnarray}
\Sigma(r) = \Sigma_0 \left(\frac r{r_0}\right)^{-p} \quad \quad T(r) = T_0 \left(\frac r{r_0}\right)^{-q}.
\end{eqnarray}
Typical normalisation values at $r_0=1\,\mathrm{AU}$ are $\Sigma_0=1700\,\mathrm{g\,cm}^{-2}$ and $T_0=280\,\mathrm{K}$. Standard values of $p$ range from $0$ to $5/3$, those of $q$ from $1/2$ to $3/4$ \citep{Hayashi1981,Cuzzi1993}.
The typical mass of the disc is about one percent of the stellar mass.
Although the choice of disc model is essential for most aspects of planet formation, little emphasis has been placed on alternatives \citep[see e.g.][]{Desch2007,Crida2009}.

To form planets and planetary cores, dust is an important ingredient (see also below).
In the model of \citeauthor{Hayashi1981} ($q=1/2$), the radius at which temperatures drop below 170 K is around 2.7 AU. 
At larger radii, the temperatures are low enough for water ice to exist. 
This transition radius is called the snow line.
Recent studies, including more detailed models of accretional heating and radiative transfer show that the snow line could come as close as 1 AU \citep{SasselovLecar2000}.

\subsection{Planet formation mechanism}
Planets are believed to form in proto-stellar discs as a natural by-product of star formation. 
These discs are made of the same material as the star itself: gas and dust. 
Current theories give two possible explanations of the formation of giant planets inside the disc.
Both models favour planet formation at large radii, beyond the snowline. 
However, it is important to keep in mind that the process of planet formation itself is not directly observable (yet), leaving theory and numerical simulations to fill in the blanks between observations of hot circumstellar discs around young stars and planets orbiting main sequence stars.

\subsubsection{Core accretion}
One of the most important unanswered questions in the theory of planet formation is to find out what the mechanism for planetesimal formation is, i.e., the process by which the building blocks of planets are formed. 
In the core accretion model, solid components of the disc stick together, forming bigger and bigger objects until a core of about 15 Earth masses is formed \citep{BodenheimerPollack86,Pollack96}. 
The proto-planet can then start to accrete gas and form a gas giant. 

Again, there are two main theories for planetesimal formation via the core accretion model: mutual collisions \citep[e.g.,][]{Hayashi77} and gravitational instability in the dust layer \citep{GoldreichWard1973}. In the first hypothesis, dust particles grow as the result of accretion-dominated collisions. Although the formation of planetesimals by mutual collisions is consistent with meteoritic evidence, the collision speed between dust particles (or aggregates) must be much slower than the typical velocity dispersion in a standard proto-stellar disc to avoid destructive collisions \citep{BlumWurm2008}. In addition, the planetesimal formation process is so slow that metre-sized particles are in danger of spiralling into the star before growing large enough to decouple from the gas \citep{Weidenschilling1977}. Even if km-sized planetesimals were able to form, they would be in danger of being ground down again by mutual collisions \citep{Stewart2009}. 

Gravitational instability is often considered to be a solution to most of these problems because the intermediate sizes are avoided all together. In this theory, the dust layer becomes dense enough for the Keplerian shear and velocity dispersion of the dust particles to be unable to support the dust against its own self gravity. The dust then collapses into clumps that eventually cool via drag forces and mutual collisions into planetesimals. 
Many different groups have been working on this subject. Until recently, the focus has been on quiet, non-turbulent, and low density regions of the accretion disc \citep[see e.g.,][]{Michikoshi2009,Michikoshi2007,Tanga2004}.

However, the turbulent gas in the proto-planetary nebula stirs the dust, which increases the velocity dispersion of the dust particles. Several ideas have been proposed to overcome the turbulence-induced mixing of the dust particles and create localised clumps. For example, \citet{Cuzzi2008} suggest that the same turbulence that stirs the dust on larger scales may also collect the dust particles on small scales. A similar idea was proposed by \citet{Johansen2007}, in which dust particles are localised into clumps promoted by both turbulence and the streaming instability \citep{YoudinGoodman2005}. These dense clumps then become gravitationally unstable.  A third hypothesis suggests that large structures, such as vortices, may be able to collect and protect dust particles from the turbulent background \citep[][]{BargeSommeria95,Lyra2009}.

In chapter \ref{ch:planetesimals}, we focus on the gravitational collapse in a very dense and turbulent region of the proto-planetary disc. 
We look carefully at the numerical requirements of modelling gravitational instability accurately and test the validity of using super-particles in a high density region. 
The results suggest that some of the early results of graviational collapse from other authors may have been too optimistic.

\subsubsection{Gravitational fragmentation}
In this model, gas planets form directly as a result of gravitational instability within the disc \citep{Boss01}; no solid core is needed. 
However, the planet can accrete dust particles later on, and hence form a core. 

For the gravitational instability to occur, it is necessary to have a Toomre $Q$ parameter of order unity \citep{Toomre1964}. 
The precise criterion depends strongly on the thermodynamics on the disc, especially the cooling time.
However, it is unlikely that the physical conditions in a proto-planetary disc are compatible with this constraint. 
If so, this is most likely to occur at large radii ($\sim100$ AU). It therefore cannot be the formation mechanism for most discovered exo-planets.

\section{Evolution of planetary systems}

\subsection{Migration}\label{sec:introduction:migration}
Many observed planets are very close to their host star. It is implausible that they have formed at such small radii (see section \ref{sec:introduction:snowline}),
even taking into account the strong selection effect of discovering close in exo-planets (see section \ref{sec:introduction:methods}).

Tidal interactions with the proto-planetary disc give the planet some radial mobility and might therefore be the solution to this problem \citep{GoldreichTremaine1980}. 
The mutual angular momentum exchange depends on many parameters of both the planet and the disc. 
Several population synthesis simulations were able to reproduce the observed mass-period distribution using migration \citep[e.g.][]{IdaLin2004}.
Planet migration can be classified in the following four main categories.

\subsubsection{Type I}
Planets are completely embedded in a proto-planetary disc for sufficiently low mass. 
Density waves are excited in the disc, both interior and exterior to the planet, at so-called Lindblad resonances. The waves carry angular momentum and produce a torque on the planet that leads to planetary migration \citep{GoldreichTremaine79}. 
The direction of migration depends on parameters of the disc model such as the gradients of the surface density, sound speed and scale height \citep{Ward97}.
In most cases, the outer torque is bigger than the inner one and the planet migrates inward. This effect is called type I migration. 

If the migration rate is very fast, low mass planets are in danger of spiralling into the star within the disc lifetime.  
The precise speed and direction in a realistic disc model are still subject to debate, where recently the focus has been on non-isothermal equations of state and the effects of co-rotation torques \citep{Paardekooper2010,PaardekooperPapaloizou2009}.

\subsubsection{Type II}
If the planet is massive enough, it can open a gap in the disc. For most disc models about a Jupiter mass is needed to clear a clean gap. The process is similar to the gaps opening in Saturn's rings with moons orbiting within the gap. 
The gap establishes a flow barrier to the disc material, effectively locking the planet to the viscous disc evolution \citep{LinPapaloizou1986}.
Thus, the planet still migrates, the regime being called type II migration. However, the migration timescale is set by the disc evolution timescale which is in general much longer than the type I migration timescale. 
 
\subsubsection{Type III}
In a disc with a relatively large surface density (several times the MMSN), the density distribution in the co-orbital region of the planet can be asymmetric.
This leads to a self-sustained torque that is proportional to the migration rate. 
In this regime, called type III migration, the planet can fall inward on a timescale much shorter than the disc evolution time obtained for type II migration, even shorter than the timescales associated with type I migration. 
The planet moves faster than the disc can respond to its perturbation, thus maintaining an asymmetry.
The net torque can cause both inward or outward migration \citep{Peplinski2008,Peplinski2008b,Peplinski2008c}.

In chapter \ref{ch:threetwo} we present the first observable indication that type III migration is indeed responsible for shaping planetary systems.

\subsubsection{Type IV}
The different migration regimes have all been studied in quiet non-turbulent discs. However, a proto-planetary disc is thought to be at least partially turbulent (see section \ref{sec:introduction:mri}). Assuming the simplest scenario in which the effects of turbulence and the net migration are separable, one can describe the stochastic migration as a diffusion process in a distribution of planets on top of the net orbital migration, described above. We call this regime type IV.

This idea leads to a Fokker-Planck description \citep{AdamsBloch2009}. There are two limits to this approach. The first, in which the net migration dominates and the effects of turbulence are negligible (mostly for heavy planets). The second, in which the net migration can be ignored and the turbulence driven diffusion dominates on short timescales (mostly for small mass planets and planetesimals). We are in an intermediate regime if the migration and diffusion timescales are comparable. It is unlikely that the interplay of stochastic migration and net migration plays an important role in determining the final semi major axis of planets because the associated timescales would have to be very similar. 
However, resonant systems are more sensitive to stochastic forces and might therefore provide observational constraints on the strength of turbulence that was present at early times.

So far, only preliminary work has been done on studying the effect of the turbulence on resonances \citep{Adams08, NelsonPapaloizou04}. 
If the turbulence is active long enough, it will eventually kick the planets out of resonance. This might happen even if the stochastic forces are not strong
enough to cause significant orbital migration. 
However, the timescale needed for destroying a resonance is not well constrained.  New results on this issue are presented in chapter \ref{ch:randwalk}.

\section{Mean motion resonances}\label{sec:introduction:resonance}
Resonances in celestial mechanics can occur when there is a simple numerical relation between two frequencies \citep{solarsystemdynamics}. For example, a planet could feel a periodic gravitational force from another planet that is a multiple of its own orbital frequency. This can either lead to an unstable situation in which angular momentum is exchanged until the resonance ceases to exist or a planet gets ejected from the system, or a stable configuration. 

Resonances can form when dissipative forces act on the planets, for example in a proto-planetary disc (see e.g. \cite{Malhotra1993} and appendix \ref{app:leepeale}).
When there are two (or more) planets in the disc, the migration rates might differ for various reasons. One possibility is that the planets are in different migration regimes (see section \ref{sec:introduction:migration}). 
The resulting differential migration changes the ratio of orbital periods~$P_i$ and the ratio of semi major axes~$a_i$. 
These ratios are related by Kepler's third law, $\left({T_1}/{T_2}\right)^2 = \left( {a_1}/{a_2} \right)^{3}$.

The planets can become locked into a mean motion resonance (MMR) if the ratio of orbital periods gets close to a rational number $q/p$ with small integers $q$ and $p$, e.g. $ 1/2, 1/3$ or $2/3$. 
The numbers $p$ and $q$ can be interpreted as the number of completed orbits after the same time.  
Some of these resonances are stable and planets can stay in resonance over a long period of time, migrating together and keeping the ratio of their orbital periods constant. 
Beside the presence of Hot Jupiters, resonance capturing is admitted as strong evidence that planets undergo a phase of orbital migration. 
This idea was successful in explaining the observed resonant multi-planet systems GJ876 and 55 Cancri \citep{LeePeale01,SnellgrovePapaloizouNelson01} which are in a 2:1 resonance. 
In chapter \ref{ch:threetwo}, we discuss the first successful formation scenario for a system that is in a 3:2 resonance.

\chapter{The dynamical origin of the multi-planetary system HD45364} \label{ch:threetwo}
\setlength{\epigraphwidth}{9cm}
\epigraph{
Truth is ever to be found in simplicity, and not in the multiplicity and confusion of things. 
}{\tiny Issac Newton, unpublished manuscript, Frank E. Manuel 1974}

\noindent The recently discovered planetary system HD45364, which consists of Jupiter- and Saturn-mass planets, is
very likely in a 3:2 mean motion resonance. 
The standard scenario for forming planetary commensurabilities involves convergent migration of two planets embedded in a
proto-planetary disc. However, when the planets are initially separated by a period ratio larger than two,
convergent migration will most likely lead to a stable 2:1 resonance, incompatible with current observations.

Rapid type III migration of the outer planet crossing the 2:1 resonance is one possible way around this problem. 
Here, we investigate this idea in detail. 
We present an estimate of the required convergent migration rate in section \ref{sec:threetwo:estimate} and confirm this with N-body simulations in section \ref{sec:threetwo:formationnbody}  and hydrodynamical simulations in section \ref{sec:threetwo:formationhydro}. 
If the dynamical history of the planetary system had a phase of rapid inward migration that 
forms a resonant configuration as we suggest here, then we predict that the orbital parameters of the two planets
will always be very similar and should show evidence of that. 

We use the orbital parameters from our simulation to calculate a radial velocity curve and compare it to observations in section \ref{sec:threetwo:observation}.
Our model provides a fit that is as good as the previously reported one. 
However, the eccentricities of both planets are considerably smaller and the libration pattern is different. 
Within a few years, it will be possible to observe the planet-planet interaction directly and thus distinguish
between these different dynamical states. 

This is the first prediction of orbital parameters for a specific extra-solar planetary system 
derived from planet migration theory alone. 
It provides strong evidence on how the system formed.

\section{The planetary system HD45364}
\subsection{The standard formation scenario and its problems}
The planets in this system have masses of $m_1=0.1906M_{\text{Jup}}$ and $m_2=0.6891M_{\text{Jup}}$ and are orbiting the star 
at a distance of $a_1=0.6813~\text{AU}$ and $a_2=0.8972~\text{AU}$, respectively \citep{CorreiaUdry2008}. 
The period ratio is close to $1.5$. 
This alone does not imply that the planets are in a 3:2 mean motion resonance. 
However, a stability analysis shows that the three body system (two planets and a star) is only stable if the planets are  
in a 3:2 mean motion resonance. Furthermore, the region of greatest stability also contains the best statistical, Keplerian fit to the radial velocity measurements \citep{CorreiaUdry2008}.

In the core accretion model, a solid core is firstly formed by dust aggregation. 
After a critical core mass is attained \citep{Mizuno1980}, the proto-planet accretes a gaseous envelope 
from the nebula \citep{BodenheimerPollack86}. 

As explained in more detail in section \ref{sec:introduction:snowline}, the planets have most likely formed further out in cooler 
regions of the proto-stellar disc as water ice, which is an important ingredient for dust 
aggregation, can only exist beyond the snow line.  The snow line is generally assumed to be at radii 
larger than $2~\text{AU}$ \citep{SasselovLecar2000}.

When the planets have obtained a substantial part of their mass, they migrate due to planet disc interactions (see section \ref{sec:introduction:migration}).
Although the details of this process are still being hotly debated, the existence 
of many resonant multiplanetary systems and hot Jupiters supports this idea and suggests that planets preferably migrate inwards.
Both planets in the HD45364 system are inside the snow line, implying that they should have migrated inwards significantly.

The migration rate depends on many parameters of the disc such as surface density, viscosity, and the mass of the planets. The planets are therefore in general expected to 
have different migration rates, which leads to the possibility of convergent migration. 
In this process the planets approach orbital commensurabilities.
If they do this slowly enough, resonance capture may occur \citep{Gold65},
after which they migrate together maintaining a constant period ratio thereafter. 

Studies by several authors have shown that when two planets, either of equal 
mass or with the outer one more massive, undergo differential convergent migration,
capture into a mean motion commensurability is expected provided that
the convergent migration rate is not too fast \citep{SnellgrovePapaloizouNelson01}. 
The observed inner and outer planet masses are such that, if (as is commonly assumed for multiplanetary systems of this kind)
the planets are initially separated widely enough that their period ratio exceeds $2,$
a 2:1 commensurability is expected to form at low migration rates 
\cite[e.g.][]{NelsonPapaloizou2002,KleyPeitz04}.

\cite{PierensNelson2008} have studied a similar scenario where the goal was to resemble 
the 3:2 resonance between Jupiter and Saturn in the early solar system. 
They also find that the 2:1 resonance forms in early stages;
however, in their case the inner planet had the higher mass, whereas the planetary system
that we are considering has the heavier planet outside. 
In their situation the 2:1 resonance is unstable, enabling the formation
of a 3:2 resonance later on, and the migration rate can stall or even reverse \citep{MassetSnellgrove2001}. 

For the planetary system HD45364, this standard picture poses a new problem. Assuming that the 
planets have formed far apart and were not much smaller during the migration phase, the outcome is almost always a 2:1 
mean motion resonance, not 3:2 as observed. The 2:1 resonance that forms is found to be extremely stable. 
One possible way around this is a very rapid convergent migration phase that passes quickly through the 2:1 resonance.

\subsection{Avoiding the 2:1 mean motion resonance}
\label{sec:threetwo:formationnbody2:1}
\label{sec:threetwo:estimate}
We found that, if two planets with masses of the observed system are in a 2:1 mean motion resonance, 
which has been formed via convergent migration, this resonance is very stable.
An important constraint arises, because  at the slowest migration rates, the 2:1 resonance is expected to form
rather than the observed 3:2 commensurability provided the planets start migrating
outside any low-order commensurability.

We can estimate the critical relative migration timescale $\tau_{a,\text{crit}}$
above which a 2:1 commensurability forms with the condition that
the planets spend at least one libration period migrating through the resonance.
The resonance's semi-major axis width $\Delta a$ associated with the 2:1 resonance can be estimated from
the condition that two thirds of the mean motion difference across $\Delta a$
be equal in magnitude to $2\pi$ over the libration period. This gives
\begin{eqnarray}
\Delta a &=& \frac{\omega_{lf}a_2}{n_2} \label{eq:reswidth}
\end{eqnarray}
where $a_2$ and $n_2$ are the semi major axis and the mean motion of the outer planet, respectively. 
The libration period $2\pi/\omega_{lf}$ can be expressed in terms of the orbital parameters 
\citep[see e.g.][and also chapter \ref{ch:randwalk}]{Gold65, ReinPapaloizou2009} but is here measured numerically, for convenience. 
If we assume the semi-major axes of the two planets
evolve on constant (but different) timescales $ |a_1/{\dot a_1}| = \tau_{a,1}$ 
and $|a_2/{\dot a_2}| = \tau_{a,2}$, the condition that the resonance width
is not crossed within a libration period gives
\begin{eqnarray}
\tau_{a,\text{crit}} \equiv \left| \frac{1}{1/\tau_{a,1}-1/\tau_{a,2}}\right| 
 & \gtrsim & 2\pi \frac{a_2}{\omega_{lf}\Delta a } 
 = 2 \pi \frac{n_2}{\omega_{lf}^2 }
\end{eqnarray}
to pass through the 2:1 MMR. 

If the planets of the HD45364 system are placed in a 2:1 resonance with the inner planet 
located at $1~\text{AU}$, the libration period $2\pi /\omega_{lf}$ is found to be 
approximately $75~\text{yrs}$. Thus, a relative migration timescale shorter than 
$\tau_{a,\text{crit}}\approx810~\text{yrs}$ is needed  to pass through the 2:1 resonance. 
For example, if we assume that the inner planet migrates on a timescale of 2000 years, then the outer 
planet has to migrate with a timescale of
\begin{eqnarray}
\label{eq:threetwo:analyticestimate} \tau_{a,2,crit} \lesssim 576~\text{yrs}. 
\end{eqnarray}

\section{N-body simulations}\label{sec:threetwo:nbody}
\label{sec:threetwo:formationnbody}

We ran $N$-body simulations to explore the large parameter space and confirm the estimate from the previous section.
In an N-body simulation all objects (stars, planets) are treated as point masses interacting only gravitationally with each other. 
To model the planet-disc interaction, one can explicitly add dissipative and stochastic forces. 
Thus, the total force acting on an object $i$ is a sum of the following terms
\begin{eqnarray}
	\vec F_i &=& \vec F_{i\text{ gravity}} +\vec F_{i\text{ indirect}} + \vec F_{i\text{ a-damping}}+ \vec F_{i\text{ e-damping}}+\vec F_{i\text{ stochastic}}.\label{eq:threetwo:nbodyforce}
\end{eqnarray}
The first term is due to the gravitational interaction. In units where $G=1$ we have
\begin{eqnarray}
	 \vec F_{i\text{ gravity}} = \sum_{j \neq i} m_i m_j \frac{ \vec r_j- \vec r_i }{\left| \vec r_j - \vec r_i\right|^{3} },
\end{eqnarray} 
where we sum over all particles in the system, except the $i$-th particle.
In the heliocentric coordinate system which is used here, the star is located at a fixed position at $\vec r=0$. 
That brings about an additional indirect term as we are in an accelerated, non-inertial  frame,
\begin{eqnarray}
	 \vec F_{i\text{ indirect}} = - \sum_{j} m_i m_j \frac{\vec  r_j }{\left|  \vec r_j\right|^{3} }.
\end{eqnarray}
The next two terms in eqation \ref{eq:threetwo:nbodyforce} are due to the laminar planet-disc interaction. 
In a first approximation the interaction damps the semi-major axis $a$ and the eccentricity $e$ on time scales $\tau_a$ and $\tau_e$, respectively. 
The exact terms depend on the orbital parameters of the planet and are given in appendix \ref{app:leepeale}. 
It is common practice to define the ratio between the timescales $K\equiv \tau_a /  \tau_e$. 
To obtain those timescales, we have to compare the N-body simulations to full hydrodynamical simulations which will be done in the following section.
An N-body simulation is much faster than a hydrodynamical simulation and thus can be integrated over a longer time span. 
This is essential because resonance capturing and migration act on timescales $\approx 1000-10000$ years and we might even want to integrate over millions of years. 
The last term in equation \ref{eq:threetwo:nbodyforce} adds stochastic forces, simulating the turbulent nature of proto-planetary discs. 
The implementation is discussed in chapter \ref{ch:randwalk}.

Newton's second law together with \Eq~\ref{eq:threetwo:nbodyforce} forms a system of ordinary differential equations.
To solve them, a new N-body code has been developed which is highly modular and easily expandable. It incorporates different modules for stochastic forces, migration, data output and time-stepping. 	

The choice of integrator depends on the problem. 
The following algorithms have been implemented:
\begin{description}
\item[Runge-Kutta Method (RK4)]
The classical, fourth order Runge-Kutta method is a widely-used standard integrator. It is an explicit method that needs four function evaluations per time-step. 
The main disadvantage of the RK4 method is the fixed time-step. We have to set a specific value at the beginning of the computation that is not refined later on. 
\item[Runge-Kutta-Fehlberg Method (RKF45)]
The Runge-Kutta-Fehlberg method is a fifth order explicit method with an embedded fourth order method \citep{fehlberg69}. We can use the two different results to estimate the numerical error and a new time-step. We repeat the step with a smaller time-step if the error is larger than a specified limit $eps$.
\item[Midpoint Method]
The midpoint method belongs to the class of second order Runge-Kutta methods. Due to the low order it is not efficient to use it on its own, but it is used as a sub-timestep integrator by the Bulirsch-Stoer Method. 
\item[Bulirsch-Stoer Method (BS)] This method \citep{StoerBulirsch02,nr} is based on a stepwise extrapolation. For each time-step we calculate the new positions and velocities several times with different sub-time-steps $\Delta t_i$ using the modified midpoint method.
We then perform an extrapolation to a \textit{perfect} sub-step in the limit where $\Delta t\rightarrow 0$. This allows the use of very large time-steps while still obtaining a high accuracy. We can also use the extrapolation to estimate the error and thus make the method adaptive.
\end{description}

The code has been used in \cite{ReinPapaloizou2009} and \cite{ReinPapaloizouKley2010}, usually with the Bulirsch-Stoer integrator and a precision of $\epsilon=10^{-11}$.
A symplectic integrator has not been implemented, although it would have better long term conservation properties. This is because the integrator would formally not be symplectic any more once velocity dependent forces have been added. This is the case in all simulations presented here, which include either migration or stochastic forces. Note that in some cases it is possible to find a symplectic integrator for systems with velocity dependent forces, for example for Hill's equations \citep{Quinn2010}. Such an integrator (a modified version of the leap-frog algorithm) has been used in shearing sheet calculations of planetary rings, presented in chapter \ref{ch:moonlet}.

\subsection{Parameter space survey}
\begin{figure}[p]
\centering
\subbottom[Period ratio $P_2/P_1$ as a function of time ($y$-axis) and migration timescale of the outer planet $\tau_{a,2}$ ($x$-axis).]{
\begin{pspicture}(0,0)(0.8,0.50) 
\rput(0.4,0.25){\scalebox{0.8}{\includegraphics[]{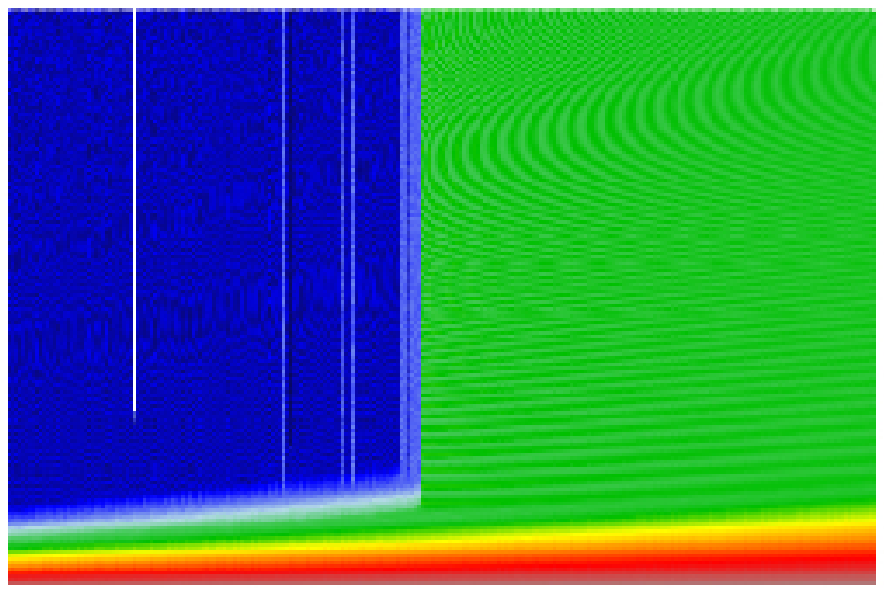}}}
\rput(0.4,0.25){\scalebox{0.8}{\includegraphics[]{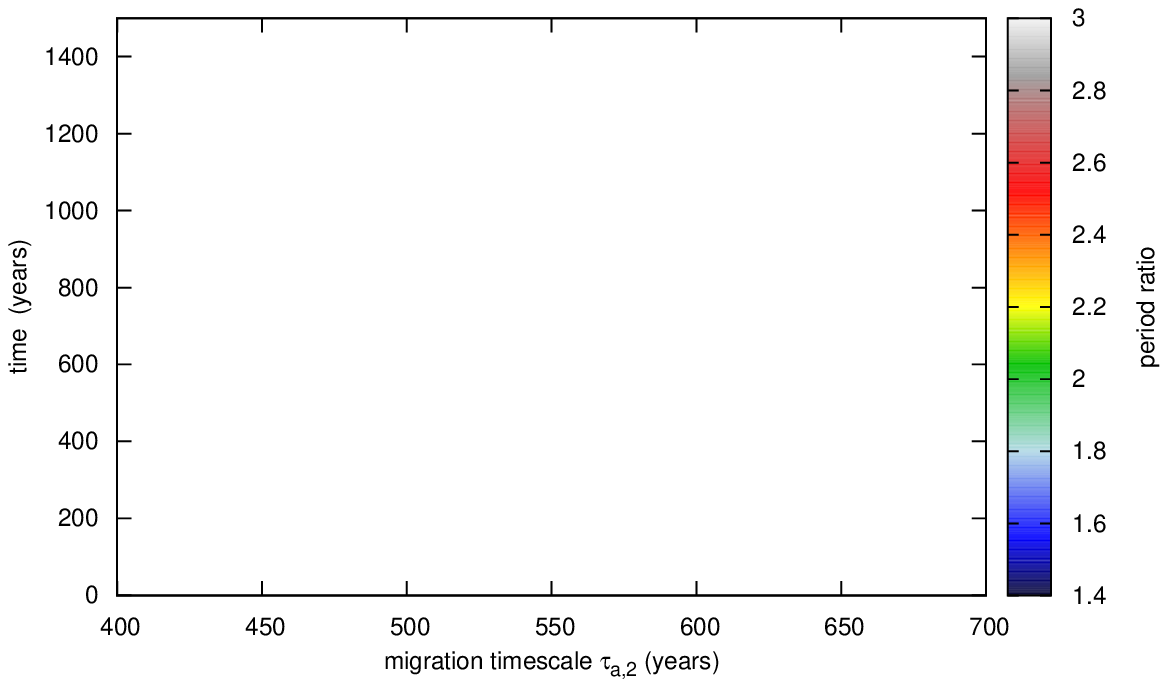}}}
\end{pspicture}
}
\subbottom[Eccentricity of the inner planet $e_1$ as a function of time ($y$-axis) and migration timescale of the outer planet $\tau_{a,2}$ ($x$-axis).]{
\begin{pspicture}(0,0)(0.8,0.50) 
\rput(0.4,0.25){\scalebox{0.8}{\includegraphics[]{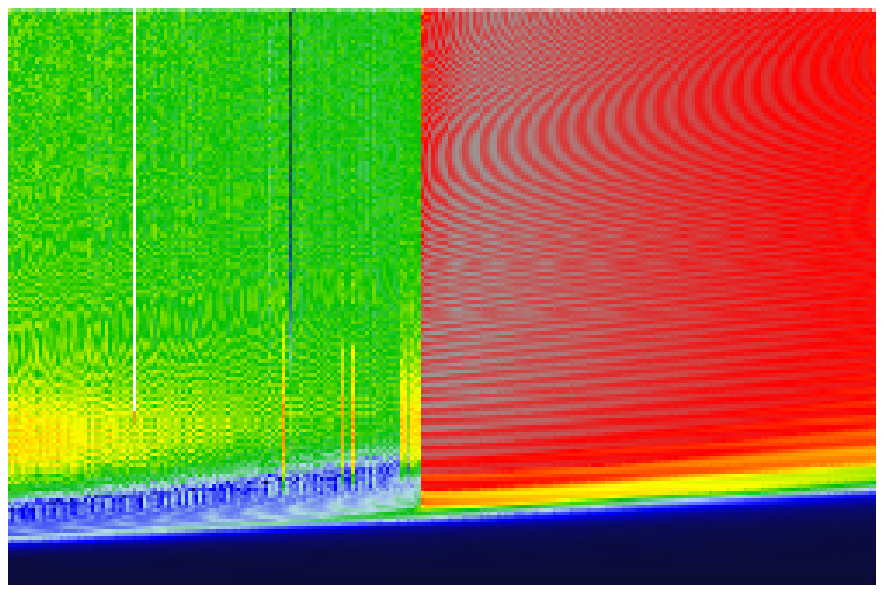}}}
\rput(0.4,0.25){\scalebox{0.8}{\includegraphics[]{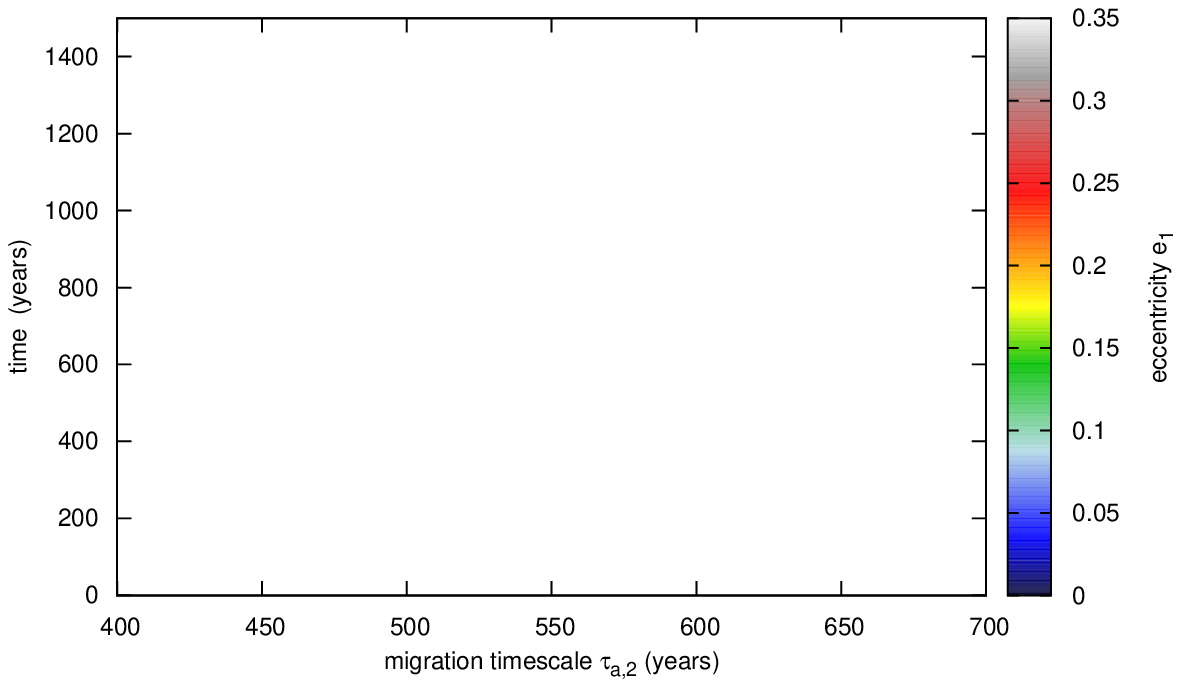}}}
\end{pspicture}
}
\subbottom[Eccentricity of the outer planet $e_2$ as a function of time ($y$-axis) and migration timescale of the outer planet $\tau_{a,2}$ ($x$-axis).]{
\begin{pspicture}(0,0)(0.8,0.50) 
\rput(0.4,0.25){\scalebox{0.8}{\includegraphics[]{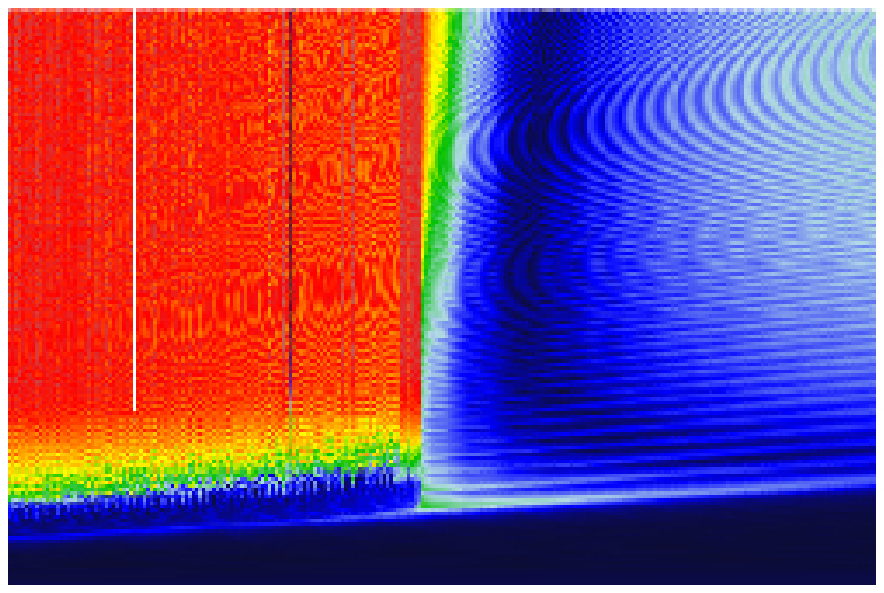}}}
\rput(0.4,0.25){\scalebox{0.8}{\includegraphics[]{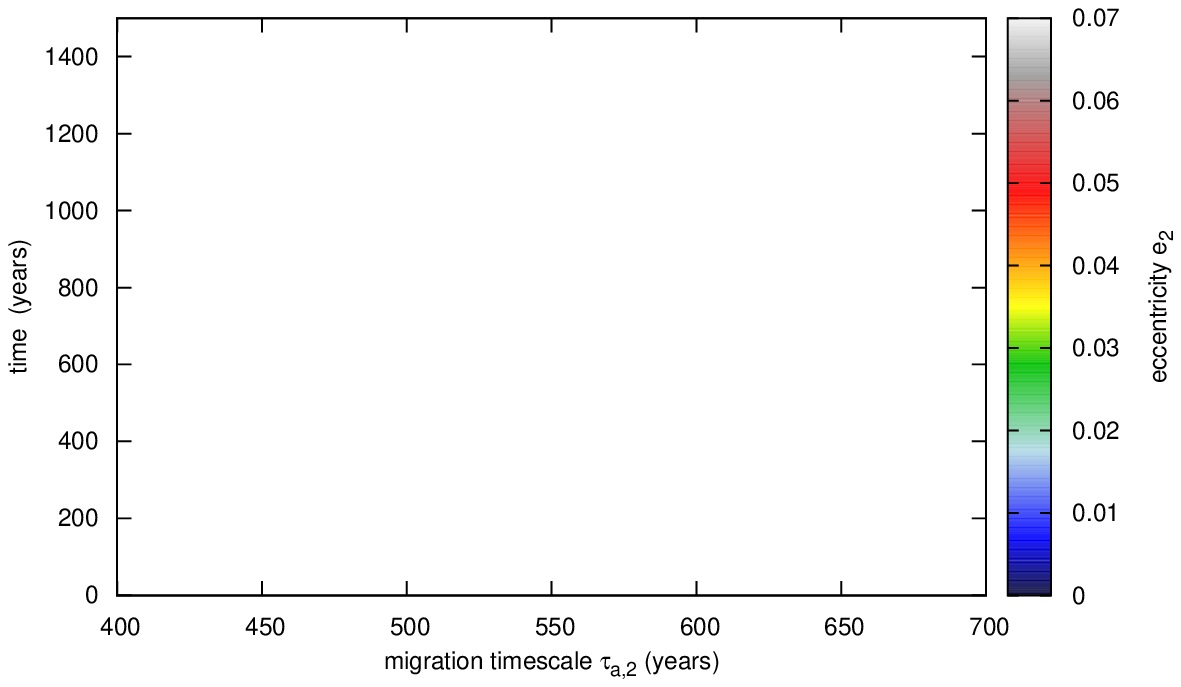}}}
\end{pspicture}
}
\caption{Parameter space survey. The migration timescale of the inner planet is $\tau_{a,1}=2000~\text{yrs}$. 
The eccentricity damping is given through 
$K\equiv \tau_a / \tau_e =10$.	\label{fig:threetwo:K10}}
\end{figure}

\begin{figure}[p]
\centering
\subbottom[Period ratio $P_2/P_1$ as a function of time ($y$-axis) and migration timescale of the outer planet $\tau_{a,2}$ ($x$-axis).]{
\begin{pspicture}(0,0)(0.8,0.50) 
\rput(0.4,0.25){\scalebox{0.8}{\includegraphics[]{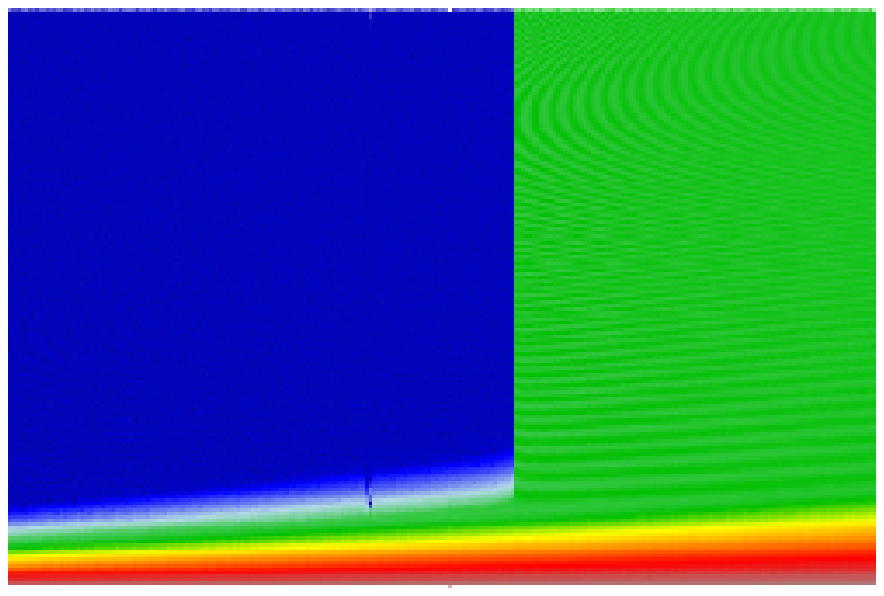}}}
\rput(0.4,0.25){\scalebox{0.8}{\includegraphics[]{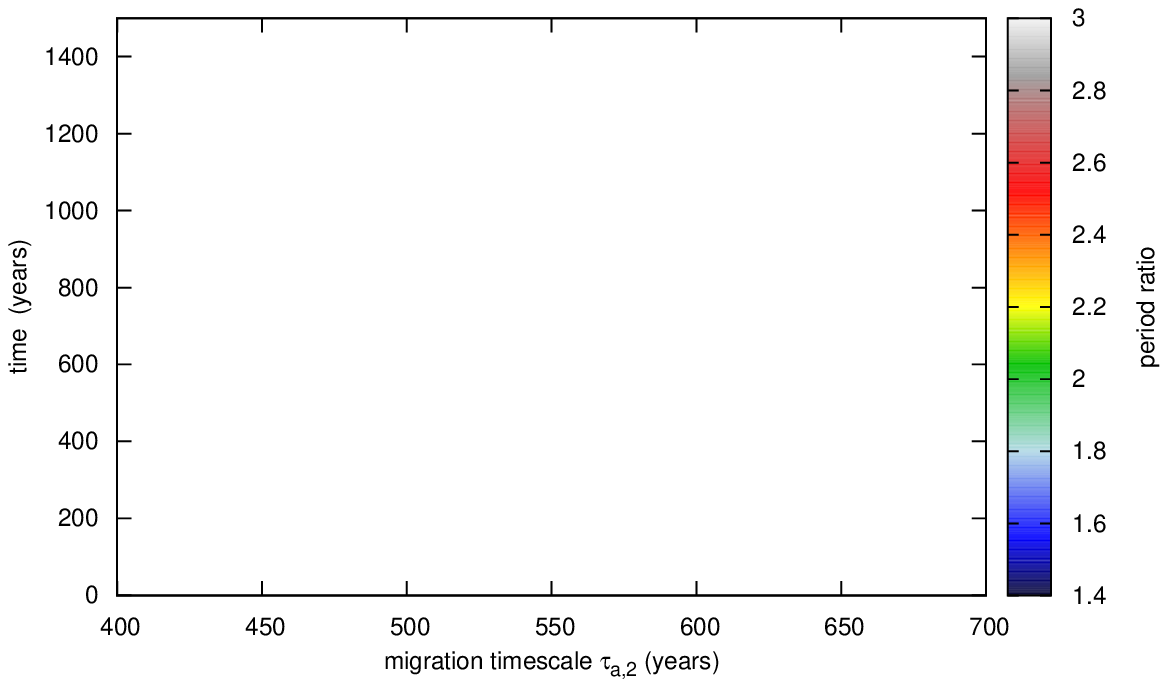}}}
\end{pspicture}
}
\subbottom[Eccentricity of the inner planet $e_1$ as a function of time ($y$-axis) and migration timescale of the outer planet $\tau_{a,2}$ ($x$-axis).]{
\begin{pspicture}(0,0)(0.8,0.50) 
\rput(0.4,0.25){\scalebox{0.8}{\includegraphics[]{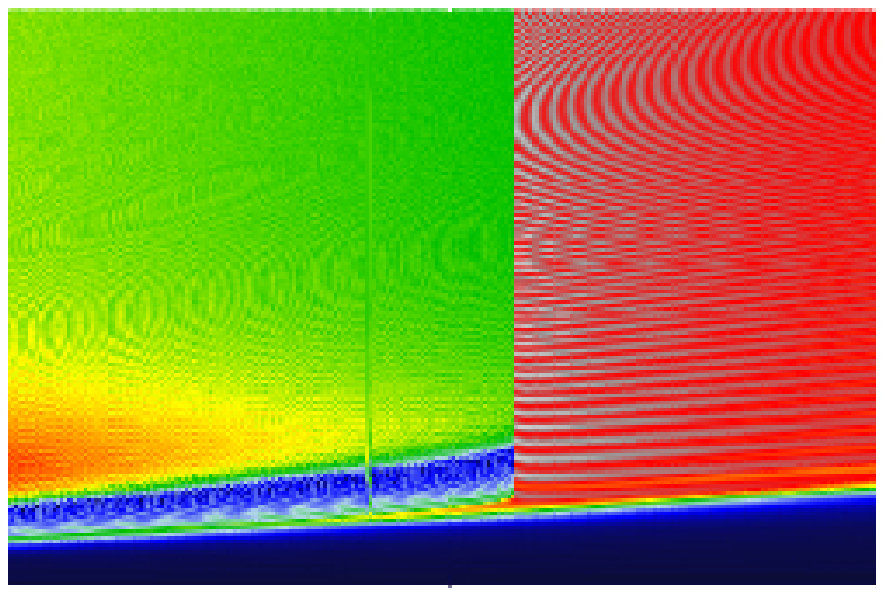}}}
\rput(0.4,0.25){\scalebox{0.8}{\includegraphics[]{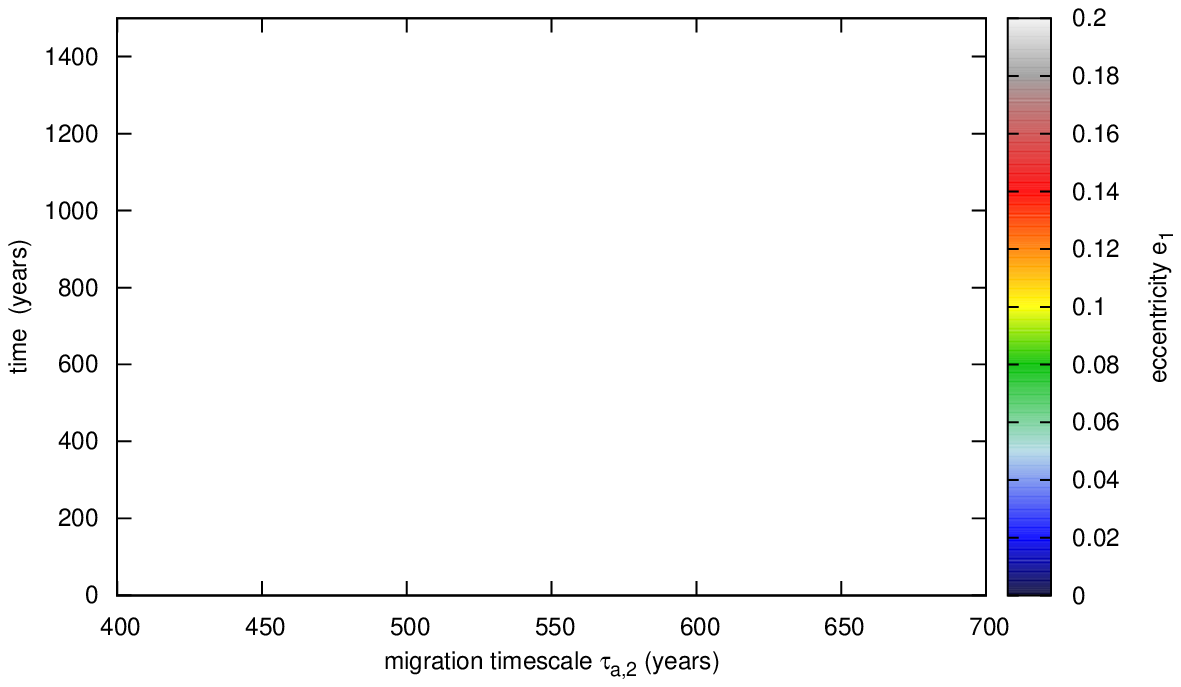}}}
\end{pspicture}
}
\subbottom[Eccentricity of the outer planet $e_2$ as a function of time ($y$-axis) and migration timescale of the outer planet $\tau_{a,2}$ ($x$-axis).]{
\begin{pspicture}(0,0)(0.8,0.50) 
\rput(0.4,0.25){\scalebox{0.8}{\includegraphics[]{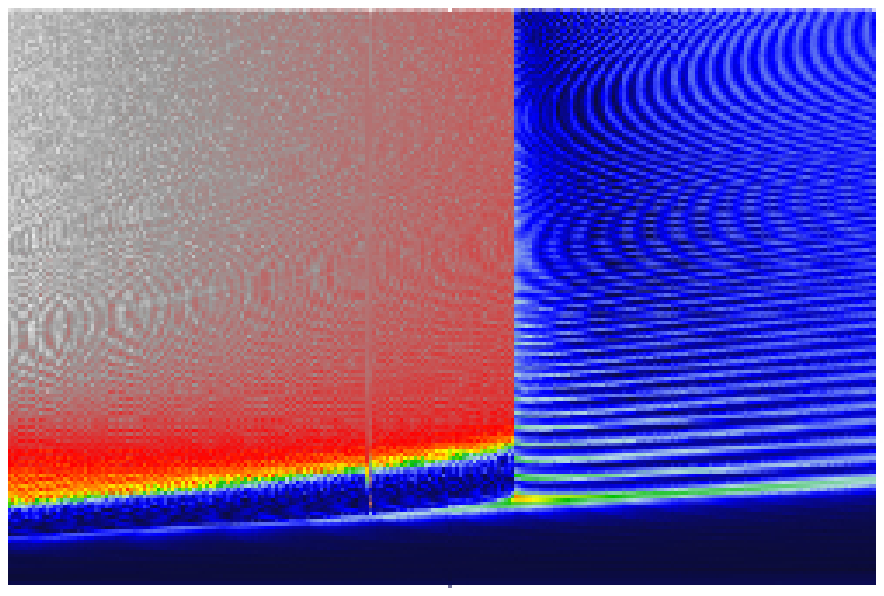}}}
\rput(0.4,0.25){\scalebox{0.8}{\includegraphics[]{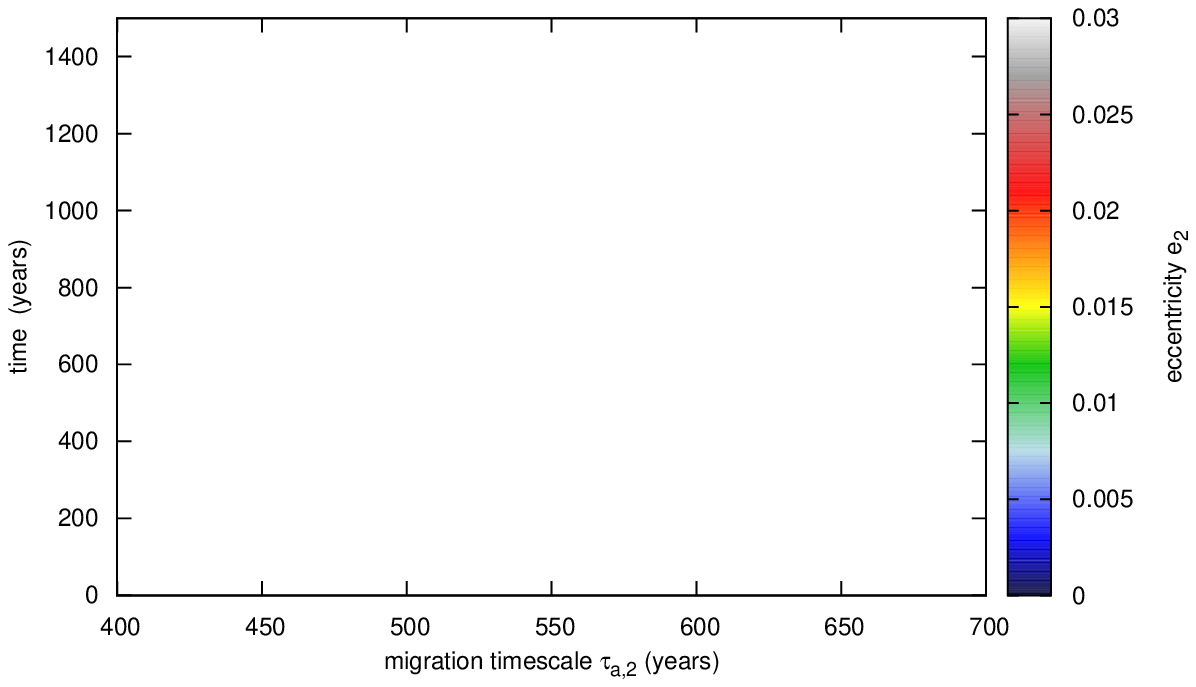}}}
\end{pspicture}
}
\caption{Parameter space survey. Same parameters as in \Fig~\ref{fig:threetwo:K10} except $K\equiv \tau_a / \tau_e =30$.	\label{fig:threetwo:K30}}
\end{figure}

\begin{figure}[p]
\centering
\subbottom[Period ratio $P_2/P_1$ as a function of time ($y$-axis) and migration timescale of the outer planet $\tau_{a,2}$ ($x$-axis).]{
\begin{pspicture}(0,0)(0.8,0.50) 
\rput(0.4,0.25){\scalebox{0.8}{\includegraphics[]{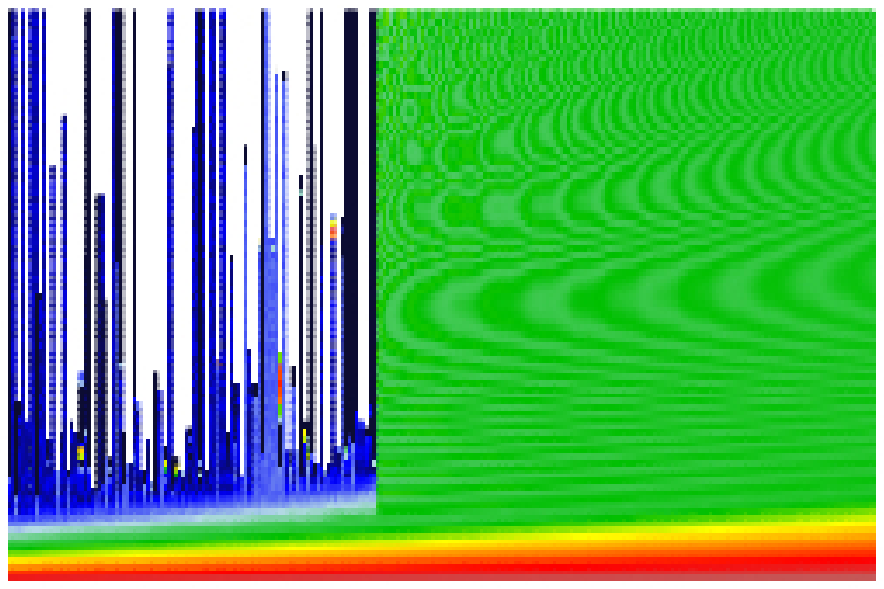}}}
\rput(0.4,0.25){\scalebox{0.8}{\includegraphics[]{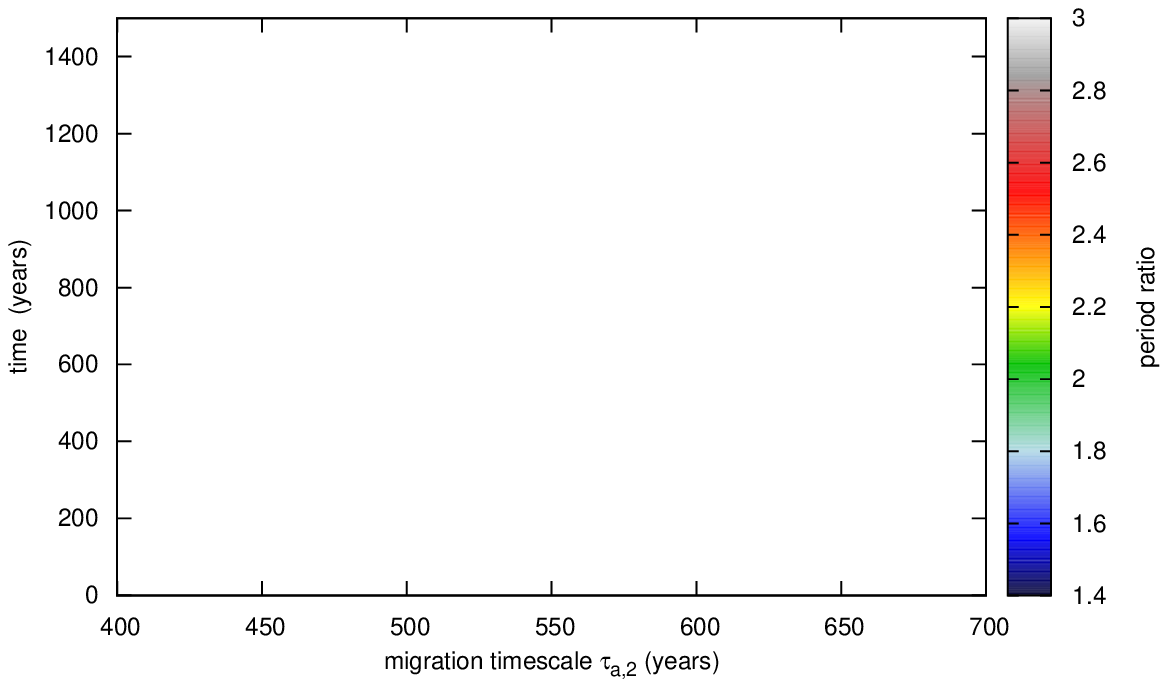}}}
\end{pspicture}
}
\subbottom[Eccentricity of the inner planet $e_1$ as a function of time ($y$-axis) and migration timescale of the outer planet $\tau_{a,2}$ ($x$-axis).]{
\begin{pspicture}(0,0)(0.8,0.50) 
\rput(0.4,0.25){\scalebox{0.8}{\includegraphics[]{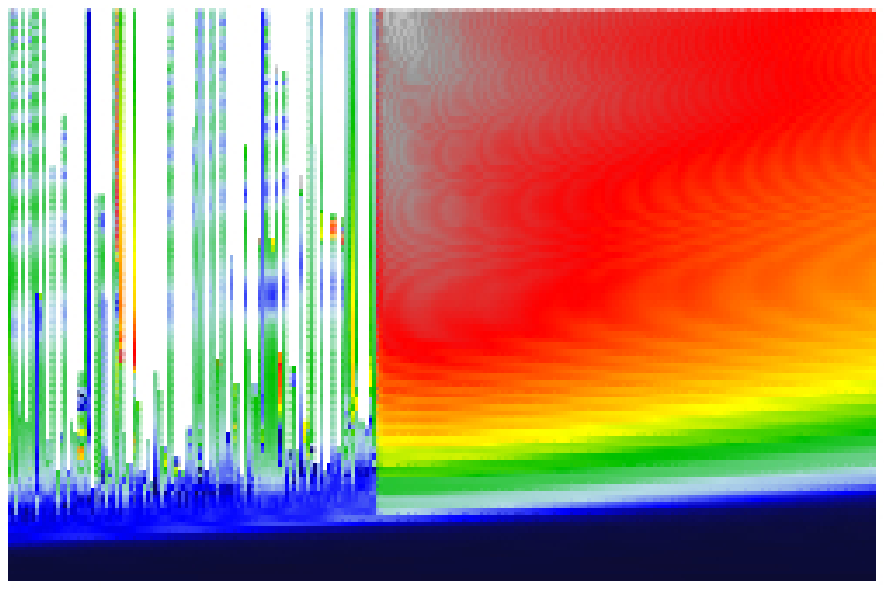}}}
\rput(0.4,0.25){\scalebox{0.8}{\includegraphics[]{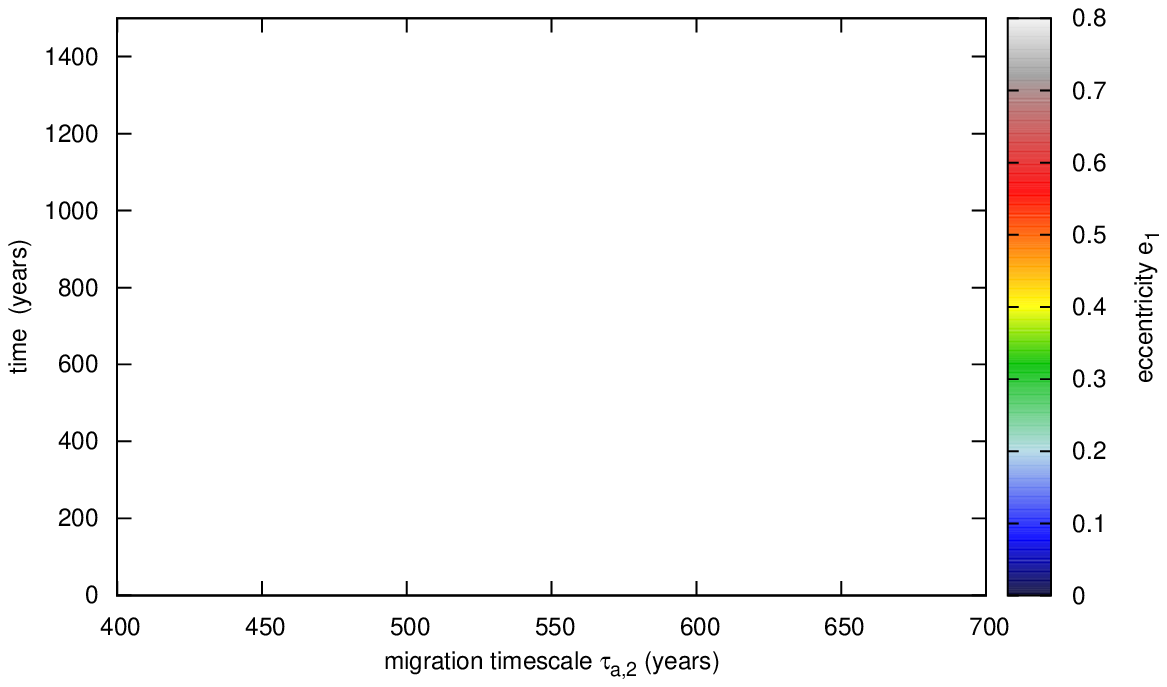}}}
\end{pspicture}
}
\subbottom[Eccentricity of the outer planet $e_2$ as a function of time ($y$-axis) and migration timescale of the outer planet $\tau_{a,2}$ ($x$-axis).]{
\begin{pspicture}(0,0)(0.8,0.50) 
\rput(0.4,0.25){\scalebox{0.8}{\includegraphics[]{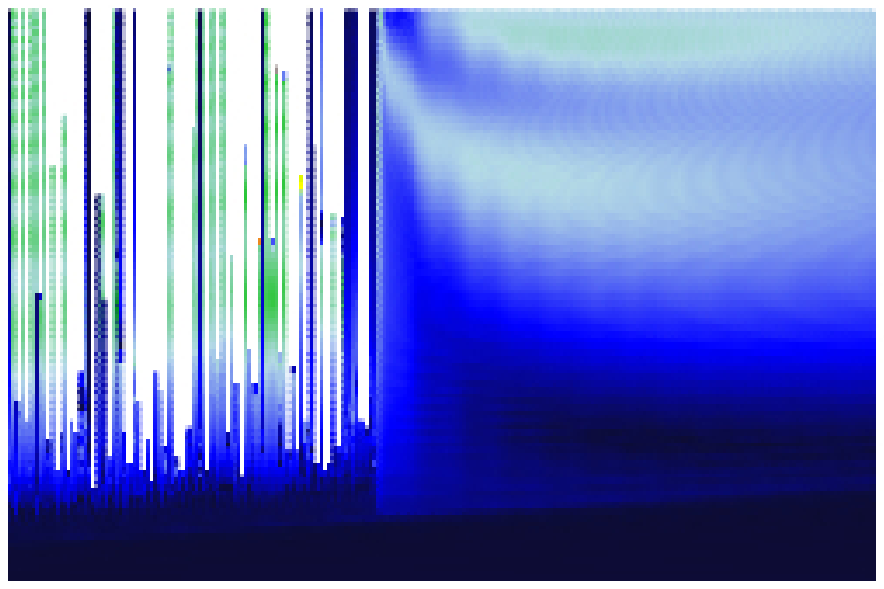}}}
\rput(0.4,0.25){\scalebox{0.8}{\includegraphics[]{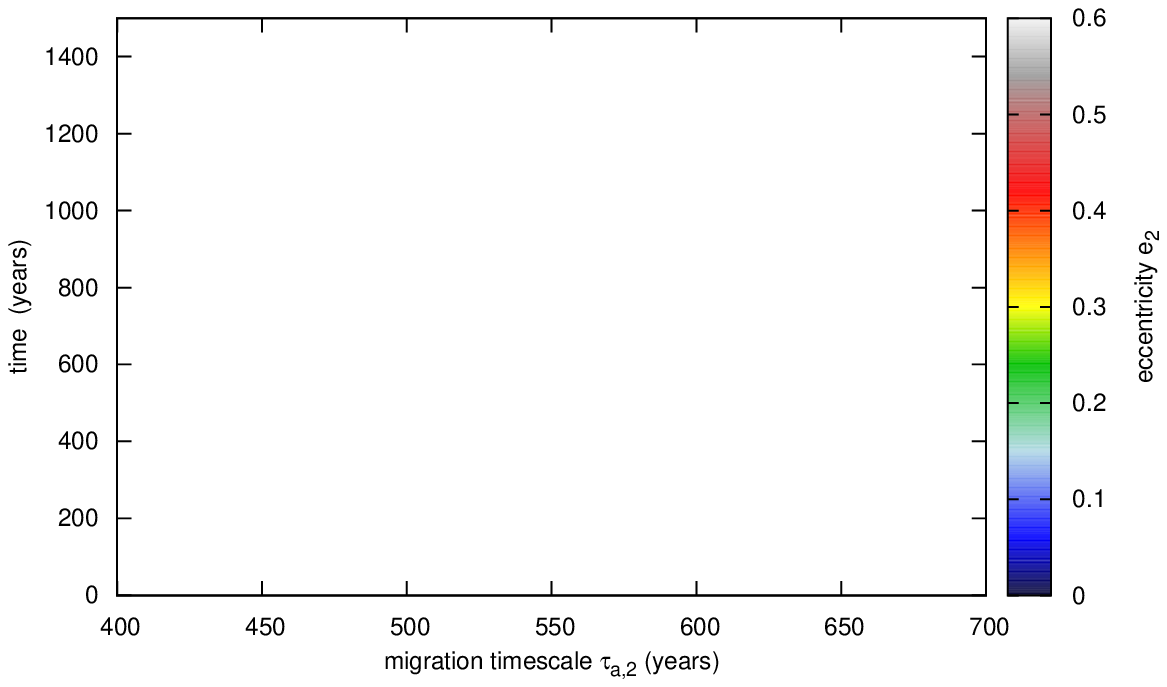}}}
\end{pspicture}
}
\caption{Parameter space survey. Same parameters as in \Fig~\ref{fig:threetwo:K10} except $K\equiv \tau_a / \tau_e =1$.	Many systems eject one or more planets, indicated by the white colour (on the left). \label{fig:threetwo:K1}}
\end{figure}

With the new N-body code, we can now easily scan the parameter space and confirm the analytic estimate of the critical migration rate needed to capture into the 3:2 resonance. 

We place both planets on circular orbits at~$a_1~=~1~\text{AU}$ and $a_2~=~2~\text{AU}$ initially. 
The migration timescale for the inner planet is fixed at~$\tau_{a,1}=2000~\text{yrs}$, while the migration 
timescale for the outer planet is varied. In \Figs~\ref{fig:threetwo:K10}, \ref{fig:threetwo:K30} and \ref{fig:threetwo:K1} we plot the period 
ratio~$P_2/P_1$ and the eccentricities of both planets as a function of time~$t$ for different migration timescales~$\tau_{a,2}$.
In all cases, there is a sharp transition of the final resonant configuration from 2:1 to 3:2 at around $\tau_{a,2}\approx 550~\text{yrs}$. 
This value agrees closely with the analytic estimate given by \Eq~\ref{eq:threetwo:analyticestimate}.
Note that once the planets are in resonance, the eccentricities quickly reach an equilibrium value as the planets migrate adiabatically.

In \Fig~\ref{fig:threetwo:K30} the eccentricity damping corresponds to $K=30$ and is therefore three times stronger than in \Fig~\ref{fig:threetwo:K10}.
A value of $K=30$ is rather high and probably unphysical. It can be seen that a smaller eccentricity damping timescale  
pushes the boundary towards a longer migration time-scale. Although this effect is rather weak (a three times stronger eccentricity damping shifts the boundary by only~4\%), it can be easily understood within the toy model presented above. The eccentricities have to rise quickly, while the planets are within the resonance width. This is more difficult for short eccentricity damping timescales. 

In \Fig~\ref{fig:threetwo:K1} the eccentricity damping corresponds to $K=1$ and is therefore ten times weaker than in \Fig~\ref{fig:threetwo:K10}.
Again, a value of $K=1$ is probably unphysical. 
This results in a rapid rise of eccentricities as soon as the planets are in resonance. 
Systems that are in a 3:2 resonance become unstable within a few thousand years. 
On the other hand, systems that are in a 2:1 resonance remain stable, although eccentricities are high ($e_1\sim0.5$). 

The results show clearly that, if the planets begin with a period ratio exceeding two, to get them 
into the observed 3:2 resonance, the relative migration time has to be shorter than what is obtained from 
the standard theory of type II migration applied to these planets in a standard model disc \citep{Nelson2000}.
In that case one expects this timescale to be larger than $10^4$~years, being effectively the disc evolution timescale. However, it is possible to obtain the 
required shorter migration timescales in a massive disc in which the planets migrate in a type III regime 
\cite[see e.g.][]{MassetPapaloizou2003, Peplinski2008}. In this regime, the surface density distribution in the 
co-orbital region is asymmetric, leading to a large torque which is able to cause the planet to fall 
inwards on a much shorter timescale than the disc evolution time obtained for type II migration. 
Other parameters such as the eccentricity damping timescale do not have a strong effect on capture probabilities. 

We explore the feasibility of this scenario in more detail in the following sections.

\section{Hydrodynamical simulations}
\label{sec:threetwo:formationhydro}
 
\begin{table*}[tb]
\begin{center}
\begin{tabular}{l|lll|ll|l}
\hline
\hline
   Run &   $H/R$ & $b/H$& $\Sigma$   & $N_r$ & $N_\phi$ & Result   \\ \hline
\texttt{F1} & 0.05 &0.6 & 0.001   & 768 & 768 &3:2\\
\texttt{F2} & 0.05 &0.6 & 0.0005  & 768 & 768 &3:2 D\\
\texttt{F3} & 0.05 &0.6 & 0.00025 & 768 & 768 &2:1 D\\
\texttt{F4} & 0.04 &0.6 & 0.001   & 768 & 768 &3:2 D\\
\texttt{F5} & 0.07 &0.6 & 0.001   & 768 & 768 &3:2\\
\texttt{R1} & rad  &1.0 & 0.0005  & 300 & 300 &3:2\\\hline
\hline
\end{tabular}
\end{center}
\caption{Parameters for some of the hydrodynamic simulations. 
The second column gives the disc aspect ratio, the third column the softening length in units of the
local scale height and the fourth column gives the initial surface density.
The fifth and sixth columns give the number of radial and azimuthal grid points used and finally 
the simulation outcome is indicated in the last column. 
 \label{tab:sun1}}
\end{table*}

Two-dimensional, grid-based hydrodynamic simulations of
an accretion disc with two gravitationally interacting planets were performed to test the rapid migration hypothesis. 
The simulations performed here are similar in concept to those performed by \citet{SnellgrovePapaloizouNelson01} 
of the resonant coupling in the GJ876
system that may have been induced by orbital migration resulting from interaction with the proto-planetary disc.
We performed studies using the FARGO code \citep{Masset00} with a modified locally isothermal equation of state. 
Those runs are indicated by the letter~\texttt{F}.

Willy Kley also ran simulations including viscous heating and radiative transport using the {\texttt{RH2D}} code. Those runs are indicated by the letter~\texttt{R} and more details can be found in \cite{ReinPapaloizouKley2010}.

Further simulations have been performed with a new hydrodynamics code called Prometheus, which is described in appendix \ref{app:dpmhd3d}. The code is similar to FARGO, using operator splitting, a staggered grid and fast orbital advection. However, it it can be run in two dimensions (cylindrical grid) and in three dimensions (spherical coordinates). Furthermore, it includes a particle and an MHD module.

\subsection{Initial configuration and computational set up}
\label{sec:initcon}
We use a system of units in which the unit of mass is the central mass $M_*,$ the unit of 
distance is the initial semi-major axis of the inner planet, $r_{1},$ and the unit of time 
is $(GM_*/r_{1}^3)^{-1/2}$. Thus the orbital period of the initial orbit of the inner planet 
is $2\pi$ in these dimensionless units.
The parameters for some of the simulations we conducted, as well as their outcomes,
are given in table \ref{tab:sun1}. 

In all simulations presented here, the mass ratio of the inner and outer planet is $q_1=2.18\cdot10^{-4}$ and $q_2=7.89\cdot10^{-4}$, respectively. 
The mass ratios adopted are those estimated for HD45364. 
The initial separation of the planets is $r_2/r_1=1.7$. The viscosity is chosen to be $\nu=10^{-5}$. 
The simulations have the inner and outer boundary of the 2D grid at $r_i=0.25$ and $r_o=3.0$, respectively. 
In all simulations, a smoothing length $b=0.6~H$ has been adopted, where $H$ is the disc 
thickness at the planet's position. This corresponds to about 4 zone widths in a simulation 
with a resolution of $768\times768$. The role of the softening parameter $b$ that is used
in two-dimensional calculations is to account for the smoothing that would
result from the vertical structure of the disc in three dimensional calculations
(e.g. Masset et al 2006). We find that the migration rate of the outer planet is 
only independent of the smoothing length if the sound speed is given by \Eq~\ref{eq:cs} (see next section). 

The planets are initialised on circular orbits, slowly turning on their mass in the first 5 orbits. 
We assume there is no mass accretion, so that these 
remain fixed in the simulation \cite[see discussion in][and also Sect. \ref{sec:threetwo:otherformationscenarios}]{Peplinski2008}.
The initial surface density profile is constant, and tests indicate that
varying the initial surface density profile does not change the outcome very much.
The total disc mass in the simulations listed in table \ref{tab:sun1} ranges from 0.014 to 0.055 solar masses.
Non-reflecting boundary conditions have been used throughout this chapter.

\subsection{Equation of state}
The outer planet is likely to undergo rapid type III migration, and the co-orbital region will be very asymmetric. 
We find, in accordance with \cite{Peplinski2008}, that the standard softening description does not lead to convergent results 
\citep[for a comparative study see][]{Cridaetal2009}. 
Because of the massive disc, a high density spike develops near the planet. 
Any small asymmetry will then generate a large torque, leading to erratic results. 
We follow the prescription of \cite{Peplinski2008} and increase the sound speed near the outer planet since the locally 
isothermal model breaks down in the circum-planetary disc. The new sound speed is given by
\begin{eqnarray}
\label{eq:sound}
	c_s = \left[ \left( h_s r_s \right)^{-n} +\left( h_p r_p \right)^{-n} \right]^{-1/n} \cdot \sqrt{\,\Omega^2_s+\Omega^2_p\,}\label{eq:cs},
\end{eqnarray}
where $r_s$ and $r_p$ are the distance to the star and the outer planet, respectively. 
$\Omega_s$ and $\Omega_p$ are the Keplerian angular velocity and $h_s$ and $h_p$ are the aspect ratio, 
both of the circumstellar and circum-planetary disc, respectively. The parameter $n$ is chosen to be~$3.5$, 
and the aspect ratio of the circum-planetary disc is~$h_p=0.4$.
The sound speed has not been changed in the vicinity of the inner, less massive planet as it will not undergo 
type III migration and the density peak near the planet is much lower. 

In radiative simulations (\texttt{R}), we go beyond the locally isothermal approximation
and include the full thermal energy equation, which takes the
generated viscous heat and radiative transport into account.
In this radiative formulation, the sound speed is a direct outcome of the simulation,
so we do not use \Eq \ref{eq:sound} \citep{ReinPapaloizouKley2010}.
 
\subsection{Simulation results}

\begin{figure}[p]
\centering
\includegraphics[width=0.9\columnwidth]{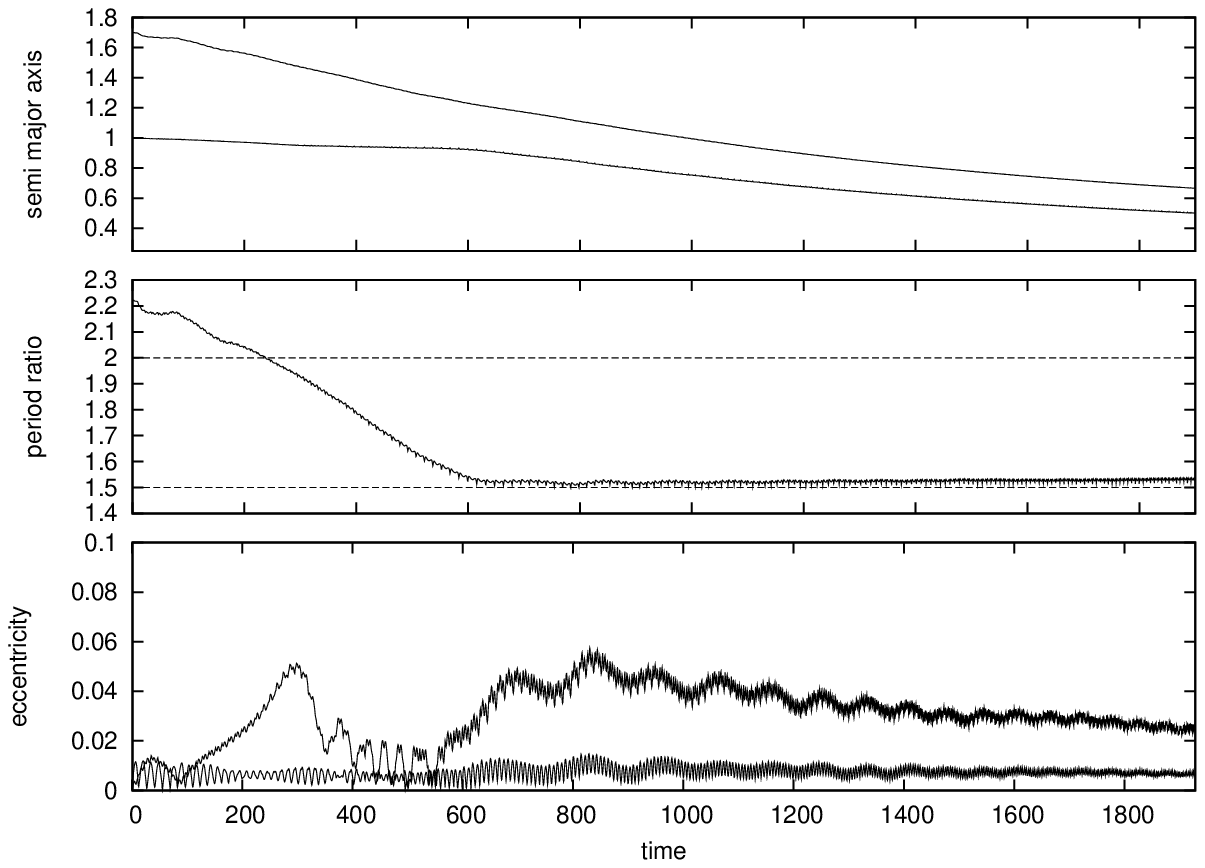}
\caption{The semi-major axes (top), period ratio $P_2/P_1$ (middle), and eccentricities (bottom) 
of the two planets plotted as a function of time 
in dimensionless units for run \texttt{F5} with a disc aspect ratio of $h=0.07$. 
In the bottom panel, the upper curve corresponds to the inner planet. 
\label{fig:conf_b0.60_h0.07_sigma0.0010_1cut_evolution}}
\includegraphics[width=0.9\columnwidth]{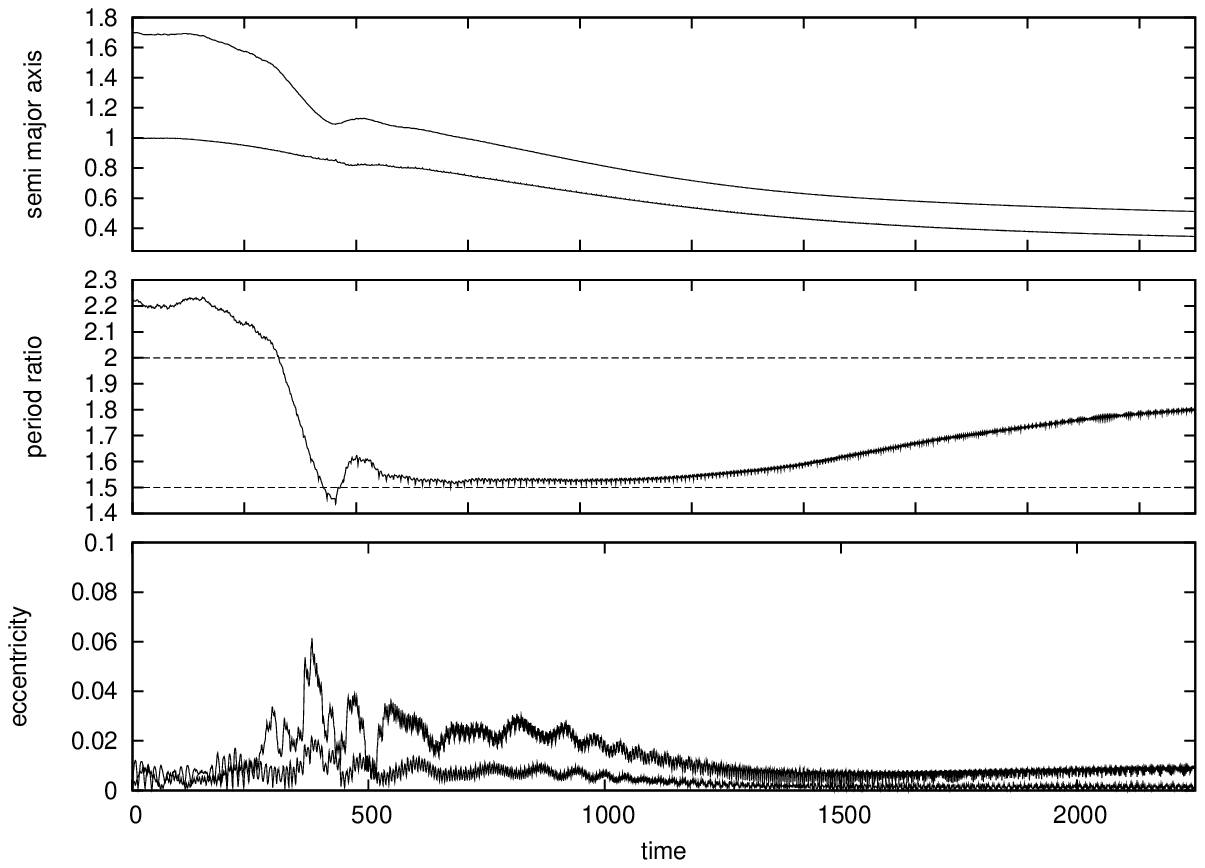}
\caption{Same plot as above for simulation \texttt{F4} with a disc aspect ratio of $h=0.04$.
\label{fig:conf_b0.60_h0.04_sigma0.0010_1cut_evolution}} 
\end{figure}

We find several possible simulation outcomes.
These are indicated in the final column of table \ref{tab:sun1}, which 
gives the resonance obtained in each case with a final letter \texttt{D} denoting that
the migration became ultimately divergent, resulting in a loss of the commensurability. 
We find that convergent migration could lead to a 2:1 resonance that was set up in the initial 
stages of the simulation (\texttt{F3}).
Cases \texttt{F1,F2,F4,F5,R1} provide more rapid and consistent convergent 
migration than \texttt{F3} and can attain a 3:2 resonance directly.

Thus avoidance of the attainment of sustained 2:1 commensurability
and the effective attainment of 3:2 commensurability required a rapid convergent
migration, as predicted by the N-body simulations and the analytic estimate 
(see section \ref{sec:threetwo:formationnbody2:1}). That is apparently helped initially 
by a rapid inward migration phase of the outer planet which shows evidence of type III migration.
The outer planet went through that phase in all simulations with a surface density 
higher than $0.00025$, and a 3:2 commensurability was obtained.
Thus, a surface density comparable to the minimum solar nebula \citep[MMSN,][]{Hayashi1981}, 
as used in simulation \texttt{F3}, is not sufficient to allow the planets undergo rapid enough type III migration.
As soon as the planets approach the 3:2 commensurability, type III migration 
stops due to the interaction with the inner planet and the outer planet starts to 
migrate in a standard type II regime. 

This imposes another constraint on the long-term sustainability of the resonance. The inner planet 
remains embedded in the disc and thus potentially undergoes a fairly rapid inward type I migration. 
If the type II migration rate of the outer planet is slower than the type I migration rate of the inner planet, 
then the planets diverge and the resonance is not sustained. A precise estimate of the 
migration rates is impossible in late stages, as the planets interact strongly with the 
density structure imposed on the disc by each other.

Accordingly, outflow boundary conditions at the inner boundary that prevent
the build up of an inner disc are more favourable to the maintenance of 
a 3:2 commensurability because the type I migration rate scales linearly with the disc 
surface density. However, those are not presented here, as the effect is weak and 
we stop the simulation before a large inner disc can build up near the boundary.

The migration rate for the inner planet depends on the aspect ratio $h$. 
It is decreased for an increased disc thickness \citep{Tanakaetal2002}. 
That explains why models with a large disc thickness (\texttt{F5,R1}) tend to stay in 
resonance, whereas models with a smaller disc thickness (\texttt{F2,F4}) tend towards divergent migration at late times.
The larger thickness is consistent with radiative runs (\texttt{R1}) which give a thickness of $h\sim0.075$ \citep[see][]{ReinPapaloizouKley2010}.

\begin{figure}[!p]
\centering
\begin{pspicture}(0,0)(1,1.2) 
\rput(0.5,0.7){\includegraphics{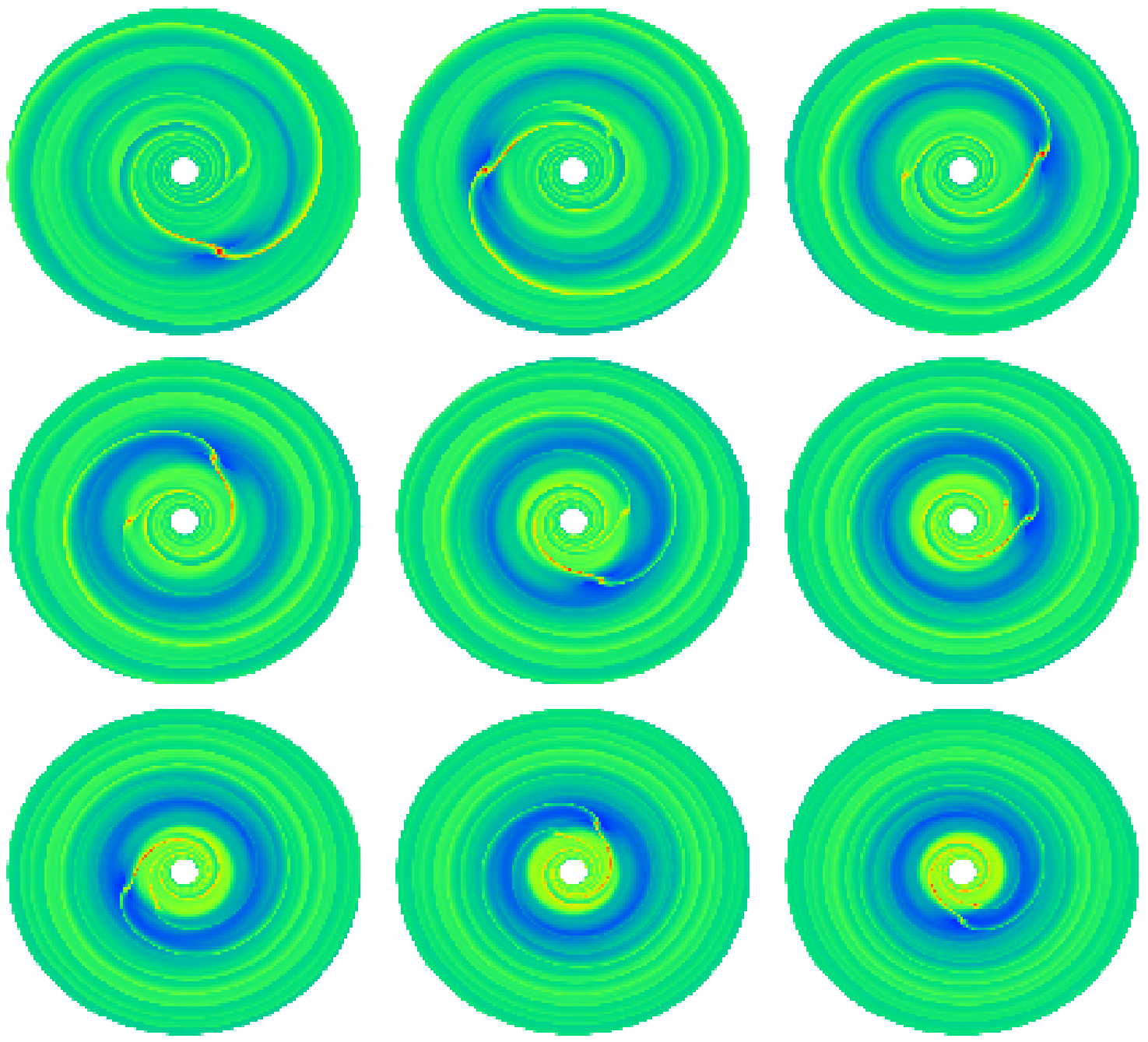}}
\rput(0.5,0.7){\includegraphics{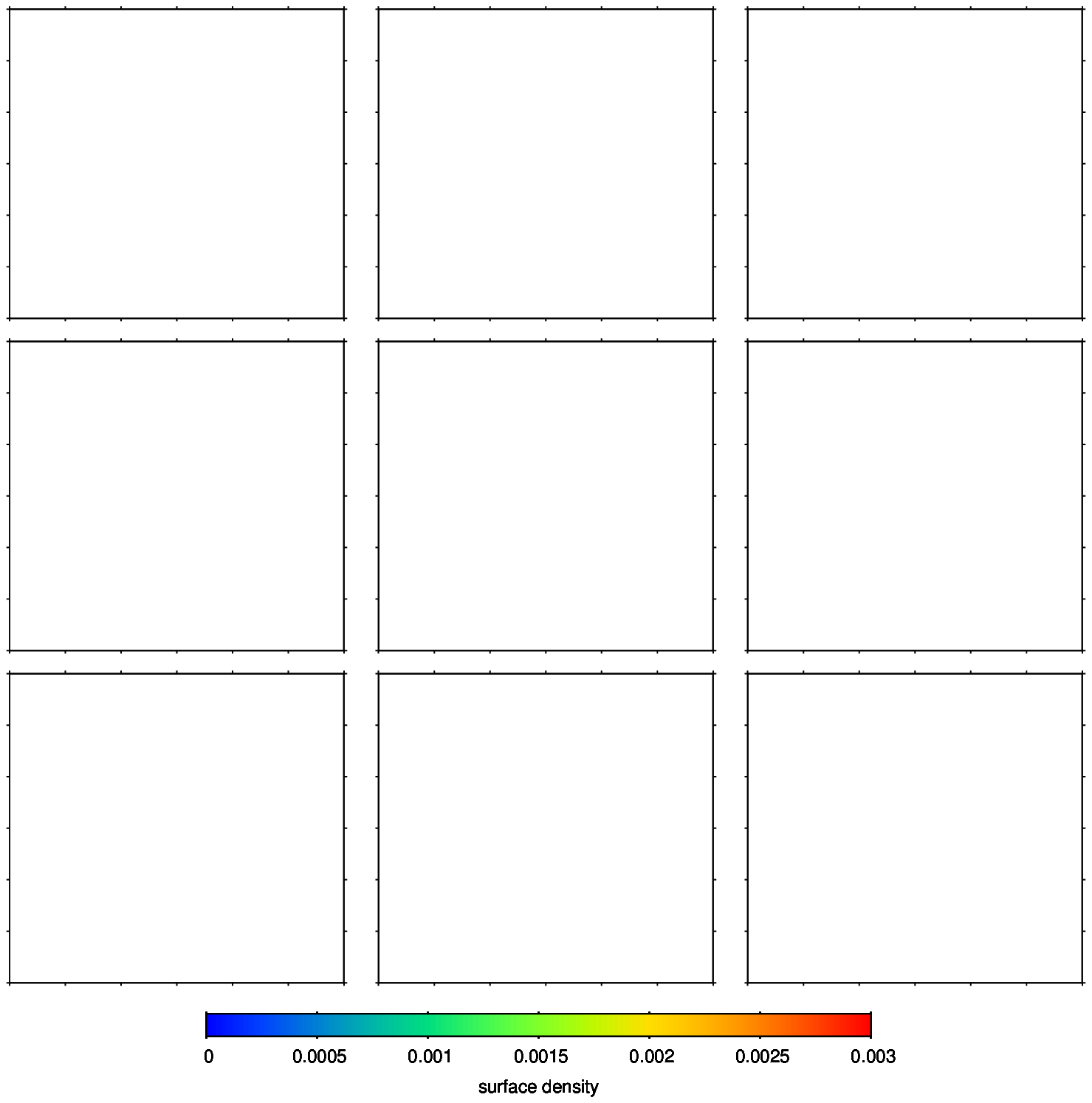}}
\end{pspicture}
\caption{Surface-density contour plots for simulation \texttt{F5} after 
20, 40, 60, 80, 100, 120, 140, 160 and 180 inner planet orbits. At the end of the type III migration phase,
the outer planet establishes a definite gap, while the inner planet
remains embedded at the edge of the outer planet's gap. 
\label{fig:conf_b0.60_h0.07_sigma0.0010_1cut_gasdens10}}
\end{figure}

\begin{figure}[!p]
\centering
\begin{pspicture}(0,0)(1,1.2) 
\rput(0.5,0.7){\includegraphics{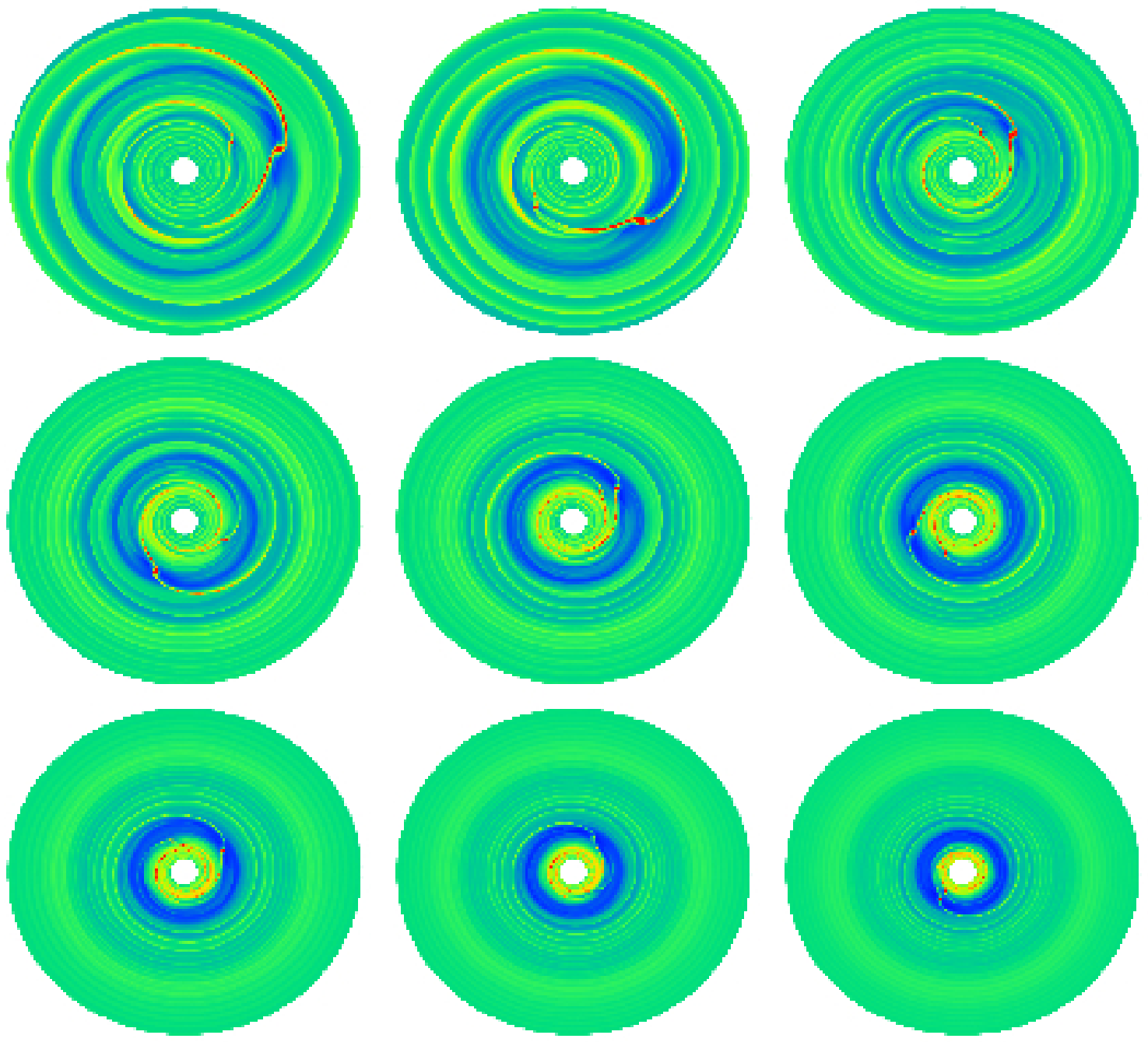}}
\rput(0.5,0.7){\includegraphics{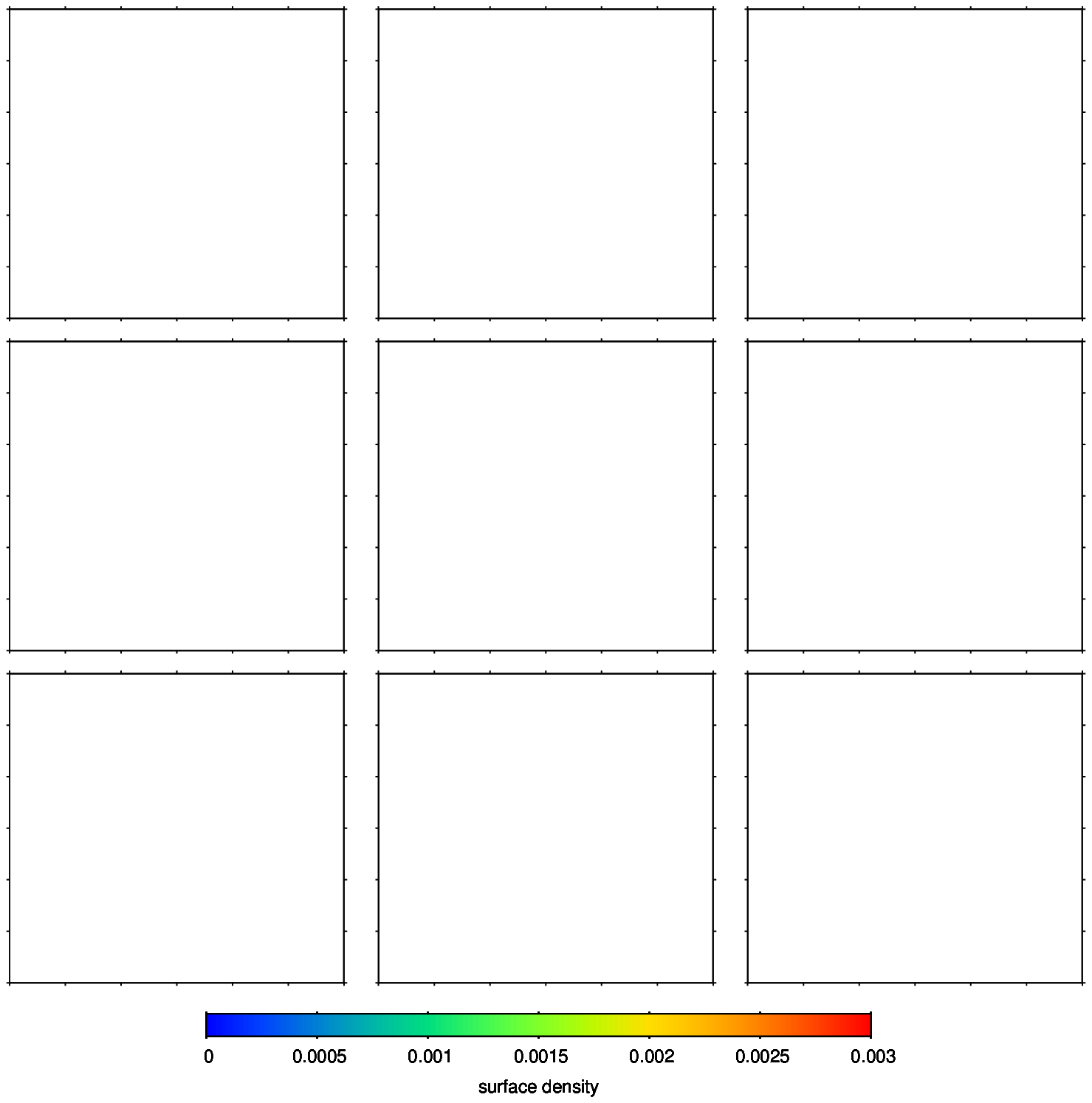}}
\end{pspicture}
\caption{Surface-density contour plots for simulation \texttt{F4} after 
20, 40, 60, 80, 100, 120, 140, 160 and 180 inner planet orbits. 
The extended 1D grid is not shown. 
\label{fig:conf_b0.60_h0.04_sigma0.0010_1cut_gasdens15}}
\end{figure}

As an illustration of the evolution of typical configuration (\texttt{F5}) that forms and maintains
a 3:2 commensurability during which the orbital radii contract by a factor of at least
$\sim 2$ we plot the evolutions of the semi-major axes, the period ratio, and eccentricities in \Fig~\ref{fig:conf_b0.60_h0.07_sigma0.0010_1cut_evolution}. We also provide surface density
contour plots in \Fig~\ref{fig:conf_b0.60_h0.07_sigma0.0010_1cut_gasdens10} after 20, 40, 60, 80, 100, 120, 140, 160 and 180 initial inner planet orbits.
The eccentricity peak in \Fig~\ref{fig:conf_b0.60_h0.07_sigma0.0010_1cut_evolution} at $t\sim300$ comes from
passing through the 2:1 commensurability. At $t\sim800$, the 3:2 commensurability is reached and maintained 
until the end of the simulation. 
The surface density in this simulation is approximately 5 times higher than the MMSN at $1~\text{AU}$ \citep{Hayashi1981}.

We also present an illustration of the evolution of the semi major axes and eccentricities, 
as well as surface density plots from a simulation (\texttt{F4}) that does form a 3:2 commensurability, 
but loses it because the inner planet is migrating too fast in figures
\ref{fig:conf_b0.60_h0.04_sigma0.0010_1cut_evolution} and 
\ref{fig:conf_b0.60_h0.04_sigma0.0010_1cut_gasdens15}.
One can see that a massive inner disc has piled up. 
This and the small aspect ratio of $h=0.04$ make 
the inner planet go faster than the outer planet, which has opened a clear gap. 
The commensurability is lost at $t\sim1200$.

In the radiative run \texttt{R1}, the initial type III migration rate is slower when compared to run \texttt{F5} because the
surface density is lower. After the the 2:1 resonance is passed, the outer planet migrates 
in a type II regime, so that the capture into the 3:2 resonance appears later. However, the orbital parameters
measured at the end of the simulations are very similar to any other run that we performed. 
This indicates that the parameter space that is populated by this kind of planet-disc simulation is very generic.

We tested the effect of 
disc dispersal at the late stages in our models to evolve the system self-consistently to the present day. 
In model \texttt{F5} after $t=2000$ we allow the disc mass to exponentially decay on 
a timescale of $\tau_{dis}\sim2000$. This timescale is shorter than the 
photo-evaporation timescale \citep{AlexanderClarke2006}. However, this scenario is expected to give a stronger 
effect than a long timescale \citep{SandorKley06}. In agreement with those authors, we
found that the dynamical state of the system does not change for the above parameters. At a late stage, the resonance
is well established and the planets undergo a slow inward migration. Strong effects are only expected if the disc dispersal 
happens during the short period of rapid type III migration, which is very unlikely. We observed that the 
eccentricities show a trend toward decreasing and the libration amplitudes tend toward slightly increasing during the dispersal phase.  
However, these changes are not different than what has been observed in runs without a disappearing disc.

Even though we use a high surface density in our simulations, the Toomre $Q$ parameter is still larger than unity at the outer boundary (typically 1.8). 
This gives us confidence that we do not need to include the effects of self-gravity in the calculation. 
Furthermore, it is not expected that self-gravity plays an important role for the migration rate of the outer planet which undergoes type III migration (M.K. Lin, private communication). 

\clearpage
\section{Other scenarios for the origin of HD45364}
\label{sec:threetwo:otherformationscenarios}

In the above discussion, we have considered the situation when the planets
attain their final masses while having a wider separation than required 
for a 2:1 commensurability, and found that convergent migration scenarios can 
be found that bring them into the observed 3:2 commensurability by disc planet interactions.
However, it is possible that they could be brought to their
current configuration in a number of different ways as considered below.
It is important to note that, because the final commensurable
state results from disc planet interactions,
it should have similar properties to those described above when making
comparisons with observations.

It is possible that the solid cores of both planets
approach each other more closely than 
the 2:1 commensurability before entering the rapid gas accretion phase 
and attaining their final masses prior to entering the 3:2 commensurability.
Although it is difficult to rule out such possibilities entirely, we note 
that the cores would be expected to be in the super earth 
mass range, where in general closer commensurabilities than 
2:1 and even 3:2 are found for typical type I migration rates
\citep[e.g.][]{PapaloizouSzuszkiewicz2005, CresswellNelson2008}.
One may also envisage the possibility that the solid cores grew in situ 
in a 3:2 commensurability, but this would have to survive expected strongly 
varying migration rates as a result of disc planet interactions as the 
planets grew in mass.

Another issue is whether the embedded inner planet is in a rapid accretion phase.
The onset of the rapid accretion phase (also called phase  3) occurs when the core
and envelope mass are about equal \citep{Pollacketal1996}. The total planet
mass depends at this stage on the boundary conditions of the circum-planetary disc.
When these allow the planet to have a significant convective envelope, the transition
to rapid accretion may not occur until the planet mass exceeds $60 \text{M}_{\oplus}$
\citep{Wuchterl1993}, which is the mass of the inner planet (see also model J3 of 
\citeauthor{Pollacketal1996} \citeyear{Pollacketal1996}, 
and models of \citeauthor{PapaloizouTerquem1999} \citeyear{PapaloizouTerquem1999}). 
Because of the above results, it is reasonable 
that the inner planet is not in a rapid accretion phase.

Finally we remark that proto-planetary discs are believed to maintain turbulence in some part
of their structure. Using the prescription of \cite{ReinPapaloizou2009}, we can simulate the turbulent 
behaviour of the disc by adding stochastic forces to an N-body simulation (see also chapter \ref{ch:randwalk}). 
These forces will ultimately eject the planets from the 2:1 resonance should that form. 
Provided they are strong enough, this can happen within the lifetime of the disc, 
thus making a subsequent capture into the observed 3:2 resonance possible. 

We confirmed numerically that such cases can occur for moderately large diffusion coefficients 
\citep[as estimated by][]{ReinPapaloizou2009}. However, this outcome seems to be the exception rather than the rule. 
Should the 2:1 resonance be broken, a planet-planet scattering event appears to be more likely. 
In all the simulations we performed, we find that only a small 
fraction of systems ($1\%$~-~$5\%$) eventually end up in a 3:2 resonance.

\section{Comparison with observations} \label{sec:threetwo:observation}

Table \ref{tab:orbit} lists the orbital parameters that \cite{CorreiaUdry2008} obtained 
from their best statistical fit of two Keplerian orbits to the radial velocity data. 
The radial velocity data collected so far is insufficient for detecting
any interactions between the planets. 
This is borne out by the fact that the authors did not obtain any improvement in terms of
minimising $\chi^2$ when a 3-body Newtonian 
fit rather than a Keplerian fit was carried out. 
However, a stability analysis supports the viability of the determined
parameters, as these lie inside a stable region of orbital parameter space.
The best-fit solution shows a 3:2 mean motion commensurability,
although no planet-planet interactions have been observed from which 
this could be inferred directly. 
It is important to note that the large minimum $\chi^2$ associated with 
the best-fit that could be obtained indicates large uncertainties 
that are not accounted for by the magnitudes of the errors quoted in their paper. 

There are two main reasons why we believe that the effective
errors associated with the observations are large.
First, many data points are clustered and clearly do not 
provide a random time sampling. This could result in 
correlations that effectively reduce the total number of independent measurements.
Second, the high minimum $\chi^2$ value that is three times the quoted
observational error indicates that there are additional effects
(e.g. sunspots, additional planets, etc.) that produce a jitter in the central star
(M. Mayor, private communication), which then enhances the effective observational error. 

No matter what process generates the additional noise, 
we can assume in a first approximation that 
it follows a Gaussian distribution. We then have to 
conclude that the effective error of each observation is a factor 
$\sim3$ larger than reported in order to account for the 
minimum $\chi^2$, given the quoted number
of $\sim 60$ independent observations. Under these circumstances,
we would then conclude that any fit with $\sqrt{\chi^2}$
in the range $2-4$ would be an equally valid possibility.
 
In the following, we show that indeed a variety of orbital 
solutions match the observed RV data with no statistically significant difference
when compared to the quoted best fit by \cite{CorreiaUdry2008}. 
As one illustration, we use the simulation \texttt{F5} obtained in section \ref{sec:threetwo:formationhydro}. 
We take the orbital parameters at a time when the orbital period 
of the outer planet is closest to the observed value 
and integrate them for several orbits with our N-body code. 
The solution is stable for at least one million years. 
There are only two free parameters available
to fit the reflex motion of the central star to the observed radial velocity: 
the origin of time (epoch) and the angle between the line of sight and the pericentre of the planets. 

\begin{table*}[tbp]
\begin{minipage}[t]{\textwidth}
\renewcommand{\footnoterule}{}
\begin{center}
\begin{tabular}{ll|ll|ll}
\hline\hline
& &\multicolumn{2}{l|}{\cite{CorreiaUdry2008}} & \multicolumn{2}{l}{Simulation \texttt{F5}}  \\
Parameter& Unit & b & c  & b & c \\\hline
$M \sin i$ & $[\mbox{M}_{\text{Jup}}]$ 		& 0.1872 & 0.6579 			& 0.1872 & 0.6579 		 		\\
$M_*$ &  $[M_{\odot}]$ 				&\multicolumn{2}{c|}{0.82} 		& \multicolumn{2}{c}{0.82} 			\\
$a$ &$[\mbox{AU}]$				& 0.6813 & 0.8972 			& 0.6804 & 0.8994  	 	\\
$e$ 	&		& 			$ 0.17\pm 0.02$ & $0.097 \pm0.012$ 	& 0.036  & 0.017  		\\
$\lambda$ & $[\mbox{deg}]$ 			& $105.8 \pm1.4$ & $269.5\pm0.6$ 	& 352.5 & 153.9 		\\
$\varpi$\footnote{Note that $\varpi$ is not well constrained for nearly circular orbits.} & $[\mbox{deg}]$ 			& $162.6 \pm6.3$ & $7.4\pm4.3$ 		& 87.9& 292.2		\\
$\sqrt{\chi^2}$ &				& \multicolumn{2}{c|}{2.79} 		& \multicolumn{2}{c}{2.76\footnote{Radial velocity amplitude was scaled down by $8\%$.} (3.51)} 	\\
Date  &[JD]					& \multicolumn{2}{c|}{2453500} 		& \multicolumn{2}{c}{2453500}  	\\
\hline
\hline
\end{tabular}
\end{center}
\caption{Orbital parameters of HD45364b and HD45364c from different fits.  \label{tab:orbit}}
\end{minipage}
\end{table*}

\begin{figure*}[t]
\centering
\includegraphics[width=1.0\textwidth]{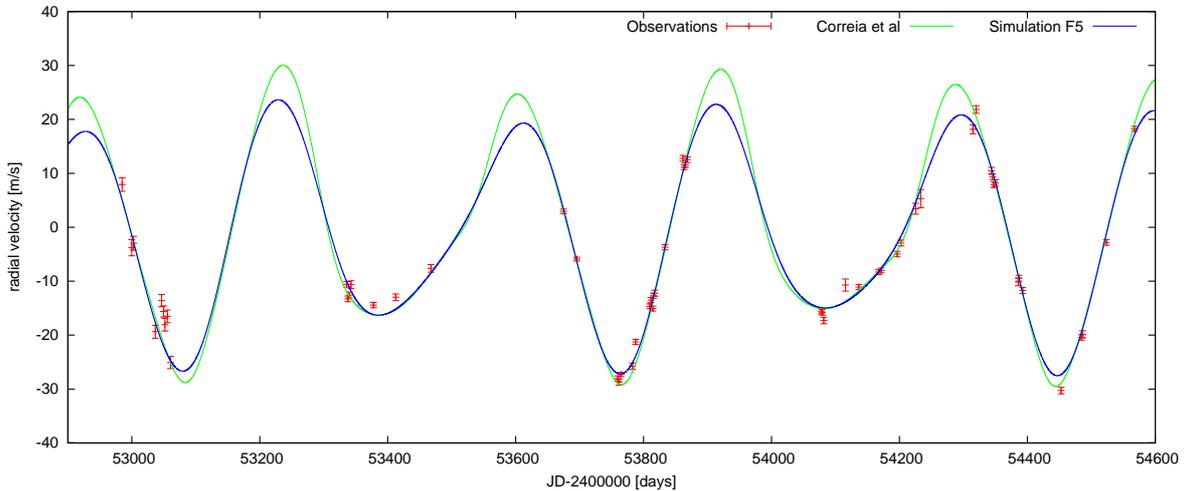}
\caption{Comparison of different orbital solutions. 
See text for a description of the different models. 
Radial velocity measurements (red points) and the published orbital 
solution (green curve) are taken from \cite{CorreiaUdry2008}.
\label{fig:nbodym_rv1}}
\end{figure*}

We can safely assume that the planet masses and periods are measured with high accuracy and that 
only the shape of the orbit contains large errors. Our best fit results in an unreduced 
$\sqrt{\chi^2}$ value of $3.51$. According to the above discussion, this solution is statistically 
indistinguishable from a solution with $\sqrt{\chi^2}\sim2.8$.

It is possible to reduce the value of $\chi^2$ even further, when assuming that the planet masses
are not fixed. In that case, we have found a fit corresponding to $\sqrt{\chi^2}=2.76$ where we have reduced the
radial velocity amplitude by $8\%$ (effectively adding one free parameter to the fit). 
This could be explained by either a heavier star, less massive planets, or a less inclined orbit.
The orientation of the orbital plane has been kept fixed ($i=90^\circ$) in all fits.

The results are shown in figure \ref{fig:nbodym_rv1}. 
The blue curve corresponds to the outcome of simulation \texttt{F5} and 
we list the orbital parameters in table \ref{tab:orbit}. 
We also plot the best fit given by \cite{CorreiaUdry2008} for comparison (green curve). 
In accordance with the discussion given above, it is very 
difficult to see any differences in the quality of these fits,
which indeed suggests that the models are statistically indistinguishable. 

However, there is an important difference between
the fits obtained from our simulations and that of \cite{CorreiaUdry2008}.
Our models consistently predict lower values for both 
eccentricities $e_1$ and $e_2$ (see table \ref{tab:orbit}). 
Furthermore, the ratio of eccentricities $e_1/e_2$ is higher than the previously reported value of $1.73$. 
The eccentricities are oscillating and the ratio is on average $\sim3$.
 
This in turn results in a different libration pattern: the slow libration mode 
\citep[see figure 3a in][]{CorreiaUdry2008} that is associated
with oscillations of the angle between the two apsidal lines is absent, as shown 
in figure \ref{fig:nbody_angles}. 
Thus there is a marked difference in the form the interaction between the
two planets takes place.

\begin{figure}[tbp]
\centering
\includegraphics[angle=270,width=0.95\columnwidth]{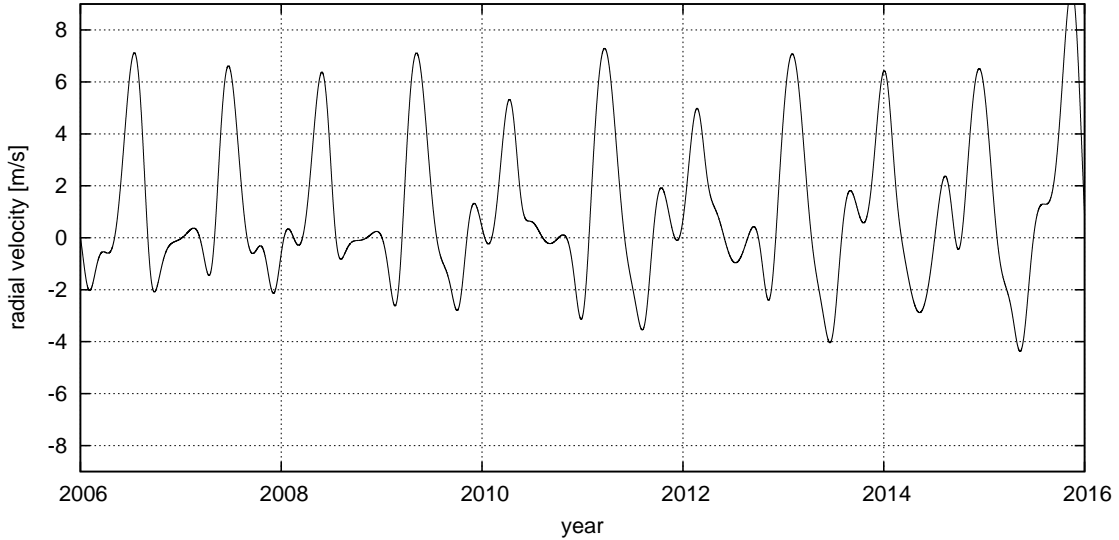}
\caption{Difference in radial velocity between the previously reported fit by 
\cite{CorreiaUdry2008} and the new fit obtained from simulation \texttt{F5}.
\label{fig:longtermdiff}}
\end{figure}

\begin{figure}[tbp]
\centering
\includegraphics[angle=270,width=0.95\columnwidth]{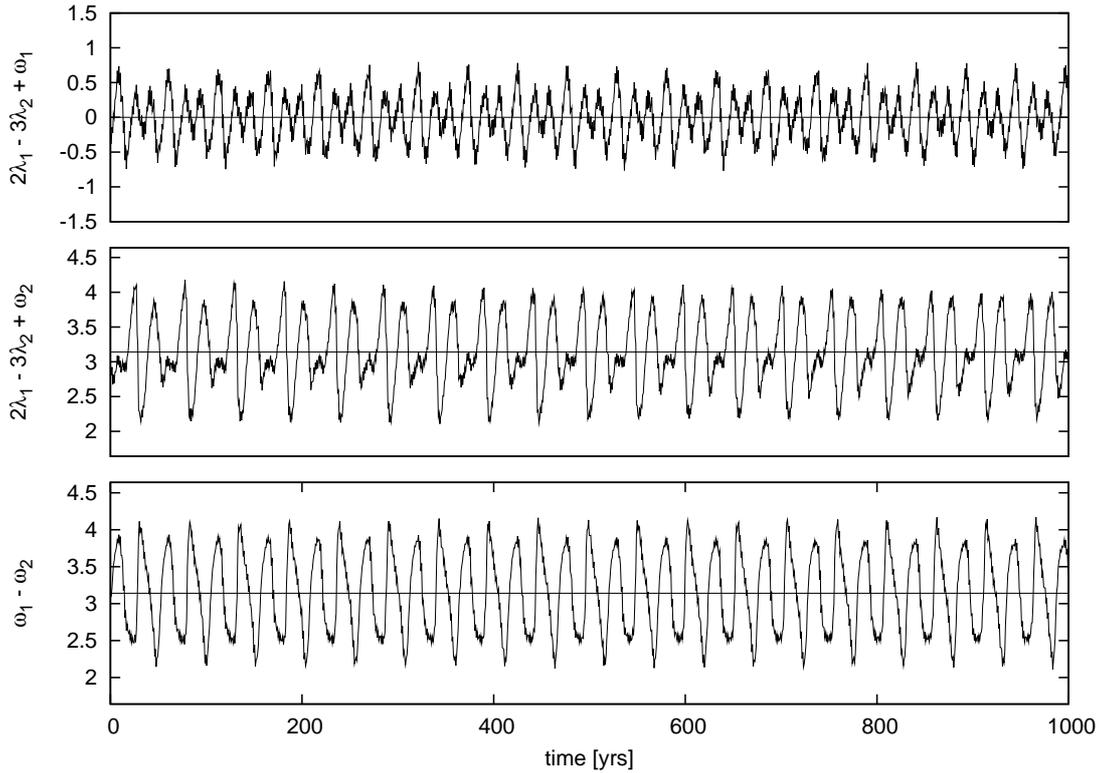}
\caption{Evolution of the resonant angles 
$\phi_1=2\lambda_1-3\lambda_2+\varpi_1$, $\phi_2=2\lambda_1-3\lambda_2+\varpi_2$, $\varpi_1-\varpi_2$ 
for simulation \texttt{F5} using the parameters given in table \ref{tab:orbit}. The angles are measured in radians.
\label{fig:nbody_angles}}
\end{figure}

It is hoped that future observations will be able to resolve this issue. 
The evolution of the difference in radial velocity over the next few years between our fit from simulation \texttt{F5} 
and the fit found by \cite{CorreiaUdry2008} is plotted in figure \ref{fig:longtermdiff}.
There is approximately one window each year that will allow further 
observations to distinguish easily between the two models. The large difference 
at these dates is not caused by the secular evolution of the system, but is simply a
consequence of smaller eccentricities.  


All runs that yield a final configuration with a 3:2 MMR are very
similar in their dynamical behaviour, as found for model \texttt{F5}. They are characterised
by an antisymmetric ($\varpi_2 - \varpi_1 \sim \pi$) state with a relatively small
libration of $\phi_1$ and $\phi_2$. This is a generic outcome of
all convergent disc-planet migration scenarios that we performed.

\section{Conclusions}
The planets in the multi-planetary system HD45364 are most likely in a 3:2 mean motion resonance. 
This poses interesting questions on its formation history. 
Assuming that the planets form far apart and migrate with a 
moderate migration rate, as predicted by standard planet formation and 
migration theories, the most likely outcome is a 2:1 mean motion resonance, 
contrary to the observation of a 3:2 MMR.

In this chapter, we investigated a possible way around this problem by letting 
the outer planet undergo a rapid inward type~III migration.
We presented an analytical estimate of the required migration rate and performed both N-body and hydrodynamical simulations.

We find that it is indeed possible to form a 3:2 MMR and avoid the 2:1 resonance, 
thus resembling the observed planetary system using reasonable disc parameters. 
Hydrodynamical simulations suggest that the system is more likely to sustain 
the resonance for high aspect ratios, as the migration of the inner planet is slowed down, 
thus avoiding divergent migration. 

Finally, we used the orbital configuration found in the hydrodynamical 
formation scenario to calculate a radial velocity curve. 
This curve was then compared to observations and the resulting fit 
has an identical $\chi^2$ value to the previously reported \textit{best fit}.

Our solution is stable for at least a million years. It is in a dynamically 
different state, both planets having lower eccentricities and a different libration pattern.
This is the first time that planet migration theory can predict a precise orbital 
configuration of a multiplanetary system. 
This might also be the first direct evidence for type III migration if 
this scenario turns out to be true. 

The system HD45364 remains an interesting object for observers, as the 
differences between the two orbital solutions can be measured in radial velocity within a couple of years.

\chapter{The survival of mean motion resonances in a turbulent disc}
\label{ch:randwalk}
\epigraph{
Every body continues in its state of rest, or of uniform motion in a right line, unless it is compelled to change that state by forces impressed upon it.
}{\tiny Isaac Newton, Principia Mathematica, 1687}

\noindent As shown for the system HD45364 in chapter \ref{ch:threetwo}, resonant configurations can be established as a result of dissipative forces acting on the planets which lead to convergent migration. 
The previous chapter assumes a laminar disc. However, in general the disc is thought to be turbulent because, as discussed in more detail in chapter \ref{ch:introduction}, the molecular viscosity is not large enough to drive the observed outward angular momentum flux. 
The most promising mechanism to drive turbulence is the magneto-rotational instability, MRI \citep{BalbusHawley91}, although other possibilities are still being discussed \citep[see e.g.][]{LesurPapaloizou2009}. 

In this chapter, we study the effects of density perturbations in the disc resulting from any kind of turbulence on planetary systems. 
We do not solve the full 3D magneto hydrodynamics equations, but rather assume a parametrisation for the forces acting on planets. 
This has two main advantages. First, we can understand the relevant dynamical processes much more easily. 
Second, there is a large diversity of simulations of the MRI in the literature which exhibit different diffusion coefficients, varying by several orders of magnitude. 
By parametrising the forces, we avoid these difficulties altogether.

Using this approach, we study the response of both a single planet and planetary systems in a 2:1 mean motion commensurability to stochastic forcing.
The forcing could for example result from MRI turbulence.
We first develop an analytic model from first principles. 
We can isolate two distinct libration modes for the resonant angles which react differently to stochastic forcing. 

Systems are quickly destabilised if the magnitude of the stochastic forcing is large. 
The growth of libration amplitudes is parametrised as a function of the diffusion coefficient and other relevant physical parameters.
We also perform numerical N-body simulations with additional stochastic forcing terms to represent the effects of disc turbulence.
These are in excellent agreement with the analytic model. 

Stochastic forcing due to disc turbulence may have played a role in shaping the configurations of observed systems in mean motion resonance.
It naturally provides a mechanism for accounting for the HD128311 system, for which the fast mode librates and the slow mode is apparently 
near the borderline between libration and circulation.

\section{Basic equations}
We begin by writing down the equations of motion for a single planet moving in a fixed plane under a general Hamiltonian $H$ in the form
\citep[see e.g.][]{SnellgrovePapaloizouNelson01, PapaloizouResonances}
\begin{eqnarray}
\dot E &=& -n\frac{\partial H}{\partial \lambda}\label{eqnmo1}\\
\dot L &=& -\left(\frac{\partial H}{\partial \lambda}+\frac{\partial H}{\partial \varpi}\right)\\
\dot \lambda &=& \frac{\partial H}{\partial L} + n \frac{\partial H}{\partial E}\label{eq:randwalk:eqnmo3}\\
\dot \varpi &=& \frac{\partial H}{\partial L}.\label{eqnmo4}
\end{eqnarray} 
Here the angular momentum of the planet is $L$ and the energy is $E$.
For a Keplerian orbital motion around a central point mass $M$ we have 
\begin{eqnarray}
L &=& m\sqrt{\G Ma(1-e^2)} \hspace{0.5cm} {\mathrm{ and}} \\
E &=& -\frac{{\G Mm}}{{2a}},
\end{eqnarray}
where $\G$ is the gravitational constant, $a$ the semi-major axis and $e$ the eccentricity (see appendix \ref{app:orbit}).
For an elliptical orbit around a point mass, equation \ref{eq:randwalk:eqnmo3} results in a linear growth of the mean longitude 
\begin{eqnarray}
\lambda = n (t-t_0) +\varpi,
\end{eqnarray}
where $n$ is the mean motion, $t_0$ being the time of periastron passage and $\varpi$ being the longitude of periastron. 
All other equations of motion are trivial as the right hand side equates to zero.

\subsection{Additional forcing of a single planet}\label{AdF}
In order to study phenomena such as stochastic forcing, we need to consider
the effects of an additional external force per unit mass ${\mathbf F}$, which may or may not be described using a Hamiltonian formalism.
However, as may be seen by considering general coordinate transformations starting from a Cartesian representation,
the equations of motion are linear in the components of ${\mathbf F}$.
Because of this we may determine them by considering forces of the form ${\mathbf F} = (F_x, F_y)$ for which the Cartesian components are constant. 
Having done this we may then suppose that these vary with coordinates and time in an arbitrary manner.
Following this procedure we note that when ${\mathbf F},$ as in the above form is constant, we can derive the equations
of motion by replacing the original Hamiltonian with a new Hamiltonian defined through
\begin{eqnarray}
H &\rightarrow& H - m\left( F_x\;x+F_y\;y\right) = H - m\left(\mathbf{r}\cdot\mathbf{F}\right).
\end{eqnarray} 
The additional terms proportional to the components of ${\mathbf F}$ correspond to the Gaussian
form of the equations of motion \citep{BrouwerClemence1961}.

The various derivatives involving $\mathbf r$ can be calculated by elementary means and expressed in terms of $E, L, \lambda$ and $\varpi.$ 
One thus finds additional contributions to the equations of motion \ref{eqnmo1}~-~\ref{eqnmo4}, indicated with a subscript $F,$ in the form
\begin{eqnarray}
\dot L_F &=&m\left( \frac{\partial }{\partial \lambda } + \frac{\partial }{\partial\varpi } \right)
\;\left(\mathbf{ r \cdot F}\right) =
m\;\mathbf{\left( r\times F \right) \cdot \hat e}_z \label{feqnmo1}\\
\dot E_F &=& m n\frac{\partial }{\partial \lambda} \left( \mathbf{ r \cdot F} \right) = m\;(\mathbf {v \cdot F} )\label{feqnmo2} \\
\dot \varpi_F &=& - m\frac{\partial }{\partial L} \left( \mathbf{ r \cdot F} \right) \ \ \ {\mathrm{ or \ \ equivalently}} \\
\dot \varpi_F &=& \frac{ \sqrt{(1-e^2)} }{n a\,e} 
\left[F_\theta \left( 1 + \frac{1}{{1-e^2}} \frac {r}{a} \right) \sin f
- F_r \cos f \right] \label{feqnmo3}\\
\dot \lambda_F &=& - m\left(\frac{\partial }{\partial L} 
+ n \frac{\partial}{\partial E} \right) \left( \mathbf{ r \cdot F} \right) 
= 
\left( 1- \sqrt{1-e^2}\right) \dot \varpi_F 
+\frac{2an}{\G M} \;(\mathbf{r\cdot F}) , \label{feqnmo4}
\end{eqnarray}
where the true anomaly $f$ is defined as the difference between the true longitude and the longitude of periastron, $f = \theta - \varpi$ (see also appendix \ref{app:orbit}).
Note that from equation \ref{feqnmo2}
we obtain
\begin{eqnarray}
\dot a_F \ \ = \ \ -\frac{2 a \dot n}{ 3 n} 
&=& \frac {2( {F_r} e \sin f + {F_{\theta}}(1+ e \cos f))}
	{n\sqrt{1-e^2}}.\label{feqnmo6}
\end{eqnarray}
and from equation \ref{feqnmo1} together with equation \ref{feqnmo2}
we obtain
\begin{eqnarray}
	\dot e_F &=& \frac{L(2 E \dot L + L \dot E) }{\G^2m^3M^2 e}.
\end{eqnarray}
In the limit $e \ll 1$ this becomes (ignoring terms $O(e)$ and smaller)
\begin{eqnarray}
	\dot e_F &=& {F_r} \frac 1 {an} \sin f + {F_{\theta}} \frac 1{an} 2 \cos f.\label{feqnmo5}
\end{eqnarray}
Furthermore in this limit we may replace $f$ by $f= \lambda -\varpi = n(t-t_0).$

Note that that the above formalism results in \Eq~\ref{feqnmo3} for $\dot \varpi_F$, which diverges for small $e$ as ${1/ e}$. 
This comes from the choice of coordinates used and is not associated with any actual singularity or instability in the system.
This is readily seen if one uses $h=e\sin\varpi,$ and $k=e\cos\varpi$ as dynamical variables rather than $e$ and $\varpi$. 
The former set behave like Cartesian coordinates, while the latter set are the corresponding cylindrical polar coordinates. 
When the former set is used, potentially divergent terms $\propto 1/e$ do not appear.
This can be seen from \Eqs \ref{feqnmo3} and \ref{feqnmo5} which give in the small $e$ limit 
\begin{eqnarray}
\dot h_F &=& -{F_r} \frac 1 {an} \sin \lambda + {F_{\theta}} \frac 1{an} 2 \cos \lambda
\label{feqnmo7} \\
\dot k_F &=& {F_r} \frac 1 {an} \sin \lambda + {F_{\theta}} \frac 1{an} 2 \cos \lambda
.\label{feqnmo8}
\end{eqnarray}
Abrupt changes to $\varpi$ may occur when $h$ and $k$ pass through the origin in the $(h,k)$ plane. 
But this is clearly just a coordinate singularity rather than a problem with the physical system which changes smoothly on transition through the origin. 
The abrupt changes to the $\varpi$ coordinate occur because very small perturbations to very nearly circular orbits produce large changes to this angle.

\subsection{Multiple planets}
Up to now we have considered a single planet. 
We now generalise the formalism so that it applies to a system of two planets. 
We follow closely the discussions in \cite{PapaloizouResonances} and \cite{PapaloizouSzuszkiewicz2005}. 
Excluding stochastic forcing for the time being, we start from the Hamiltonian formalism describing their mutual interactions using Jacobi coordinates \citep{Sinclair1975}.
In this formalism the motion of the outer planet is described around the centre of mass of the star plus the inner planet rather then the star alone.
Thus, the radius vector ${\mathbf r}_2,$ of the inner planet of mass $m_2$ is measured from the star with mass $M$ and that of the outer planet, 
${\mathbf r }_1,$ of mass $m_1$ is referred to the centre of mass of $M$ and $m_2$. The reduced masses of $m_2$ and $m_1$ are then defined as $\mu_2=m_2\cdot M/(M+m_2)$ and $\mu_1=m_1 \cdot (M+m_2)/(M+m_2+m_1)$.
We also define $M_{2}=M+m_2$ and $M_{1}= M + m_2+m_1$.
From now on we consistently adopt subscripts $1$ and $2$ for coordinates related to the outer and inner planets respectively. 
Note that this is different from the previous chapter.

The required Hamiltonian, is given by \citep[][p.440ff]{solarsystemdynamics}: 
\begin{eqnarray} H & = & 
  \frac12 \mu_2| \dot {\mathbf r}_2|^2 
+ \frac12 \mu_1| \dot {\mathbf r}_1|^2 
- \frac{GM_2\mu_2}{|\mathbf{r}_2|}
- \frac{GMm_1}{|\mathbf{r}_{01}|}
- \frac{G m_{1}m_2}{ | {\mathbf r}_{12}|}
. \label{hamil}
\end{eqnarray}
Here $\mathbf{r}_{ij}$ is the relative position vector between the objects with subscript $i$ and $j$, where 0 denotes the star. 
The Hamiltonian can be expressed in terms of $E_i,L_i,\varpi_i, \lambda_i, i=1,2$ and the time $t$. 
The energies are given by $E_i = -G i\mu_iM_{i}/(2a_i)$ and 
the angular momenta $L_i = \mu_i\sqrt{G M_{i}a_i(1-e_i^2)}$ with $a_i$ and $e_i$ denoting the semi-major axes and eccentricities respectively. 
The mean motions are $n_i = \sqrt{G M_{i}/a_i^3}$. 
As $m_i\ll M$, we replace $M_{i}$ by $M$ and $\mu_i$ by $m_i$ from now on.

The interaction Hamiltonian (the term $\propto m_1m_2$ in equation \ref{hamil}) can be expanded in a Fourier series involving linear combinations of the three angular differences
$\lambda_i - \varpi_i, i=1,2$ and $\varpi_1 -\varpi_2$ \citep[e.g.][]{BrouwerClemence1961}.
Near a first order $p+1:p$ resonance, we expect that both
$\phi_1 = (p+1)\lambda_1-p\lambda_2-\varpi_2, $ and
$\phi_2 = (p+1)\lambda_1-p\lambda_2-\varpi_1,$
will be slowly varying.
Following standard practice \citep[see e.g.][]{PapaloizouResonances, PapaloizouSzuszkiewicz2005} 
only terms in the Fourier expansion involving linear 
combinations of $\phi_1$ and $\phi_2$
as argument are retained because 
only these are expected to lead to large long term perturbations. 

The resulting Hamiltonian may then be written in the general form $H=E_1+E_2+ H _{12},$ where the interaction Hamiltonian is given by
\begin{equation}
H _{12}= -\frac{G m_1m_2}{a_1}\sum C_{k,l}\left( \frac{a_1}{a_2}, e_1,e_2 \right) \cos (l\phi_1 +k\phi_2), \label{Hamil} 
\end{equation}
where in the above and similar summations below, the sum ranges over all positive and negative integers $(k,l)$ and the dimensionless coefficients $C_{k,l}$
depend on $e_1,e_2$ and the ratio of semi-major axes $a_1/a_2$ only.

\subsubsection{Equations of motion for two planets}
The equations of motion for each planet can now be easily derived, that take into account
the contributions due to their mutual interactions \citep[see][]{PapaloizouResonances, PapaloizouSzuszkiewicz2005} and contributions from \ref{feqnmo1} - \ref{feqnmo4}. 
The latter terms arising from external forcing are indicated with a subscript $F$. 
We obtain to lowest order in the perturbing masses:
\begin{eqnarray}
\frac{dn_1}{ dt} & = & \frac{3(p+1)n_1^2 m_2}{ M} \sum C_{k,l}(k+l)\sin (l\phi_1 +k\phi_2) + \left(\frac{dn_1}{dt}\right)_F\label{first}\\
\frac{dn_2}{dt} & = & -\frac{3pn_2^2 m_1 a_2}{ M a_1 }\sum C_{k,l}(k+l)\sin (l\phi_1 +k\phi_2) 
+ \left(\frac{dn_2}{dt}\right)_F\label{first1}
 \end{eqnarray}
\begin{eqnarray}
\frac{de_1}{ dt} &=& - \frac{m_2 n_1 \sqrt{1-e_1^2} }{ e_1 M } \cdot \sum C_{k,l}\sin (l\phi_1 +k\phi_2) \label{first2} \\ 
&& \cdot
\left[k-(p+1)(k+l)\left(1-\sqrt{1-e_1^2}\right)\right] + \left(\frac{de_1}{dt}\right)_F\nonumber\\ 
\frac{de_2}{dt} &=& - \frac{m_1 a_2 n_2 \sqrt{1-e_2^2}}{ a_1 e_2 M }\cdot \sum C_{k,l}\sin (l\phi_1 +k\phi_2) \nonumber \\
&& \cdot 
\left[l+p(k+l)\left(1-\sqrt{1-e_2^2}\right)\right] + \left(\frac{de_2}{ dt}\right)_F \label{first3}
 \end{eqnarray}
\begin{eqnarray}
\frac{d\phi_2}{dt} & =& (p+1)n_1- pn_2 -
\sum (D_{k,l} +E_{k,l})\cos(l\phi_1 +k\phi_2)
+ \left(\frac{d\phi_2}{ dt}\right)_F\label{last2} \\
\frac{d\phi_1 }{ dt} &=& (p+1)n_1- pn_2 -
\sum ( D_{k,l} + F_{k,l} )\cos(l\phi_1 +k\phi_2)
+ \left(\frac{d\phi_1}{ dt}\right)_F \label{last} .
 \end{eqnarray}
Here
 \begin{eqnarray} 
D_{k,l} &=& \frac{2(p+1)n_1a_1^2m_2}{ M}{\frac{\partial}{\partial a_1}}\left( C_{k,l}/a_1 \right)
- \frac{2pn_2a_2^2m_1 }{ M}{\frac{\partial}{ \partial a_2}}\left( C_{k,l}/a_1 \right),\\
E_{k,l} & = & \frac{n_1m_2 \left((p+1)(1 -e_1^2)-p\sqrt{1-e_1^2}\right) }{ e_1 M}\frac{\partial C_{k,l}}{ \partial e_1}\nonumber \\
&& +\; \frac{pn_2a_2m_1\left(\sqrt{1-e_2^2}-1+e_2^2\right) }{ a_1 e_2 M}\frac{\partial C_{k,l}}{ \partial e_2} 
\end{eqnarray}
and
\begin{eqnarray}
F_{k,l} & = & \frac{(p+1)n_1m_2 \left(1 -e_1^2-\sqrt{1-e_1^2}\right) }{ e_1 M}\frac{\partial C_{k,l}}{ \partial e_1} \nonumber \\
&& +\; \frac{n_2a_2m_1\left((p+1)\sqrt{1-e_2^2}-p(1-e_2^2)\right)}{ a_1 e_2 M}\frac{\partial C_{k,l}}{ \partial e_2}.
\end{eqnarray}
Note that $\phi_2 - \phi_1 = \varpi_2 - \varpi_1\equiv \zeta $
is the angle between the two apsidal lines of the two planets.
We also comment that, up to now, we have not assumed that the eccentricities are small and that, in addition to stochastic contributions, the external forcing terms may in general
contain contributions from very slowly varying disc tides.

\subsubsection{Modes of libration}\label{FSmodes}
We first consider two planets in resonance with no external forces, to identify the possible libration modes. 
In the absence of external forces, \Eqs \ref{first} - \ref{last} can have a solution for which $a_i$ and $e_i$ are constants and the angles $\phi_1$ and $\phi_2$ are zero.
In general, other values for the angles may be possible but such cases do not occur for the numerical examples presented below. 
When the angles are zero, \Eqs \ref{last2} and \ref{last} provide a relationship between $e_1$ and $e_2$ \citep[see e.g.][]{PapaloizouResonances}.

We go on to consider small amplitude oscillations or librations of the angles about their equilibrium state. 
Because two planets are involved there are two modes of oscillation, which we find convenient to separate and describe as fast and slow modes. 
Assuming the planets have comparable masses, the fast mode has a libration frequency $\propto \sqrt{m}$ and the slow mode has a libration frequency $\propto m$.
These modes clearly separate as the planet masses are changed while maintaining fixed eccentricities.

\subsubsection{Fast mode}
To obtain the fast mode we linearise \Eqs \ref{first}~-~\ref{last} and neglect second order terms in the planet masses. 
This is equivalent to neglecting the variation of $D_{k,l}, E_{k,l}$ and $ F_{k,l}$ in \Eqs \ref{last2} and \ref{last} which then require that $\phi_1 = \phi_2$ very nearly for this mode.
Noting that for linear modes of the type considered here and in the next section, \Eqs \ref{first}~-~\ref{first3} 
imply that $\dot n_i$ and $\dot e_i$ are proportional to linear combinations of the librating angles, 
differentiation of either of \Eqs \ref{last2} or \ref{last} with respect to time then gives for small amplitude oscillations
\begin{eqnarray}
\frac{d^2\phi_i }{ dt^2} +\omega_{lf}^2 \phi_i &=& 0, \ \ \ ( i= 1,2 ), 
\label{fast} \end{eqnarray}
where
\begin{equation}
\omega_{lf}^2= - \frac{3p^2n_2^2 m_2 }{ M } \left(1+\frac{ a_2 m_1}{ a_1m_2}\right)\cdot \sum C_{k,l}(k+l)^2\label{fastp},
\end{equation}
and we have used the resonance condition that $(p+1)n_1 = p n_2$ which is satisfied to within a correction of order $\sqrt{m_1/M}$.
Note that for this mode the fact that $\phi_1 = \phi_2$ very nearly, implies that $\varpi_2 -\varpi_1$ is small. 
Thus that quantity does not participate in the oscillation.

\subsubsection{Slow mode}
Now, we look for low frequency librations with frequency $\propto m_1$.
Equations \ref{last2} and \ref{last} imply that, for such oscillations, to within a small relative error of order $m_1/M$, $(p+1)n_1 = pn_2$ throughout. 

Equations \ref{first} and \ref{first1} then imply that the two angles are related by $\phi_2=\beta \phi_1$, where $\beta = - \sum C_{k,l}(k+l)l)/(\sum C_{k,l}(k+l)k)$.
Subtracting \Eq \ref{last} from \Eq \ref{last2}, differentiating with respect to time and using this condition results in an equation
for $\zeta = \phi_ 2- \phi_1 = \varpi_2 -\varpi_1$ 
\begin{eqnarray}
\frac{d^2\zeta}{ dt^2} +\omega_{ls}^2 \zeta &=& 0,
\label{slow} \end{eqnarray}
where
\begin{equation}
\omega_{ls}^2= \alpha_1 \frac{n_1m_2\sqrt{1 -e_1^2} 
}{ e_1 M (1-\beta)}\frac{\partial W }{ \partial e_1} + \alpha_2 \frac{n_2a_2m_1\sqrt{1 -e_2^2} 
}{ a_1 e_2 M (1-\beta)}\frac{\partial W }{ \partial e_2}\label{slowp}
\end{equation}
with
\begin{eqnarray}
\alpha_1 &= & 
\sum C_{k,l}\left[k-(p+1)(k+l)\left(1-\sqrt{1-e_1^2}\right)\right](k\beta+l),\nonumber\\
\alpha_2 & =& 
\sum C_{k,l}\left[l+p(k+l)\left(1-\sqrt{1-e_2^2}\right)\right](k\beta+l)\nonumber
\end{eqnarray}
and
\begin{eqnarray}
 W & = & \left( \frac{n_1a_2m_2\sqrt{1 -e_2^2} }{ a_1 e_2 M}{\frac{\partial}{ \partial e_2}}
-\frac{n_2m_1 \sqrt{1 -e_1^2}}{ e_1 M}{\frac{\partial }{ \partial e_1}} \right)\nonumber \cdot \sum C_{k,l}. \nonumber
 \end{eqnarray}
Although the expressions for the mode frequencies are complicated, the fast frequency scales as the square root of the planet mass
and the slow frequency scales as the planet mass. Both frequencies scale with the characteristic mean motion of the system.

\subsubsection{Librations with external forcing}
When external forcing is included source terms appear on the right hand sides of \Eqs \ref{fast} and \ref{slow}.
We assume that the forcing terms are small so that terms involving products of these and both the libration amplitudes and the planet masses may be neglected.
Then, in the case of the slow mode, repeating the derivation given above including the forcing terms, we find that \Eq \ref{slow} becomes
\begin{eqnarray}
\frac{d^2\zeta}{ dt^2} +\omega_{ls}^2 \zeta &=& \frac{d}{ dt}\left(\dot \varpi_{2F} - \dot \varpi_{1F} \right). \label{slowf} 
\end{eqnarray}
The quantities $\dot\varpi_{iF}$ are readily obtained for each planet from \Eq \ref{feqnmo3}.
From this we see that for small eccentricities, $\dot\varpi_{iF} \propto 1/e_i$, indicating large effects when $e_i$ is small. 
As already discussed, this feature arises from a coordinate singularity rather than physically significant changes to the system.

A similar description may be found for the fast mode. 
In this case, neglecting terms of the order of the square of the planet masses or higher, one may use \Eqs \ref{last2} and \ref{last}
to obtain an equation for $\phi_1 $ in the form
\begin{eqnarray}
\frac{d^2 \phi_1}{ dt^2} +\omega_{lf}^2 \phi_1&=&\frac{d}{ dt}\left( \dot \phi_{1F} \right )+(p+1)\dot n_{1F}-p\dot n_{2F}. \label{fastf} 
\end{eqnarray}

Equations \ref{fastf} and \ref{slowf} form a pair of equations for the stochastically forced fast and slow modes respectively.
This mode separation is not precise.
However, it can be made so by choosing appropriate linear combinations of the above modes. 
Numerical results confirm that $\phi_1$ predominantly manifests the fast mode and $\zeta$ the slow mode, so we do not
expect such a change of basis to significantly affect conclusions.

We further comment that because $\dot\phi_{1F}$ contains $\dot\varpi_{2F}$ but not $\dot\varpi_{1F}$,
 for small eccentricities there are only potential forcing terms $\propto$~$1/e_2$ that occur when forcing is applied to the inner planet.
As this planet has the larger eccentricity for the situations we consider, small eccentricities are not found to play any significant role in this case.

Each mode responds as a forced oscillator. 
The forcing contains a stochastic component which tends to excite the respective oscillator and ultimately convert libration into circulation. 
But we stress that the above formulation as well as developments below assume small librations, so we may only assess the initial growth of oscillation amplitude. 
However, inferences based on the structure of the non linear governing equations and an extrapolation of the linear results enable successful comparison with numerical results.

\subsection{Stochastic forces}\label{StochF}
We assume that turbulence causes the external force per unit mass $(F_r, F_{\theta})$ acting on each planet to be stochastic.
For simplicity we shall adopt the simplest possible model. 
Each component of the force satisfies the relation 
\begin{equation}
F_i(t)F_i(t')= \left< F_i^2 \right> g(|t-t'|)
\end{equation}
where the auto-correlation function $g(x)$ is such that $\int^{\infty}_0 g(x) dx = \tau_c$, 
where $\tau_c$ is the correlation time and $\sqrt{\left< F_i^2 \right>}$ is the root mean square value of the $i$ component. 
Note that an ensemble average is implied on the left hand side. 
Again, for simplicity we assume that different components acting on the same planet as well as components acting on different planets are uncorrelated. 
The root mean square values as well as $\tau_c$ may in general depend on $t$, but we shall not take this into account here and simply assume that these
quantities are constant and the same for each force component. 

The stochastic forces make quantities they act on undergo a random walk. 
Thus, if for example $\dot A = F_i$ for some quantity\footnote{Constants or slowly varying quantities originally multiplying $F_i$ 
may be absorbed by a redefinition of $A$ and so do not affect the discussion given below.} $A$, the square of the change of $A$ occurring after a time interval $t$ is given by
\begin{eqnarray}
(\Delta A)^2 &=& \int^t_0 \int^t_0 F_i(t')F_i(t'')dt'dt'' \nonumber \\
&=& \int^t_0 \int^t_0 \left< F_i^2 \right>g(|t'-t''|) dt'dt'' \rightarrow D_i t.
\label{stoch}
\end{eqnarray}
Here we take the limit where $t/\tau_c $ is very large corresponding to an integration time of many correlation times 
and $D_i = 2\left< F_i^2 \right>\tau_c$ is the diffusion coefficient. 

When the evolution of a stochastically forced planetary orbit is considered, it is more appropriate to consider a model governing equation for $A$ of the generic form
\begin{equation}
\dot A = F_i \sin(nt),\label{generic} 
\end{equation} 
where we recall that $2\pi/n$ is the orbital period (but note that a different value could equally well be considered).
We note in passing that, by shifting the origin of time, an arbitrary phase may be added to the argument without changing the results given below. 
One readily finds that \Eq \ref{stoch} is replaced by 
\begin{eqnarray}
(\Delta A)^2&=& \int^t_0 \int^t_0 F_i(t')F_i(t'')\sin(n t') \sin(n t'') dt'dt''\nonumber \\
&=&\int^t_0 \int^t_0\left< F_i^2 \right>g(|t'-t''|) \sin(n t')\sin(n t'')dt' dt'' \nonumber \\
&\rightarrow& \frac{\gamma Dt}{ 2},
\label{stochreduced}
\end{eqnarray}
where we introduce a new correction factor $\gamma(n) = \int^{\infty}_0 g(x)\cos(n x) dx$. 
If $n\tau_c\ll 1$, corresponding to the correlation time being much less than the orbital period, then $\gamma \rightarrow 1$.
For larger $\tau_c,$ $\gamma <1$ gives a reduction factor for the amount of stochastic diffusion. 
For example if we adopt, an exponential form for the auto-correlation function such that
\begin{eqnarray*}
	g(|t'-t''|) = \exp \left(- \frac{|t'-t''|}{\tau_c}\right),
\end{eqnarray*}
we find 
\begin{eqnarray}
 \gamma = \frac 1{1+n^2 \tau_c^2}\label{eq:reduction},
\end{eqnarray}
and for the purposes of simplicity and comparison with numerical work we shall use this from now on.

\subsubsection{Stochastic forcing of an isolated planet}
In the case of a stochastically forced single planet we may obtain a statistical estimate for the 
characteristic growth of the orbital parameters as a function of time by directly integrating 
\Eqs \ref{feqnmo2}, \ref{feqnmo3}, \ref{feqnmo5}, \ref{feqnmo7} and \ref{feqnmo8} with respect to time. 
We may then apply the formalism leading to the results expressed in generic form through 
\Eqs \ref{stoch}~-~\ref{eq:reduction} to obtain estimates for the stochastic diffusion of the orbital elements in the limit of small eccentricity in the form
\begin{eqnarray}
(\Delta a)^2 &=& 4 \frac { D t }{ n^2 }\label{eq:growtha}\\
(\Delta n)^2 &=& 9 \frac { D t }{ a^2 }\label{eq:growthn}\\
(\Delta e )^{2} &=& 2.5 \frac {\gamma D t }{ n^2 a^2 } \label{eq:growthe}\\
(\Delta \varpi)^2 &=& \frac{2.5}{e^2} \frac{\gamma Dt}{n^2 a^2} \\ \label{eq:growthw}
(\Delta h )^{2} &=& 2.5 \frac {\gamma D t }{ n^2 a^2 } \label{eq:growthh}\\
(\Delta k )^{2} &=& 2.5 \frac {\gamma D t }{ n^2 a^2 } \label{eq:growthk}.
\end{eqnarray}
As mentioned before, the $1/e^2$ dependence of $(\Delta \varpi)^2$ which arises from the coordinate singularity does not appear for $(\Delta h )^{2}$ and $ (\Delta k )^{2}$.
Note that the definitions of $h$ and $k$ imply, consistently with the above, that
\begin{eqnarray}
 (\Delta h )^{2} &=& (\Delta e )^{2}\sin^2\varpi+ e^{2} (\Delta \varpi)^2\cos^2\varpi \label{eq:growthh1}\\
 (\Delta k )^{2} &=& (\Delta e )^{2}\cos^2\varpi+ e^{2} (\Delta \varpi)^2\sin^2\varpi \label{eq:growthk1}.
\end{eqnarray}

\subsubsection{The evolution of the longitude}
Of special interest is also the evolution of the planet's mean longitude. 
In observations of Saturn's rings, for example, it is much easier to measure a shift of the longitude of an embedded moonlet compared to a direct
measurement of the change in its semi-major axis.
The time derivative of the mean motion is given by
\begin{eqnarray}
\dot n = - \frac32 \sqrt{\frac{\G M}{a^5} }\dot a = - \frac{3\;  F_\theta }{a}.
\end{eqnarray}
Assuming a nearly circular ($e\ll 1$) orbit, the mean longitude is thus given by the double integral
\begin{eqnarray}
\lambda(t) 
	&=& \int_0^t \left(n+\Delta n(t')\right)\;dt'\\
	&=& nt - \int_0^t \int_0^{t'} 		\frac{3F_\theta(t'')}a		\;dt''dt'.
\end{eqnarray}
The root mean square of the difference compared to the unperturbed orbit is
\begin{eqnarray}
\left(\Delta \lambda\right)^2 &=&
\left<\left(\lambda(t)-nt\right)^2 \right>^{0.5}\\
	&=&  \int_0^t \int_0^{t'} \int_0^t \int_0^{t'''}		\frac{9F_\theta(t'')F_\theta(t'''')}{a^2}		\;dt''''dt'''dt''dt'\\
	&=&  \frac{9\left< F_\theta^2 \right>}{a^2}  \int_0^t \int_0^{t'} \int_0^t \int_0^{t'''}	        g(|t''-t''''|) 			\;dt''''dt'''dt''dt' \label{eq:randwalk:notdoable}\\
	&=&  \frac{9\left< F_\theta^2 \right>}{a^2} \left(
	-2\tau_c^4+
	\left( 2\tau_c^3t+2\tau_c^4+\tau_c^2t^2 \right ) e^{-t/\tau_c}
	+\frac 13 \tau_c t^3\label{eq:growthlambdaall}
	\right) \\
	&\overset{t\gg\tau_c}{=}& 
	\frac{9\left< F_\theta^2 \right> \tau_c }{a^2}   \;t^3
	\;= \;\frac32 \frac{D}{a^2} \;t^3 \;=\; \left(\Delta n\right)^2 \frac{t^2}6,\label{eq:growthlambda}
\end{eqnarray}
where we have taken the limit of $t\gg\tau_c$ in the last line. 
We also assume an exponentially decaying auto-correlation function $g$ because the integral in \Eq~\ref{eq:randwalk:notdoable} can not be solved analytically for an arbitrary function $g$. 
Note that on short timescales, other terms in equation \ref{eq:growthlambda} may dominate the growth of $\lambda$.

Instead of a stochastic migration, let us also assume a laminar migration with constant migration rate $\tau_a$, given by
\begin{eqnarray}
\tau_a&=&\frac{a}{\dot a}=-\frac32 \frac{n}{\dot n}.
\end{eqnarray}
As above, we can calculate the longitude as a function of time as the following double integral
\begin{eqnarray}
\lambda(t) 
	&=& \int_0^t \left(n+\Delta n(t')\right)\;dt'\\
	&=& nt - \int_0^t \int_0^{t'} 		\frac32 \frac{n}{\tau_a}		\;dt''dt'
	= nt - \frac34 \frac{n}{\tau_a}\;t^2		
\end{eqnarray}
and thus
\begin{eqnarray}
\left(\Delta \lambda\right)^2 &=&\frac98 \frac{n^2}{\tau_a^2}\;t^4 = \left( \Delta n\right)^2 \frac{t^2}2.\label{eq:growthlambdalaminar}
\end{eqnarray}
Note that \Eq \ref{eq:growthlambdalaminar} is an exact solution, whereas \Eq \ref{eq:growthlambda} is a statistical quantity, describing the root mean square value in an ensemble average. 
In the stochastic and laminar case, $\left(\Delta\lambda\right)^2$ grows like $\sim t^3$ and  $\sim t^4$, respectively. This allows to discriminate the different mechanisms by observing $\Delta \lambda$ over an extended period of time.

\subsubsection{Stochastic variation of the resonant angles in the two planet case} \label{STF2}
We now consider the effects of stochastic forcing on the resonant angles. 
The expressions are more complicated than in the previous section because there are more variables involved. 
However, we basically follow the same formalism.
 
While the libration amplitude is small enough for linearisation to be reasonable, the evolution is described by \Eqs \ref{slowf} and \ref{fastf}.
These may be solved by the method of variation of parameters. Assuming the amplitude is zero at $t=0$, the solution of \Eq \ref{slowf} is given by 
\begin{eqnarray}
\zeta  &=& \sin(\omega_{ls}t)\int^t_0\frac{S_\zeta\cos(\omega_{ls}t)}{\omega_{ls}}dt 
 - \cos(\omega_{ls}t)\int^t_0\frac{S_\zeta\sin(\omega_{ls}t)}{\omega_{ls}}dt,
\label{zetasol}
\end{eqnarray}
where $S_\zeta= {d}(\dot\varpi_{2F}- \dot \varpi_{1F})/dt$.
Equation \ref{zetasol} may be regarded as describing a harmonic oscillator whose amplitude varies in time such that the square of the amplitude after a time interval $t$ is given by
\begin{equation}
(\Delta \zeta )^2 = \left(\int^t_0\frac{S_\zeta\sin(\omega_{ls}t)}{\omega_{ls}}dt\right)^2
+\left(\int^t_0\frac{S_\zeta\cos(\omega_{ls}t)}{ \omega_{ls}}dt\right)^2.
\label{Dzeta} \end{equation}
The corresponding expression for \Eq \ref{fastf} is
\begin{equation}
(\Delta \phi_1)^2 = \left(\int^t_0\frac{S_\phi\sin(\omega_{lf}t)}{ \omega_{lf}}dt\right)^2
+\left(\int^t_0\frac{S_\phi\cos(\omega_{lf}t)}{ \omega_{lf}}dt\right)^2,
\label{DQ} \end{equation}
where $S_\phi= d(\dot \phi_{1F})/dt +(p+1)\dot n_{1F} - p\dot n_{2F}.$

We now evaluate the expectation values of these using the formalism of the previous section.
For simplicity we specialise to the case when stochastic forces act only on the outer planet. 
This is also what has been considered numerically later.
Taking \Eq \ref{Dzeta}, we perform an integration by parts, neglecting the end point contributions as these are sub-dominant (they increase less rapidly than $t$ for large $t$), to obtain
\begin{equation}
(\Delta \zeta )^2 = \left(\int^t_0 \dot \varpi_{1F}
\sin(\omega_{ls}t)dt\right)^2
+\left(\int^t_0 \dot \varpi_{1F}\cos(\omega_{ls}t)dt\right)^2.
\label{Dzetaf}
\end{equation}
In dealing with equation \ref{DQ} we neglect $\dot \phi_{1F}$ in $S_\phi$ because after integration by parts 
this leads to a contribution on the order $\omega_{lf}/n$ smaller than that derived from $\dot n_{1F}$. 
Thus, we simply obtain
\begin{eqnarray}
\frac{(\Delta \phi_1)^2}{ (p+1)^2}
&=& \left(\int^t_0\frac{\dot n_{1F}\sin(\omega_{lf}t)}{ \omega_{lf}}dt\right)^2 \nonumber
+\left(\int^t_0\frac{\dot n_{1F}\cos(\omega_{lf}t)}{ \omega_{lf}}dt\right)^2.
\end{eqnarray}
We again follow the procedures outlined in the previous section to obtain the final growth estimates 
\begin{eqnarray}
(\Delta e_1e_2\sin\zeta )^2 &=& 
2.5 \frac{De_2^2 \gamma_s t}{ 2a_1^2 n_1^2 }\label{eq:growths}\quad \quad \text{and} \\
\frac{(\Delta \phi_1)^2}{ (p+1)^2}
&=& \frac{9 D \gamma_ft}{ a_1^2\omega_{lf}^2}, 
\label{eq:growthf}
\end{eqnarray}
where the corrections factors are given by 
\begin{eqnarray}
\gamma_f &=& \frac1{1+\omega_{lf}^2\tau_c^2} \quad \quad \text{and}\\
\gamma_s &=& \frac1{1+(n_1+\omega_{ls})^2\tau_c^2} +\frac1{1+(n_1-\omega_{ls})^2\tau_c^2 }.
\end{eqnarray}
The correction factors $\gamma_f$ and $\gamma_s$ compare the relevant timescales of the system (here, the libration periods) to the correlation time of the stochastic force.

\subsubsection{Growth of libration amplitudes}\label{GRLIB}
Equations \ref{eq:growths} and \ref{eq:growthf} express the expected growth of the resonant angle libration amplitudes as a function of time.
These expressions can be simply related to those obtained for a single planet. 
Thus \Eqs \ref{eq:growtha} and \ref{eq:growthf} applied to the outer planet imply that 
\begin{equation}
(\Delta \phi_1)^2 /(\Delta a_1)^2= 9(p+1)^2 n_1^2\gamma_f /(4 a_1^2\omega_{lf}^2). \nonumber
\end{equation}
As we are interested in the case $p=1$, the width of the libration zone is $\sim a_1\omega_{lf}/n_1$, 
we see that the time for $(\Delta \phi_1)^2$ to reach unity is comparable for the semi-major axis to diffuse through the libration zone.

For fixed eccentricity of the inner planet $e_2$, \Eq \ref{eq:growths} indicates that $\zeta$, being the angle between the apsidal 
lines of the two planets, diffuses in the same way as for an isolated outer planet subject to stochastic forces. 
Thus, in the small amplitude regime, the way this diffusion occurs would appear to be essentially independent of whether the planets are in resonance.

An important consequence of \Eq \ref{eq:growths} is the behaviour of $\zeta$ for small $e_1$. 
The latter quantity was assumed constant in the analysis.
As already discussed earlier, abrupt changes to $\zeta$ are expected when the eccentricity of one planet gets very small. 
Then, even an initially small amplitude libration is converted to circulation.
Thus if $e_1$ is small, then \Eq \ref{eq:growths} indicates that a time $t \sim 4a_1^2 n_1^2 e_1^2 / (5 D \gamma_s \zeta^2 )$, is required to convert libration to circulation.
This can be small if $e_1$ is small. Even if $e_1$ is not small initially, it is important to note that it also undergoes stochastic diffusion 
(see \Eq \ref{eq:growthe}) as well as oscillations through its participation in libration. 
Should $e_1^2$ attain very small values through this process, then from \Eq \ref{eq:growths} we expect the onset of a rapid evolution of $\zeta$. 
This is particularly important when $e_1^2$ already starts from relatively small values.

In fact, application of \Eqs \ref{eq:growths} and \ref{eq:growthf} to the numerical examples discussed below, adopting the initial orbital elements, indicate that the diffusion
of $\zeta$ is significantly smaller than $\phi_1$ unless $e_1$ starts out with a very small value. 
This would suggest that $\phi_1$ reaches circulation before $\zeta$.
However, this neglects the coupling between the angles that occurs once the libration amplitudes become significant.
It is readily seen that it is not expected that $\phi_1$ could circulate while $\zeta$ remains librating as it was initially.
One expects to recover standard secular dynamics for $\zeta$ from the governing \Eqs \ref{first}~-~\ref{last2}, 
when these are averaged over an assumed $\phi_1$ circulating with constant $\dot \phi_1$.
As a libration of the initial form would not occur under those conditions, we expect, and find, large excursions or increases in the libration amplitude of $\zeta$ to be correlated with
increases in the libration amplitude of $\phi_1$. 
This in turn increases the oscillation amplitude of the eccentricity $e_1$, allowing it to approach zero. 
The consequent rapid evolution of $\zeta$ then enables it to pass to circulation.
Thus, the breaking of resonance is ultimately found to be controlled by the excitation of large amplitude librations for $\phi_1$,
which induce $\zeta$ to pass to circulation somewhat before $\phi_1$ itself.

\section{Numerical simulations}

We have performed numerical simulations of one and two planet systems that allow for the incorporation of additional stochastic forces with the properties described above. 
These forces provide a simple prescription for estimating the effects of stochastic gravitational forces produced by density fluctuations associated with disc turbulence.
The results have been obtained using the N-body code described in chapter \ref{ch:threetwo}. 
We have checked that results are converged and do not depend on the integrator used.

We first discuss the expected scaling of the stochastic forces with the physical parameters of the disc and their implementation in N-body integrations. 
In order to clarify the physical mechanisms involved, and to check the analytic predictions for stochastic diffusion given by \Eqs \ref{eq:growtha}~-~\ref{eq:growthw}, we 
consider simulations of a single planet undergoing stochastic forcing first. 
We then move on to consider two planet systems with and without stochastic forcing. 
We focus on the way a 2:1 commensurability is disrupted and highlight the various evolutionary stages a system goes through as it evolves
from a state with a strong commensurability affecting the interaction dynamics, to one where the commensurability is completely disrupted. We find that in some cases a strong scattering occurs. 
We consider a range of different planet masses and eccentricities (see table \ref{table:initorb}).

\subsection{Stochastic forces}\label{sec:scaling}
In order to mimic the effects of turbulence, for example produced by the MRI, it is necessary to calibrate these forces with reference to MHD simulations. 
As described above, the basic parameters characterising the prescription for stochastic forcing that we have implemented are the root mean square value of the force components 
per unit mass (in cylindrical coordinates) $\sqrt{\left< F_i^2 \right>}$ and the auto correlation time $\tau_c$. 

From our analytic considerations, we conclude that stochastic forces make the orbital parameters undergo a random walk 
that is dependent on the force model primarily through the diffusion coefficient $D = 2\left< F_i^2 \right>\tau_c$ but also through the dimensionless correction factors $\gamma, \gamma_s$ and $\gamma_f$.
\begin{figure}[tb]
\center
\resizebox{0.7\columnwidth}{!}{
\input{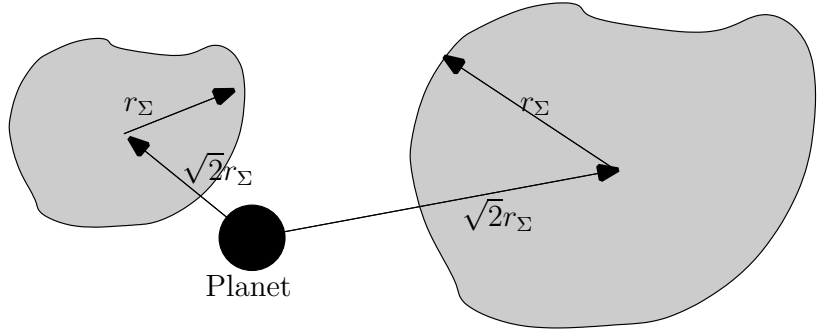}
}
\caption{The planet is embedded in a proto-planetary disc with random density fluctuations. The force exerted by both clumps in the picture on the planet has the same strength. It is independent of the clump size in a two dimensional system, assuming that the clumps are randomly distributed.
\label{fig:randwalk:discforce}}
\end{figure}

For planets under the gravitational influence of a proto-planetary disc, the natural scale for the force per unit mass components,~$F_i$, is 
$F_0(r) = \pi \G\Sigma(r)/2$, where $\Sigma$ is the characteristic disc surface density \citep[see e.g.][]{papaloizouterquem06}.
$F_0(r)$ is the gravitational force per unit mass due to a small circular disc patch of radius $r_{\Sigma}$ at a distance $\sqrt{2} \; r_{\Sigma}$ from its centre assuming
that all its mass is concentrated there. See also \Fig~\ref{fig:randwalk:discforce}.
This is independent of $r_{\Sigma}$.
The natural correlation time $\tau_c$ is the inverse of the orbital angular frequency $\tau_{c,0} = \Omega^{-1}$. 
We adopt a minimum mass solar nebula model \citep[MMSN, see][]{Weidenschilling1977} with 
\begin{eqnarray}
\Sigma(r) &=& 4200 \frac{\mbox{g}}{\mbox{cm}^2} \left(\frac{r}{\mbox{1 AU}}\right)^{-3/2}.
\end{eqnarray}
This provides a natural scale for $D$ as a function of the local disc radius and the central stellar mass through\footnote{The scaling in \cite{ReinPapaloizou2009} is incorrect due to a missing factor of $\pi$. The value in this thesis has been corrected.} 
\begin{eqnarray}
D_0 &=& 2 C F_0^2 \tau_{c,0} = 1.95 \; C \; \frac{ \mbox{cm}^2}{\mbox{s}^{3}} \;\left( \frac{r}{1\mbox{ AU}}\right)^{-\frac32} \left(\frac{M_*}{1 \mbox{ M}_\odot}\right)^{-\frac 12},
\end{eqnarray}
where $C$ is a dimensionless constant.
There are several factors which contribute to the value of~$C$: 
\begin{itemize}
\item 
The density fluctuations found in MRI simulations ${\delta \rho}/{\rho} \sim {\delta \Sigma}/{\Sigma}$ are typically of the order $0.1$ \citep[e.g.][]{Nelson2005}. 
\item 
The presence of a dead zone in the mid plane regions of the disc, where the MRI is not active, has been found to cause reductions 
in the magnitude of $F_0$ by one order of magnitude or more, as compared to active cases \citep{oishi2007}.
\item
Massive planets open a gap in the disc.
\cite{oishi2007} found that most of the contribution to the stochastic force comes from density fluctuations within a distance of one scale-height from the planet. 
When a gap forms, this region is cleared of material leading to a substantial decrease in the magnitude of turbulent density fluctuations. 
Consequently $F_0$ should be reduced on account of a lower ambient surface density. 
A factor of $0.1$ seems reasonable, although it might be even smaller \citep{DiscComp2006}. 
\item
The correlation time $\tau_c$ is found to be slightly shorter than an orbital period, namely~$0.5\;\Omega^{-1}$ \citep{NelsonPapaloizou04,oishi2007}.
\end{itemize}

\noindent If it is appropriate to include reduction factors to account for all of the above effects, one finds $C= 5\cdot 10^{-7}$ and we expect a natural scale 
for the diffusion coefficient to be specified through
\begin{equation}
D_0 = 10^{-6}\left( \frac{r}{1\mbox{ AU}}\right)^{-3/2}
\left(\frac{M_*}{1 \mbox{ M}_\odot}\right)^{-1/2} \frac{ \mbox{cm}^2}{\mbox{s}^{3}}.\label{eq:scalingd} 
\end{equation}
The same value of D may be equivalently scaled to the orbital parameters of the planets without reference to the disc by writing
\begin{eqnarray}
D_0 &=& 3.5 \cdot 10^{6}\left( \frac{r}{1\mbox{ AU}}\right)^{-5/2}\left( \frac{M_*}{1M_{\odot}}\right)^{3/2} 
\cdot \left(\frac{ r^4 \left< F_i^2 \right>\Omega \tau_c}{(GM_*)^2}\right) \frac{ \mbox{cm}^2}{\mbox{s}^{3}}. \label{DCOEFF}
\end{eqnarray}
Thus, a value $D_0=10^{-5}$ in cgs units corresponds to a ratio of the root mean square stochastic force component to that due to the central star 
of about $\sim 10^{-6}$ for a central solar mass at $1$~AU.
It is a simple matter to scale to other locations. 

Of course we emphasise that the value of this quantity is very uncertain, a situation that is exacerbated by its proportionality to the 
square of the magnitude of the stochastic force per unit mass. 
For this reason we perform simulations for a range of $D$, covering many orders of magnitude.

\begin{table}[bt]
  
\begin{minipage}{\columnwidth}
\renewcommand{\footnoterule}{}
\renewcommand{\thempfootnote}{\arabic{mpfootnote}}
\centering 
\begin{tabular}{ r | c c c c c } 
\hline 
\hline
Planet & $m$ ($\mathrm{M}_J$) & $a$ (AU) & $P$ (days) &$e$ & $ \zeta$ \\  
\hline  
GJ876 c & 0.790 & 0.131 &30.46& 0.263 & $10\degree $ \\ 
      b & 2.530 & 0.208 &60.83& 0.031 & \\
\hline    
GJ876 LM c & 0.13 & 0.131 &30.46& 0.263 & $10\degree $ \\ 
       b & 0.42 & 0.208 &60.83& 0.031 &  \\
\hline    
GJ876 SE c & 0.013 & 0.131 &30.46& 0.263 & $10\degree $ \\ 
       b & 0.042 & 0.208 &60.83& 0.031 &  \\
\hline    
GJ876 E c & 0.0013 & 0.0131 &30.46& 0.263 & $10\degree $ \\ 
       b & 0.0042 & 0.208 &60.83& 0.031 &  \\
\hline    
GJ876 LM HE c & 0.13 & 0.131 &30.46& 0.41 & $10\degree $ \\ 
       b & 0.42 & 0.208 &60.83& 0.09 &  \\
\hline    
GJ876 SE HE c & 0.013 & 0.131 &30.46& 0.41 & $10\degree $ \\ 
       b & 0.042 & 0.208 &60.83& 0.09 &  \\
\hline    
HD128311 b & 1.56 & 1.109 & 476.8 & 0.38 & $58 \degree$ \\
         c & 3.08 & 1.735 & 933.1 & 0.21 & \\
\hline
\hline
\end{tabular}
\caption{
Parameters of the model systems considered.
\label{table:initorb}
Orbital elements of the system GJ876 are based on a fit by \cite{Rivera2005}.
LM, SE and E indicate that planet masses have scaled down by a factor of 6, 60 and 600, respectively. 
This is equivalent to a Jovian mass planet, a super-Earth and Earth mass planet around a solar mass star.
HE indicates higher eccentricities than in the observed system.
Orbital elements of the system HD128311 are taken from \cite{Vogt2005}.}
\end{minipage}
\end{table}

\subsection{Numerical implementation of correlated noise}
The procedure we implemented, uses a discrete first order Markov process to generate a correlated, continuous noise that is added as an additional force. 
This statistical process is defined by two parameters, the root mean square of the amplitude and the correlation time $\tau_c$ \citep{KASDIN}. 
It has a zero mean value and has no memory. 
This has the advantage that previous values do not need to be stored in memory. 
The auto-correlation function decays exponentially and thus mimics the auto-correlation function measured in MHD simulations by \cite{oishi2007}. 
It is also consistent with the assumptions made in the previous sections.

The procedure for generating noise in a variable \texttt{x} with an auto-correlation time \texttt{tau} and root mean square value of 1 is as follows. 
The update of \texttt{x} from \texttt{t} to \texttt{t+dt} in pseudo code is
\begin{verbatim}
    x *= exp( -dt / tau) 
    x += Xi * sqrt( 1 - exp( 2 * dt / tau ) )
\end{verbatim}
where \texttt{Xi} is a random variable with normal distribution and a variance of unity.

\subsection{Stochastic forces acting on a single planet}
\begin{figure}[p]
\centering
\includegraphics[width=0.9\columnwidth]{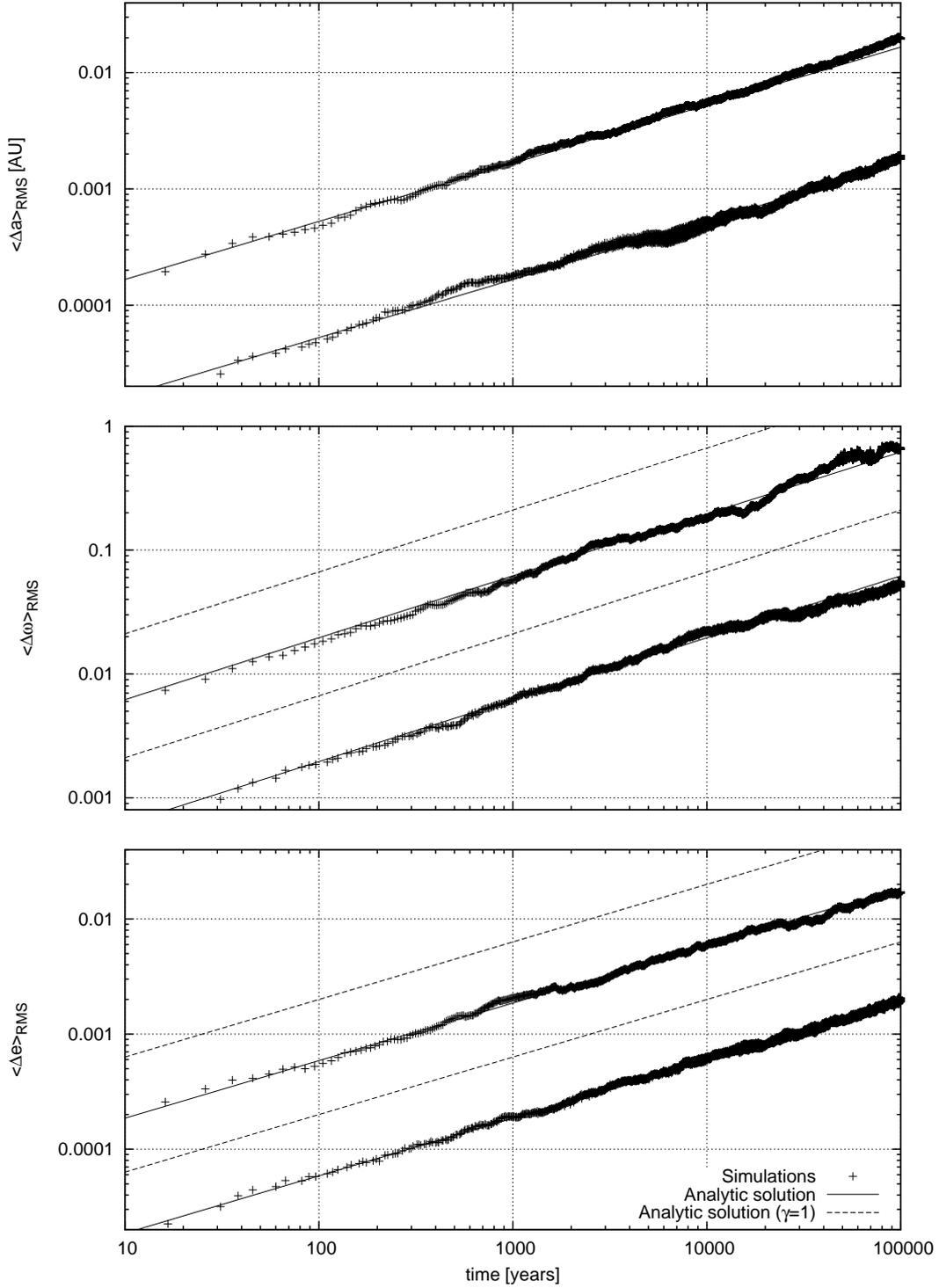}
\caption{Growth of the semi-major axis $\Delta a$, the periastron $\Delta \varpi$ and the eccentricity $\Delta e$ for a single planet as a function of time. 
The initial orbital parameters of the planet are taken to be those of GJ876~b (see table \ref{table:initorb}). 
One hundred simulations starting from the same initial conditions are averaged for each value of the diffusion coefficient.
The diffusion coefficient $D$ is $8.2\cdot 10^{-3} { \mbox{cm}^2}/{\mbox{s}^{3}}$ for the upper
and $8.2 \cdot 10^{-5} { \mbox{cm}^2}/{\mbox{s}^{3}}$ for the lower curves, respectively. 
Solid lines correspond to the analytic predictions, i.e., these are not fits. 
Dashed lines correspond to the analytic predictions for uncorrelated, random kicks.
\label{fig:singleplanet} }
\end{figure}

\begin{figure}[t]
\centering
\includegraphics[width=0.9\columnwidth]{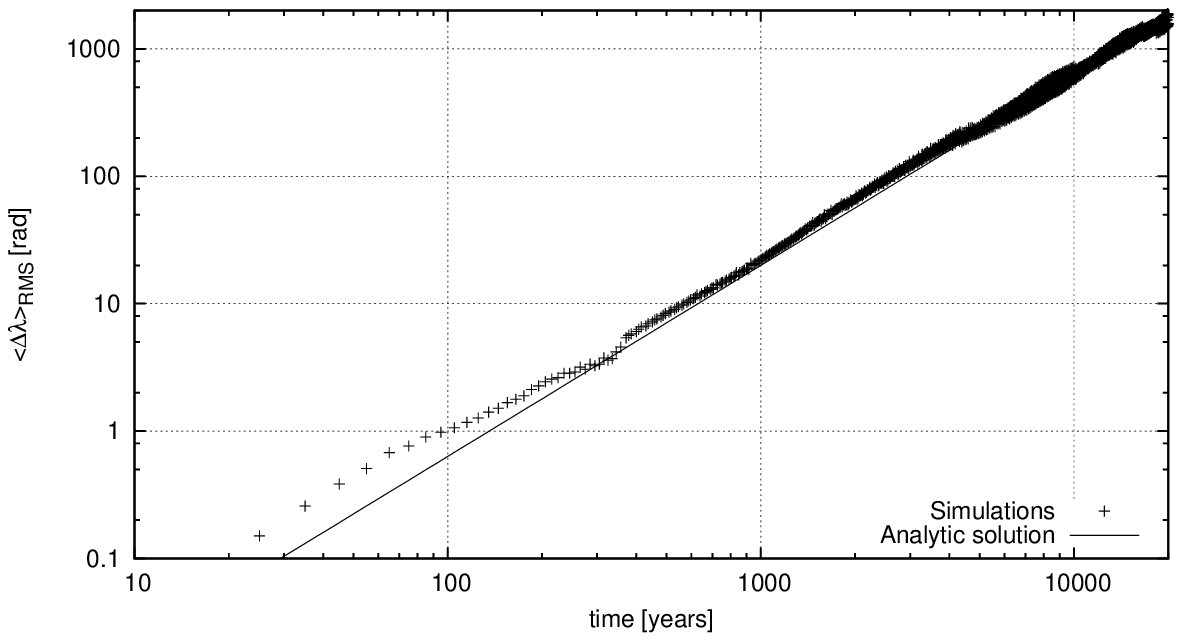}
\caption{Growth of the difference in longitude $\Delta \lambda$ as a function of time for a single planet. 
The initial orbital parameters of the planet are taken to be those of GJ876~b (see table \ref{table:initorb}). 
One hundred simulations starting from the same initial conditions are averaged. 
Here, the diffusion coefficient is $D = 8.2 \cdot 10^{-5} { \mbox{cm}^2}/{\mbox{s}^{3}}$.
The solid line corresponds to the analytic prediction, given by \Eq \ref{eq:growthlambda}.
\label{fig:singleplanetlambda} }
\end{figure}

We first investigate the long-term effect of stochastic forces on a single isolated planet.
The initial orbital parameters are taken to be the observed parameters of GJ876~b (see table \ref{table:initorb}).
In this simulation, we use the reduced diffusion coefficients
\begin{eqnarray}
 D= 8.2 \cdot 10^{-5} \frac{ \mbox{cm}^2}{\mbox{s}^{3}}, \quad \text{and}\quad
 D= 8.2 \cdot 10^{-3} \frac{ \mbox{cm}^2}{\mbox{s}^{3}}, \nonumber
\end{eqnarray}
which can be represented by a correlation time of half the orbital period and a specific force with root mean square values 
$\sqrt{\left<F^2\right>}~=~4.05\cdot10^{-6}~\mbox{cm}/\mbox{s}^{2}$, and 
$\sqrt{\left<F^2\right>}~=~4.05\cdot10^{-5}~\mbox{cm}/\mbox{s}^{2}$, respectively (see also \Eq \ref{DCOEFF}).
In that case the correction factor $\gamma$ evaluates to $\gamma= 0.0866$.
 
The growth rates of the root mean square values for $\Delta a$, $\Delta \varpi$ and $\Delta e$ in 200 simulations are plotted in \Fig \ref{fig:singleplanet}. 
We also plot the analytic predictions (\Eqs \ref{eq:growtha}, \ref{eq:growthw} and \ref{eq:growthe}) as solid lines. 
The numerical simulations and the analytic model are in excellent agreement and scale as expected. 
We have performed simulations for a variety of diffusion coefficients and get results that are fully consistent in all cases. 

We also plot the analytic growth rates coming from \Eqs \ref{eq:growthw} and \ref{eq:growthe}
while setting $\gamma=1$, which corresponds to uncorrelated noise.
The resulting curves (dashed lines) do not agree with simulation results.
This shows clearly that correlation effects are important, and have to be taken into account. 
The model presented by \cite{Adams08} makes use of such an uncorrelated force (random kicks), 
leading to an overestimated diffusion coefficient by almost a factor $\sim 16$.
Note that only orbital parameters involving the geometry of the orbit are affected,
i.e., not the semi-major axis.

In \Fig~\ref{fig:singleplanetlambda}, we plot the root mean square of the difference in longitude. 
One can clearly seen that  $\left(\Delta\lambda\right)^2 $ does grow like $\sim t^3$, as expected from \Eq \ref{eq:growthlambda}.

\subsection{Illustration of the modes of libration in a two planet system}
\begin{figure}[tbp]
\centering
\includegraphics[width=0.8\columnwidth]{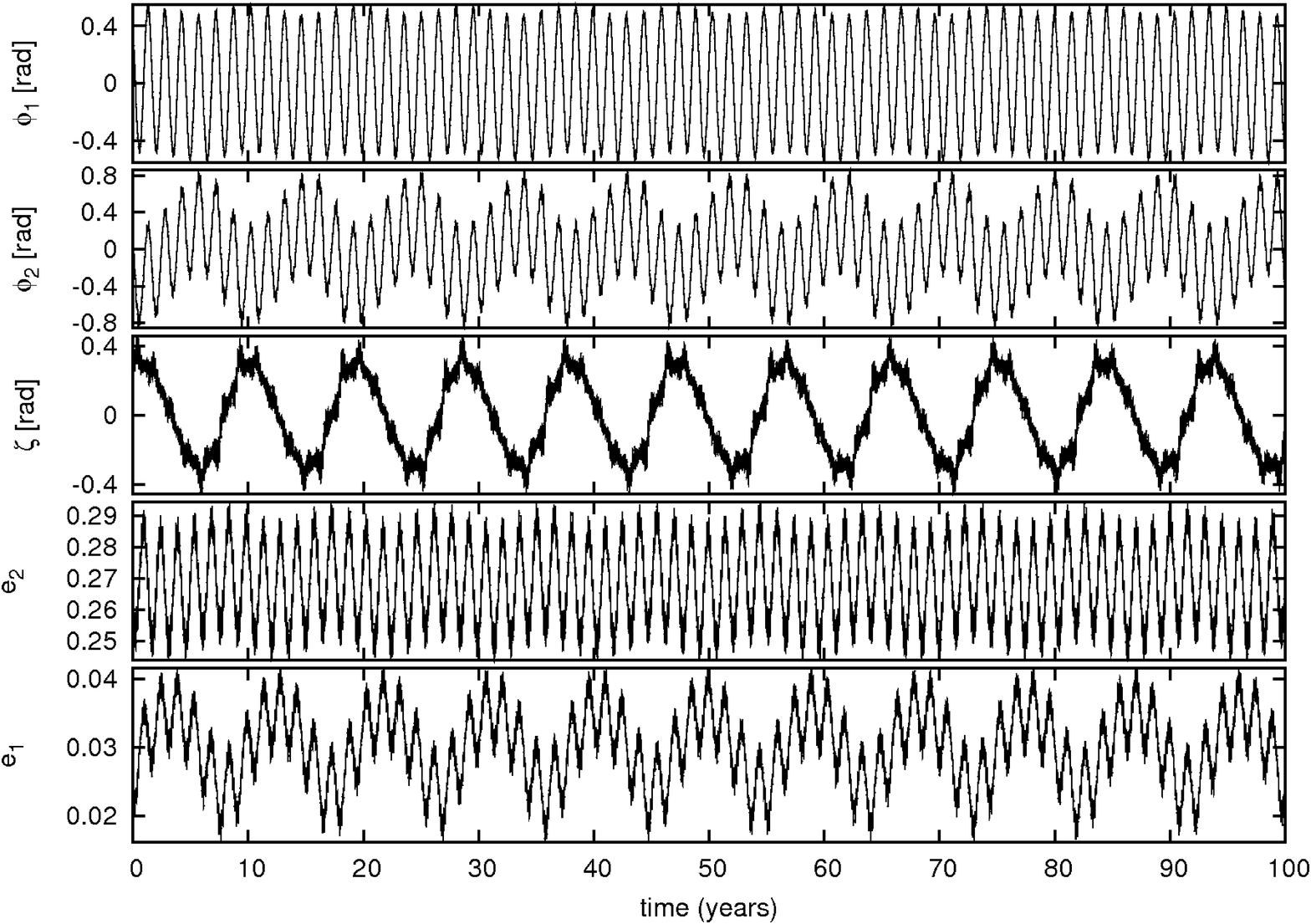}
\caption{Time evolution of the resonant angles and the eccentricities in the system GJ876 without turbulent forcing. 
The dominance of the fast mode with period $\sim 1.4$~years in the oscillations of $\phi_1$, 
and the dominance of the slow mode with period $\sim 10$~years in $\zeta$ can be clearly seen.\label{fig:twoplanets}}
\end{figure}
We now go on to consider two planet systems.
As an illustrative example we discuss the GJ876 system (see table \ref{table:initorb}). 
We begin by considering the evolution of the system without stochastic forcing in order to characterise the modes of libration of the resonant angles and other orbital parameters.
In particular, we identify the fast and slow modes discussed in section \ref{FSmodes}.

The time evolution of the resonant angles and the eccentricities of the unperturbed system is plotted in \Fig \ref{fig:twoplanets}.
Clearly visible are the slow and fast oscillation modes. 
The fast mode, which is seen to have a period of about 1.4 years, dominates the librations of $e_2$ and $\phi_1$ while also being present in those of $\phi_2$. 
On the other hand the slow mode, which is seen to have a period of about 10 years, dominates the librations of $\zeta$ while also being present in those of $e_1$ and $\phi_2.$ 

We emphasise the fact that the eccentricities of the two planets participate in the librations, and are therefore not constant. 
In particular, the eccentricity of GJ876~b oscillates around a mean value of $0.03$ with an amplitude $\Delta e \approx 0.01 - 0.02$. 
This behaviour involving the attainment of smaller values of the eccentricity has important consequences for stochastic 
evolution as discussed above (section \ref{GRLIB}) and also below. 

We remark that similar behaviour occurs for all the systems we have studied, these having a wide range of eccentricities and planet masses. 
The fast and slow mode periods scale with the planet masses as given by \Eqs \ref{fastp} and \ref{slowp} respectively.
These indicate, as confirmed by our simulations, that if both masses are reduced by a factor $\Lambda$ 
then the period of the fast mode scales as $\sqrt{\Lambda}$ and the period of the slow mode scales as $\Lambda.$

\begin{figure}[tbp]
\centering
\includegraphics[angle=270,width=1.0\columnwidth]{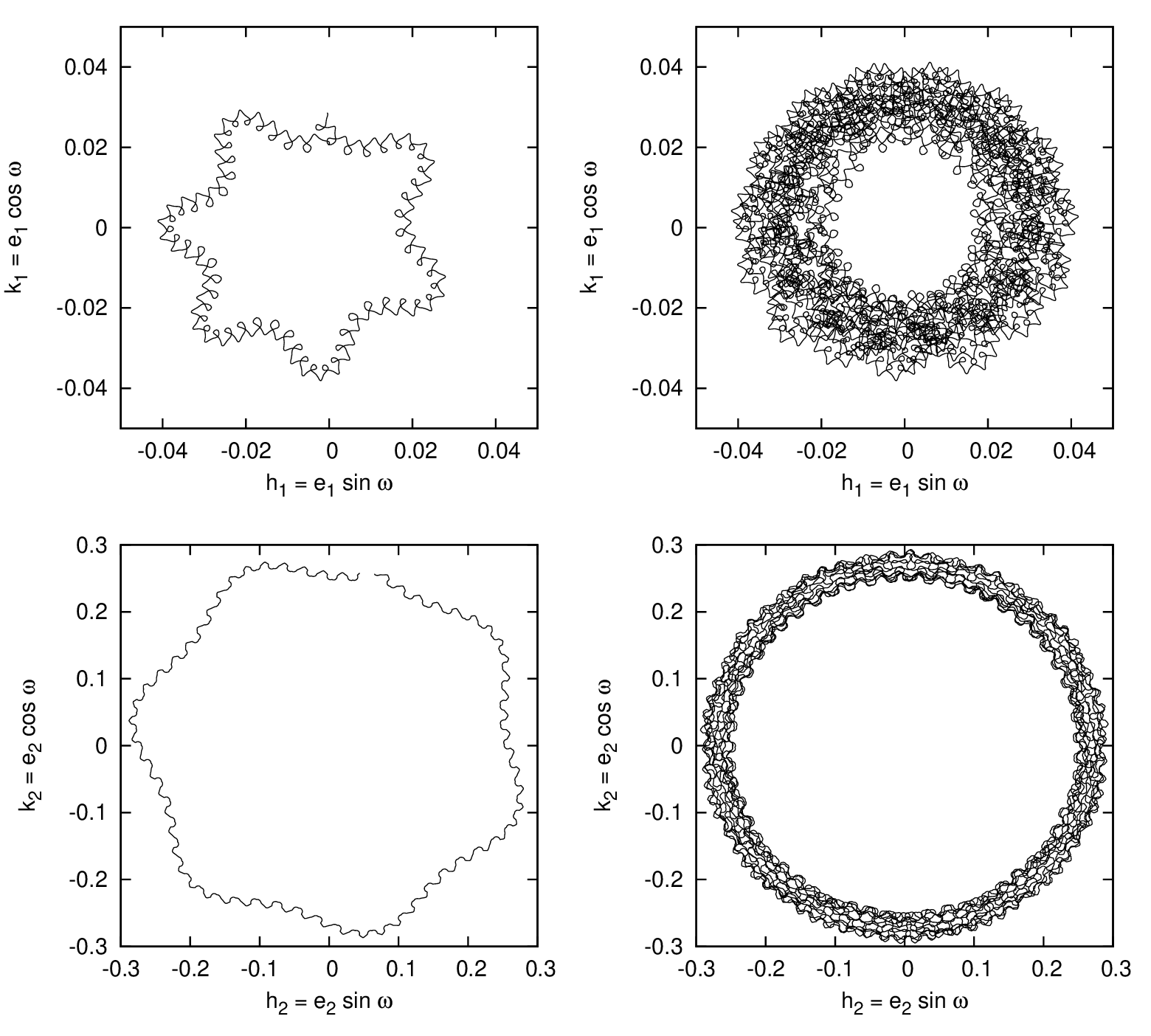}
\caption{ Librations in the GJ876 system, visualised using $h$ and $k$ variables. The initial conditions are the same as in figure \ref{fig:twoplanets}. The left plots show the evolution during one slow mode period, which is equivalent to about 6 fast mode periods. The right plots show the evolution over many libration periods.
\label{fig:randwalk:hk}}
\end{figure}

In \Fig \ref{fig:randwalk:hk} we plot the librations in $h$ and $k$ variables. The left plots show one period of the slow mode libration, corresponding to one complete circle around the origin. Super-imposed on that are slow mode librations with higher frequency. The frequency ratio is $\sim6$, thus the hexagon shape. The even higher frequency modes correspond to the orbital period itself.
The right plots show many libration periods.

\section{Two planets with stochastic forcing}
We now consider systems of two planets with stochastic forcing.
For simplicity we begin by applying additional forces only to the outer planet.
We have found that adding the same form of forcing to the inner planet tends to speed up the evolution by approximately a factor of two without changing qualitative details.
For illustrative purposes we again start with the GJ876 system and consider two diffusion coefficients 
$D = 8.2~\cdot~10^{-5} \text{cm}^2/\text{s}^{3}$ and $D = 8.2~\cdot~10^{-4} \text{cm}^2/\text{s}^{3}$. 
In all of our simulations a constant value of $D=2\langle F_i\rangle ^2\tau_c$ is used by
maintaining both parameters $\left< F_i\right>$ and $\tau_c$ constant.
For the cases considered here, there is little orbital migration so that this effect can be ignored.

\begin{figure}[p]
\centering
\includegraphics[width=0.9\columnwidth]{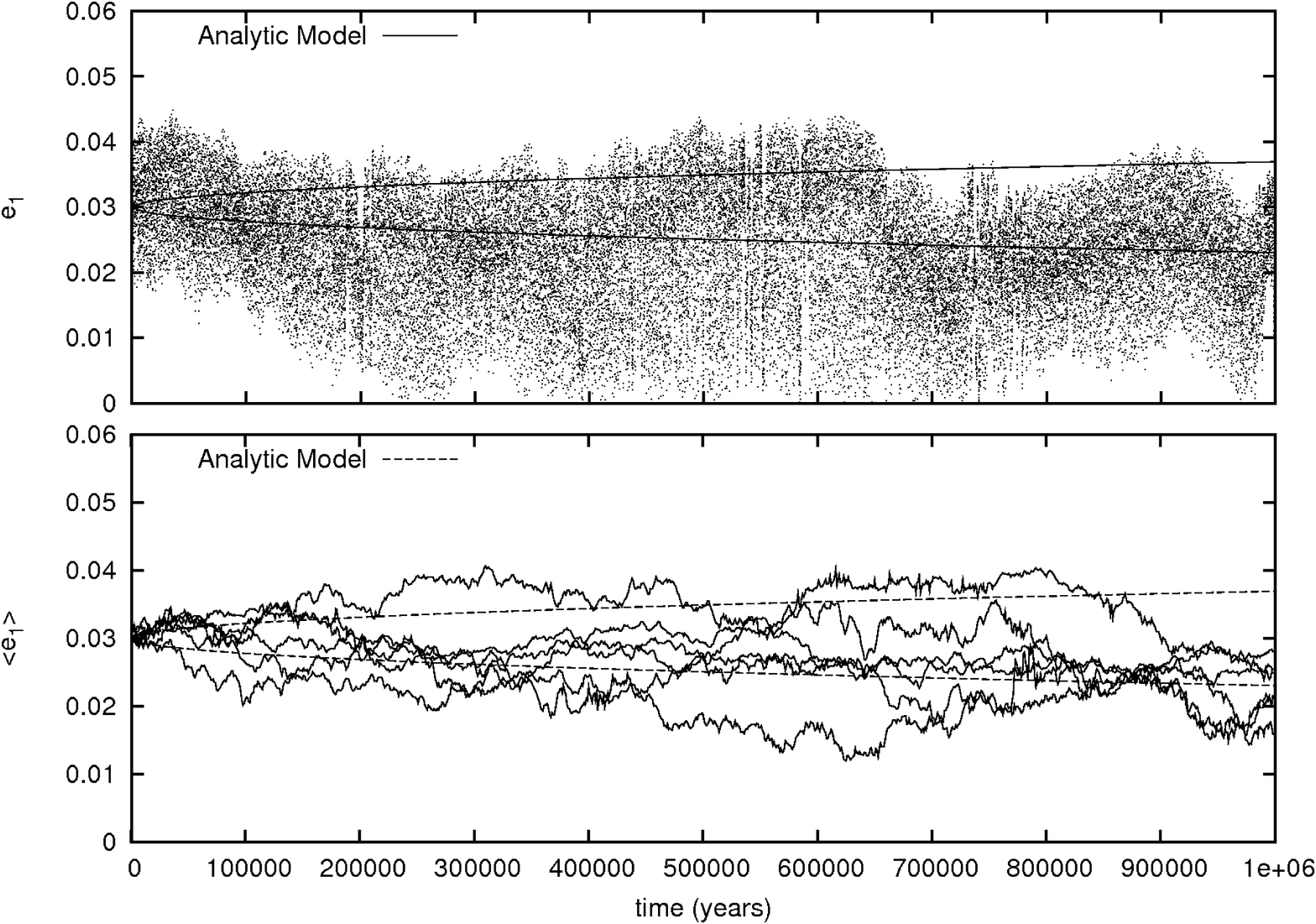}
\includegraphics[width=0.9\columnwidth]{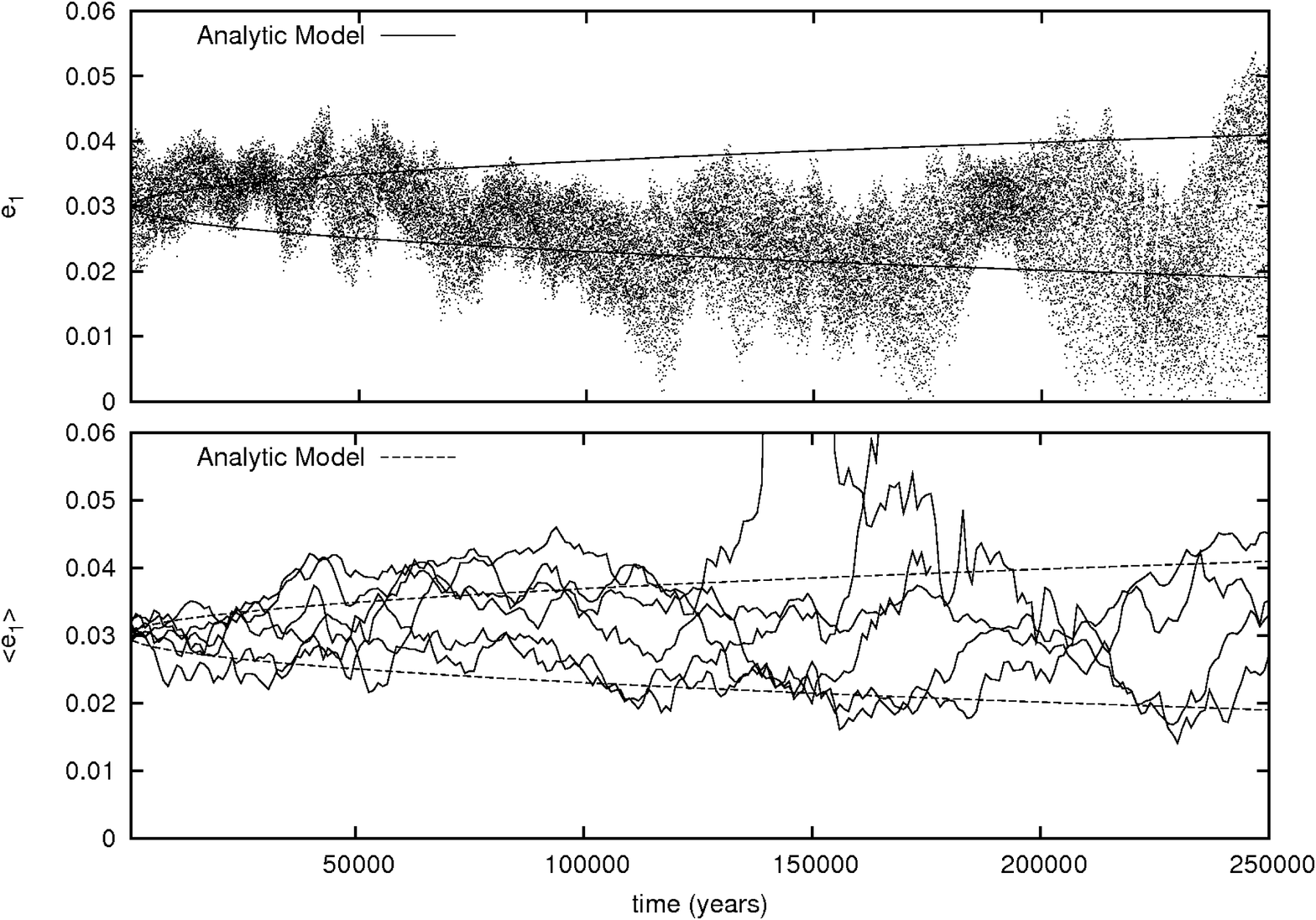}
\caption{Time evolution of the eccentricities in the GJ876 system with turbulent forcing included. 
The diffusion coefficient is $D = 8.2\cdot10^{-5} \text{cm}^2/\text{s}^{3}$ for the two upper plots 
and $D = 8.2\cdot10^{-4} \text{cm}^2/\text{s}^{3}$ for the two lower plots. 
The first (uppermost) and the third plots each show a single run. 
The second and fourth (lowermost) plots show the time averaged eccentricities for six different runs. 
The averaging interval is 1000~years, corresponding to $\sim 100$ slow mode periods.
\label{fig:twoplanets_turb}}
\end{figure}

The time evolution of the eccentricities is plotted for two runs in the first and third plot in \Fig \ref{fig:twoplanets_turb}. 
We see fast oscillations superimposed on a random walk. 
The amplitude of the oscillations, as well as the mean value, change with time. 

Our simple analytic model assumes slowly changing background eccentricities and semi-major axes. Accordingly, it does not incorporate the 
oscillations of the eccentricity due to the resonant interaction of the planets. 
In order to make a comparison, we perform a time average over many periods to get smoothed quantities whose behaviour we can compare with that expected from 
\Eqs \ref{eq:growtha}~-~\ref{eq:growthw}. 
When this procedure is followed (see second and forth plot in \Fig \ref{fig:twoplanets_turb}), the evolution is in reasonable accord with that expected from the analytic model provided that 
allowance is made for the importance of small values of $e_1$ in determining the growth of the libration amplitude of the angle between 
the apsidal lines of the two planets (see \Eq~\ref{eq:growthw}).
The presence of this feature results in the behaviour of the libration amplitude being more complex than that implied 
by a process governed by a simple diffusive random walk. This is discussed in the following sections.

\subsection{Disruption of a resonance in stages}
Systems with mean motion commensurabilities can be in many different configurations. 
Here we describe the important evolutionary stages as they appear in a stochastically forced system 
starting from a situation in which all the resonant angles show small amplitude libration. 
Observations of GJ876 suggest that the system is currently in such a state with the ratio of the orbital periods $P_1/P_2$ oscillating about a mean value of $2.$

\subsubsection {Attainment of circulation of the angle between the apsidal lines}
When the initial eccentricity of the outer planet, $e_1$, is small, the excitation of the libration amplitudes of the 
resonant angles readily brings about a situation where $e_1$ attains even smaller values.
This then causes the periastron difference $\zeta$ to undergo large oscillations and eventually reach circulation (see \Eq \ref{eq:growthw}).
Should stochastic forces cause $e_1$ to reach zero, on account of the coordinate singularity, $\varpi_1$ becomes undefined.
Note that this occurs without a large physical perturbation to the system.
Thus, the occurrence of this phenomenon does not imply the system ceases to be in a commensurability.

In this context we note that the eccentricity of GJ876~b initially is such that $e_1 \sim 0.03$ with values $e_1\sim 0.01$ often being attained during libration cycles.
Thus only a small change may cause the above situation to occur.
In all cases that we have considered, we find that $\zeta$ enters circulation prior to the fast angle $\phi_1$,
which may remain librating until that too is driven into circulation.

\subsubsection{Attainment of circulation of the fast angle}
Both before and after $\zeta$ enters circulation, stochastic forces increase the libration amplitude of the fast mode $\phi_1$.
This mode dominates both the librations of the resonant angle $\phi_1$ and the semi-major axes.
Eventually $\phi_1$ starts circulation.  Shortly afterwards, commensurability is lost and $P_1/P_2$ starts to undergo a random walk with a centre that drifts away from 2.
Note that it is possible for some realisations to re-enter commensurability. 
For systems with the masses of the observed GJ876 system, the most likely outcome is a scattering event that causes complete disruption of the system. 

\begin{figure}[tbp]
\centering
\includegraphics[angle=270,width=1.0\columnwidth]{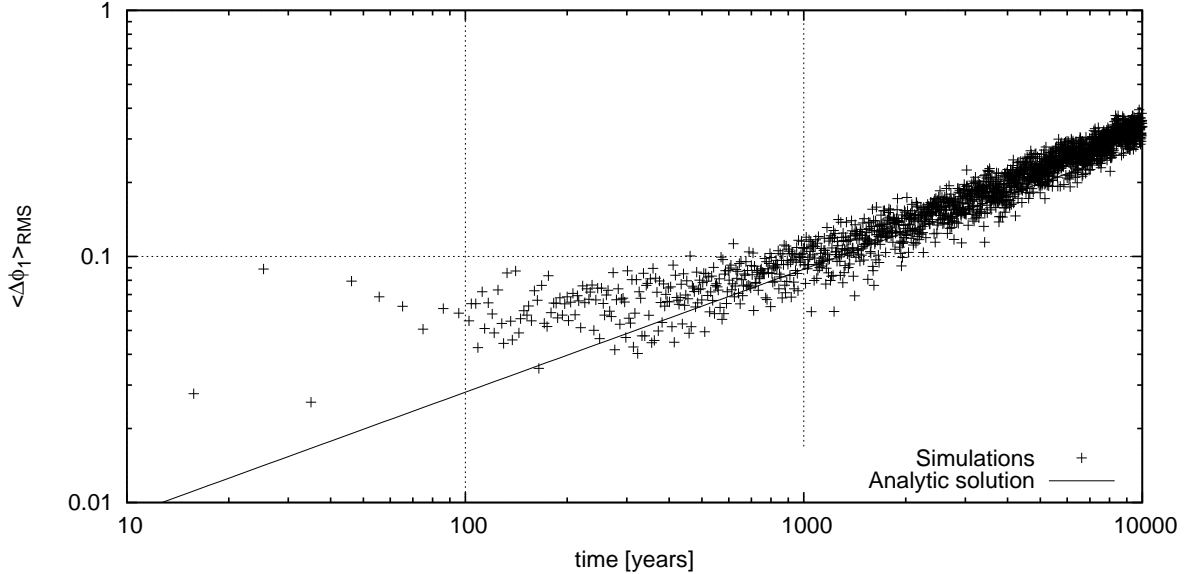}
\caption{Growth of the difference in the resonance angle $\phi_1$ as a function of time. The initial orbital parameters are those of GJ876 (see table \ref{table:initorb}). Here, the diffusion coefficient is  $D = 8.2\cdot 10^{-3} \text{cm}^2/\text{s}^{3}$. The solid line corresponds to the analytic prediction, given by \Eq \ref{eq:growthf}.
\label{fig:randwalk:twoplanet_rms}}
\end{figure}

Because the eccentricities do not play an important role in the attainment of circulation of the fast angle,
we can use \Eq \ref{eq:growthf} to calculate the growth rate. The root mean square values of $\Delta \phi_1$ as a function of time, taken from both, analytical and numerical results, are plotted in \Fig \ref{fig:randwalk:twoplanet_rms}. Here, we use $D = 8.2~\cdot~10^{-3} \text{cm}^2/\text{s}^{3}$. 
Note that as mentioned above, $\zeta$ is circulating first (in the presented case this happens after $\sim1000$~years). However, this does not influence the growth of $\Delta \phi_1$.

\subsection{A numerical illustration}
\begin{figure}[p]
\centering
\includegraphics[width=0.8\columnwidth]{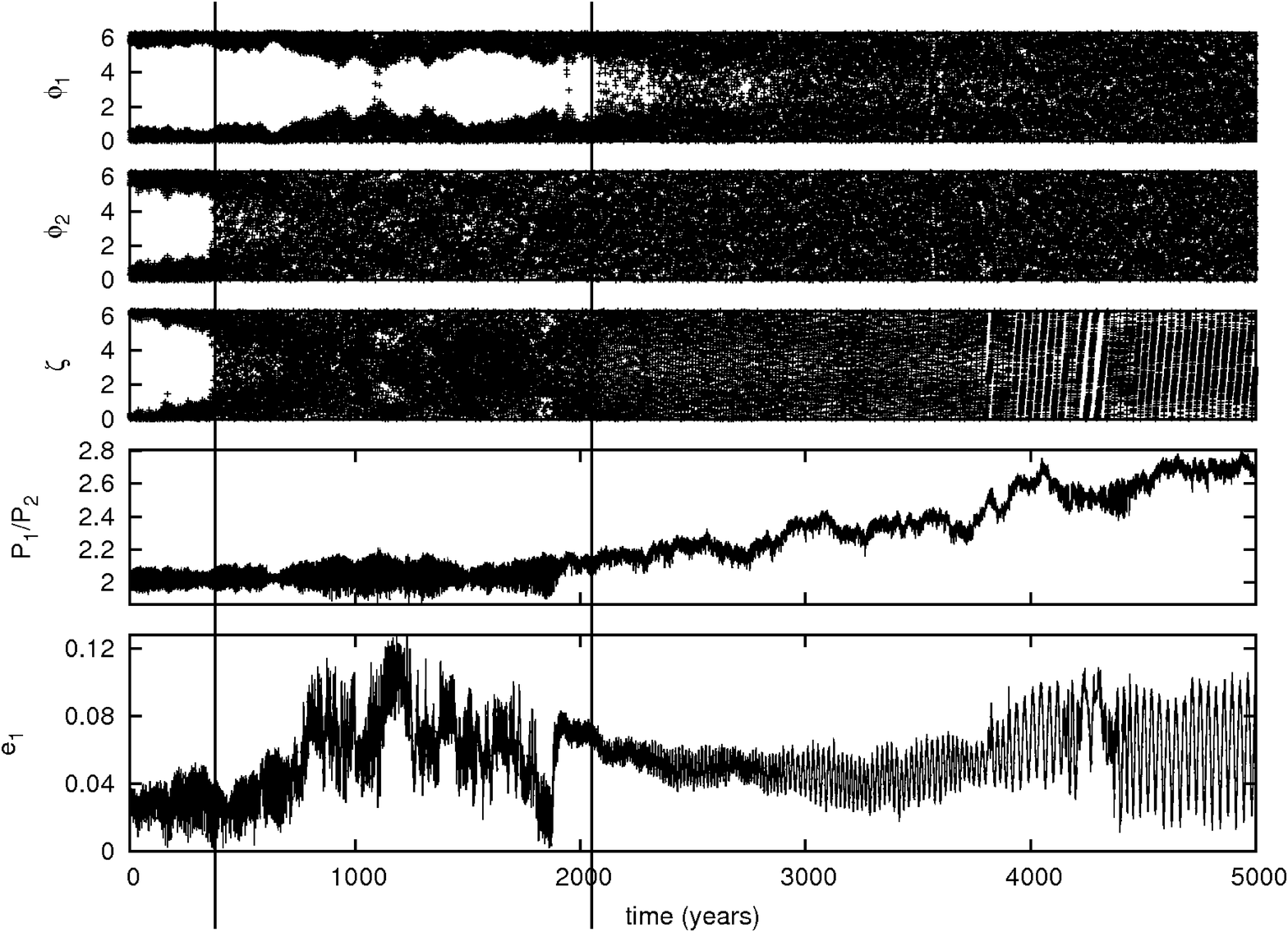}
\includegraphics[width=0.8\columnwidth]{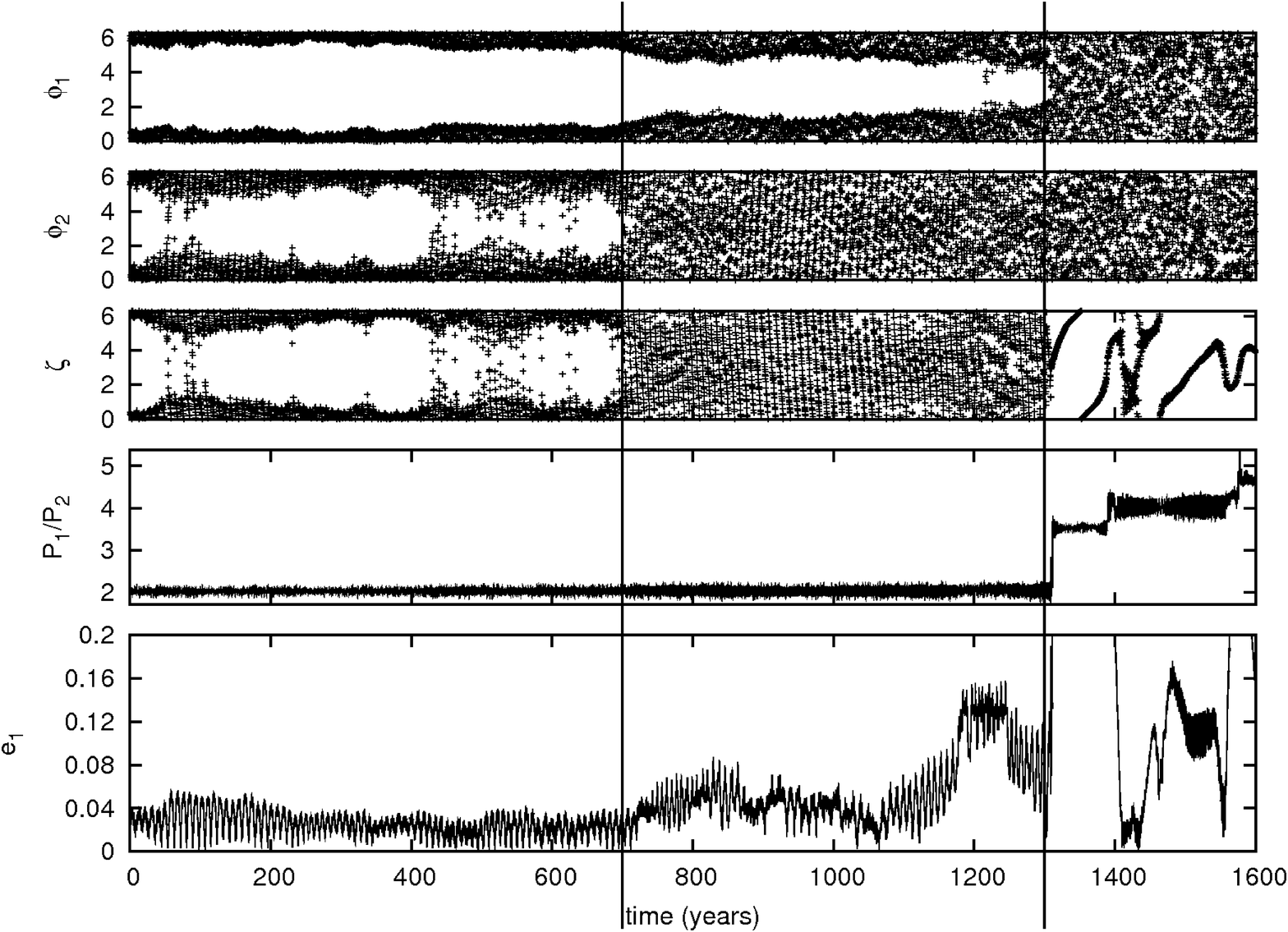}
\caption{Time evolution of the resonant angles, the period ratio $P_2/P_1$ and the eccentricity, $e_1$, 
in the GJ876 system with stochastic forcing corresponding to $D = 0.42 \text{cm}^2/\text{s}^{3}$. 
The vertical lines indicate when the angles enter circulation for a prolonged period.
The realisation illustrated in the lower panel scatters shortly after $\phi_1$ goes into circularisation. 
\label{fig:twoplanets_turb_break}}
\end{figure}

In order to illustrate the evolutionary sequence described above we plot results for 
two realisations of the evolution of the GJ876 system in figure \ref{fig:twoplanets_turb_break}.
For these runs we adopted the diffusion coefficient $D = 0.42 \text{cm}^2/\text{s}^{3}$. 
Note that increasing $D$ only decreases the evolutionary time which has been found, both analytically and numerically to be~$\propto~1/D$ (see below).

The times at which the transition from libration to circulation occurs for both the slow and fast angles, 
are indicated by the vertical lines in figure \ref{fig:twoplanets_turb_break}.

Several of the features discussed above and in section \ref{GRLIB} can be seen in figure \ref{fig:twoplanets_turb_break}.
In particular the tendency for the occurrence of very small values of $e_1$ to be associated with transitions to circulation
of $\zeta$ can be seen for the realisation plotted in the lower panels at around $t=80$~years and $t= 500$~years.
Such episodes always seem to occur when the libration amplitude of the fast angle $\phi_1$ is relatively boosted, 
indicating that this plays a role in boosting the slow angle. 
If the duration for which $e_1$ attains small values is small and $\phi_1$ recovers small amplitude librations, the slow angle returns to circulation.
Thus, the attainment of long period circulation for the slow angle is related to the diffusion time for $\phi_1$.
We also see from figure \ref{fig:twoplanets_turb_break} that the angle $\phi_2$, which has a large contribution from the slow mode, 
behaves in the same way as $\zeta $ as far as libration/circularisation is concerned.
We have verified by considering the results from the simulations of GJ876 LM HE, which started with a larger value of $e_1,$ that, as expected, 
the attainment of circulation of the slow angle takes relatively longer in this case, the time approaching more closely the time when $\phi_1$ attains circulation. 
Also as expected, the time when $\phi_1$ attains circulation is not affected by the change in $e_1$.

\subsection{Dependence on the diffusion coefficient}
We now consider the stability of the systems listed in table \ref{table:initorb} as a function of $D$.
These systems have a variety of masses and orbital eccentricities.
In particular, in view of the complex interaction between the resonant angles discussed above, we wish to investigate 
whether the mean amplitude growth at a given time is indeed proportional to $D$. 
As also mentioned above, the value of the diffusion parameter $D$ that should be adopted, is very uncertain. 
We have therefore considered values of $D$ ranging over five orders of magnitude. 
The correlation time $\tau_c$ is always taken to be given by $\tau_c = 0.5\Omega^{-1}$ while the RMS value of the force is changed. 
In order to determine the \textit{lifetime} of a resonant angle, we monitor whether it is librating or circulating. 
Numerically, libration is defined to cease when the angle is first seen to reach absolute values larger than $2$. 
We note that the angle can in general regain small values afterwards.
However, this is a transient effect and changes the lifetime by no more than a factor of $\sim 2$ in all our simulations.

In this context we consider the fast angle $\phi_1$ and the slow angle $\zeta$. As we saw above, 
the latter can be replaced by $\phi_2$ which exhibits the same behaviour. 
Because $\phi_1$ is the last to start circulating, the resonance is defined to be broken at that point.

\begin{figure}[p]
\centering
\includegraphics[width=0.8\columnwidth]{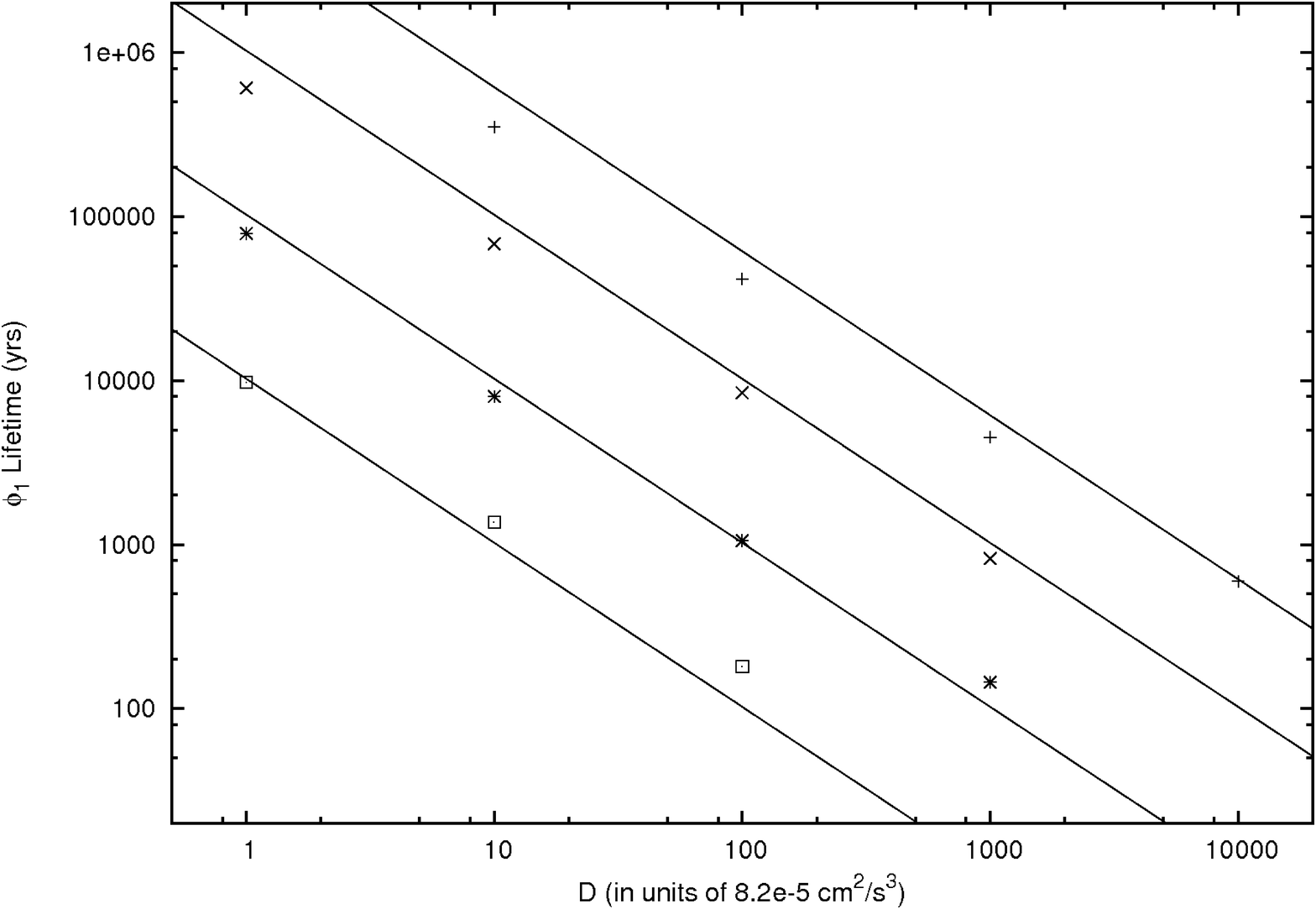}
\includegraphics[width=0.8\columnwidth]{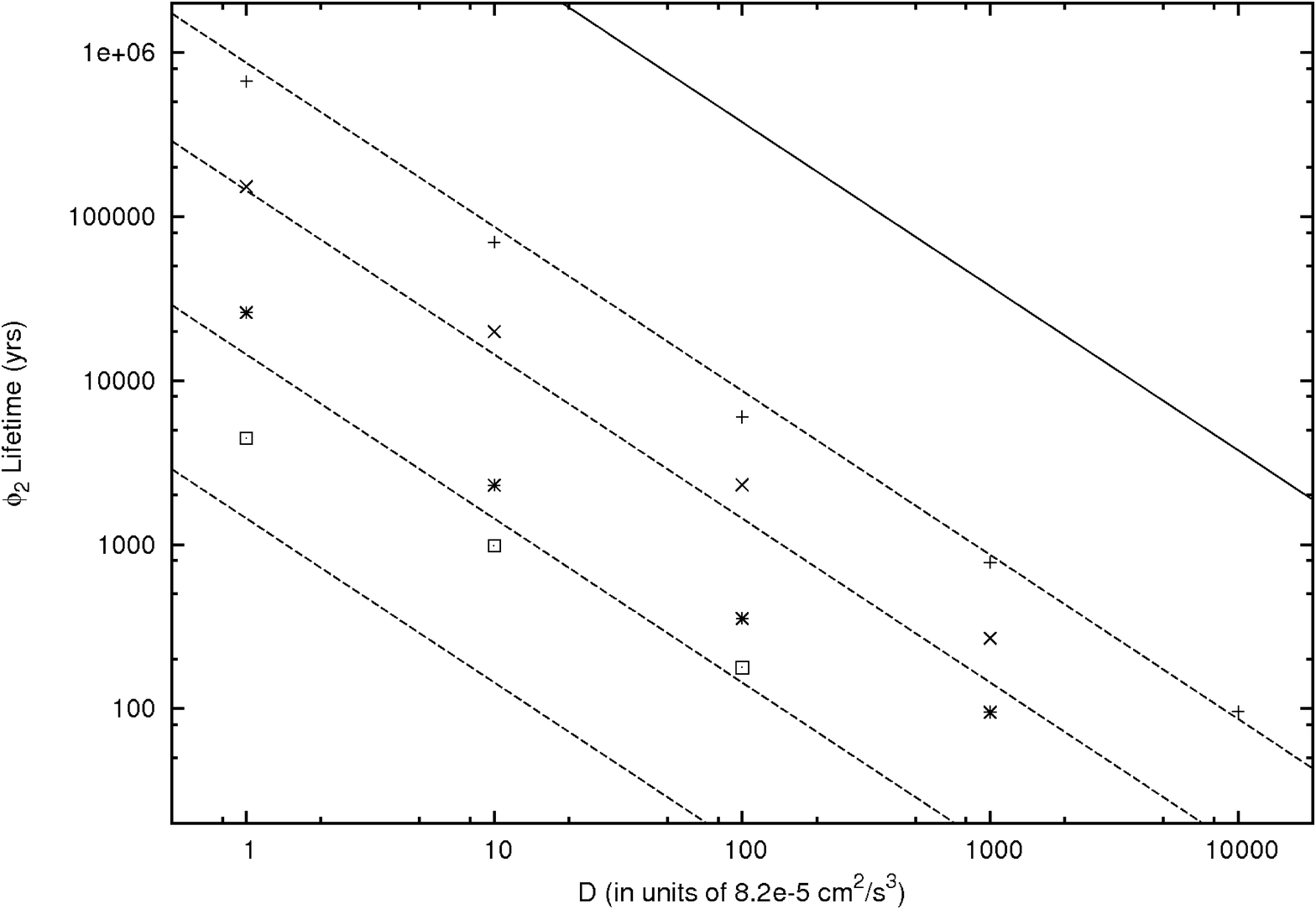}
\caption{Average time until circulation of the resonant angles $\phi_1$ (top) and $\phi_2$ (bottom) in the systems GJ876
(indicated with $+$) , GJ876LM (indicated with $\times$), GJ876SE (indicated with $\star$) and GJ876E (indicated with $\Box$)
as a function of the stochastic forcing diffusion coefficient $D$.
In the case of the upper panel, the analytic curves explained in the text, are from top to bottom for the GJ876, GJ876LM, GJ876SE and GJ876E systems, respectively. 
In the lower panel the analytic curves, as explained in the text, are from top to bottom for the GJ876 (solid curve and top dashed curve), 
GJ876LM, GJ876SE and GJ876E systems, respectively. 
\label{fig:lifetime1}}
\end{figure}

Equations \ref{eq:growthf} and \ref{eq:growths} estimate the spreading of the resonant angles as a function of time. 
We plot both the numerical and analytical results in figure \ref{fig:lifetime1}. 
To remove statistical fluctuations and obtain a mean spreading time, the numerical values for a particular value of $D$ were obtained by averaging over $60$ realisations.

From \Fig \ref{fig:lifetime1} it is apparent that the evolutionary times scale is $\propto 1/D$ for $D$ varying by many orders of magnitude. 
The analytic estimates for the libration survival times of $\phi_1$, dominated by the fast mode, obtained using \Eq \ref{eq:growthf}
adopting the initial orbital elements and with the fast libration frequency determined from the simulations, are plotted in the upper panel of figure \ref{fig:lifetime1}. 
These are in good agreement with the numerical results.
However, the estimate for $\phi_2$ based on \Eq~\ref{eq:growths} using the initial value of $e_1$ overestimates the lifetime by a factor of 
at least $\sim 40$ (see solid line in the lower plot of \Fig \ref{fig:lifetime1}).
Furthermore \Eq \ref{eq:growths} has no dependence on planet mass which is in clear conflict with the numerical results.
As discussed above, this is due to the temporary attainment of small eccentricities. 
This causes the disruption of the libration of $\zeta$ earlier than predicted assuming $e_1$ being constant. 
In fact this disruption occurs at times that can be up to $10$ times shorter than those required to disrupt $\phi_1$. 
We estimate the lifetime of librations of $\phi_2$ by calculating the time $\phi_1$ needs to reach values close to $1$ 
(see dashed lines in the lower plot of \Fig \ref{fig:lifetime1}).
As explained above in section \ref{GRLIB}, large amplitude variations of $\phi_1$ are expected to couple to and excite the slow mode. 
Variations induced in the eccentricity $e_1$ allow $e_1$ to reach zero and we lose libration of $\phi_2$. 
These simple estimates are in good agreement with the numerical results and accordingly support the idea that $\phi_1$ and $\phi_2$ are 
indeed coupled in the non-linear regime.
For low mass planets this simple analytic prediction underestimates the lifetime of $\phi_2$ by a factor of $\sim 2$, 
suggesting that the coupling between the two modes depends slightly on the planet masses.

To confirm our understanding of these processes, we have repeated the calculations with systems that are in the same state as the system discussed above but have higher 
eccentricities (see GJ876~LM~HE and GJ876~SE~HE in table \ref{table:initorb}). 
The lifetime of the librating state of $\phi_1$ is unchanged, as this does not depend significantly on the eccentricity. 
However, the lifetime of the librating state of $\phi_2$ is a factor of $2-3$ longer. 
This is due to the fact that a larger excitation of $\phi_2$ is needed to make $e_1$ reach small values in these simulations. 

The random walk description breaks down completely when the anticipated disruption time becomes shorter than the libration period. 
This is expected to happen for small enough masses as can be verified from \Fig \ref{fig:lifetime1}.
It is because the disruption time decreases linearly with the planet masses while the libration period increases as the square root of the planet masses.
Then we cannot average over many libration periods.
This situation is apparent in \Fig \ref{fig:lifetime1} for very short disruption times of the order 100-1000~years where survival times cease to vary linearly with $1/D$.

\section{Formation of HD128311}
Having understood the effects of stochastic forces on resonances, we
now go on to discuss the application of these ideas to the formation of the orbital configuration of the planetary system HD128311.
This is (with 99\% confidence) in a 2:1 mean motion resonance, with the angle $\phi_1$ librating and the angle $\zeta$ circulating 
so that there is no apsidal co-rotation \citep{Vogt2005}. 
However, the orbital parameters are not well constrained. 
The original Keplerian fit by \cite{Vogt2005} is such that the planets undergo a close encounter after only 2000 years. 
The values in table \ref{table:initorb} have been obtained from a fit to the observational data that includes interactions between the planets. 
The values quoted have large error bars. 
For example the eccentricities $e_1$ and $e_2$ have an uncertainty of 33\% and 21\%, respectively. 
Although the best fit doesn't manifest apsidal co-rotation, the system could undergo large amplitude librations and still be stable.

According to \cite{LeePeale2002} the planets should have apsidal co-rotation if the commensurability was formed by the two planets undergoing sufficiently slow convergent 
inward migration, and if they then both migrated inwards significantly while maintaining the commensurability \citep[see also][]{SnellgrovePapaloizouNelson01}.
Accordingly \cite{SandorKley06} suggested a possible formation scenario with inward migration as described above, but with an additional perturbing event, 
such as a close encounter with an additional third planet occurring after the halting of the inward migration. 
This perturbation is needed to alter the behaviour of $\zeta$, so that it undergoes circulation rather than libration and thus producing orbital 
parameters similar to the observed ones. 

We showed above that stochastic forcing possibly resulting from turbulence driven by the MRI readily produces systems with commensurabilities 
without apsidal co-rotation if the eccentricities are not too large. 
This suggests that a scenario that forms the commensurability through disc-induced inward convergent migration might readily produce commensurable systems
without apsidal co-rotation if stochastic forcing is included.
Such scenarios are investigated in this section.

We present simulations of the formation of HD128311 that include migration and stochastic forcing due to turbulence 
but do not invoke artificial perturbation events involving additional planets. 
We find that model systems with orbital parameters resembling the observed ones are readily produced and are better matches than those provided by \cite{SandorKley06}.
The planets in this system are of the order of several Jupiter masses and the eccentricity of one planet can get very small during one libration period. 
The observed orbital parameters are given in table \ref{table:initorb} and their time evolution (due to planet-planet interactions) is plotted on the right hand side of 
\Figs \ref{fig:hd128311}, \ref{fig:hd128311_2} and \ref{fig:hd128311_3}. 
The mass of the star is $0.8~{\mathrm M}_{\odot}$.

\subsection{Migration forces}
We incorporate the non-conservative forces exerted by a proto-stellar disc that lead to inward migration by following the approach outlined in appendix \ref{app:leepeale}.
The migration process is characterised by migration and eccentricity damping timescales, $\tau_a = {a}/{\dot a}$ and $\tau_e = {e}/{\dot e}$, respectively. 
This procedure is now widely used \citep[e.g.][]{SandorKley06, Crida08}. 
We also use the ratio of the above timescales, $K=\left|\tau_a\right|/\tau_e$, as defined in chapter \ref{ch:threetwo}. 
This ratio determines the eccentricities in the state of self-similar inward migration of the commensurable system that is attained after large times. 

In this work we allow the planets to form a commensurability through convergent inward migration with stochastic forcing included. 
The disc is then removed through having its surface density reduced to zero on a 2000~year timescale. 
This reduces both the migration forces and the stochastic forces simultaneously. 
This procedure is very similar to that adopted by the above authors but we have included stochastic forcing and removed the disc on longer, more realistic, time-scales.

The stochastic forces need to have the right balance with respect to the migration rate.
We have found that inward migration imparts stability to the resonant system.
If the migration rate is too fast relative to the stochastic forcing the migration keeps down the libration amplitudes and we do not get circulation. 
On the other hand large eccentricity damping favours broken apsidal co-rotation. 
This might look counter intuitive at first sight, but remember that the diffusion of $\zeta$ depends on $1/e^2_1$ and $1/e^2_2$. 

Due to the stochastic nature of the problem, it is hard to present a continuum of solutions so we restrict ourselves to the discussion of three representative examples. 
However, we comment that we are able to obtain similar end states for a wide range of migration parameters. 

As above, the parameters $\left< F_i\right>$ and $\tau_c$ are kept constant, so maintaining $D$ constant, 
for the duration of the simulation, with $\tau_c$ being determined for the initial location of the outer planet. 
As discussed above, these values may scale with the radius of the planet and we thus expect them to change during migration. 
However, the semi-major axis of the outer planet changes only by $\sim 30$~\% during the simulation. Consequently, we ignore this effect.

\subsection{Model 1}
\begin{figure}[tbp]
\centering
\includegraphics[width=0.8\columnwidth]{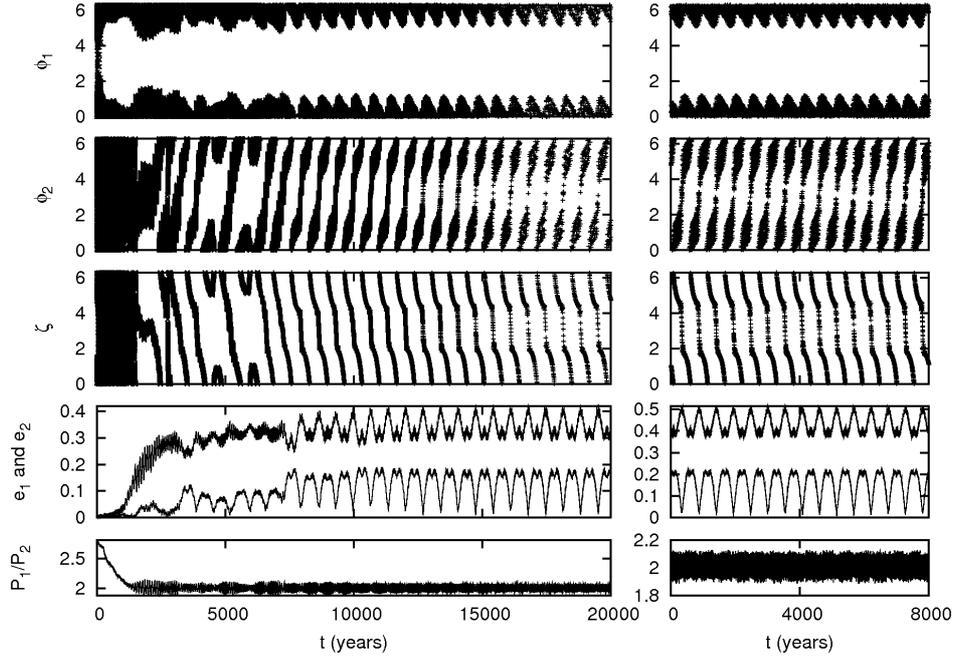}
\caption{The plots on the left hand side show a possible formation scenario for the system HD128311 including turbulence and migration. 
We plot the observed system on the right hand side as a comparison (see table~\ref{table:initorb} and text). 
The plots show, from top to bottom, the resonant angles $\phi_1, \phi_2, \zeta$, the eccentricities $e_1, e_2$ and 
the period ratio $P_1/P_2$.
Resonance capturing occurs after 2000~years.
\label{fig:hd128311}}
\end{figure}

The planets are initially on circular orbits at radii $a_1=4$~AU and $a_2=2$~AU.
Stochastic forcing is applied to the outer planet only, with a diffusion coefficient of $D=6.4\cdot 10^{-3} \text{cm}^2/\text{s}^3$.
Note that although this value is 64~times larger than the scaling given by \Eq \ref{eq:scalingd}, corresponding to a force that is 8~times larger than 
the simple estimate given in section \ref{sec:scaling}, corresponding to smaller reduction factors resulting from a gap or dead zone, 
it is too small for the angle $\phi_1$ to be brought to circulation during our runs. 
The results given above and summarised in \Fig \ref{fig:lifetime1}, and further tests, indicate that similar results would be obtained if $D$ is reduced, 
but on a longer time-scale $\propto 1/D,$ provided the migration rate is also ultimately reduced appropriately so that the system can survive 
for long enough to enable $\zeta$ to be driven into circulation.

The outer planet is made to migrate inwards on a timescale $\tau_{a_1}=8000$~years. 
The inner planet migrates slowly outwards on a timescale of $\tau_{a_2}=-20000$~years. 
This results in convergent migration.
For both planets we use an eccentricity to semi-major axis damping ratio of $K=8$. 
The resulting evolutionary timescales are significantly larger than those used by \cite{SandorKley06} and more easily justified by hydrodynamical simulations. 

The time evolution is shown in the left plot of \Fig \ref{fig:hd128311}. 
After resonance capturing all angles are either initially librating or on the border between libration and circulation.
The slow mode has a period of $\sim 700$~years, whereas the fast mode period is 20~times shorter.

It can be seen from \Fig \ref{fig:hd128311} and other similar figures below, that while migration continues, libration amplitudes 
tend to be controlled apart from when $e_1$ becomes either zero or very close to zero. 
Then, due to stochastic forcing, $\phi_2$ and $\zeta$ start circulating. 
Subsequently, libration is recovered over time intervals for which $e_1$ does not attain very small values but eventually additional stochastic forcing
together with the repeated attainment of small values for $e_1$ causes $\phi_2$ to remain circulating for the remainder of the simulation.
After $13000$~years, both the forces due to migration and turbulence are reduced smoothly on a timescale of $2000$~years. 
The result is a stable configuration that resembles the observed system very well. 

\subsection{Model 2}
\begin{figure}[tbp]
\centering
\includegraphics[width=0.8\columnwidth]{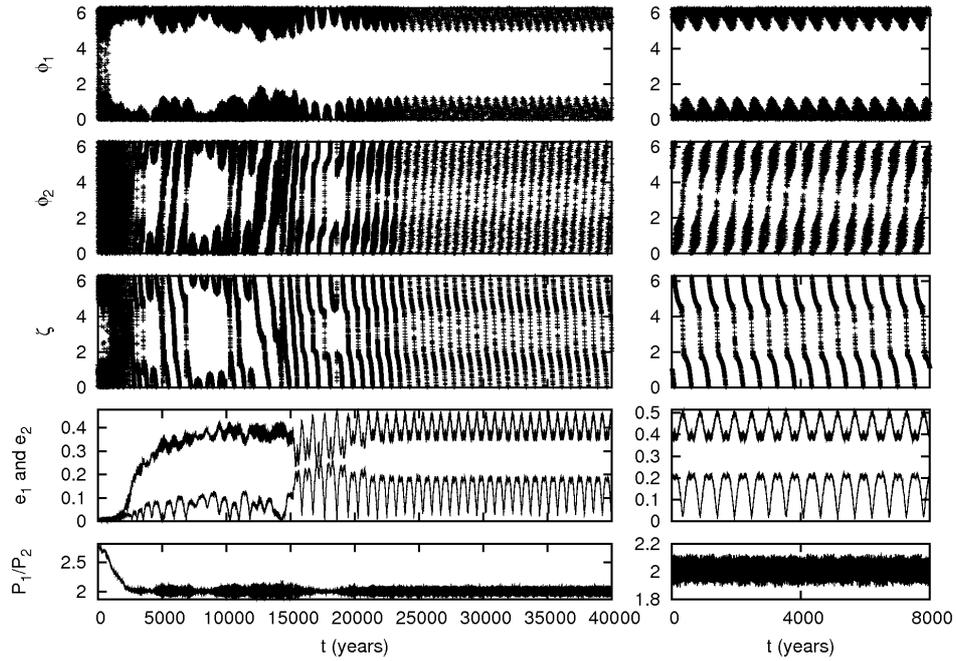}
\caption{The plots on the left hand side show another possible formation scenario for the system HD128311. 
Again, we plot the observed system on the right hand side as a comparison (see table~\ref{table:initorb} and text). 
The migration timescales in this run are $\tau_{a,1}=16000$, $\tau_{a,2}=-40000$ and $K=5.5$. 
Note that the eccentricities are larger compared to figure \ref{fig:hd128311} because $K$ is smaller. 
\label{fig:hd128311_2}}
\end{figure}

For this simulation, we lengthen the migration timescales by a factor of 2 to show that the results are generic.
We also decrease the value of $K$ to $5.5$. 
All other parameters are the same as for model 1. 
This results in larger eccentricities and the final state better resembled our representation of the observed system. 
However, it should be kept in mind that the eccentricities are not well constrained by the observations. 

The time evolution is plotted in \Fig \ref{fig:hd128311_2}.
The resonant angles librate immediately after the capture into resonance, as predicted for sufficiently slow migration.
However, in the same way as for model 1 described above, stochastic forces make $\phi_2$ circulate soon afterwards. 
Once the migration forces and stochastic forces are removed between $20000$ and $22000$~years, the system stays in a stable configuration with no apsidal co-rotation. 

\subsection{Model 3}
\begin{figure}[tbp]
\centering
\includegraphics[width=0.8\columnwidth]{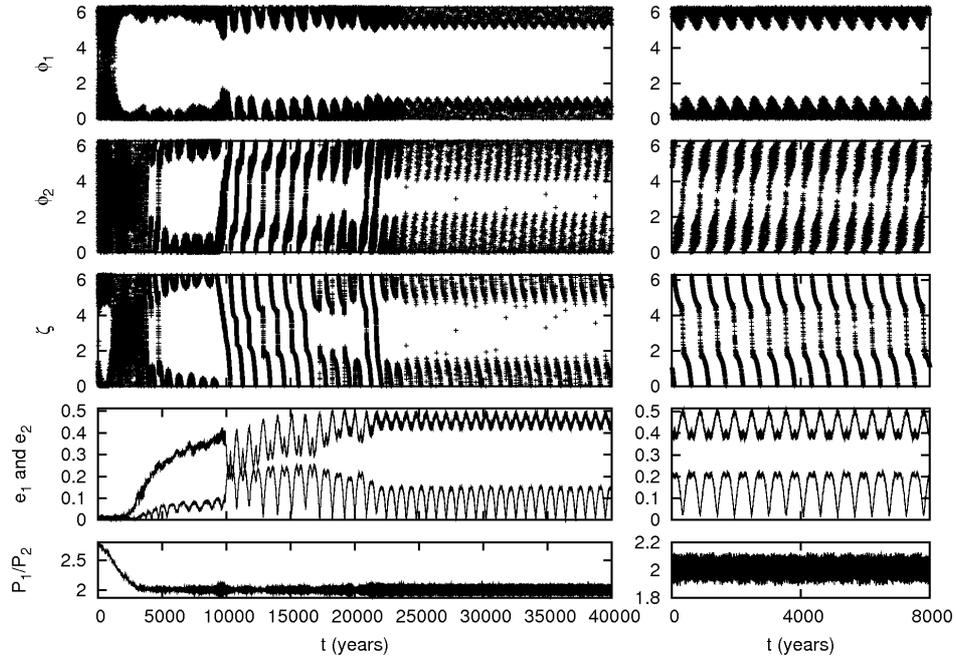}
\caption{ These plots show another possible formation scenario of HD128311. 
The damping parameters are the same as in figure \ref{fig:hd128311_2}.
\label{fig:hd128311_3}}
\end{figure}
All parameters are the same as for model 2. 
Accordingly, model 3 represents another statistical realisation of that case.
The time evolution is plotted in figure \ref{fig:hd128311_3}. 
The final state is very close to the boundary between libration and circulation of $\zeta$.
As discussed above, large uncertainties in the orbital parameters mean that the actual system could be in such a state. 

Due to stochastic forces, which may result from local turbulence, we are able to generate a broad spectrum of model systems.
Some of them undergo a strong scattering during the migration process. 
However, the surviving systems resemble the observed system very well. 
Although the orbital parameters are not well constrained yet, we showed that values in the right range are naturally produced by a turbulent disc.

\section{Conclusions} \label{sec:discussion}
In this chapter, we have presented a self-consistent analytical model of both a single planet, and two planets in a mean motion resonance, being 
subject to external stochastic forcing. 
The stochastic forces could result from MRI-driven turbulence within the proto-planetary disc but our treatment is equally applicable to any other source, for example from a 
gravitoturbulent disc such as Saturn's A-ring (see chapter~\ref{ch:moonlet}).

We considered the evolution of a stochastically forced two planetary system that is initially deep inside a MMR 
(i.e., the two independent resonant angles librate with small amplitude).
Stochastic forces cause libration amplitudes to increase in the mean with time until all resonant angles are eventually driven into 
circulation, at which point the commensurability is lost.
Often a strong scattering occurs soon afterwards for systems composed of planets in the Jovian mass range.

We isolated a fast libration mode which is associated with oscillations of the semi-major axes and a slow libration mode which is mostly associated with the
motion of the angle between the apsidal lines of the two planets.
These modes respond differently to stochastic forcing, the slow mode being more readily converted to circulation than the fast mode. 
This slow mode is sensitive to the attainment of small eccentricities, which cause rapid transitions between libration and circulation.
The amplitude of the fast mode grows more regularly in the mean, with the square of the libration amplitude in most cases increasing linearly with time
and being proportional to the diffusion coefficient $D$. 
Of course this discussion is simplified and there are limitations. 
For example if the total mass of the system is reduced, the disruption time eventually becomes comparable to the libration period. 
In that case the averaging process that we use in the derivation is no longer valid and the lifetime no longer scales as $1/D.$ 

The analytic model was compared to numerical simulations which incorporate stochastic forces. 
Those forces, parametrised by the mean square values of each component in cylindrical coordinates and the auto-correlation time, 
are applied in a continuous manner giving results that could be directly compared with the analytic model.
The simulations are in broad agreement with analytic predictions and we presented illustrative examples of the disruption process. 
We performed the simulations for a large range of diffusion parameters, planet masses and initial eccentricities to verify the scaling law for the
commensurability disruption time summarised hereafter.

To summarise our results, recalling that the slow angle is driven into circulation before the fast angle, so that the ultimate lifetime is determined
by the time taken for the fast angle to achieve circulation, we determine the lifetime, $t_f$, using \Eq \ref{eq:growthf},
setting $t=t_f,$ $(\Delta \phi_1)^2 = (\Delta \phi_1)^2_0 = 4,$ together with $\gamma_f =p =1$, we obtain
\begin{eqnarray}
t_f&=&\frac {a_1^2 \omega_{lf}^2(\Delta \phi_1)^2_0}{36D}.
\end{eqnarray}
We showed above that this gives good agreement with our numerical results.
Using $D=2\langle F_i^2\rangle \tau_c,$ we can express this result in terms of the relative magnitude of the stochastic forcing in the form
\begin{eqnarray}
t_f&=&2.4\times 10^{-4}\left(\frac {a_1^2n_1^4}{\langle F_i^2\rangle}\right) \left(\frac{(\Delta \phi_1)^2_0}{8n_1\tau_c}\right) \;\;\cdot
\left( \frac{8.5\omega_{lf}\sqrt{q_{GJ}}}{n_1\sqrt{q}}\right)^2\frac{q}{q_{GJ}} P_1,\label{RESEQ}
\end{eqnarray}
where $P_1$ is the orbital period of the outer planet.
Here the first quantity in brackets represents the ratio of the square of the central force per unit mass to the mean 
square stochastic force per unit mass acting on the outer planet.
The other quantities in brackets, scaled to the GJ876 system are expected to be unity, while the last factor $q/q_{GJ}$ is the 
ratio of the total mass ratio of the system to the same quantity for GJ876. 
Here, it is assumed that the two planets in the system have comparable masses.

From \Eq \ref{RESEQ}, we see that a non-migrating system such as GJ876 could survive in resonance for $t_f \sim 10^6$~years if the stochastic force amplitude
is $\sim 10^{-5}$ times the central force. 
This expression enables scaling to other systems at other disc locations for other stochastic forcing amplitudes.
Inference of survival probabilities for particular systems depends on many uncertain aspects, such as the proto-planetary disc model and the strength of the turbulence. 
However, the mass ratio dependence in \Eq \ref{RESEQ} indicates that survival is favoured for more massive systems. 
At the present time the number of observed resonant systems is too small for definitive conclusions to be made.
However, the fact that several systems exhibiting commensurabilities have been observed indicates that resonances are not always 
completely disrupted by stochastic forces due to turbulence, but rather may be modified as in our study of HD128311.

The most likely orbital configuration of the HD128311 system is such that the fast mode librates with the slow mode being near the borderline between libration and circulation.
We found that such a configuration was readily produced in a scenario in which the commensurability was formed through a temporary period of convergent migration,
with the addition of stochastic forcing. 
During a migration phase moderate adiabatic invariance applied to the libration modes together with eccentricity damping leads to 
increased stabilisation and a longer lifetime for the resonance.
However, the time evolution of the eccentricity, and in particular the attainment of small values, plays an 
important role in causing the slow mode to circulate. This corresponds to the loss of apsidal co-rotation. 
Thus we expect that a large eccentricity damping rate does not necessarily stabilise the apsidal co-rotation of a system. 
Additional simulations have shown that this is lost more easily for large damping rates ($K\gg 10$).
On the other hand, it should be noted that this has less significance when one of the orbits becomes nearly circular.

Further observations of extra-solar planetary systems leading to better statistics may lead to an improved situation for assessing the role of stochastic forcing.

\chapter{Stochastic migration of small bodies in Saturn's rings}\label{ch:moonlet}
\epigraph{
{
Have not the small particles of bodies certain powers, virtues, or forces, by which they act at a distance, not only upon the rays of light for reflecting, refracting, and inflecting them, but also upon one another for producing a great part of the Phenomena of nature?
}}{\tiny Isaac Newton, Optiks, Query 31, 1729}

\noindent In the previous chapter, we discussed the stochastic migration of planets, embedded in a circum-stellar disc. 
A very similar process also occurs in the circum-planetary disc known as Saturn's rings.
However, both the length and time scales involved are three orders of magnitude smaller.
The bodies of interest are small moonlets. 
These $20\,\mathrm{m}\,-100\,\mathrm{m}$ sized bodies have been predicted both analytically and numerically \citep{Spahn2000,Sremcevic2002,Seiss2005}. 
They create propeller shaped structures due to their disturbance of the rings and have been observed only recently by the Cassini spacecraft in the A and B ring \citep{Tiscareno2006, Tiscareno2008}.

They can migrate within the rings, similar to proto-planets which migrate in a proto-stellar disc. 
Depending on the disc properties and the moonlet size, this can happen in either a smooth or in a stochastic (random walk) fashion. 
We refer to those migration regimes as type I and type IV, respectively, in analogy to the terminology in disc-planet interactions (see section \ref{sec:introduction:migration}). 
\cite{Crida2010} showed that there is a laminar type I regime that might be important on very long timescales.
This migration is qualitatively different to the migration in a pressure supported gas disc.
However, the migration of moonlets in the A ring is generally dominated by type IV migration, at least on short timescales. 

We here study their dynamical evolution in the type IV regime analytically, including important effects such as collisions which were not present in the discussion in chapter \ref{ch:randwalk}.
We also perform realistic three dimensional collisional N-body simulations with up to a quarter of a million particles.
A new set of \emph{pseudo} shear periodic boundary conditions is used which reduces the computational costs by an order of magnitude compared to previous studies. 
The numerical simulations confirm our analytic estimates to within a factor of two.

For low ring surface densities, the main effects on the evolution of the eccentricity and the semi-major axis are found to be due to collisions and the gravitational interaction with particles in the vicinity of the moonlet. 
For large surface densities, the gravitational interaction with self-gravitating wakes becomes important. These stochastic forces are very similar to what has been discussed earlier, in chapter \ref{ch:randwalk}.

On short timescales the evolution is always dominated by stochastic effects caused by collisions 
and gravitational interaction with self-gravitating ring particles.
These result in a random walk of the moonlet's semi-major axis. 
The eccentricity of the moonlet quickly reaches an equilibrium value due to collisional damping.
The average change in semi-major axis of the moonlet after 100 orbital periods is 10-100m. 
This translates to an offset in the azimuthal direction of several hundred kilometres. 
We expect that such a shift is easily observable.

We start be reviewing the basic equations governing the moonlet and the ring particles in a shearing box approximation in section \ref{moonlet:sec:analytic}.
Then, in section \ref{dampe}, we estimate the eccentricity damping timescale due to ring particles colliding with the moonlet and ring particles on horseshoe orbits as well as the effect of particles on circulating orbits.
In section \ref{moonlet:sec:excitee} we estimate the excitation of the moonlet eccentricity caused by stochastic 
particle collisions and gravitational interactions with ring particles.
This enables us to derive an analytic estimate of the equilibrium eccentricity.

In section \ref{moonlet:sec:randa} we discuss and evaluate processes, such as collisions and gravitational interactions
with ring particles and self-gravitating clumps, that lead to a random walk in the semi-major axis of the moonlet. 

We describe our numerical setup and the initial conditions used in section \ref{moonlet:sec:numerical}. 
The results are presented in section \ref{moonlet:sec:results}. 
All analytic estimates are confirmed both in terms of qualitative trends
and quantitatively to within a factor of about two in all simulations that we performed. 
We also discuss the long term evolution of the longitude of the moonlet and its observability,
before summarising our results in section \ref{moonlet:sec:conclusions}.

\section{Basic equations governing the moonlet and ring particles}\label{moonlet:sec:analytic}
\subsection{Epicyclic motion} \label{moonlet:sec:coordinates}
\begin{figure*}
\center
\resizebox{\textwidth}{!}{
\begin{picture}(0,0)%
\includegraphics{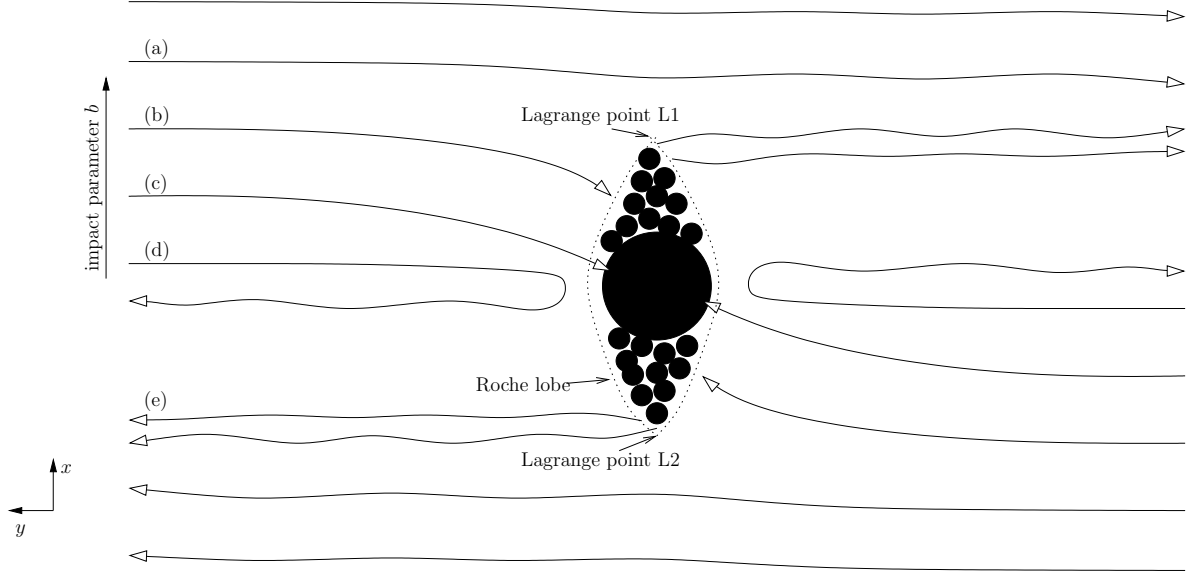}%
\end{picture}%
\setlength{\unitlength}{3947sp}%
\begingroup\makeatletter\ifx\SetFigFont\undefined%
\gdef\SetFigFont#1#2#3#4#5{%
  \reset@font\fontsize{#1}{#2pt}%
  \fontfamily{#3}\fontseries{#4}\fontshape{#5}%
  \selectfont}%
\fi\endgroup%
\begin{picture}(11724,5724)(-611,-7198)
\put(-74,-6211){\makebox(0,0)[lb]{\smash{{\SetFigFont{12}{14.4}{\familydefault}{\mddefault}{\updefault}{\color[rgb]{0,0,0}$x$}%
}}}}
\put(-524,-6811){\makebox(0,0)[lb]{\smash{{\SetFigFont{12}{14.4}{\familydefault}{\mddefault}{\updefault}{\color[rgb]{0,0,0}$y$}%
}}}}
\put(751,-2011){\makebox(0,0)[lb]{\smash{{\SetFigFont{12}{14.4}{\familydefault}{\mddefault}{\updefault}{\color[rgb]{0,0,0}(a)}%
}}}}
\put(751,-2686){\makebox(0,0)[lb]{\smash{{\SetFigFont{12}{14.4}{\familydefault}{\mddefault}{\updefault}{\color[rgb]{0,0,0}(b)}%
}}}}
\put(751,-3361){\makebox(0,0)[lb]{\smash{{\SetFigFont{12}{14.4}{\familydefault}{\mddefault}{\updefault}{\color[rgb]{0,0,0}(c)}%
}}}}
\put(751,-4036){\makebox(0,0)[lb]{\smash{{\SetFigFont{12}{14.4}{\familydefault}{\mddefault}{\updefault}{\color[rgb]{0,0,0}(d)}%
}}}}
\put(4501,-2686){\makebox(0,0)[lb]{\smash{{\SetFigFont{12}{14.4}{\familydefault}{\mddefault}{\updefault}{\color[rgb]{0,0,0}Lagrange point L1}%
}}}}
\put(4051,-5386){\makebox(0,0)[lb]{\smash{{\SetFigFont{12}{14.4}{\familydefault}{\mddefault}{\updefault}{\color[rgb]{0,0,0}Roche lobe}%
}}}}
\put(751,-5536){\makebox(0,0)[lb]{\smash{{\SetFigFont{12}{14.4}{\familydefault}{\mddefault}{\updefault}{\color[rgb]{0,0,0}(e)}%
}}}}
\put(4501,-6136){\makebox(0,0)[lb]{\smash{{\SetFigFont{12}{14.4}{\familydefault}{\mddefault}{\updefault}{\color[rgb]{0,0,0}Lagrange point L2}%
}}}}
\put(301,-4186){\rotatebox{90.0}{\makebox(0,0)[lb]{\smash{{\SetFigFont{12}{14.4}{\familydefault}{\mddefault}{\updefault}{\color[rgb]{0,0,0}impact parameter $b$}%
}}}}}
\end{picture}%
}
	\caption{Trajectories of ring particles in a frame centred on the moonlet. Particles accumulate near the moonlet and fill its Roche lobe. Particles on trajectories labelled a) are on circulating orbits. Particles on trajectories labelled b) collide with other particles in the moonlet's vicinity.
	Particles on trajectories labelled (c) collide with the moonlet directly.
	Particles on trajectories labelled (d) are on horseshoe orbits. Particles on trajectories
	labelled (e) leave the vicinity of the moonlet. \label{fig:trajectory}}
\end{figure*}

We adopt a local right handed Cartesian coordinate system with its origin being in circular Keplerian orbit with semi-major axis $a$ and 
rotating uniformly
with angular velocity $\Omega.$
This orbit coincides with that of the moonlet when it is assumed to be 
unperturbed by ring particles. The $x$ axis coincides with the line joining the 
central object of mass $M_p$ and the origin. The unit vector in the $x$ direction, $\mathbf{e}_x,$ points away
from the central object. The unit vector in the $y$ direction $ \mathbf{e}_y$ points in the direction
of rotation and the unit vector in the $z$ direction, $\mathbf{e}_z$ points in the vertical direction
being normal to the disc mid-plane (see also appendix \ref{app:gravtree}). 

In general, we shall consider a ring particle of
mass $m_1$ interacting with the moonlet which has a much larger mass $m_2.$ 
A sketch of three possible types of particle trajectory in the vicinity of the moonlet is shown in figure \ref{fig:trajectory}. 
These correspond to three distinct regimes, a) denoting circulating orbits, b) and c) denoting orbits that result in a 
collision with particles in the vicinity of the moonlet
and directly with the moonlet respectively, and d) denoting horseshoe orbits. All these types of trajectory occur for particles
that are initially in circular orbits, both interior and exterior to the moonlet, when at large distances from it.

Approximating the gravitational force due to the central object by its first order Taylor expansion
about the origin leads to Hill's equations (see equations \ref{eq:app:gravtree:hills1}-\ref{eq:app:gravtree:hills3}), governing the motion of a particle of mass $m_1$ of the form 
\begin{eqnarray}
\ddot {\mathbf r} &=& -2\Omega {\mathbf e}_z\times \dot {\mathbf r} + 3 \Omega^2 ({\mathbf r} \cdot {\mathbf e}_x) {\mathbf e}_x - {\mathbf \nabla} \Psi/m_1, \label{eq:hills}
\end{eqnarray}
where $\mathbf{r} = (x,y,z)$ is position of a particle with mass $m_1$ and $\Psi$ is the gravitational potential
acting on the particle due to other objects of interest such as the moonlet. 

The square amplitude of the epicyclic motion ${\mathcal E}^2$ can be defined through
\begin{eqnarray}
{\mathcal E}^2 = {\Omega^{-2}{\dot x}^2+(2\Omega^{-1}\dot y+3x )^2} \label{eq:defe2}.
\end{eqnarray}
Note that neither the eccentricity $e$, nor the semi-major axis $a$ are formally defined in the local coordinate system. 
However, in the absence of interaction with other masses $(\Psi=0),$ ${\mathcal E}$ is conserved,
and up to first order in the eccentricity we may make the identification ${\mathcal E}=e\,a.$ 
We recall the classical definition of the eccentricity $e$ in a coordinate
system centred on the central object
\begin{eqnarray}
e = \left| \frac{\mathbf w \times \left( \mathbf s \times \mathbf w \right)}{\Omega^2a^3} - \frac{\mathbf s}{|\mathbf s |}\right|. \label{eq:defe1}
\end{eqnarray}
Here, $\mathbf{s}$ is the position vector $(a,\,0,\,0)$ and $\mathbf w$ is the velocity vector of the particle relative to the mean shear 
associated with circular Keplerian motion as viewed in the local coordinate system
plus the circular Keplerian velocity corresponding to the orbital frequency $\Omega$ 
of the origin. Thus
\begin{eqnarray}
\mathbf w &=& \left(\dot x,\,\,\dot y + \frac32 \Omega x+ a \Omega,\,\,\dot z\right).
\end{eqnarray}
All eccentricities considered here are 
very small ($\sim 10^{-8}-10^{-7}$) so that the difference between the quantities defined through use of equations \ref{eq:defe2} and \ref{eq:defe1} is negligible, allowing us to adopt $\mathcal E$ as a measure of the eccentricity throughout this chapter.

Let us define another quantity, $\mathcal A$, that is also conserved for non interacting particle motion, $\Psi=0,$ which
is the $x$ coordinate of the centre of the epicyclic motion and is given by 
\begin{eqnarray}
{\mathcal A} &=& 2 \Omega^{-1}{\dot y} + 4 x.
\end{eqnarray}
We identify a change in ${\mathcal A}$ as a change in the semi-major axis $a$ of the particle, again, under the assumption that the eccentricity is small.

\subsection{Two interacting particles}
We now consider the motion of two gravitationally interacting particles with position 
vectors ${\bf r}_1= (x_1,y_1,z_1)$ and ${\bf r}_2=(x_2,y_2,z_2).$ Their corresponding masses are $m_1$ and $m_2$, respectively. 
The governing equations of motion are
\begin{eqnarray}
\ddot {\mathbf r}_1 &=& -2\Omega {\mathbf e}_z\times \dot {\mathbf r}_1 + 3 \Omega^2 ({\mathbf r}_1 \cdot {\mathbf e}_x) {\mathbf e}_x - {\mathbf \nabla}_{{\mathbf r}_1}\Psi_{12}/m_1\\
\ddot {\mathbf r}_2 &=& -2\Omega {\mathbf e}_z\times \dot {\mathbf r}_2 + 3 \Omega^2 ({\mathbf r}_2 \cdot {\mathbf e}_x) {\mathbf e}_x + {\mathbf\nabla}_{{\mathbf r}_1}\Psi_{12}/m_2,
\end{eqnarray}
where the interaction gravitational potential is
$\Psi_{12}= -Gm_1m_2 / \left|\mathbf{r}_1-\mathbf{r}_2\right|$.
The position vector of the centre of mass of the two particles is given by
\begin{eqnarray}
\mathbf{\bar r}&=&\frac{m_1 \mathbf{r}_1 + m_2\mathbf{r}_2 }{m_1+m_2}.
\end{eqnarray}
The vector $\mathbf{\bar r}$ also obeys equation \ref{eq:hills} with $\Psi=0$, which applies to an isolated particle. 
This is because the interaction potential does not affect the motion of the centre of mass.
We also find it useful to define the vector
\begin{eqnarray}
{\mbox{\boldmath${\mathcal E}$}}_i = (\Omega^{-1}{\dot x}_i,\;2\Omega^{-1}\dot y_i+3x_i )\ \ \ \ i=1,2 \label{eq:defevec}.
\end{eqnarray}
Then, consistently with our earlier definition of ${\mathcal E}$, we have ${\mathcal E}_i =| {\mbox{\boldmath${\mathcal E}$}}_i|.$
The amplitude of the epicyclic motion of the centre of mass $\bar {\mathcal E}$ is given by
\begin{eqnarray}
(m_1+m_2)^2\bar{\mathcal E}^2 &=& m_1^2{\mathcal E}_1^2+m_2^2{\mathcal E}_2^2 + 2m_1m_2{\mathcal E}_1{\mathcal E}_2 \cos(\phi_{12}).
\end{eqnarray}
Here $\phi_{12}$ is the angle between $ {\mbox{\boldmath${\mathcal E}$}}_1$ and $ {\mbox{\boldmath${\mathcal E}$}}_2.$
It is important to note that $\bar {\mathcal E}$ is conserved even if the two
particles approach
each other and become bound. This is as long as frictional forces are internal to the
two particle system and do
not affect the centre of mass motion.

\section{Effects leading to damping of the eccentricity of the moonlet}\label{dampe}
We begin by estimating the moonlet eccentricity damping rate associated with ring particles that either collide
directly with the moonlet, particles in its vicinity, or interact only gravitationally with the moonlet.

\subsection{The contribution due to particles impacting the moonlet} \label{moonlet:sec:edampcollision}
Particles impacting the moonlet in an eccentric orbit exchange momentum with it.
Let us assume that a ring particle, identified with $m_1$ has zero epicyclic amplitude, so that ${\mathcal E}_1=0$ far away from the moonlet. 
The moonlet is identified with $m_2$ and has an initial epicyclic amplitude ${\mathcal E}_2$. 
The epicyclic amplitude of the centre of mass is therefore 
\begin{eqnarray}
\bar{\mathcal E} = \frac{m_2}{m_1+m_2} {\mathcal E}_2 \simeq \left(1-\frac{m_1}{m_2}\right) {\mathcal E}_2 \label{eq:ecm},
\end{eqnarray}
where we have assumed that $m_1\ll m_2$. 

The moonlet is assumed to be in a steady state in which there is no net accretion of ring particles. 
Therefore, for every particle that either collides directly with the moonlet or nearby particles bound to it (and so itself becomes temporarily bound to it), one particle must also escape from the moonlet. 
This happens primarily through slow leakage from locations close to the $\mathrm{L}_{1}$ and $\mathrm{L}_{2}$ points 
such that most particles escape from the moonlet with almost zero velocity (as viewed from the centre of mass frame)
and so do not change its orbital eccentricity.
However, after an impacting particle becomes bound to the moonlet, conservation of the epicyclic amplitude associated
with the centre of mass motion together with equation \ref{eq:ecm} imply that
each impacting particle will reduce the eccentricity by a factor $1-m_1/m_2.$ 

It is now an easy task to estimate the eccentricity damping timescale by determining the number of particle
collisions per time unit with the moonlet or particles bound to it. 
To do that, a smooth window function $W_{b+c}(b)$ is used, being unity 
for impact parameters $b$ that always result in an impact with the moonlet or particles nearby that are bound to it, being zero for impact parameters that never result in an impact. 
	
To ensure that the impact band is well defined, we assume that the epicyclic amplitude is smaller than the moonlet's Hill radius, which is the case in all simulations discussed below. 
However, if this were not the case, one could simply calculate the collisions per time units by other means, for example by using the velocity dispersion and the geometrical cross section of the moonlet. 

The number of particles impacting the moonlet per time unit, $dN/dt$, is obtained by integrating over the impact parameter
with the result that
\begin{eqnarray}
\frac{dN}{dt} &=& \frac{1}{m_1}{\int_{-\infty}^{\infty} \frac32 \, \Sigma\, \Omega \,|b| \; W_{b+c}(b)\, \mathrm{d}b}\label{Window}.
\end{eqnarray}
We note that allowing $b$ to be negative enables impacts from both sides of the moonlet to be taken into account.

Therefore, after using equation \ref{eq:ecm} we find that the rate of change of the moonlet's eccentricity $e_2$, or equivalently of it's amplitude of epicyclic motion ${\mathcal E}_2$, is given by
\begin{eqnarray}
\frac{d{\mathcal E}_2}{dt} = -\frac{{\mathcal E}_2}{\tau_{e,\mathrm{collisions}}} &=& -\frac{{\mathcal E}_2}{m_2}{\int_{-\infty}^{\infty} \frac32 \, \Sigma\, \Omega \,|b| \; W_{b+c}(b)\, \mathrm{d}b}, \label{eq:d13}
\end{eqnarray}
where $\tau_{e,\mathrm{collisions}}$ defines the circularisation time arising from collisions with the moonlet. We remark that the natural unit
for $b$ is the Hill radius of the moonlet, $r_H = (m_2/(3M_p))^{1/3}a,$ so that the dimensional scaling for
$\tau_{e,\mathrm{collisions}}$ is given by 
\begin{eqnarray}
\tau_{e,\mathrm{collisions}}^{-1} &\propto& G\;\Sigma\; r_H^{-1}\;\Omega^{-1} \label{eq:scaling}
\end{eqnarray}
which we find to also apply to all the processes for modifying the moonlet's eccentricity discussed below.
If we assume that $W_{b+c}(|b|)$ can be approximated by a box function, being unity in the interval $[1.5r_H,\,2.5r_H]$ and zero elsewhere, we get
\begin{eqnarray}
\tau_{e,\mathrm{collisions}}^{-1} &=& 2.0 \;G\;\Sigma\; r_H^{-1}\;\Omega^{-1} \label{eq:taucol}.
\end{eqnarray}

\subsection{Eccentricity damping due the interaction of the moonlet with particles on horseshoe orbits}\label{moonlet:sec:edamphorseshoe}
The eccentricity of the moonlet manifests itself in a small oscillation of the moonlet about the origin.
Primarily ring particles on horseshoe orbits will respond to that oscillation and damp it.
This is because only those particles on horseshoe orbits spend enough time in the vicinity of the moonlet, i.e. many epicyclic periods.

In appendix \ref{app:response}, we calculate the amplitude of epicyclic motion ${\mathcal E}_{1f}$ (or equivalently the eccentricity $e_{1f}$) that is induced in a single ring particle in a horseshoe orbit undergoing a close approach to a moonlet which is assumed to be in an eccentric orbit. 
In order to calculate the circularisation time, we have to consider all relevant impact parameters. 
Note that each particle encounter with the moonlet 
is conservative and is such that for each particle,
the Jacobi constant, applicable when the moonlet is in circular orbit, is increased
by an amount $m_1\Omega^2{\mathcal E}_{1f}^2/2$ by the action of
the perturbing force, associated with the eccentricity of the moonlet, as the particle passes by.
Because the Jacobi constant, or energy in the rotating frame, for the moonlet and the particle together is conserved,
the square of the epicyclic amplitude associated with the moonlet alone changes by ${\mathcal E}_{1f}^2\,m_1/m_2.$
Accordingly the change in the amplitude of epicyclic motion of the moonlet ${\mathcal E}_2$, consequent
on inducing the amplitude of epicyclic motion ${\mathcal E}_{1f}$ in the horseshoe particle, is given by 
\begin{eqnarray}
\Delta {\mathcal E}_{2}^2 &=& -\frac{m_1}{m_2} {\mathcal E}_{1f}^2.
\end{eqnarray}
Note that this is different compared to equation \ref{eq:ecm}. 
Here, we are dealing with a second order effect. 
First, the eccentric moonlet excites eccentricity in a ring particle.
Second, because the total epicyclic motion is conserved, the epicyclic motion of the moonlet is reduced. 

Integrating over the impact parameters associated with ring particles
and taking into account particles streaming by the moonlet from both directions
by allowing negative impact parameters gives
\begin{eqnarray}
\left.\frac{d{\mathcal E}_2^2}{dt}\right|_{\mathrm{horseshoe}}&=& -\int_{-\infty}^{\infty} \frac{3\Sigma {\mathcal E}_{1f}^2 \Omega \,|b| \,W_d(b)\,db}{2m_2} \equiv -\frac{2{\mathcal E}_2^2}{\tau_{e,\mathrm{horseshoe}}},
\end{eqnarray}
where $\tau_{e,\mathrm{horseshoe}}$ is the circularisation time and, as above, we have inserted a window function, which
is unity on impact parameters that lead to horseshoe orbits, otherwise being zero.
Using ${\mathcal E}_{1f}$ given by equation \ref{inducede}, we obtain
\begin{eqnarray}
\tau_{e,\mathrm{horseshoe}}^{-1} =\frac{9}{128}\left(\frac{\Sigma r_H^2\Omega}{m_2}\right)\int_{-\infty}^{\infty}{\mathcal I}^2\eta^{4/3} W_d\left((2\eta r_h)^{1/6}\right)\,d\eta.
\end{eqnarray}
For a sharp cutoff of $W_d(|b|)$ at $b_m =1.5r_H$ we find the value of the integral in the above to be $2.84$.
Thus, in this case we get 
\begin{eqnarray}
\tau_{e,\mathrm{horseshoe}}^{-1}= 0.13 \;G \;\Sigma\;r_H^{-1}\;\Omega^{-1}.\label{eq:damphorse}
\end{eqnarray}
However, note that this value is sensitive to the value of $b_m$ adopted. For $b_m= 1.25r_H,$ $\tau_e$ is a factor of $4.25$ larger.

\subsection{The effects of circulating particles}\label{moonlet:sec:dampecirc}
The effect of the response of circulating particles to the gravitational
perturbation of the moonlet on the moonlet's eccentricity can be estimated from the work of 
\cite{GoldreichTremaine1980} and \cite{GoldreichTremaine1982}.
These authors considered a ring separated from a satellite such that co-orbital effects
were not considered. Thus, their expressions may be applied to estimate
effects due to circulating particles. However, we exclude their corotation torques as they 
are determined by the ring edges and are absent in a local model with uniform azimuthally averaged surface density.
Equivalently, one may simply assume that the corotation torques are saturated.
When this is done only Lindblad torques act on the moonlet. These tend to excite the moonlet's eccentricity rather than damp it.

We replace the ring mass in equation 70 of \cite{GoldreichTremaine1982} by the integral over the impact parameter and switch to our notation to obtain
\begin{eqnarray}
\left.\frac{d{\mathcal E}_2^2}{dt}\right|_{\mathrm{circ}}&=& 9.55 \int_{-\infty}^{\infty} m_2 \, \Sigma \, G^2 \, \Omega^{-3} \,  |b|^{-5} \;{\mathcal E}_2^2\;W_a(b) \,\mathrm{d}b,\label{Goldcirc}
\end{eqnarray}
where $W_a(b)$ is the appropriate window function for circulating particles. 
Assuming a sharp cutoff of $W_a(|b|)$ at $b_m$, we can evaluate the integral in equation \ref{Goldcirc} to get 
\begin{eqnarray}
\frac{1}{{\mathcal E}_2}\left.\frac{d{\mathcal E}_2}{dt}\right|_{\mathrm{circ}}=-\frac{1}{\tau_{e,circ}}&=&  0.183 \;G \;\Sigma\;r_H^{-1}\;\Omega^{-1},
\label{Goldcirc2} \end{eqnarray}
where we have adopted $b_m=2.5r_H$, consistently with the simulation results presented below.
We see that, although $\tau_{e,circ}<0$ corresponds to growth rather than damping of the eccentricity, it scales in the same manner
as the circularisation times in sections \ref{moonlet:sec:edampcollision} and \ref{moonlet:sec:edamphorseshoe} scale (see equation \ref{eq:scaling}). 

This timescale and the timescale associated with particles on horseshoe orbits, $\tau_{e,\mathrm{horseshoe}}$, are significantly larger than the timescale associated with collisions, given in equation \ref{eq:taucol}.
We can therefore ignore those effects for most of the discussion in this chapter.

\section{Processes leading to the excitation of the eccentricity of the moonlet} \label{moonlet:sec:excitee}
In section \ref{dampe} we assumed that the moonlet had a small eccentricity but neglected
the initial eccentricity of the impacting ring particles.
When this is included, collisions and gravitational interactions of ring particles with the moonlet may also excite its eccentricity.

\subsection{Collisional eccentricity excitation}
To see this, assume that the moonlet initially has zero eccentricity,
or equivalently no epicyclic motion, but the ring particles do.

As above we consider the conservation of the epicyclic motion of the centre of mass in
order to connect the amplitude of the
final epicyclic motion of the combined moonlet and ring particle
to the initial amplitude of the ring particle's epicyclic motion.
This gives the epicyclic amplitude of the centre of mass after one impact as
\begin{eqnarray}
\bar{\mathcal E} = \frac{m_1}{m_1+m_2} {\mathcal E}_1 \simeq \left(\frac{m_1}{m_2}\right) {\mathcal E}_1 \label{eq:ecmexc}.
\end{eqnarray}
Assuming that successive collisions are uncorrelated and occur stochastically
with the mean time between consecutive encounters being $\tau_{ce}$, the evolution of $\bar{\mathcal E}$ is governed by the equation
\begin{eqnarray}
\left.\frac{d \bar{\mathcal E}^2}{dt}\right|_{\mathrm{collisions}} = \left( \frac{m_1}{m_2}\right)^{2} \langle {\mathcal E}_1^{2} \rangle \tau_{ce}^{-1},\label{eq:edot}
\end{eqnarray}
where $\tau_{ce}^{-1} = dN/dt$ can be expressed in terms of the surface density and an impact window function $W_{b+c}(b)$ (see equation \ref{Window}).
The quantity $\langle {\mathcal E}_1^{2} \rangle$ is the mean square value of ${\mathcal E}_1$ for the ring particles.
Using equation \ref{eq:defe2}, this may be written in terms of the mean squares of the components of the velocity dispersion
relative to the background shear, in the form
\begin{eqnarray}
\langle {\mathcal E}_1^{2} \rangle = \Omega^{-2}\langle ({\dot x_1}^2+4({\dot y_1}+3\Omega x_1/2 )^2\rangle .
\end{eqnarray}
We find in numerical simulations $\langle e_1\rangle\sim10^{-6}$ for almost all ring parameters. This value is mainly determined by the coefficient of restitution \citep[see e.g. figure 4 in][]{MorishimaSalo2006}.

\subsection{Stochastic excitation due to circulating particles}\label{circpart}
Ring particles that are on circulating orbits, such as that given by path (a) in figure \ref{fig:trajectory}, exchange energy and angular momentum with the moonlet and therefore change its eccentricity. 
\cite{GoldreichTremaine1982} calculated the change in the eccentricity of a ring particle due to a satellite. 
We are interested in the change of the eccentricity of the moonlet induced by a ring of particles and therefore swap the satellite mass with the mass of a ring particle.
Thus, rewriting their results (equation 64 in \cite{GoldreichTremaine1982}) in our local notation without reference to the semi-major axis, we have
\begin{eqnarray}
\Delta {\mathcal E}_2^2 &=& 5.02 \;m_1^2\;G^2\;\Omega^{-4}\; {b}^{-4}.
\end{eqnarray}
Supposing that the moonlet has very small eccentricity, it will receive stochastic impulses that cause its eccentricity to undergo
a random walk that will result in ${\mathcal E}_2^2$ increasing linearly with time, so that we may write
\begin{eqnarray}
\left.\frac{d{\mathcal E}_2^2}{dt}\right|_{\mathrm{circulating\;particles}} 
&=&
\int_{-\infty}^{\infty} W_a(b) \Delta {\mathcal E}_2^2 d(1/t_b),
\end{eqnarray}
where $d(1/t_b)$ is the mean encounter rate with particles which have an impact parameter in the interval $(b,b+db)$.
$W_a(b)$ is the window function describing the band of particles in circulating orbits.
Setting $d(1/t_b)=3\Sigma\Omega |b| db/(2m_1),$ we obtain
\begin{eqnarray}
\left.\frac{d{\mathcal E}_2^2}{dt}\right|_{\mathrm{circulating\;particles}} 
&=& 7.53\int_{-\infty}^{\infty} W_a(b) m_1 |b|^{-3}\Sigma G^2 \Omega^{-3}\mathrm{d}b.
\label{circpe}
\end{eqnarray}
For high surface densities an additional effect can excite the eccentricity when gravitational wakes occur.
The Toomre parameter $Q$ is defined as 
\begin{eqnarray}
Q=\frac{\bar{v} \Omega}{\pi G\Sigma}, \label{eq:moonlet:toomre}
\end{eqnarray}
where $\bar v$ is the velocity dispersion of the ring particles\footnote{The Toomre criterion used here was originally derived for a flat gaseous disc.
To make use of it we replace the sound speed by the radial velocity dispersion of the ring particles.}. 
When the surface density is sufficiently high such that $Q$ approaches unity, transient clumps will form in the rings. 
The typical length scale of these structures is given by the critical Toomre wave length \citep{Daisaka2001} 
\begin{eqnarray}
\lambda_\mathrm{T} &=& {4 \pi^2 G \Sigma}{\Omega^{-2}}, \label{eq:moonlet:toomrelambda}
\end{eqnarray}
which can be used to estimate a typical mass of the clump:
\begin{eqnarray}
m_\mathrm{T} 	&\sim& \pi \left(\frac{\lambda_\mathrm{T}}2\right)^2 \Sigma
	= 	{4\pi^5G^2\Sigma^3}{\Omega^{-4}}.
\end{eqnarray}
Whenever strong clumping occurs, $m_\mathrm{T}$ should be used in the above calculation instead of the mass of a single particle $m_1$:
\begin{eqnarray}
\left.\frac{d{\mathcal E}_2^2}{dt}\right|_{\mathrm{circulating\;clumps}}&=& 7.53 \int_{-\infty}^{\infty} W'_a(b) m_T |b|^{-3}\Sigma G^2 \Omega^{-3} \mathrm{d}b,
\label{clumpe}
\end{eqnarray}
where $W'_a(b)\approx W_a(b)$ is the appropriate window function.
For typical parameters used in section \ref{moonlet:sec:results}, this transition occurs at $\Sigma\sim200\mathrm{kg/m^2}$.

\subsection{Equilibrium eccentricity}
Putting together the results from this and the previous section, we can estimate an equilibrium eccentricity of the moonlet. 
Let us assume the eccentricity $e_2$, or the amplitude of the epicyclic motion ${\mathcal E}_2$, evolves under the influence of excitation and damping forces as follows
\begin{eqnarray}
\frac{d{\mathcal E}_2^2}{dt}&=& -2 \left(
 \tau_{e,\mathrm{collisions}}^{-1} 
 + \tau_{e,\mathrm{horseshoe}}^{-1} 
 +\tau_{e,\mathrm{circ}}^{-1} \right) {\mathcal E}_2^2 \nonumber \\
&&+\left.\frac{d{\mathcal E}_2^2}{dt}\right|_{\mathrm{collisions}}
 +\left.\frac{d{\mathcal E}_2^2}{dt}\right|_{\mathrm{circulating\;particles}}
+\left.\frac{d{\mathcal E}_2^2}{dt}\right|_{\mathrm{circulating\;clumps}}
.\label{equile}
\end{eqnarray}
The equilibrium eccentricity is then found by setting the above equation equal to zero and solving for ${\mathcal E}_2$. 

To make quantitative estimates we need to specify
the window functions $W_a(b)$, $W_{b+c}(b)$ and $W_d(b)$ that determine in which impact parameter
bands the particles are in (see figure \ref{fig:trajectory}). 
To compare our analytic estimates to the numerical simulations presented below, we measure the window functions numerically.
Alternatively, one could simply use sharp cutoffs at some multiple of the Hill radius (see also section \ref{sec:moonlet:window}). 
We already made use of this approximation as a simple estimate in the previous sections. 
The results may vary slightly, but not significantly. 

However, as the window functions are dimensionless, it is possible to obtain the dimensional scaling of ${\mathcal E}_2$ by adopting the length scale applicable to the impact parameter to be the Hill radius $r_H$ and simply assume that the window functions are of order unity. 
As already indicated above, all of the circularisation times scale as $\tau_{e} \propto \Omega \,r_H\,G^{-1}\,\Sigma^{-1}$, or equivalently $\propto m_2/(\Sigma r_H^2\Omega)$.
Assuming the ring particles have zero velocity dispersion, the eccentricity excitation is then due to circulating clumps or particles and the scaling of $d{\mathcal E}_2^2/{dt}$ due to this cause is given by equations \ref{circpe} and \ref{clumpe} by
\begin{eqnarray}
\frac{d{\mathcal E}_2^2}{dt}&\propto& m_i r_H^{-2}\Sigma G^2 \Omega^{-3},
\end{eqnarray}
where $m_i$ is either $m_1$ or $m_T.$ We may then find the scaling
of the equilibrium value of ${\mathcal E}_2$ from consideration of equations \ref{equile} and \ref{eq:d13} as
\begin{eqnarray}
{\mathcal E}^2_2&\propto & \frac{m_i}{m_2} r_H^{2}.
\end{eqnarray}
This means that the expected kinetic energy in the non circular motion of the moonlet
is $\propto m_i r_H^{2}\Omega^2$ which can be viewed as stating that
the non circular moonlet motion scales in equipartition with the mass $m_i$ moving 
with speed $r_H\Omega.$ This speed applies when the dispersion velocity associated 
with these masses is zero indicating that the shear across a Hill radius
replaces the dispersion velocity in that limit. 

In the opposite limit when the dispersion velocity exceeds the shear across a Hill
radius and the dominant source of eccentricity excitation is due to collisions,
equation \ref{equile} gives 
\begin{eqnarray}
m_2{\mathcal E}_2^2&=& m_1\langle {\mathcal E}_1^{2} \rangle 
\end{eqnarray}
so that the moonlet is in equipartition with the ring particles.
Results for the two limiting cases can be combined to give
an expression for the amplitude of the epicyclic motion excited in the moonlet
of the form
\begin{eqnarray}
m_2\Omega^2{\mathcal E}_2^2&=& m_1\Omega^2 \langle {\mathcal E}_1^{2} \rangle + C_im_i\Omega^2 r_H^{2},
\end{eqnarray}
where $C_i$ is a constant of order unity.
This indicates the transition between the shear dominated and the velocity dispersion
dominated limits.

\section{Processes leading to a random walk in the semi-major axis of the moonlet}\label{moonlet:sec:randa}
We have established estimates for the equilibrium eccentricity of the moonlet in the previous section. 
Here, we estimate the random walk of the semi-major axis of the moonlet. 
In contrast to the case of the eccentricity, there is no tendency
for the semi-major axis to relax to any particular value, so that there are no damping terms
and the deviation of the semi-major axis from its value at time $t=0$
grows on average as $\sqrt{t}$ for large $t.$

Depending on the surface density, there are different effects that dominate the contributions
to the random walk of the moonlet. For low surface densities, collisions and horseshoe orbits are most important.
For high surface densities, the random walk is dominated by the stochastic gravitational force of
circulating particles and clumps. In this section, we try to estimate the strength of each effect.

\subsection{Random walk due to collisions}\label{moonlet:sec:randacol}
Let us assume, without loss of generality, that the guiding centre of the epicyclic motion of a ring particle is displaced from the
orbit of the moonlet by ${\mathcal A}_1$ in the inertial frame, whereas in the co-rotating frame the moonlet is initially located at the origin with ${\mathcal A}_2=0$. 
When the impacting particle becomes bound to the moonlet, the guiding centre of the epicyclic motion of the centre of mass of the combined object, as viewed in the inertial
frame, is then displaced from the original
moonlet orbit by 
\begin{eqnarray}
\Delta \bar{\mathcal A} = \frac{m_1}{m_1+m_2} {\mathcal A}_1 \sim \frac{m_1}{m_2} {\mathcal A}_1 .
\end{eqnarray}
This is the analogue of equation \ref{eq:ecmexc} for the evolution of the semi-major axis. For an
ensemble of particles with impact parameter $b,$ 
the average centre of epicyclic motion is $\langle {\mathcal A}_1\rangle=b$. Thus we can write the evolution of $\bar{\mathcal A}$ due to consecutive encounters as
\begin{eqnarray}
\left.\frac{d\bar{\mathcal A}^2}{dt}\right|_{\mathrm{collisions}} &=& \left(\frac{m_1}{m_2}\right)^2 \langle {\mathcal A}_1^2\rangle\; \tau_{ce}^{-1}\\
&=& \frac{m_1}{m_2^2}\;\int_{-\infty}^\infty \frac32\, \Sigma\, \Omega \;\left|b\right|^3\; W_{b+c}(b) \, \mathrm{d}b, \label{eq:randwalka:collisions}
\end{eqnarray}
which should be compared to the corresponding expression for the eccentricity in equation \ref{eq:edot}.

\subsection{Random walk due to stochastic forces from circulating particles and clumps}
Particles on a circulating trajectory (see path (a) in figure \ref{fig:trajectory}) that come close the moonlet will exert a stochastic gravitational force.
The largest contributions occur for particles within a few Hill radii.
Thus, we can crudely estimate the magnitude of the specific gravitational force (acceleration) due to a single particle as
\begin{eqnarray}
f_{cp} &=& \frac{G\,m_1}{(2r_H)^2}.
\end{eqnarray}
When self-gravity is important, gravitational wakes (or clumps) have to be taken into account, as was done in section \ref{circpart}. 
Then a rough estimate of the largest specific gravitational force due to self gravitating clumps is 
\begin{eqnarray}
f_{cc} &=& \frac{G\,m_\mathrm{T}}{\left(2r_H+\lambda_T\right)^2},
\end{eqnarray}
where $2r_H$ has been replaced by $2r_H+\lambda_T$. This allows for the fact 
that $\lambda_T$ could be significantly larger than $r_H$, in which case the approximate distance of the clump to the moonlet is $\lambda_T$ (see also figure \ref{fig:randwalk:discforce} and discussion in section \ref{sec:scaling}). 

Following the formalism of chapter \ref{ch:randwalk}, we define a diffusion coefficient as 
$D= 2\tau_c \langle f^2\rangle,$ where $f$ is an acceleration, $\tau_c$ is the correlation time
and the angle brackets denote a mean value.
We here take the correlation time associated with
these forces to be the orbital period, $2\pi\Omega^{-1}.$
This is the natural dynamical timescale of the system and has been found to be a reasonable assumption from an analysis of the simulations described below.
Using eqation \ref{eq:growtha},  we may estimate the random walk in $\bar {\mathcal A}$ due to circulating particles and self-gravitating clumps 
to be given by
\begin{eqnarray}
\left.\frac{d\bar{\mathcal A}^2}{dt}\right|_{\mathrm{circulating\;particles}} &=& 4 D_{cp}\,\Omega^{-2}
= 16\pi\, \Omega^{-3}  f_{cp}^2 \\
&=&16 \pi \, \Omega^{-3} \left(\frac{G\,m_1}{(2r_H)^2}\right)^2\label{eq:randwalka:circulatingparticles}\\
\left.\frac{d\bar{\mathcal A}^2}{dt}\right|_{\mathrm{circulating\;clumps}} &=& 4 D_{cc}\,\Omega^{-2}
= 16\pi\, \Omega^{-3}  f_{cc}^2\\
&=&16 \pi\,\Omega^{-3}\left(
\frac{G\,m_\mathrm{T}}{\left(2r_H+\lambda_T\right)^2}\right)^2. \label{eq:randwalka:circulatingclumps}
\end{eqnarray}
This is only a crude estimate of the random walk undergone by the moonlet. In reality several additional effects might also play a role. For example, circulating particles and clumps are clearly correlated, the gravitational wakes have a large extent in the azimuthal direction and particles that spend a long time in the vicinity of the moonlet have more complex trajectories. 
Nevertheless, we find that the above estimates are correct up to a factor 2 for all the simulations that we performed (see below).

\subsection{Random walk due to particles in horseshoe orbits}
Finally, let us calculate the random walk induced by particles on horseshoe orbits.
Particles undergoing horseshoe turns on opposite sides of the planet produce changes of opposite sign.
Encounters with the moonlet are stochastic and therefore the semi-major axis will undergo a random walk.
A single particle with impact parameter $b$ will change the semi-major axis of the moonlet by
\begin{eqnarray}
\Delta\bar{\mathcal A} = 2 \frac{m_1}{m_1+m_2} b.
\end{eqnarray}
Analogous to the analysis in section \ref{moonlet:sec:randacol}, the time evolution of $\bar {\mathcal A}$ is then governed by
\begin{eqnarray}
\left.\frac{d\bar{\mathcal A}^2}{dt}\right|_{\mathrm{horseshoe}} &=& 4 \left(\frac{m_1}{m_2}\right)^2 b^2\; \tau_{he}^{-1}\\
&=& 6 \frac{m_1}{m_2^2}\;\int_{-\infty}^\infty \Sigma\, \Omega \;\left|b\right|^3\; W_d(b) \, \mathrm{d}b.\label{eq:randwalka:horseshoe}
\end{eqnarray}
Note that this equation is identical to equation \ref{eq:randwalka:collisions} except a factor 4, as particles with impact parameter $b$ will leave the vicinity of the moonlet at $-b$.

\clearpage
\section{Numerical Simulations}\label{moonlet:sec:numerical}
\begin{table*}[tb]
\begin{center}
\begin{tabular}{l|rr|rrr}
Name 		&	\multicolumn{1}{c}{$\Sigma$} 		& \multicolumn{1}{c|}{$r_2$} 		& \multicolumn{1}{c}{$dt$} & \multicolumn{1}{c}{$L_x\times L_y$} 				& \multicolumn{1}{c}{$N$} \\\hline\hline
\texttt{EQ5025}	& 	$50\,\mathrm{kg/m^2}$	& $25\,$m			& 4s& $1000\,\mathrm{m}\times1000\,\mathrm{m}$	& 7.2k \\
\texttt{EQ5050}	& 	$50\,\mathrm{kg/m^2}$	& $50\,$m			& 4s& $1000\,\mathrm{m}\times1000\,\mathrm{m}$	& 7.2k \\
\texttt{EQ20025}& 	$200\,\mathrm{kg/m^2}$	& $25\,$m			& 4s& $1000\,\mathrm{m}\times1000\,\mathrm{m}$	& 28.8k \\
\texttt{EQ20050}& 	$200\,\mathrm{kg/m^2}$	& $50\,$m			& 4s& $1000\,\mathrm{m}\times1000\,\mathrm{m}$	& 28.8k \\
\texttt{EQ40025}& 	$400\,\mathrm{kg/m^2}$	& $25\,$m			& 4s& $1000\,\mathrm{m}\times1000\,\mathrm{m}$	& 57.6k \\
\texttt{EQ40050}& 	$400\,\mathrm{kg/m^2}$	& $50\,$m			& 4s& $1000\,\mathrm{m}\times1000\,\mathrm{m}$	& 57.6k \\
\texttt{EQ40050DT}& 	$400\,\mathrm{kg/m^2}$	& $50\,$m			& 40s& $1000\,\mathrm{m}\times1000\,\mathrm{m}$	& 57.6k \\
\texttt{EQ5050DTW}& 	$50\,\mathrm{kg/m^2}$	& $50\,$m			& 40s& $2000\,\mathrm{m}\times2000\,\mathrm{m}$	& 28.8k \\
\texttt{EQ20050DTW}& 	$200\,\mathrm{kg/m^2}$	& $50\,$m			& 40s& $2000\,\mathrm{m}\times2000\,\mathrm{m}$	& 115.2k \\
\texttt{EQ40050DTW}& 	$400\,\mathrm{kg/m^2}$	& $50\,$m			& 40s& $2000\,\mathrm{m}\times2000\,\mathrm{m}$	& 230.0k \\
\end{tabular}
\caption{Initial simulation parameters. The second column gives the surface density of the ring. The third gives the moonlet radius. The fourth column gives the time step. The fifth and sixth columns give the lengths of the main box as measured in the $xy$-plane
and the number of particles used, respectively. \label{tab:sim}}
\end{center}
\end{table*}

We perform realistic three dimensional simulations of Saturn's rings with an embedded moonlet and verify the analytic estimates presented above. 
The nomenclature and parameters used for the simulations are listed in table \ref{tab:sim}.

\subsection{Methods} \label{moonlet:sec:methods}
We use the GravTree code and a symplectic integrator for Hill's equations \citep{Quinn2010}. Both are described in more detail in appendix \ref{app:gravtree}.
The gravitational forces are calculated with a Barnes-Hut tree \citep{Barnes1986}. 

To further speed up the calculations, we run two coupled simulations in parallel. 
The first adopts the main box which incorporates a moonlet. The second adopts an auxiliary box
which initially has the same number of particles but which does not contain a moonlet.
This is taken to be representative of the unperturbed background ring.
We describe these so called \textit{pseudo} shear periodic boundary conditions in more detail in appendix \ref{app:gravtree}.

The moonlet is allowed to move freely in the main box. 
However, in order to prevent it leaving the computational domain, as
soon as the moonlet has left the innermost part (defined as extending one eighth of the box size), 
all particles are shifted together with the box boundaries, such that the moonlet is returned to the centre of the box. 
This is possible because the shearing box approximation is invariant with respect to translations in the $xy$ plane (see equation \ref{eq:hills}). 

Collisions between particles are resolved using the instantaneous collision model and a 
velocity dependent coefficient of restitution given by \cite{Bridges1984}:
\begin{eqnarray}
\epsilon(v) &= &\mathrm{min}\left[0.34\cdot \left( \frac{v_\parallel}{1\mathrm{cm/s}} \right)^{-0.234}, 1\right],
\end{eqnarray}
where $v_\parallel$ is the impact speed projected on the axis between the two particles.

\subsection{Initial conditions and tests}

Throughout this chapter, we use a distribution of particle sizes, $r_1$, ranging from $1$m to $5$m with a slope of $q=-3$.
Thus $dN/dr_1 \propto r_1^{-q}.$ 
The density of both the ring particles and the moonlet is taken to be $0.4 \,\mathrm{g/cm}^3$. 
This is at the lower end of what has been assumed reasonable for Saturn's A ring \citep{LewisStewart2009}. The moonlet radius is taken to be either $50\,\mathrm{m}$
or $25\,\mathrm{m}.$
We found that using a larger ring particle and moonlet density only leads to more particles being bound to the moonlet. 
This effectively increases the mass of the moonlet (or equivalently its Hill radius) and can therefore
be easily scaled to the formalism presented here. 
Simulating a gravitational aggregate of this kind
is computationally very expensive, as many more collisions have to be resolved each time-step. 

The initial velocity dispersion of the particles is set to $1\,\mathrm{mm/s}$ for the $x$ and $y$ components and $0.4\,\mathrm{mm/s}$ for the $z$ component. 
The moonlet is placed at a semi-major axis of $a=130000\,\mathrm{km}$, corresponding to an orbital period of $P=13.3\,\mathrm{hours}$.
Initially, the moonlet is placed in the centre of the shearing box on a circular orbit. 

The dimensions of each box, as viewed in the $(x,y)$ plane, are specified in table \ref{tab:sim}. 
Because of the small dispersion velocities, the vertical motion is automatically strongly confined to the mid-plane.
After a few orbits, the simulations reach an equilibrium state in which the velocity dispersion of particles does not change any more.

We ran several tests to ensure that our results are converged. 
Simulation \texttt{EQ40050DT} uses a ten times larger time step than simulation \texttt{EQ40050}. 
Simulations \texttt{EQ5050DTW}, \texttt{EQ20050DTW} and \texttt{EQ40050DTW} use a box that is twice the size of that used in simulation \texttt{EQ40050}. 
Therefore, four times more particles have been used. 
No differences in the equilibrium state of the ring particles, nor in the equilibrium state of the moonlet have been observed in any of those test cases.

\begin{figure}[t]
\centering
\includegraphics[angle=270,width=0.99\textwidth]{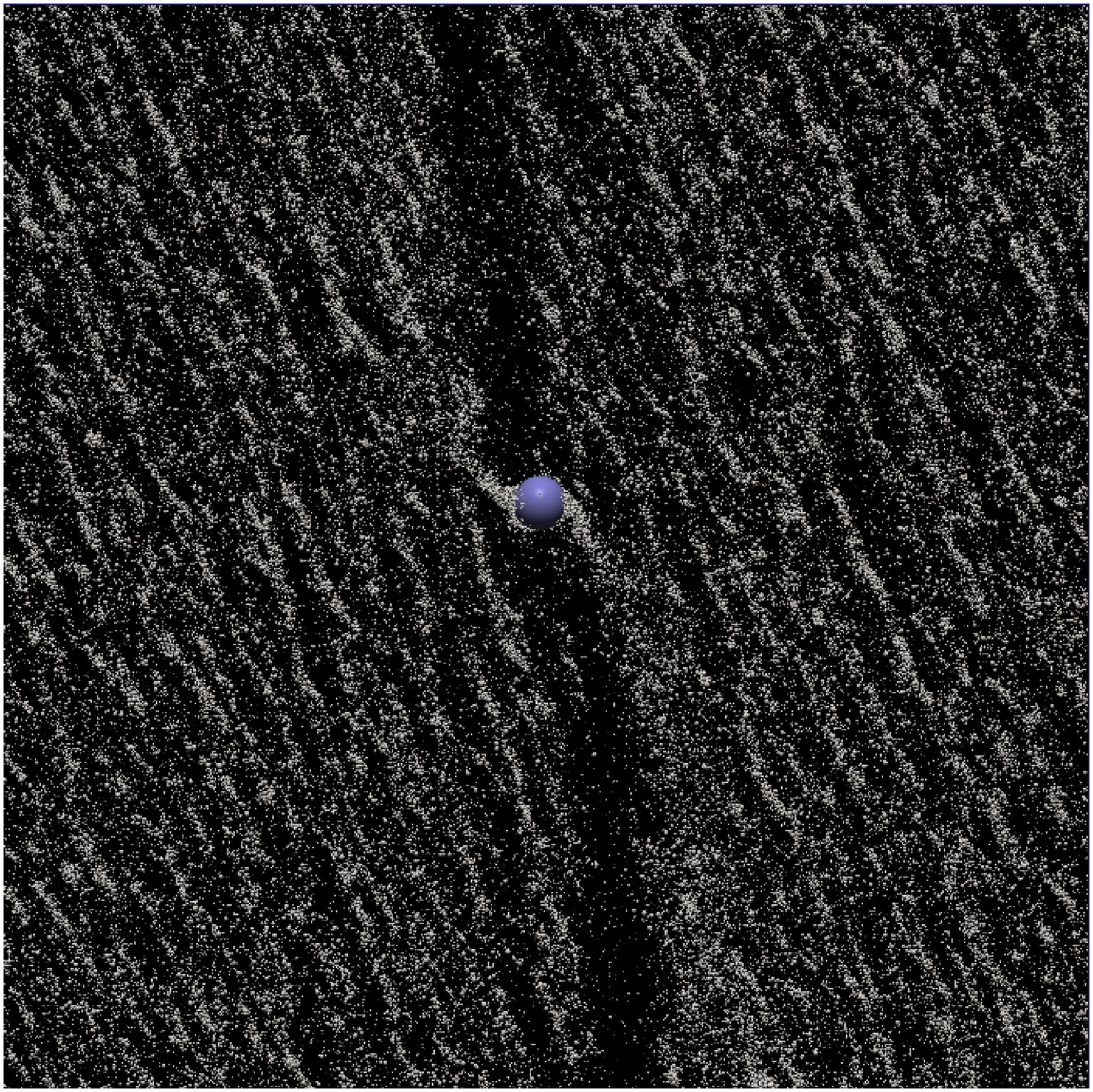}
\caption{Particle distribution in the $xy$ plane in simulation \texttt{EQ40050DTW}. The moonlet's wake is clearly visible.
\label{fig:snapdtw}}
\end{figure}

\begin{figure*}[p]
\centering
\subbottom[$\Sigma=50\,\mathrm{kg/m^2}$, $r_2=25\,\mathrm{m}$]{
\includegraphics[angle=270,width=0.65\textwidth]{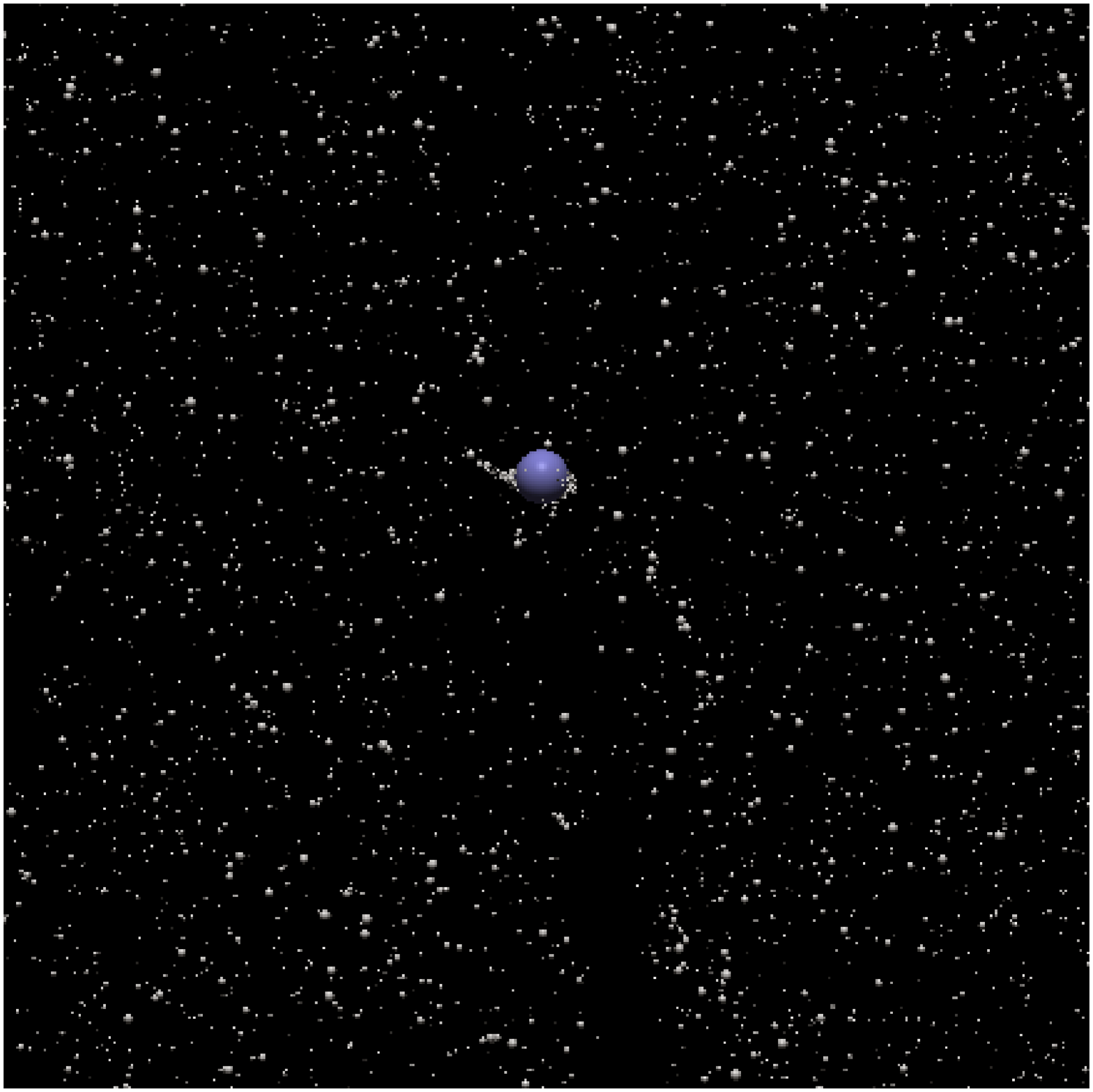}
}
\subbottom[$\Sigma=50\,\mathrm{kg/m^2}$, $r_2=50\,\mathrm{m}$]{
\includegraphics[angle=270,width=0.65\textwidth]{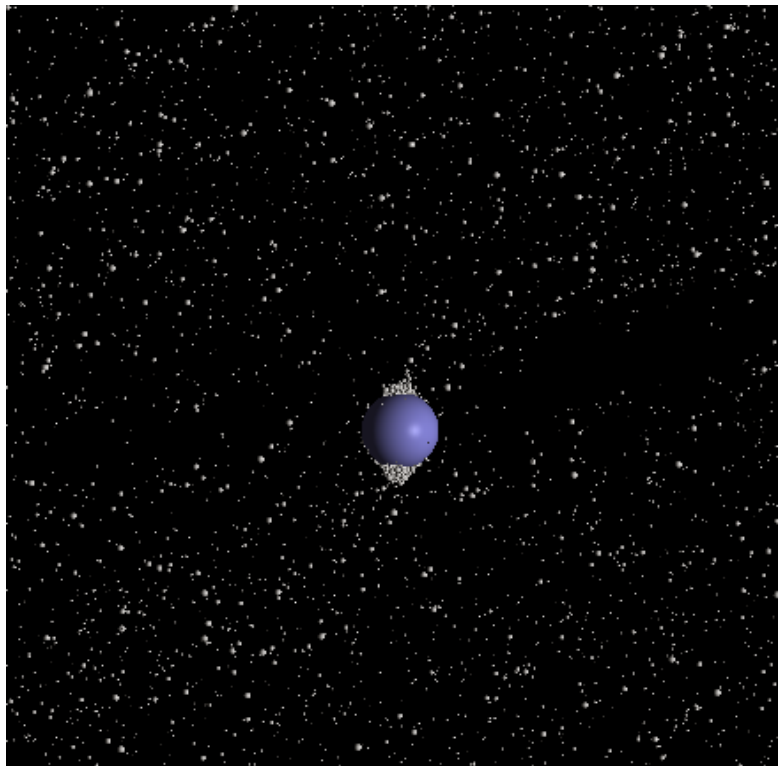}
}
\caption{Particle distribution in the $xy$ plane in simulations \texttt{EQ5025} and  \texttt{EQ5050} at a time when the moonlet and the ring particles are in equilibrium. The mean optical depth is $\tau=0.04$.
\label{fig:moonletsnap1}}
\end{figure*}

\begin{figure*}[p]
\centering
\subbottom[$\Sigma=200\,\mathrm{kg/m^2}$, $r_2=25\,\mathrm{m}$]{
\includegraphics[angle=270,width=0.65\textwidth]{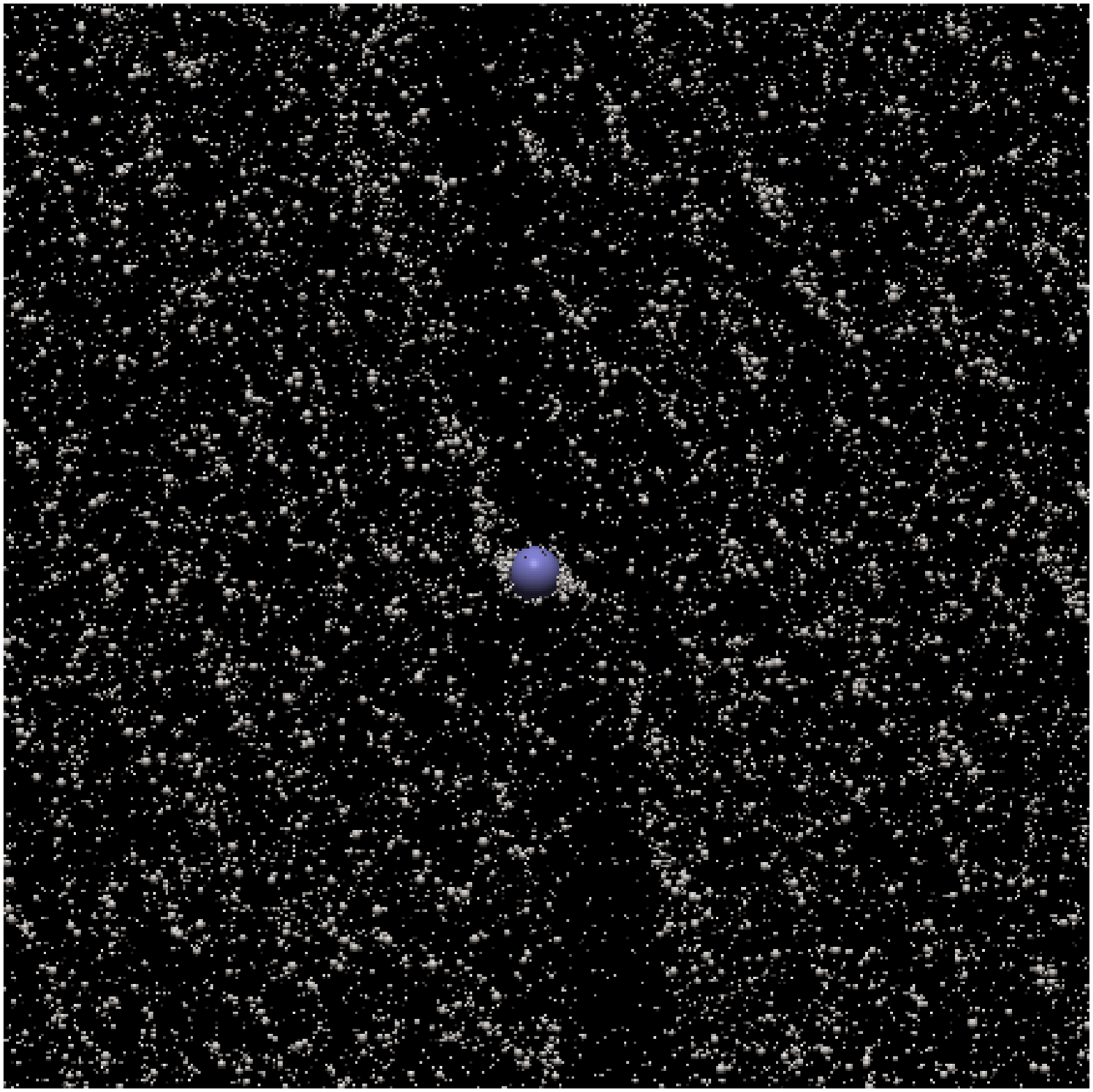}
}
\subbottom[$\Sigma=200\,\mathrm{kg/m^2}$, $r_2=50\,\mathrm{m}$]{
\includegraphics[angle=270,width=0.65\textwidth]{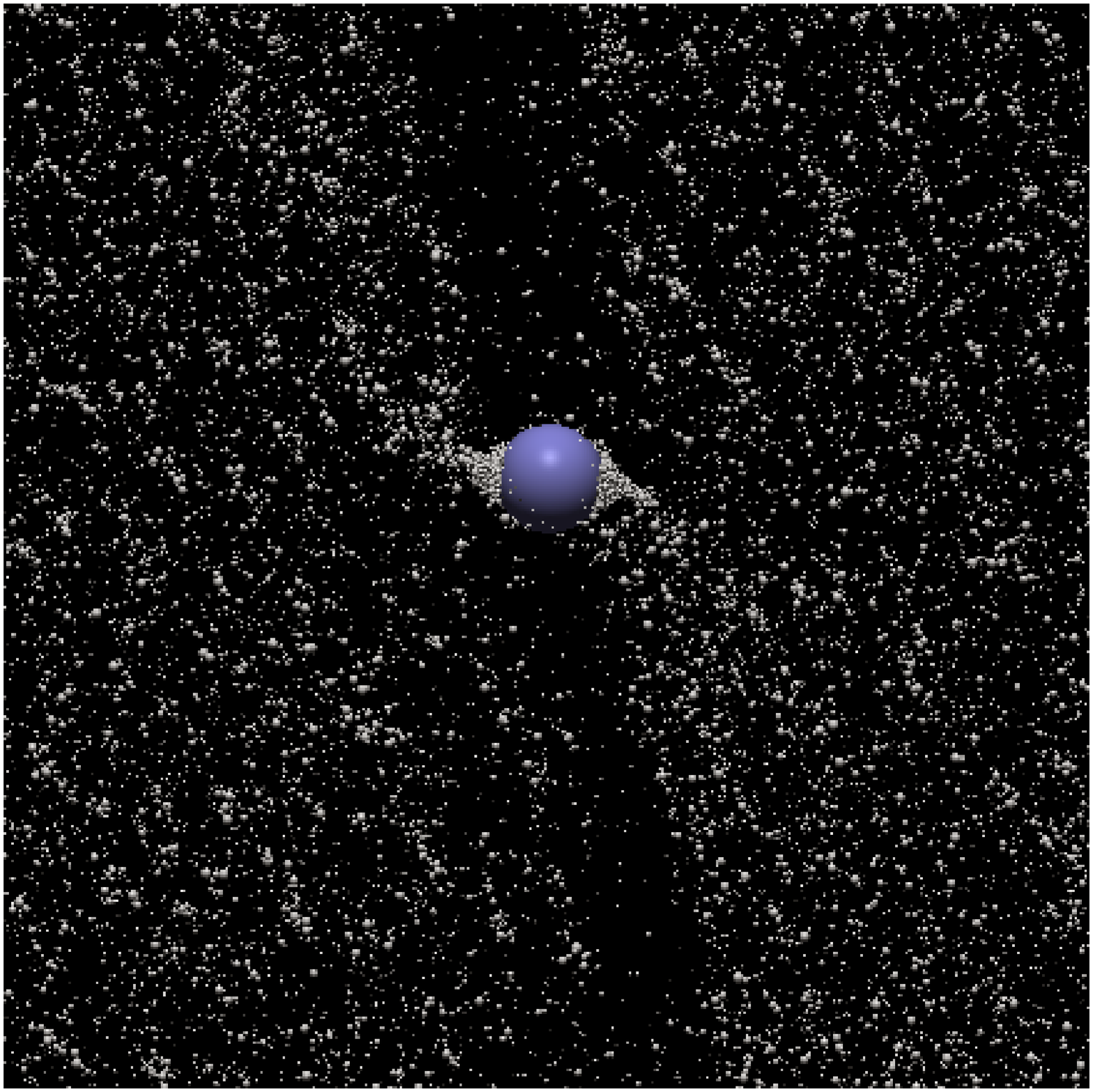}
}
\caption{Same as in figure \ref{fig:moonletsnap1} but for simulations \texttt{EQ20025} and  \texttt{EQ20050}. The mean optical depth is $\tau=0.15$.
\label{fig:moonletsnap2}}
\end{figure*}

\begin{figure*}[p]
\centering
\subbottom[$\Sigma=400\,\mathrm{kg/m^2}$, $r_2=25\,\mathrm{m}$]{
\includegraphics[angle=270,width=0.65\textwidth]{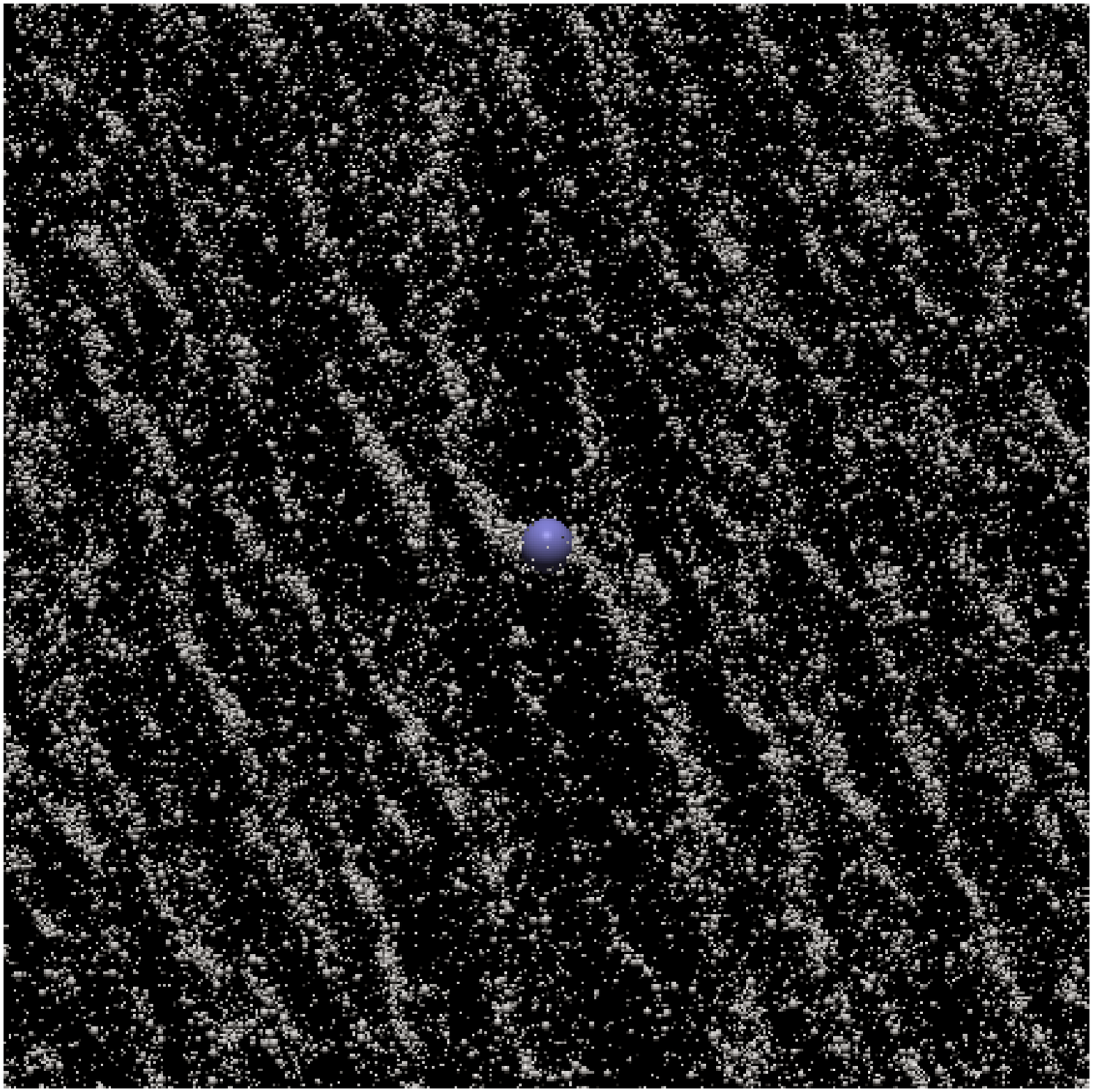}
}
\subbottom[$\Sigma=400\,\mathrm{kg/m^2}$, $r_2=50\,\mathrm{m}$]{
\includegraphics[angle=270,width=0.65\textwidth]{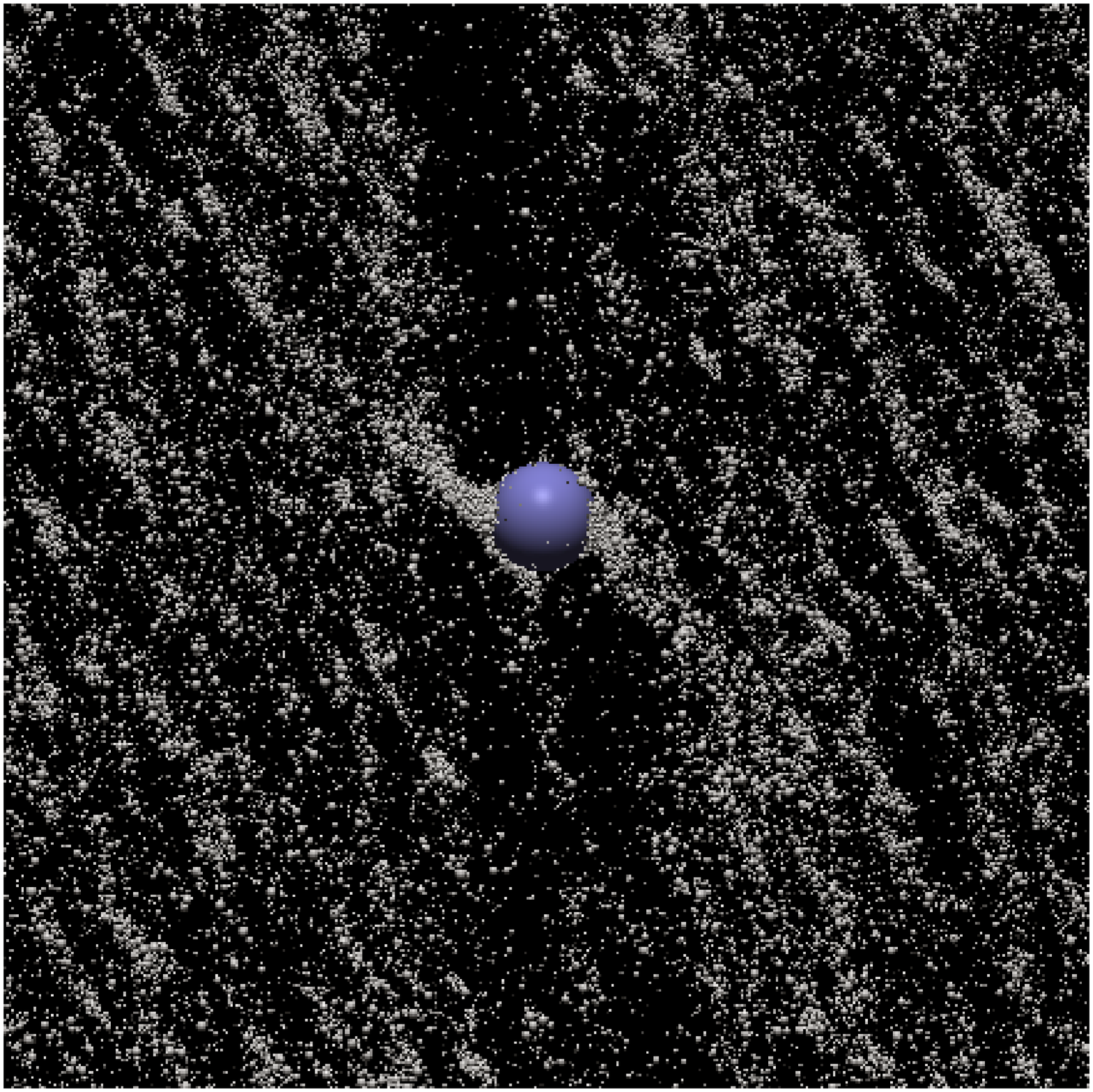}
}
\caption{Same as in figure \ref{fig:moonletsnap1} but for simulations \texttt{EQ40025} and  \texttt{EQ40050}. The mean optical depth is $\tau=0.30$.
\label{fig:moonletsnap3}}
\end{figure*}

A snapshot of the particle distribution in simulation \texttt{EQ40050DTW} is shown in figure \ref{fig:snapdtw}.
Snapshots of simulations \texttt{EQ5025}, \texttt{EQ5050}, \texttt{EQ20025}, \texttt{EQ20050}, \texttt{EQ40025} and \texttt{EQ40050}, which use 
the smaller box size, are shown in figures \ref{fig:moonletsnap1}, \ref{fig:moonletsnap2} and \ref{fig:moonletsnap3}.
In all these cases, the length of the wake that is created by the moonlet is much longer than the box size. 
Nevertheless, the pseudo shear periodic boundary conditions allow us to simulate the dynamical evolution of the moonlet accurately.


\section{Results} \label{moonlet:sec:results}

\subsection{Impact bands}
\label{sec:moonlet:window}
The impact band window functions 
$W_a(b)$, $W_{b+c}(b)$ and $W_{d}(b)$ are necessary to estimate the damping timescales and the strength of the excitation.
To measure those, each particle that enters the box is labelled with its impact parameter. 
The possible outcomes are plotted in figure \ref{fig:trajectory}: (a) being a circulating particle which leaves the box on the opposite side, (b) representing a collision with other ring particles close to the moonlet where the maximum distance 
from the centre of the moonlet 
has been taken to be twice the moonlet radius, (c) being a direct collision with the moonlet and finally (d) showing a horseshoe orbit in which the particle leaves the box on the same side that it entered. 

The impact band (b) is considered in addition to the impact band (c) because some ring particles will collide with other ring particles that are (temporarily) bound to the moonlet. Thus, these collisions take part in the transfer the energy and momentum to the moonlet. Actually, if the moonlet is simply a rubble pile of ring particles as suggested by \cite{Porco2007}, then there might be no solid moonlet core and all collisions are in the impact band (b). 

\begin{figure*}[p]
\centering
\includegraphics[angle=270,width=\textwidth]{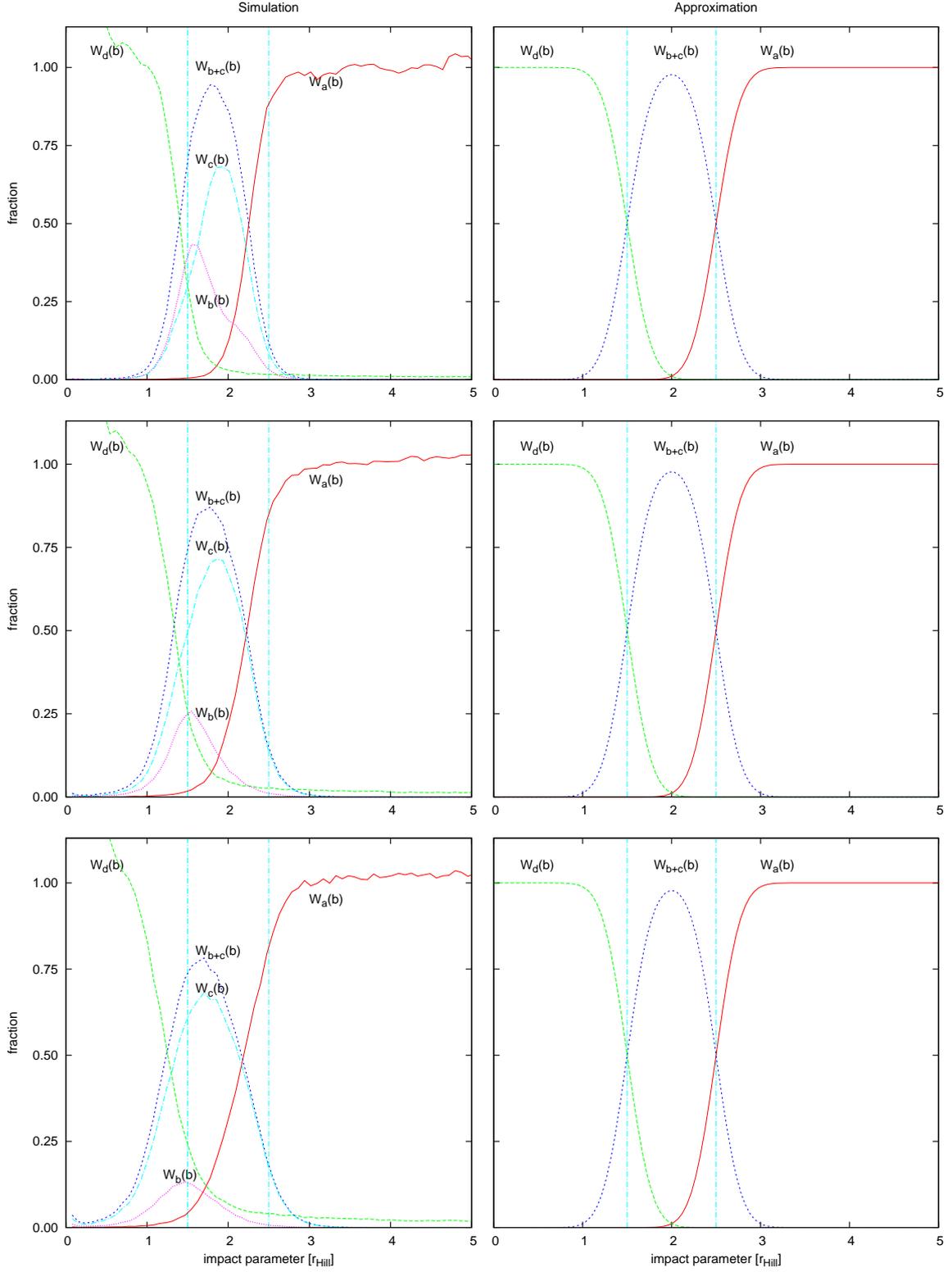}
\caption{Impact bands for simulations with a moonlet radius $r_2=25\,\mathrm{m}$. Left and right columns show the window functions measured in simulations and approximations of those, respectively (see equations \ref{eqW1}-\ref{eqW3}).
The surface density of the ring is, from top to bottom, $\Sigma=50\,\mathrm{kg/m^2}$, $\Sigma=200\,\mathrm{kg/m^2}$ and $\Sigma=400\,\mathrm{kg/m^2}$.
\label{fig:impactbands1}}
\end{figure*}

\begin{figure*}[p]
\centering
\includegraphics[angle=270,width=\textwidth]{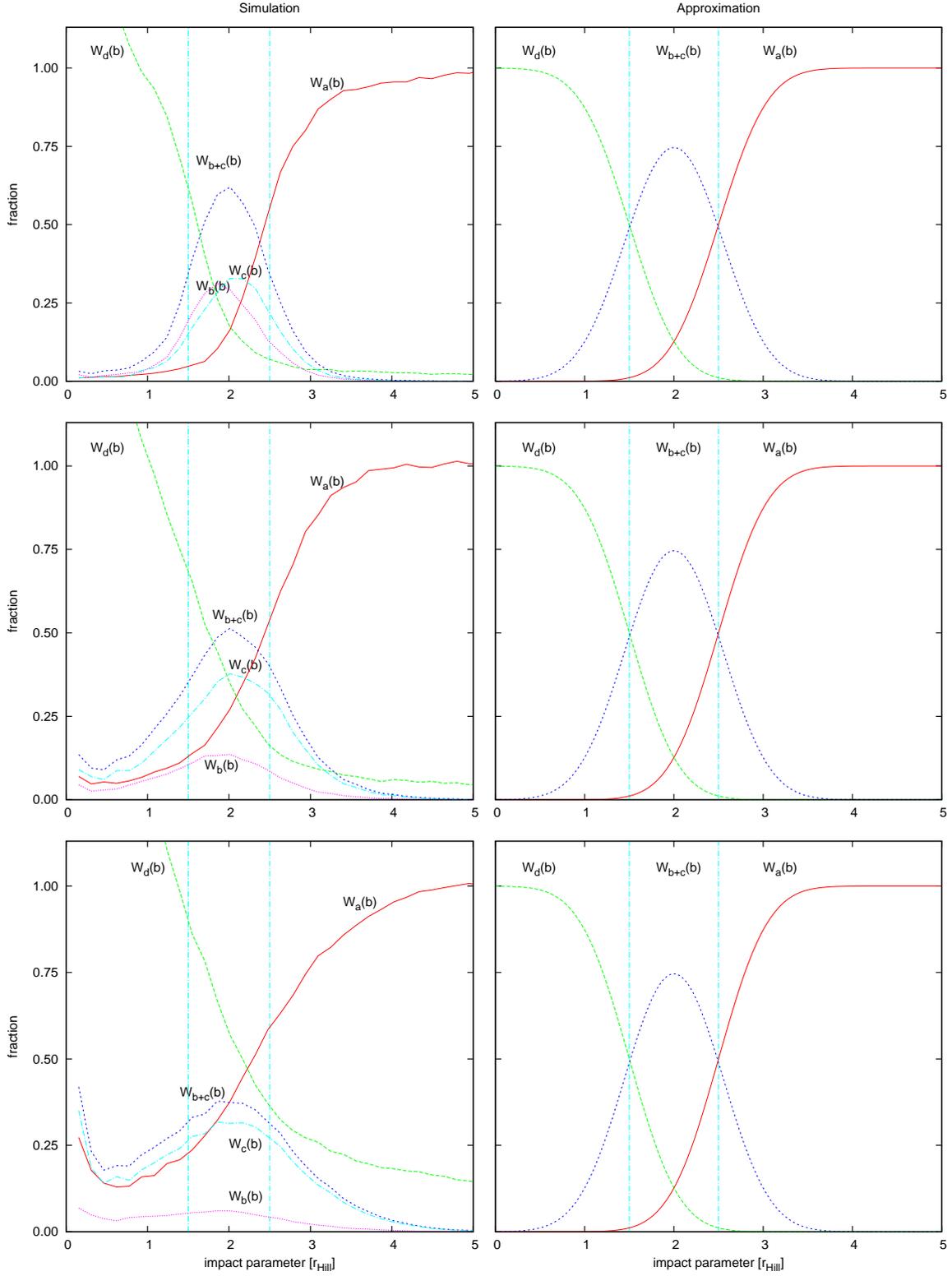}
\caption{Same as in figure \ref{fig:impactbands1} but for simulations with a moonlet radius $r_2=50\,\mathrm{m}$. 
\label{fig:impactbands2}}
\end{figure*}

Figures \ref{fig:impactbands1} and \ref{fig:impactbands2} show the impact bands, normalised to the Hill radius of the moonlet $r_{\mathrm{H}}$ for simulations \texttt{EQ5025}, \texttt{EQ20025}, \texttt{EQ40050}, \texttt{EQ5050}, \texttt{EQ20050} and \texttt{EQ40050}. 
For comparison, we also plot sharp cutoffs, at $1.5r_H$ and $2.5r_H$, as these have been used in same of the scaling laws presented above. 

Note that $\sum_{i=a,b,c,d}W_i\gtrsim1$ because ring particles are shifted from time to time when the moonlet is too far away from the origin (see section \ref{moonlet:sec:methods}).  
In this process, particles with the same initial impact parameter might become associated with the more than one impact band.

Our results show no sharp discontinuity because of the velocity dispersion in the ring particles which is $\sim 5\,\mathrm{mm/s}$. 
The velocity dispersion corresponds to an epicyclic amplitude of ${\mathcal E}_{rp}\sim20\,\mathrm{m}$ or equivalently $\sim0.4\,r_H$ and $\sim0.8\,r_H$ (for $r_2=50\mathrm{m}$ and $r_2=25\mathrm{m}$, respectively).
Note that the Hill radius is approximately equal to the physical radius of the moonlet. This is because the moonlet will always fill its Roche lobe with ring particles \citep[see also][]{Porco2007}. 

The impact bands are almost independent of the surface density and depend only on the Hill radii and the mean epicyclic amplitude of the ring particles, or, more precisely, the ratio thereof. 
Thus we expect the impact bands to have a cutoff width of ${\mathcal E}_{rp}$. 
A simple approximation of the impact bands may then be given by  
\begin{eqnarray}
W_a(b) &=& \frac12 + \frac12 \;\mathrm{erf}\left(\frac{b/r_H-2.5}{{\mathcal E}_{rp}/r_H}\right),\label{eqW1}\\
W_d(b) &=& \frac12 + \frac12 \;\mathrm{erf}\left(\frac{1.5-b/r_H}{{\mathcal E}_{rp}/r_H}\right)\quad\mathrm{and}\\
W_{b+c}(b) &=& 1-W_a(b)-W_d(b).\label{eqW3}
\end{eqnarray}
These curves are plotted on the right hand side of figures \ref{fig:impactbands1} and \ref{fig:impactbands2}, being qualitatively very similar to the measured impact bands. 
This approximation ignores the non-spherical shape of the Roche lobe and assumes $r_H=r_2$.

\clearpage

\subsection{Eccentricity damping timescale}
To measure the eccentricity damping timescale, we first let the ring particles and the moonlet reach an equilibrium and integrate them for 200 orbits. 
We then change the velocity of the moonlet and the ring particles within $2r_{\mathrm{H}}$. 
The new velocity corresponds to an eccentricity of $6\cdot 10^{-7}$, which is well above the equilibrium value. 
We then measure the decay timescale $\tau_e$ by fitting a function of the form 
\begin{eqnarray}
e_d(t)=\langle e_{eq}\rangle+\left(6\cdot10^{-7}-\langle e_{eq}\rangle\right) \; e^{-t/\tau_e}.
\end{eqnarray}
The results are given in table \ref{tab:simresults1}. 
They are in good agreement with the estimated damping timescales from section \ref{moonlet:sec:edampcollision} and \ref{moonlet:sec:edamphorseshoe}, showing clearly that the most important damping process, in every simulation considered here, is indeed through collisional damping as predicted by comparing equation \ref{eq:taucol} with equations \ref{eq:damphorse} and \ref{Goldcirc2}.

\begin{table}[t]
\begin{center}
\begin{tabular}{l|r|rrr}
Name 		&\multicolumn{1}{c|}{Numerical results}	&\multicolumn{2}{c}{Analytic results }\\	
	& 	& collisions & horseshoe \\\hline\hline
\texttt{EQ5025}		
	& 18.9	& 18.4 & 1712 	\\	
\texttt{EQ5050}		
	& 27.6 	& 33.8 & 3424	\\
\texttt{EQ20025}	
	& 9.0	& 3.9 & 428	\\
\texttt{EQ20050}	
	& 12.9	& 8.5 & 856	\\
\texttt{EQ40025}	
	& 2.8	& 2.1 & 214	\\
\texttt{EQ40050}	
	& 5.5	& 4.4 & 428
\end{tabular}
\caption{Eccentricity damping timescale $\tau_e$ of the moonlet in units of the orbital period. The second column lists the simulation results. The third and fourth column list the analytic estimates of collisional and horseshoe damping timescales, respectively. \label{tab:simresults1}}
\end{center}
\end{table}

\subsection{Mean moonlet eccentricity}

\begin{sidewaystable}[p]
\begin{center}
\begin{tabular}{l|r|rrr|r}
Name &\multicolumn{1}{c|}{Numerical results}	
		&\multicolumn{4}{c}{Analytic results}\\	
	&
	& collisions
	& circulating particles
	& \multicolumn{1}{c|}{circulating clumps}
	& \multicolumn{1}{c}{total}
		 	\\\hline\hline
\texttt{EQ5025}	&	
	$6.1\cdot10^{-8}$	&$4.5\cdot10^{-8}$ &$1.9\cdot10^{-8}$ &$0.3\cdot10^{-8}$ &$4.9\cdot10^{-8}$ \\	
\texttt{EQ5050}	& 	
	$4.3\cdot10^{-8}$ 	&$1.6\cdot10^{-8}$ &$1.4\cdot10^{-8}$ &$0.2\cdot10^{-8}$ &$2.1\cdot10^{-8}$ \\
\texttt{EQ20025}&	
	$6.4\cdot10^{-8}$	&$4.6\cdot10^{-8}$ &$1.7\cdot10^{-8}$ &$2.0\cdot10^{-8}$ &$5.3\cdot10^{-8}$ \\
\texttt{EQ20050}&	
	$4.9\cdot10^{-8}$ 	&$1.6\cdot10^{-8}$ &$1.4\cdot10^{-8}$ &$1.6\cdot10^{-8}$ &$2.7\cdot10^{-8}$ \\
\texttt{EQ40025}&	
	$11.3\cdot10^{-8}$ 	&$4.6\cdot10^{-8}$ &$2.5\cdot10^{-8}$ &$8.4\cdot10^{-8}$ &$9.9\cdot10^{-8}$ \\
\texttt{EQ40050}&	
	$7.2\cdot10^{-8}$ 	&$1.6\cdot10^{-8}$ &$1.5\cdot10^{-8}$ &$4.9\cdot10^{-8}$ &$5.4\cdot10^{-8}$
\end{tabular}
\caption{Equilibrium eccentricity $\langle e_{\mathrm{eq}}\rangle$ of the moonlet. 
The second column lists the simulation results. 
The third, fourth and fifth column list the analytic estimates of the equilibrium eccentricity assuming a single excitation mechanism. 
The last column lists the analytic estimate of the equilibrium eccentricity summing over all excitation mechanisms. \label{tab:simresults2}}
\end{center}
\vspace{1cm}
\begin{center}
\begin{tabular}{l|r|rrrr|r}
Name 		&\multicolumn{1}{c|}{Numerical results}	
		&\multicolumn{4}{c}{Analytic results }\\	
		
	& 	
	& horseshoe
	& collisions 
	& circulating particles 
	& circulating clumps
	& total 
		 	\\\hline\hline
\texttt{EQ5025}		
	&22.7m 		& 9.0m	& 10.5m 	& 17.6m 	& 0.3m		& 22.4m 	\\	
\texttt{EQ5050}		
	&12.6m 		& 4.5m	& 4.8m 		& 4.4m 		& 0.1m		& 7.9m	 	\\
\texttt{EQ20025}	
	&38.1m 		& 18.1m	& 25.1m 	& 17.6m 	& 15.7m		& 38.9m		\\
\texttt{EQ20050}	
	&22.7m 		& 13.6m	& 9.6m 		& 4.4m 		& 4.8m		& 14.7m 	\\
\texttt{EQ40025}	
	&78.1m 		& 25.6m	& 36.5m 	& 17.6m	 	& 86.6m		& 100.7m	\\
\texttt{EQ40050}	
	&45.6m 		& 12.8m	& 13.4m 	& 4.4m 		& 31.4m		& 36.7m
\end{tabular}
\caption{Change in semi-major axis of the moonlet after 100 orbits, ${\mathcal A}(\Delta t=100\,\mathrm{orbits})$. 
The second column lists the simulation results. 
The third till sixth columns list the analytic estimates of the change in semi-major axis assuming a single excitation mechanism.
The last column lists the total estimated change in semi-major axis summing over all excitation mechanisms. \label{tab:simresults3}}
\end{center}
\end{sidewaystable}
The time averaged mean eccentricity of the moonlet\footnote{Here, we are calculating the mean eccentricity, not the root-mean-square eccentricity.}
is measured in all simulations after several orbits when the ring particles and the moonlet have reached an equilibrium state. 
To compare this value with the estimates from section \ref{moonlet:sec:analytic} we set equation \ref{equile} equal to zero and use the analytic damping timescale listed in table \ref{tab:simresults1}.
The analytic estimates of the equilibrium eccentricity are calculated for each excitation mechanism separately to disentangle their effects. 
They are listed in the third, fourth and fifth column in table \ref{tab:simresults2}. 
The sixth column lists the analytic estimate for the mean eccentricity using the sum of all excitation mechanisms. 

For all simulations, the estimates are correct within a factor of about 2. 
For low surface densities, the excitation is dominated by individual particle collisions.
For larger surface densities, it is dominated by the excitation due to circulating self gravitating clumps (gravitational wakes). 
The estimates and their trends are surprisingly accurate, as we have ignored several effects (see below).

\subsection{Random walk in semi-major axis}
The random walk of the semi-major axis $a$ (or equivalently the centre of epicyclic motion ${\mathcal A}$ in the shearing sheet) of the moonlet in the numerical simulations are measured and compared to the analytic estimates presented in section \ref{moonlet:sec:randa}.
We run one simulation per parameter set. 
To get a statistically meaningful expression for the average random walk after a given time ${\mathcal A}(\Delta t)$, we average all pairs of ${\mathcal A}(t)$ and ${\mathcal A}(t')$ for which $t-t'=\Delta t$. In other words, we assume the system satisfies the Ergodic hypothesis.

${\mathcal A}(\Delta t)$ then grows like $\sqrt{\Delta t}$ and we can fit a simple square root function (see chapter \ref{ch:randwalk}).
This allows us to accurately measure the average growth in ${\mathcal A}$ after $\Delta t=100$ orbits by running one simulation for a long time and averaging over time, rather than running many simulations and performing an ensemble average.
The measured values are given in the second column of table \ref{tab:simresults3}.
The values that correspond to the analytic expressions in equations \ref{eq:randwalka:horseshoe}, \ref{eq:randwalka:collisions}, \ref{eq:randwalka:circulatingparticles} and \ref{eq:randwalka:circulatingclumps} are listed in columns three, four, five and six, respectively.

For low surface densities, the evolution of the random walk is dominated by collisions and the effect of particles on horseshoe orbits. 
For large surface densities, the main effect comes from the stochastic gravitational force due to circulating clumps (gravitational wakes). 

The change in semi-major axis is relatively small, a few tens of metres after 100 orbits ($=50$ days). 
The change in longitude $\Delta \lambda$ that corresponds to this is, however, is much larger. 

Using the results from chapter \ref{ch:randwalk}, namely equations \ref{eq:growthlambda} and \ref{eq:growthlambdalaminar}, we can predict the long term evolution of $\lambda$.
In the stochastic and laminar case, $\left(\Delta\lambda\right)^2$ grows like $\sim t^3$ and $\sim t^4$, respectively. 
This behaviour allows us to discriminate between a stochastic and laminar migration by observing $\Delta \lambda$ over an extended period of time. 

\begin{figure}[tb]
\centering
	\includegraphics[angle=270,width=0.95\columnwidth]{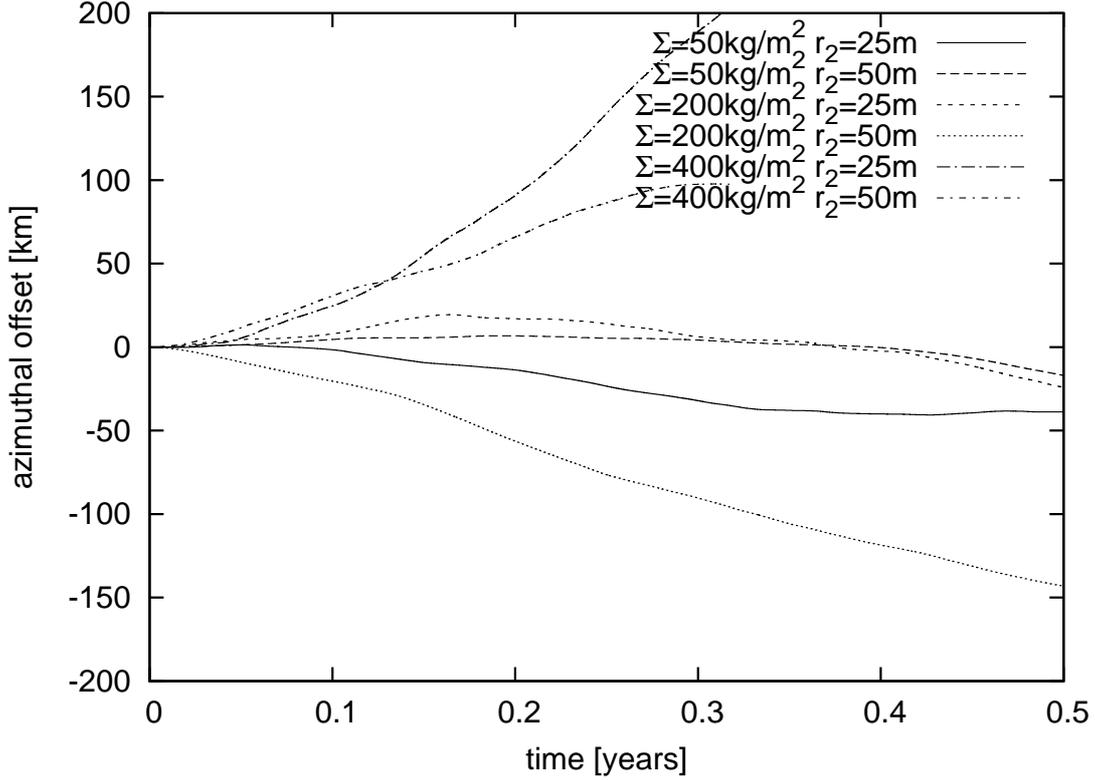}
\caption{Offset in the azimuthal distance due to the random walk in the semi-major axis $a$ measured in simulations \texttt{EQ5025}, \texttt{EQ5050}, \texttt{EQ20025}, \texttt{EQ20050}, \texttt{EQ40025} and \texttt{EQ40050}. Individual moonlets may show linear, constant or oscillatory growth. On average, the azimuthal offset grows like $t^{3/2}$.\label{fig:azimuth}}
\end{figure}

Results from individual simulations (i.e. not an ensemble average) are plotted in figure \ref{fig:azimuth}. Here, $\Delta \lambda$ is expressed in terms of the azimuthal offset relative to a Keplerian orbit. 
One can see that for an individual moonlet, the shift in azimuth can appear to be linear, constant or oscillating on short timescales (see curve for simulation \texttt{EQ20025}). 
This is partly because of the lower order terms in equation \ref{eq:growthlambdaall}, though, on average, the azimuthal offset grows very rapidly, as $\sim t^{3/2}$ for $t\gg\tau$ (see equation \ref{eq:growthlambda}).

\section{Conclusions}\label{moonlet:sec:conclusions}
In this chapter, we have discussed the dynamical response
of an embedded moonlet in Saturn's rings to interactions with ring particles both analytically and by the use of realistic three dimensional many particle simulations. 
Both the moonlet and the ring particle density were taken to be $0.4 \,\mathrm{g/cm}^3$. Moonlets of radius $25\,\mathrm{m}$ and $50\,\mathrm{m}$ were considered. 
Particle sizes ranging between $1\,\mathrm{m}$ and $5\,\mathrm{m}$ were adopted.

We estimated the eccentricity damping timescale of the moonlet due to collisions with ring particles and due to the response of ring particles to gravitational perturbations by the moonlet analytically. 
We found the effects due to the response of particles on horseshoe and circulating orbits are negligible.
On the other hand the stochastic impulses applied to the moonlet by circulating particles were found to cause the square of the eccentricity to grow linearly with time as did collisions with particles with non zero velocity dispersion.
A balance between excitation and damping processes then leads to an equilibrium moonlet eccentricity. 

We also estimated the magnitude of the random walk in the semi-major axis of the moonlet induced by collisions with individual ring particles and the gravitational interaction with particles and gravitational wakes. 
There is no tendency for the semi-major axis to relax to any particular value, so that there are no damping terms.
The deviation of the semi-major axis from its value at time $t=0$ grows on average as $\sqrt{t}$ for large $t.$

From our simulations we find that the evolution of the eccentricity is indeed dominated by collisions with ring particles. 
For large surface densities (more than $200\mathrm{kg/m^2}$) the effect of gravitational wakes also becomes important, leading to an increase in the mean steady state eccentricity of the moonlet. 
When the particle velocity dispersion is large compared to $\Omega r_H,$ we obtain approximate energy equipartition between the moonlet and ring particles as far as epicyclic motion is concerned.

Similarly, the random walk in the semi-major axis was found to be dominated by collisions for low surface densities and by gravitational wakes for large surface densities. 
In addition we have shown that on average, the difference in longitude $\Delta \lambda$ of a stochastically forced moonlet grows with time like $t^{3/2}$ for large~$t.$ 

The radial distance travelled within 100 orbits (50 days at a distance of 130000km) is, depending on the precise parameters, of the order of 10-100m. 
This translates to a shift in longitude of several hundred kilometres. 
The shift in $\lambda$ is not necessarily monotonic on short timescales (see figure \ref{fig:azimuth}).
We expect that such a shift should be easily detectable by the Cassini spacecraft.
And indeed, \cite{Tiscareno2010} report such an observation, although the data is not publicly available yet, at the time when this thesis was submitted.

There are several effects that have not been included in this study.
The analytic discussions in sections \ref{dampe}, \ref{moonlet:sec:excitee} and \ref{moonlet:sec:randa} do not model the motion of particles correctly when their trajectories take them very close to the moonlet. 
Material that is temporarily or permanently bound to the moonlet is ignored. 
The density of both the ring particles and the moonlet are at the lower end of what is assumed to be reasonable \citep{LewisStewart2009}, which has the advantage that for our computational setup only a few particles (with a total mass of less than 10\% of the moonlet) are bound to the moonlet at any given time.


\chapter{Planetesimal formation in a turbulent disc} \label{ch:planetesimals}
\epigraph{
To explain all nature is too difficult a task for any one man or even for any one age. 'Tis much better to do a little with certainty, \& leave the rest for others that come after you, than to explain all things by conjecture without making sure of any thing.
}{\tiny Isaac Newton, unpublished preface to Opticks, 1704}

\noindent In chapter \ref{ch:randwalk}, we discussed the evolution of planets and planetary systems under the effect of stochastic forces resulting from turbulence. 
The stochastic migration of moonlets in Saturn's rings was discussed in chapter \ref{ch:moonlet}. 
In this chapter, we look at another system, in which turbulence is important, namely the formation of planetesimals, the building blocks of planets.

The formation mechanism of planetesimals in proto-planetary discs is hotly debated. 
Currently, the favoured model involves the accumulation of metre-sized objects within a turbulent disc. These high spatial densities then promote the formation of planetesimals, for example via gravitational instability as discussed in more detail in chapter \ref{ch:introduction}. 
In numerical simulations that model this process, one can at most simulate a few million particles as opposed to the several trillion metre-sized particles expected in a real proto-planetary disc. Therefore, single particles are often used as super-particles to represent a distribution of many smaller particles. It is assumed that small-scale phenomena do not play a role and particle collisions are not modelled. 
However, the super-particle approximation can only be valid in a collision-less or strongly collisional system. 
In many recent numerical simulations this is not the case.

Here, a scaled system is studied that does not require the use of super-particles. 
We compare this new method to the standard super-particle approach. We find that the super-particle approach produces unreliable results that depend on numerical artifacts such as the gravitational softening in both the requirement for gravitational collapse and the resulting clump statistics.  
Our results show that short-range interactions (collisions) have to be modelled properly.

The outline of this chapter is as follows.
In section~\ref{sec:def}, we review the important timescales in the problem and show that the super-particle approach breaks down in high density regions. 
In section~\ref{sec:methods}, we discuss both the super-particle and the scaled particle method.
The numerical implementation and the initial conditions of the simulations are presented in section~\ref{sec:numerical}.
Results of numerical simulations using both the super-particle approach and our scaling method are presented in section~\ref{sec:results}. 
We conclude with presenting resolution constraints for numerical simulations and discuss the implications for the planetesimal formation process through gravitational instability in section~\ref{sec:disc}.

\section{Orders of magnitude}\label{sec:def}
\subsection{Definition of timescales}
The aim of this work is to simulate the dynamics of dust (or particles) interacting gravitationally inside a turbulent proto-planetary disc. 
Each particle is subject to five different physical processes, each operating on a typical timescale:

\begin{description}
\item \textit{Stopping time} $\tau_s$
  
Each particle of size $a$ feels the effects of the surrounding gas through a linear drag force. This force can be written as $\mathbf{F}_\text{drag}=\frac{m_p}{\tau_s} (\mathbf{v}_g-\mathbf{v}_p)$, where $\tau_s$ is the stopping time of a particle of mass $m_p$, $\mathbf{v}_g$ is the velocity of the gas, and $\mathbf{v}_p$ is the velocity of the particle \citep{W77}.
The stopping time depends on the degree of coupling to the gas \citep[e.g.][]{ChiangYoudin2009}:
\begin{eqnarray}
\tau_s= \begin{cases}
\frac{\rho_p a}{\rho_g c_s} & a\le \frac94 \lambda \quad \mathrm{Epstein\;drag}\\
\frac{4\rho_p a^2}{9\rho_g c_s \lambda} & a\ge \frac94 \lambda \quad \mathrm{Stokes\;drag},
\end{cases}
\end{eqnarray}
where $\rho_p$ and $\rho_g$ are the internal density of the particles and the density of the gas, respectively. $\lambda$ is the mean free path of gas particles and $c_s$ the sound speed. 

\item \textit{Physical collision timescale} $\tau_c$ 

Dust particles will suffer a physical collision with another dust particle on a timescale of $\tau_c=(\sigma_c \mathrm{\bar v}_p n)^{-1}$, where $\sigma_c$ is the geometrical cross-section of the particles, $\mathrm{\bar v}_p$ is the particle velocity dispersion, and $n$ is the number density of particles in a homogeneous medium. 

\item \textit{Orbital timescale} $\tau_e$ 

The timescale associated with the particles orbiting the central object in a local shearing patch is the epicyclic period $\tau_{e}=2\pi\Omega^{-1}$, where $\Omega$ is the angular velocity at the semi-major axis of the shearing box.

\end{description}

\noindent There are two limits for the gravitational interactions between dust particles, long and short range. The long-range interaction can be seen as a collective process involving interactions between clouds of particles. On the other hand, the short-range interaction is important for resolving close approaches between pairs of particles, i.e. gravitational scattering. For the sake of clarity, we separate these two processes. 

\begin{description}
\item \textit{Gravitational collapse timescale} $\tau_{Gl}$ 

On a length scale $\lambda \gg \delta_r$, where $\delta_r$ is the average distance between neighbouring dust particles, the long-range interaction of the dust particle distribution can be approximated by a continuous density $\rho$. The gravitational collapse timescale, which is of the order of the free fall time of the system, can be defined by $\tau_{Gl}=1/\sqrt{G \rho}$. \\

\item \textit{Gravitational scattering timescale} $\tau_{Gs}$ 

Short-range gravitational interactions are interactions between a pair of single particles that can result in a scattering event. As with physical collisions, the timescale for a gravitational interaction depends on the cross-section, $\tau_{Gs}=(\sigma_G \mathrm{\bar v}_p n)^{-1}$. Here, $\sigma_G\sim G^2 m_p^2/\mathrm{\bar v}_p^4$ is the gravitational cross-section of the particle. Note that the cross-section is velocity dependent, which makes it qualitatively different from physical (billard ball) collisions between two particles. 

\end{description}

\noindent Particles might also be affected by excitation from a turbulent background state on a timescale defined by the turbulence itself. 
Unlike in chapter \ref{ch:randwalk}, we here model the excitation in an even more simplified way, in which the turbulence is scale independent and has no timescale associated with it.

\subsection{The real physical system}\label{sec:real}
To identify the dominant physical processes in a proto-planetary disc, one has to quantify and compare the relevant timescales as defined above. 
In the following discussion and in the numerical simulations, we assume $R_0=1$ AU, a gas disc thickness of $H/R_0=0.01$, and a minimum mass solar nebula (MMSN) with surface gas density\footnote{Note that there are different values for MMSN in the literature. This value here is a factor 5 smaller than the one used in chapter~\ref{ch:randwalk}.}  $\Sigma=890~\text{g~cm}^{-2}$. 
One usually assumes a solid to gas ratio of~$\rho_s/\rho_g=0.01$. We, however, assume a solid to gas ratio of unity. This value is justified by numerical simulations \citep{Johansen2007,Johansen2009} that demonstrate over-densities of the order of~$10^1-10^2$ can easily occur because of the interaction with a turbulent gas disc, the streaming instability, or vertical settling. Thus, we are only interested in the late stages of the planetesimal formation process. The details of how the initial over-density has been achieved are not relevant for the discussion presented in this chapter. 
We also note that significantly different models of the solar nebula have been proposed \citep{Desch2007}. However, all results can easily be scaled to different scenarios.

We simplify the system by assuming that all solid components of the disc are in metre-sized particles. We assume that these boulders have a typical velocity dispersion caused by gas turbulence $\mathrm{\bar v}_p\sim0.05 c_s\sim 30\text{m}/\text{s}$, where $c_s$ is the local sound speed, in accordance with numerical results \citep{Johansen2007}. Since $\mathrm{\bar v}_p\ll c_s$, the particles tend to sediment toward the mid-plane, forming a finite thickness dust layer due to the non-zero velocity dispersion.

For metre-sized boulders, the physical cross-section is $\sigma_c\sim 1\,\mathrm{m}^2$, whereas the gravitational scattering cross-section is $\sigma_G\sim10^{-21}\,\mathrm{m}^2$ showing clearly that gravitational scattering is irrelevant to the dynamics of dust particles embedded in a disc. 
However, all other timescales are roughly equivalent. 
Metre-sized boulders are weakly coupled to the gas with a stopping time $\tau_s\sim \tau_e\sim \Omega^{-1}$ \citep{W77}. 
The physical collision timescale is $\tau_c=7.3 \, \Omega^{-1}$ and the long-range gravitational interaction timescales is $\tau_{Gl} \sim \Omega^{-1}$.

This physical system is expected to become gravitationally unstable, according to the Toomre criterion \citep{Toomre1964}.
The instability occurs when the gravitational collapse timescale $\tau_{Gl}$ is shorter than the transit time due to random particle motions $\lambda /\mathrm{\bar v}_p $ and the orbital timescale. 
Assuming that the particles can be modelled by an isotropic gas\footnote{This is formally not valid in the studied regime, as it is not strongly collisional.} with a sound speed $\mathrm{\bar v}_p$, the system becomes unstable when $Q\sim 1$ (equation \ref{eq:moonlet:toomre}).
In that case, the most unstable wavelength $\lambda_T$ is given by equation \ref{eq:moonlet:toomrelambda}.
In the following, we compare the physical parameters from this section to their counterpart in numerical simulations. We first summarise the super-particle approach before presenting the scaling method.

\section{Methods}
\label{sec:methods}
Two different kinds of simulation are considered in this chapter: 
\begin{enumerate}
  \item Particles are assumed to be point masses and have no physical size, the gravitational field of the particles being approximated with a smoothing length to avoid numerical divergences (see section~\ref{sec:nbody}).

  \item Particles have a physical size and, therefore, no gravitational smoothing is required but physical collisions must be included. 
\end{enumerate}
We refer to the particles as super-particles and scaled particles, respectively.

\subsection{Super-particle approximation}
Let us quickly review the standard approach using super-particles before moving on to studying our favoured method that uses scaled particles.

A super-particle represents many smaller particles. The dynamics of the small particles are not calculated exactly. It is assumed that all particles behave similarly and in a collective manner. To simulate gravitational interactions between super-particles, one has to use a softened potential. Without this, the gravitational scattering cross-section of super-particles becomes too large and super-particles undergo gravitational scattering events, which are unphysical since these events never occur in a real system. Individual particle-particle collisions are not modelled. 

There are various examples where this approach is used successfully. For example, smoothed particle hydrodynamics (SPH) uses the super-particle approximation to simulate many gas molecules \citep{Lucy1977}. These systems are often assumed to be \emph{strongly collisional} to ensure thermodynamical equilibrium inside each super-particle. One can therefore assign collective properties to clouds of particles such as pressure and temperature. Another example is the evolution of galaxies. When two galaxies collide, individual particles (stars) will usually not undergo gravitational scattering events or physical collisions. In that case, the super-particle approach models a \emph{collision-less} system in which collective dynamics are the only important physical process. In both cases, the super-particle approximation is a valid approach for simulating the system numerically, but will break down as soon as the system is \emph{marginally} collisional.

As an example, we consider two clouds of particles undergoing a ``collision''. In the strongly collisional case, the clouds will slowly merge and the thermodynamic variables (e.g., temperature) will diffuse between the clouds, the particles inside each cloud following a random walk trajectory due to numerous collisions. On the other hand, in the collision-less regime, the two clouds simply do not see each other because there is no short-range interaction present between the particles. In a marginally collisional regime, some particles will collide with particles of the other cloud, leading to a partial thermalisation of the velocity distribution, but some other particles will not have any collision at all and will follow approximately a straight line. Evidently, the outcome of this event cannot be described using a super-particle approach.

\subsection{Scaling method}
The idea of the scaling method is that one should keep all important timescales in a numerical simulation as close to those of the real physical system as possible and model all particle collisions explicitly. 

The numerical system consists of $N_\mathrm{num}$ particles in a box with length $H$, simulating a patch of the disc at a fixed radius $R_0$. 
The particle mass is $m_\mathrm{num}=\Sigma H^2/N_\mathrm{num}$ ($H$ is the box size and one scale height). Thus, the density in the box and the long-range gravitational interactions are unchanged compared to the real system.

The gravitational scattering cross-section of the numerical system is then given by 
\begin{eqnarray}
\sigma_{G,\mathrm{num}}=\frac{G^2 \Sigma^2 H^4}{N_\mathrm{num}^2 \mathrm{\bar v}_p^4} \label{eq:crossgravity}.
\end{eqnarray}
For the initial surface density and velocity dispersion used in the simulation (see above), one finds
\begin{eqnarray}
 \sigma_{G,\mathrm{num}}\simeq \frac{ H^2}{N_\mathrm{num}^2} & \simeq & \frac{0.0001\text{ AU}^2}{N_{\mathrm{num}}^2}.
\end{eqnarray}
The physical collision cross-section in the simulation is $\sigma_{c,\mathrm{num}}~=~\pi a_\mathrm{num}^2$ where $a_\mathrm{num}$, is the radius of the particles in the simulation. 
We derive two constraints from the physical and gravitational collision cross-sections:
\begin{enumerate}
\item \textit{Same mean free path in simulation and real physical systems}\nopagebreak

That means $N_{\mathrm{num}} \sigma_{c,\mathrm{num}} =N  \sigma_c $, where $\sigma_c$ and $N$ are the geometrical cross-section and the number of particles in a region of size $H^3$ in a real disc, respectively.
In other words, this condition ensures that the physical collision timescale in the simulation is exactly the same as in the real system. 

\item \textit{Negligible gravitational scattering}\nopagebreak

 When two particles approach each other, the outcome should be a physical collision, which means that $\sigma_{c,\mathrm{num}}~\gg~\sigma_{G,\mathrm{num}}$.
The gravitational scattering timescale remains long compared to the physical collision timescale.
\end{enumerate}

\noindent The first condition places a constraint on the particle size in the simulation. With $N=4.7\cdot 10^{18}$ and a particle radius of $a=1\,\mathrm{m}$ as found in a real disc within a box of volume $H^3$, using the parameters from above one finds that
\begin{eqnarray}
  a_\mathrm{num}  = \sqrt{\frac{\sigma_{c,\mathrm{num}}}\pi} =  \sqrt{\frac{N\sigma_c}{N_{\mathrm{num}}\pi}} \simeq \frac{0.014}{\sqrt{N_\mathrm{num}}}\mathrm{AU}.\label{eq:anum}
\end{eqnarray}
We note that once this condition is satisfied in the initial conditions it will be automatically satisfied at all times. In simulations, we typically find that the collision timescale is reduced by more than one order of magnitude during the gravitational collapse.

The second condition is then satisfied by changing the number of particles.
An interesting result is that if $N_\mathrm{num}>10$, the second condition is easily satisfied if the first one is satisfied. 
However, $\sigma_{G,\mathrm{num}}$ depends strongly on the velocity dispersion. 
In particular, a velocity dispersion 5 times smaller than the initial value (as found in some simulations) leads to an increase by a factor of 600 in the gravitational scattering cross-section. 
Therefore, we suggest using a large safety factor ($N_\mathrm{num}>10^5$) to ensure that the second condition is always satisfied, even for significantly smaller~$\mathrm{\bar v}_p$. This condition also allows us to estimate when our approach breaks down, namely when the number of clumps in the system is so low ($N<10\sim100$) that gravitational scattering becomes important.

Although it would be helpful as a further simplification, it is not possible to perform the above mentioned simulation in two dimensions. In that case, the filling factor, which is defined to be the ratio of the volume (or area) of all particles to the total volume (or area), is of the order of one for the above parameters. We note that in 2D an increase in particle number (and decrease in particle size as required by \Eq \ref{eq:anum}) does not decrease the filling factor if the collisional lifetime remains constant. 


\section{Numerical simulations}\label{sec:numerical}
We perform our simulations in a cubic box with shear periodic boundary conditions in the radial ($x$) and periodic boundary conditions in the azimuthal ($y$) and vertical ($z$) directions, as described in appendix \ref{app:gravtree}. 
In the local approximation, the force per mass on each particle is a sum of the contributions from Hill's equations (see \Eqs \ref{eq:app:gravtree:hills1}-\ref{eq:app:gravtree:hills3}) and interaction terms.
The interaction terms $\bm{F}_\text{int}$ are divided into components related to self gravity (including vertical gravity), physical collisions between particles, or drag and excitation forces:
\begin{eqnarray}
\bm{F}_\text{int} &=& \bm{F}_\text{grav} + \bm{F}_\text{col} + \bm{F}_\text{drag} + \bm{F}_\text{turb}.
\end{eqnarray}
We solve the resulting equations of motion with a leap frog (kick drift kick) time-stepping scheme. 
In the following subsections, we describe how each of these terms are calculated and what their physical relevance is.

The box size of $0.01~\text{AU}$ was chosen such that there are always several unstable modes $\lambda_T$ in the box, when the system is pushed into the regime of gravitational collapse, as estimated by \Eq \ref{eq:moonlet:toomrelambda}.

\subsection{Self gravity}\label{sec:nbody}
In the $N$-body problem that we consider, we have to solve Newton's equations of universal gravitation for a large number of particles $N$.
Calculating the gravity for each particle from each other particle results in $O\left(N^2\right)$ operations. To reduce the number of operations, we use an approximation, namely a Barnes-Hut (BH) tree code (see appendix \ref{app:gravtree}).

We use~$8$ \textit{rings} of ghost boxes in the radial and azimuthal direction. 
The system that we are interested in is only marginally gravitationally stable and due to the finite number of ghost boxes, we introduce a slight asymmetry that will become important after several orbits. In the beginning of our simulations, we integrate for many orbits in order to reach a stable equilibrium. We then push the system into a gravitationally unstable regime (see section ~\ref{sec:initcond} for details). During the stable phase, horizontal over-concentrations are to be expected and are indeed observed because of this asymmetry.
However, since in a real proto-planetary disc the gravitational instability is considered to appear locally (e.g., inside a vortex or a similar structure), this effect of having a preferred location for gravitational instability is not unphysical and does not affect any conclusions made. We tested this by applying a linear cut-off to the gravitational force at a distance of one box length (A. Toomre, private communication). In this case, there was no preferred location in the box and the simulation evolved in exactly the same way.

\subsection{Physical collisions}
Physical collisions between particles are treated by checking whether any two particles are overlapping after each time-step. 
Using the already existing tree structure from the gravity calculation, one can again reduce the computational costs from $O\left(N^2\right)$ to $O\left(N \log N\right)$. 
There is a small probability that we might miss a collision if the time-step is too large. To ensure that our results are converged, we tested that a reduced time-step does not alter the outcome.
See appendix \ref{app:gravtree} for more details on how collisions are resolved.

The particle radius is given by \Eq \ref{eq:anum}. In order to keep the collision timescale $\tau_c$ close to unity, which is numerically the worst case scenario because all timescales are equivalent, we increase the radius by a factor of 4 in all simulations.

\subsection{Drag force}
Each particle feels a drag force. The background velocity of the gas $\mathbf{v}_g$ is assumed to be a steady Keplerian profile
\begin{eqnarray}
\mathbf{v}_g=-\frac32 \Omega x\;\mathbf{e}_y.
\label{gasvel}
\end{eqnarray}
Although accretion flows are turbulent, we choose to use a simple velocity profile so that the behaviour of self-gravitating particles can be understood in conditions that can be easily controlled.

\subsection{Random excitation}

The system of particles described above is dynamically unstable even without self-gravity, as the coefficient of restitution and the stopping time do not depend on the particle velocities. 
The particles can either settle down in the mid-plane and create a razor-thin disc if the stopping time is too short, or they can expand vertically forever if the stopping time is above some critical value and the excitation mechanism is provided only by collisions. Dust particles in accretion discs are found in the settling regime. However, a complete settling never occurs in accretion discs because the background flow is always turbulent due to the Kelvin-Helmholtz instability \citep{JohansenHenning2006}, the MRI, or other hydrodynamic instabilities \citep[see e.g.][]{LesurPapaloizou2009} which diffuse particles vertically \citep{FromangPapaloizou2006}. This stirring process of turbulence is not present in the gas velocity field of the simulations presented here (see \Eq~\ref{gasvel}). To approximate the turbulent mixing, we add a random excitation (white noise in space and time) to the particles in the simulation. This allows us to have a well defined equilibrium in which the system is stable rather than starting from unstable initial conditions that might influence the final state.

We perturb the velocity components of each particle after each time-step by $\Delta \mathrm{v}_i =\sqrt{\delta t}\, \xi$, where $\delta t$ is the current time-step and $\xi$ is a random variable with a normal distribution around $0$ and variance $s$. This excitation mechanism heats up the particles and allows us to have a well defined three dimensional equilibrium as shown in section~\ref{sec:results}. In our simulations, we use a value of $s=1.3\e{15}\;\text{m}^2\text{s}^{-3}$ and $s=8.9\e{14}\;\text{m}^2\text{s}^{-3}$ for the simulations without and with collisions, respectively. These values were chosen such that the equilibrium state is approximately equivalent for both types of simulations.

In comparison to chapter \ref{ch:randwalk}, here we try to simplify things as far as possible. Thus, the description of stochastic forces used leads to uncorrelated and scale independent noise.

\subsection{Initial conditions}\label{sec:initcond}

All particles, which have equal mass, smoothing length, and physical size, are placed randomly inside the box in the $xy$~plane. We note that particles have either a physical size or a smoothing length associated with them, depending on the type of simulation we perform. In the z-direction, the particles are placed in a layer with an initially Gaussian distribution about the mid-plane and a standard deviation of 0.05 H. We allow the system to reach equilibrium by integrating it until $t=30\Omega^{-1}$ ($4.8\text{yrs}$). 

We tested various ways of pushing the system into the unstable regime. We here present two different scenario. First, one can switch off the excitation mechanism. The systems starts to cool down and thus becomes unstable.  Second, one can shorten the stopping time, which leads to enhanced cooling. 

Following one of the above descriptions, the system becomes gravitationally unstable and bound clumps form within a few orbits. We did not see any qualitative differences as long as we started from a well defined equilibrium state. 
This shows that, even though we simplified the treatment of turbulent stirring and gas drag, the results are generic.

\section{Results} \label{sec:results}
\subsection{Super particles}

We first present simulations without physical collisions (super-particles) which rely on the smoothing length $b$ to avoid any divergences. Although we use a tree code and a smoothed potential for the force calculation is assumed, the results are equivalent to an FFT-based method (Fast Fourier Transform)  where the grid length acts as an effective smoothing length. In general, to check the numerical resolution, the particle number $N_\text{num}$ is increased and the smoothing length $b$ is reduced independently. A simulation is resolved when the result is independent of both $N_\text{num}$ and $b$. This turns out to be impossible in the present situation. The main reason is that the smallest scale in a gravitational collapse will ultimately depend on the smoothing length. By varying both parameters at the same time, an empirical scaling of~$b\sim1/\sqrt{N_\text{num}}$ works fairly well if $N_\text{num}$ is large enough. 
The square root in this scaling comes from the fact that we are basically treating a two dimensional system. 
However, this procedure is not justified and is unphysical. The smoothing length was introduced to avoid divergent terms and not to model any small-scale physical process. Therefore, it should \emph{not} have any impact on the physical outcome of the simulation. Incidentally, the existence of this dependency indicates that short-range interactions are important for the result of these simulations and should be modelled with care.

\begin{figure}[btp]
\centering
\subbottom[Super-particles. The simulation labelled \texttt{LS} has a ten times larger smoothing length. The curves do not overlap because the simulations have not converged. \label{fig:veldisp_smooth}]{
\includegraphics[angle=270,width=0.83\columnwidth]{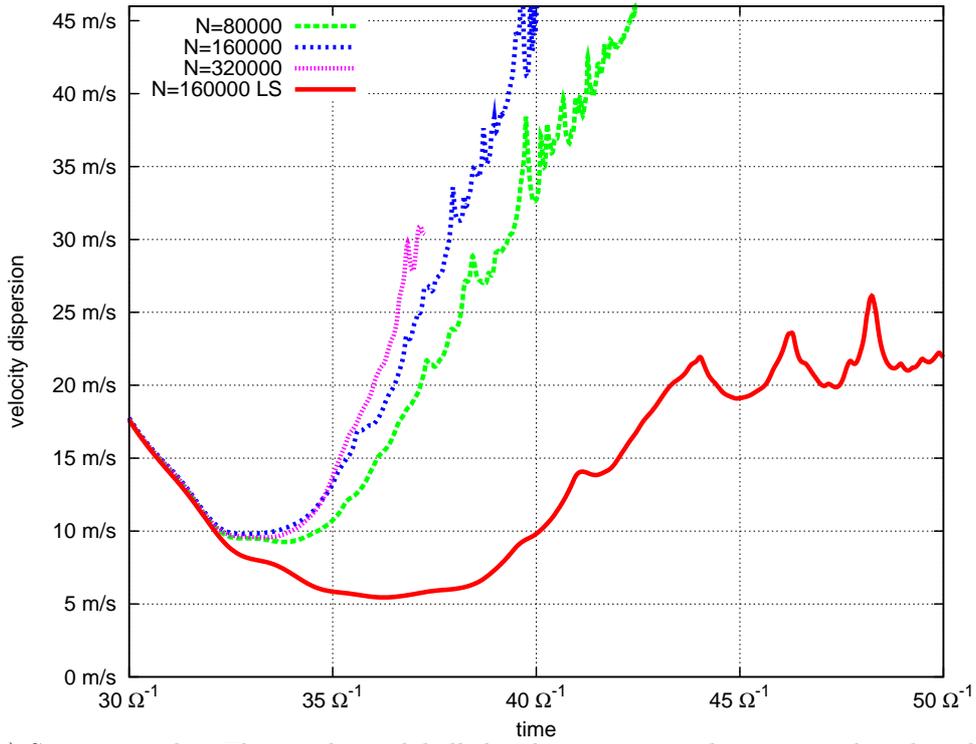}
}
\subbottom[Scaled-particles. All curves overlap at early times because the simulations are converged (see text). \label{fig:veldisp_scaled} ]{
\includegraphics[angle=270,width=0.83\columnwidth]{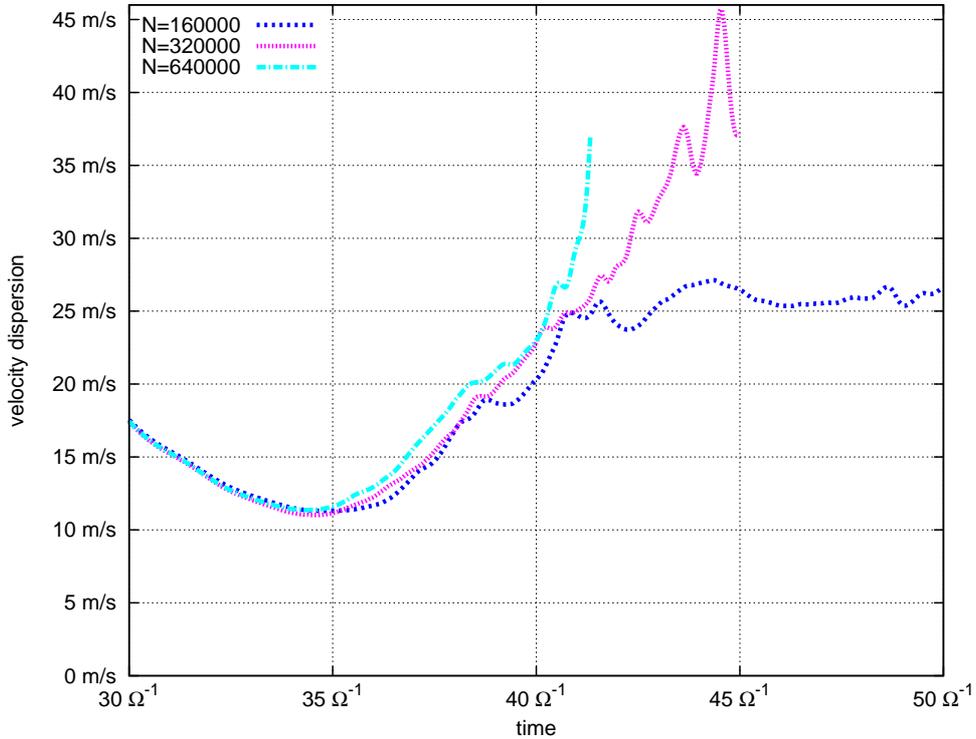} 
}
\caption{The velocity dispersion as a function of time in different runs as an indicator of convergence. Simulations use super-particles (top) and scaled-particles (bottom).\label{veldisp_noexit}}
\end{figure}

In \Fig \ref{fig:veldisp_smooth}, we plot the velocity dispersion as a function of time. The simulations begin from the stable equilibrium described in section~\ref{sec:methods}. After $t=30~\Omega^{-1}$, we switch off the excitation mechanism. The system continues to cool and it becomes gravitationally unstable within one orbit, as estimated by {\Eq \ref{eq:moonlet:toomre}}.
Once clumping occurs, the velocity dispersion begins to rise again. All simulations use the particle radius given by {\Eq~\ref{eq:anum}} as a smoothing length except the ones labelled \texttt{LS}, which use a value ten times larger. The larger softening length is approximately $128$-th of the box length and illustrates the kind of evolution expected from an FFT method using a $128^3$ grid.  
This is the resolution used in other studies, such as \cite{Johansen2007}. 

\begin{figure}[btp]
\centering
\subbottom[Super-particles. The simulation labelled \texttt{LS} has a ten times larger smoothing length. The curves do not overlap because the simulations have not converged. \label{fig:veldisp_smooth_exit}]{
\includegraphics[angle=270,width=0.83\columnwidth]{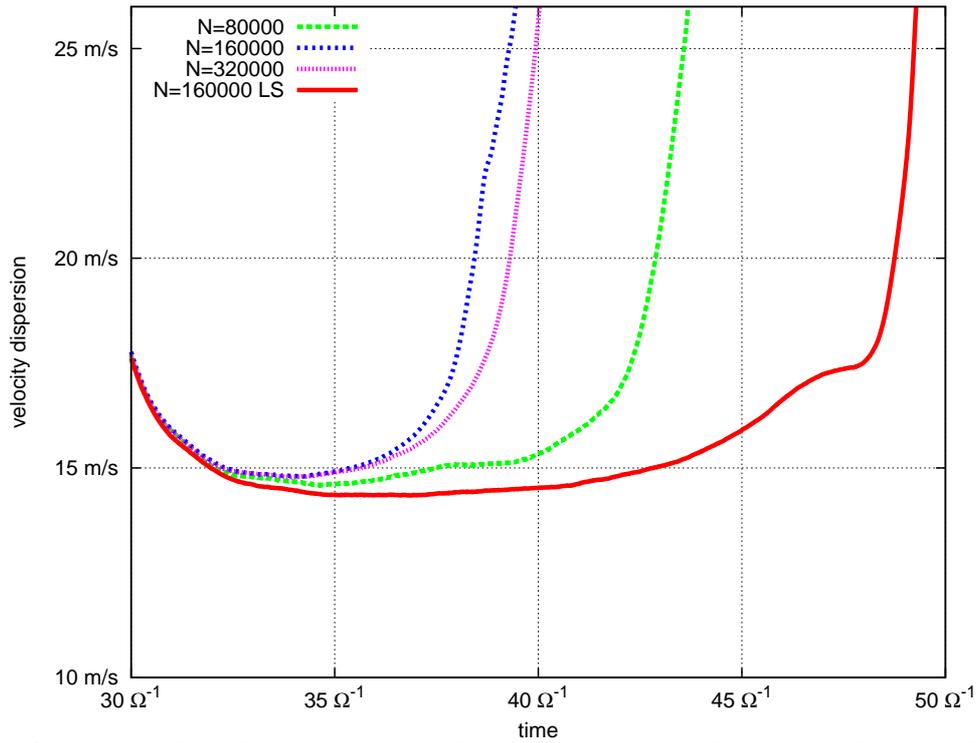}
}
\subbottom[Scaled-particles. All curves overlap because the simulations are converged. \label{fig:veldisp_scaled_exit} ]{
\includegraphics[angle=270,width=0.83\columnwidth]{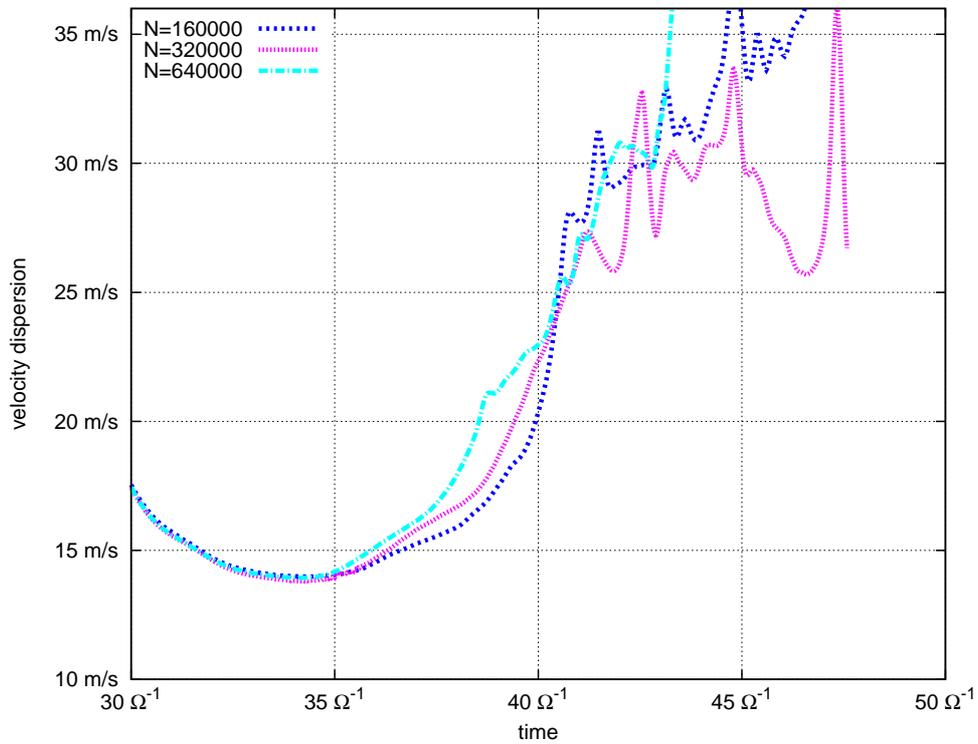} 
}
\caption{Same as in \Fig \ref{veldisp_noexit} but here the damping timescale is decreased while the excitation is unchanged. This leads to qualitatively very similar results.}
\end{figure}

The velocity dispersion evolution is also plotted in \Fig \ref{fig:veldisp_smooth_exit} for those simulations, in which the damping timescale has been decreased while keeping the excitation unchanged. The results are qualitatively very similar to those presented in \Fig \ref{fig:veldisp_smooth}.

\begin{sidewaysfigure}[p]
\centering
\includegraphics[width=0.75\textwidth]{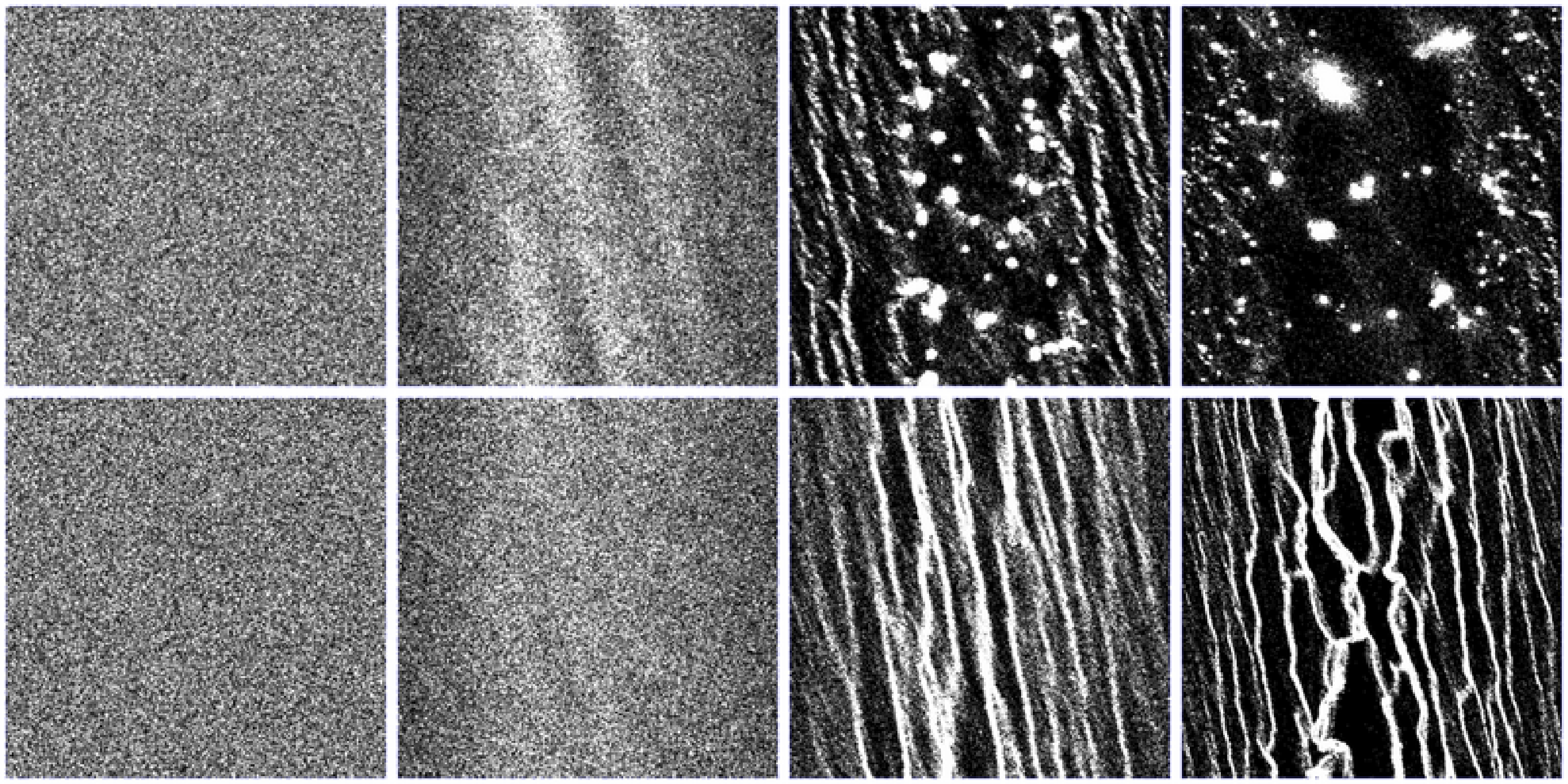}
\caption{Super-particles: Snapshots of the particle distribution in the $xy$ plane. Both simulations use 160~000 particles with different smoothing lengths. The simulation on the bottom uses smoothing length ten times larger than the one on the top. The snapshots were taken (from left to right) at $t=0, 30, 37, 40\,\Omega^{-1}$. With a large smoothing length, the outcome looks very different, the system is more stable, more stripy structure can be seen and clumps form later, if at all. \label{fig:snapsmoothing}}
\end{sidewaysfigure}

Snapshots of the particle distribution of two simulations are shown in \Fig \ref{fig:snapsmoothing}. 
Both simulations use 160~000 particles and all parameters, except the smoothing length, are the same. 
The top row is the simulation with a smoothing length given by \Eq~\ref{eq:anum}, whereas the bottom row uses a smoothing length that is ten times larger. 
The simulations correspond to the blue (medium dashed) and red (solid) curve in \Fig \ref{fig:veldisp_smooth}. 
We show the snapshots to illustrate the importance of the smoothing length in simulations without physical collisions. 
One can see that the simulations differ already before clumps form. 
Stripy structures appear on a scale given by \Eq \ref{eq:moonlet:toomrelambda}. As soon as we enter the unstable regime at~$t\sim32\Omega^{-1}$ (i.e., the velocity dispersion begins to rise, see also \Fig \ref{fig:veldisp_smooth}) the simulations evolve very differently. The simulation on the top row of \Fig \ref{fig:snapsmoothing} forms many clumps at an early time, whereas the bottom row simulation forms only a few, more massive clumps at later times. 

This can be confirmed by looking at spectra of the same two simulations as shown in \Fig \ref{fig:spectr_smooth}. 
These spectra were generated by mapping the particles onto a $128\times 128$ grid in the $xy$ plane and computing the fast Fourier transform of the mapping in the $x$ direction. 
The resulting spectra were finally averaged in the $y$ direction \citep[see][for a complete description of the procedure]{Tanga2004}. 
As suggested by the snapshots, the nonlinear dynamics of the gravitational instability strongly depends on the smoothing length used.
In particular, one observes that smaller scales are amplified more slowly for larger smoothing lengths. 
The resulting spectra at $t=40\Omega^{-1}$ also differ significantly. 
With a small enough smoothing length, the spectrum looks almost flat, whereas a large smoothing length introduces a cutoff at $k/2\pi\simeq 10$. 
We also note that the smoothing length in the latter case is of the order of the grid size ($k/2\pi\sim100$). 
The cutoff observed clearly demonstrates that the smoothing length modifies the dynamics on scales up to 10 times larger than the smoothing length itself.


\begin{sidewaysfigure}[p]
\centering
\includegraphics[width=0.75\textwidth]{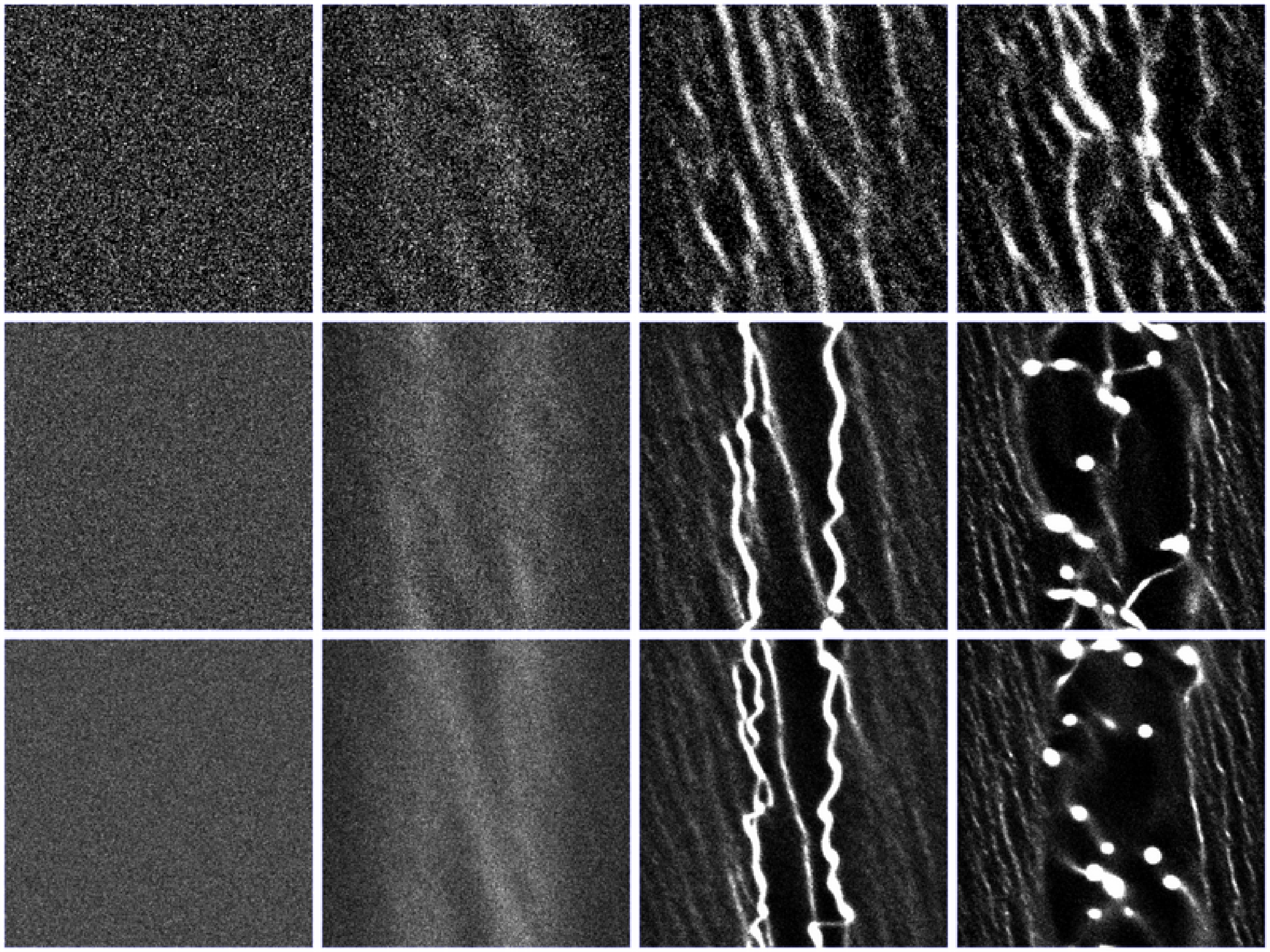}
\caption{Scaled particles: Snapshots of the particle distribution in the $xy$ plane. The simulations (from top to bottom) use 40~000, 320~000, 640~000 particles with their physical size given by \Eq~\ref{eq:anum}. In all simulations, the collision timescale was kept constant. The snapshots were taken (from left to right) at $t=0, 30, 37, 40\,\Omega^{-1}$. The simulation with 40~000 particles has a large filling factor by the last frame, which prevents clumps from forming. The intermediate resolution simulation (middle row) and the highest resolution simulation (bottom row) have more and smaller particles and thus a smaller filling factor. The results (i.e., number of clumps in the last frame) are very similar in the intermediate and high resolution simulations.   \label{fig:snapcollisions}}
\end{sidewaysfigure}

\subsection{Scaled particles}

\begin{figure}[p]
\centering
\subbottom[Super-particles: Spectrum with 160~000 particles and two different smoothing lengths that differ by a factor 10 at times $t=30$ (red, solid curve), $t=36$ (green, long dashed curve), and $t=38$ (blue, short dashed curved). The spectra at similar times are not on top of each other because the simulations have not converged.\label{fig:spectr_smooth}]{
\includegraphics[angle=270,width=0.8\columnwidth]{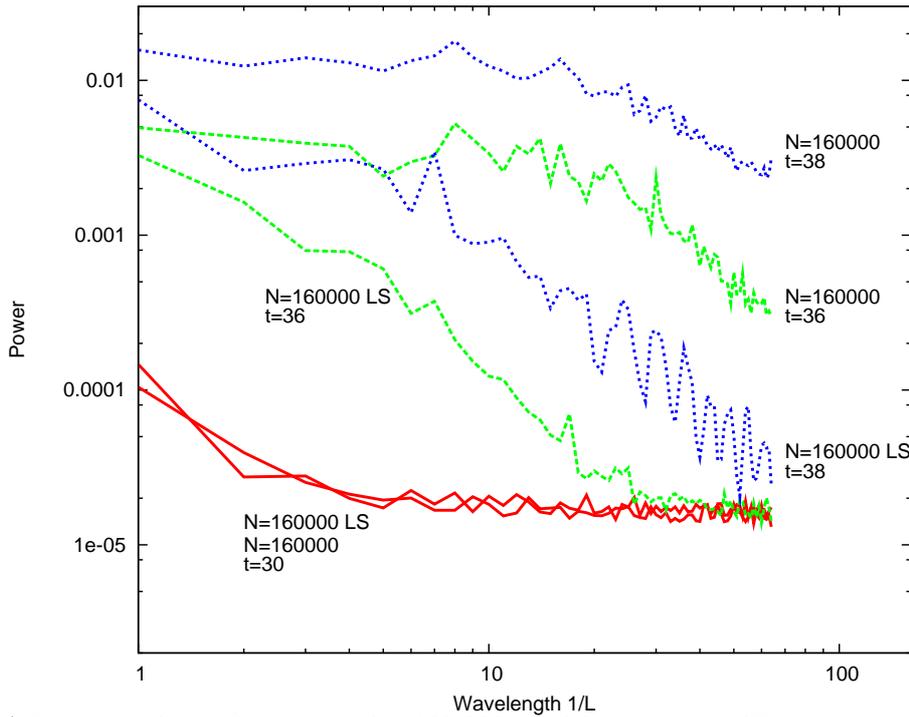}
}
\centering
\subbottom[Scaled particles: Spectrum with 160~000, 320~000, and 640~000 particles (colours are equivalent to \Fig \ref{fig:spectr_smooth}). At each time, the spectra are converged, i.e., the spectra are on top of each other and independent of the particle number, besides the noise level on very small scales. At late times ($t\gtrsim38$, blue curve), one begins to see more structure on small scales in simulations with larger number of particles (see text).  \label{fig:spectr_coll}]{
\includegraphics[angle=270,width=0.8\columnwidth]{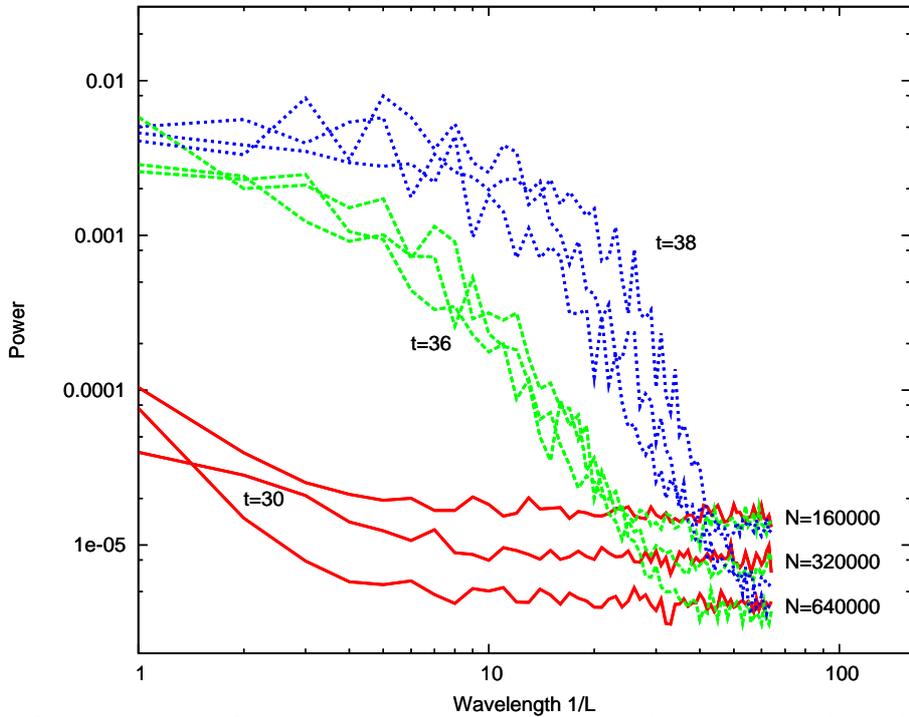}
}
\caption{Density distribution spectrum}
\end{figure}

The velocity dispersion evolution of simulations with $N=160\,000, 320\,000, 640\,000$ particles including physical collisions is presented in \Fig \ref{fig:veldisp_scaled} and \Fig \ref{fig:veldisp_scaled_exit}. In these runs, no gravitational smoothing length is needed. Again, the simulations start from a stable equilibrium and after $t=30~\Omega^{-1}$, we change the same parameters as in the non-collisional case to push the system into the gravitationally unstable regime. In \Fig \ref{fig:veldisp_scaled} we increase the damping timescale, whereas we switch off the excitation in \Fig \ref{fig:veldisp_scaled_exit}.
Snapshots of the particle distribution for three runs are plotted in \Fig \ref{fig:snapcollisions}. The middle and bottom rows of snapshots correspond to the purple (medium dashed) and dark blue (long dashed) lines of \Fig \ref{fig:veldisp_scaled}.

One can see in figures \ref{fig:veldisp_scaled}, \ref{fig:veldisp_scaled_exit} as well as \Fig \ref{fig:snapcollisions} that the intermediate and high resolution runs produce very similar results. We call those simulations converged, as a change in particle number does not change the outcome. 
An additional, lower resolution run ($N=40\,000$) has not converged because the filling factor is too large in dense regions, preventing clumps from being gravitationally bound. 
The problem does not occur in the non-collisional cases because particles are allowed to overlap. 
Since the filling factor scales with particle number as~$1/\sqrt{N_\text{num}}$, this issue is resolved for runs with more than a few hundred thousand particles. 
We note that the velocity dispersion might differ at later stages for the converged runs because the final clump radius is still determined by the physical particle radius and therefore by the particle number. 
This can be seen in particular in \Fig \ref{fig:veldisp_scaled} where the velocity dispersion of the $160\,000$ run reaches an equilibrium value before the runs with larger particle number. Note however, that up to that point, the results are converged.

We show the density distribution spectrum in \Fig \ref{fig:spectr_coll} as a function of the number of particles. The resulting spectra do not depend on the number of particles in the system, as expected. In comparison with \Fig \ref{fig:spectr_smooth}, the simulations with collisions are converged since the dynamic has not controlled by the numerical parameter $N$, which is the only free parameter in the system. 

At very late times ($t\gtrsim 38\Omega^{-1}$), the spectra begin to show a systematic trend towards more structure on smaller scales for runs with larger $N$. This is due to the filling factor, already mentioned above. When the size of a clump in the simulation is only determined by the size of the individual scaled particles that it consists of, our approach breaks down.
One can also see these clumps forming by looking at the velocity dispersion in \Fig \ref{fig:veldisp_scaled}. The particles inside a clump begin to dominate the velocity dispersion over the background after $t\sim38\Omega^{-1}$. A clear indication of this is the spiky structure with a typical correlation time of $\sim\Omega^{-1}$ corresponding to different clumps interacting and merging with each other.  One way around this issue, which will be considered in future work, is to allow particles to merge \citep{Michikoshi2009}. Using this specific accretion model, the mass of the clump can then be used as an upper limit of the clump mass. 
Eventually, one should also worry about possible effects leading to the destruction of an already formed clump.

\clearpage
\section{Discussion and implications}\label{sec:disc}
In this chapter, we have shown that convergence in $N$-body simulations of planetesimal formation via gravitational instability can be achieved when taking into account all relevant physical processes. 
It is absolutely vital to simulate gravity, damping, excitation, and physical collisions simultaneously, as the related timescales are of the same order and therefore all effects are strongly coupled. 

A set of simulations is defined to be converged when the results do not depend on the particle number or any other artificial numerical parameter such as a smoothing length.
As a test case, a box with shear periodic boundary conditions was used, containing hundreds of thousands of self-gravitating \mbox{(super-)particles} in a stratified equilibrium state. 
The particles were then pushed into a self gravitating regime that eventually led to gravitational collapse.
Simulations with and without physical collisions were studied for a range of particle numbers to test convergence.
 
In cases without physical collisions, convergence can not be achieved. However, it was possible to change multiple free simulation parameters at the same time, namely the particle number $N$ and the smoothing length $b$, such that in special cases the results did not depend on the particle number. We note however that there is a free parameter in the simulation (i.e., the smoothing length) that effects the outcome and makes it impossible to find the \textit{real} physical solution. 

In the proto-planetary disc, physical collisions dominate over gravitational scattering. In the case of planetesimal formation, the system is marginally collisional. Physical collisions will partially randomise the particle distribution, whereas a smoothing length does not because the softening length is very large compared to the gravitational cross-section. Even if the gravitational particle-particle scattering becomes important, it can be seen from \Eq~\ref{eq:crossgravity} that the velocity dependence of the cross-section $\sigma_G$ differs fundamentally from the velocity independent cross-section $\sigma_C$. We therefore argue that the super-particle approach should not be used when collisions become important.

With collisions, the simulation outcome is both quantitatively and qualitatively very different from simulations with no physical collisions. The initial size of the clumps is larger and the number of clumps is smaller. As there is no smoothing length, there is effectively one fewer free parameter. Changing the particle number while keeping the collision time constant does not change the outcome. We therefore call these simulations converged. 

Additional tests including inelastic collisions with a normal coefficient of restitution of $0.25$ have been performed to confirm that the results do not depend on the way physical collisions are modelled. As expected, the qualitative outcome in terms of number of clumps and clump size is different because the physical properties of the system have been changed. However, once individual collisions are resolved the results converge in exactly the same manner as in those simulations presented above which have a coefficient of restitution of $1$. 

This chapter focused on the numerical requirements to study gravitational instability and the formation of planetesimals in proto-planetary discs. 
The initial conditions were chosen such that the equilibrium is a well defined starting point for the convergence study. The collision, damping, collapse and orbital timescales are all of the same order, which is effectively the worst case scenario. 
To provide any constraints on planetesimal formation itself, one would have to properly simulate the turbulent background gas dynamics which is a far bigger task.

However, we were able to determine the numerical resolution needed to resolve dust particles in a proto-planetary disc.
Assuming that one wishes to simulate dust particles realistically, including collisions, and that the particles are uniformly distributed in a box of base $L^2$ and height $H$, the size of each particle is determined by the collision rate in the real disc (see \Eq~\ref{eq:anum})
\begin{eqnarray}
a_{\text{num}} = \sqrt{\frac{N}{N_{\text{num}}}} a,
\end{eqnarray}
where $a$ and $N$ are the size and number of particles in the volume $L^2H$ of the real disc. 
The number of particles in the simulation $N_{\text{num}}$ has to be large enough to ensure that the filling factor is low until clumps form. 
The requirement that the filling factor is less than unity at $t=0$ is given by
\begin{eqnarray}
\frac43 \pi  \left(a_{\text{num}}\right)^{3} \cdot N_{\text{num}} \le C_f\;  L^2 H,
\end{eqnarray}
where $C_f$ is a safety factor. Although $C_f\lesssim1$ would be enough to resolve collisions initially, it is insufficient to simulate the collapse phase. During the collapse, the filling factor rises rapidly. Assuming that one wishes to resolve collisions correctly when the particles have contracted by one order of magnitude, one has to include a safety factor of $C_f=10^{-3}$. If the collapse occurs mainly in two dimensions, as in the simulations presented here, a factor of $C_f=10^{-2}$ is sufficient. 
Furthermore, one has to ensure that the box contains at least a few unstable modes (see \Eq \ref{eq:moonlet:toomrelambda}). All this together places tight numerical constraints on numerical simulations of planetesimal formation via gravitational instability.
 
In the future, the discussion should be expanded to include more physics, namely a proper treatment and feedback of the background gas turbulence. This will allow us to further clarify the behaviour of planetesimals in the early stages of planet formation and eventually test the different formation scenarios.

\chapter{Summary}\label{ch:summary}
\epigraph{
If I have seen further it is only by standing on the shoulders of giants.
}{\tiny Issac Newton, in a letter to Robert Hooke, 1676}

\noindent Several hundred extra-solar planets have been discovered in recent years. This led to a revival of planet formation theory.
The remarkable differences of extra-solar planetary systems and our own solar system keep challenging long standing theories.

In this thesis, four main results are presented and summarised below.  
They can be categorised into two groups. 
One group is concerned with the formation of resonant multi-planetary systems. 
The other group discusses the effects of stochastic forcing on orbiting bodies such as planets, moonlets and dust.

\section{Main results}

\subsection{Most resonant planetary systems are stable}
In chapter \ref{ch:randwalk}, we develop a consistent framework for the treatment of stochastic forces in celestial mechanics. 
The growth rates of all orbital parameters are given as a function of the diffusion coefficient in the most generic way.
Furthermore, several important correction factors which take into account a finite correlation time are included.
Previous studies have overestimated the diffusion efficiency by more than an order of magnitude by ignoring these effects. 

We apply this framework to a single, stochastically forced planet and to two planets that are in resonance. 
The results show that most planetary systems are stable against the random forces originating from MRI turbulence. 
This remains true even when we take into account the considerable uncertainties in the amplitude of those forces. 

We find that resonant planetary systems have two different modes, which respond differently to stochastic forcing. 
Thus, systems that did undergo a random walk phase during their evolution should show evidence of that.
One system, in which this might have been observed, is HD128311. We present several successful formation scenarios of this system
which can reproduce the observed libration pattern. This cannot be obtained using laminar convergent migration alone.

The framework has a wide range of applications including Saturn's rings (see chapter \ref{ch:moonlet}) and promises to have a large impact on future studies involving stochastic forces of any type.

\subsection{First prediction of orbital parameters of exo-planets}
The planetary system HD45364 is in a 3:2 resonance.
The system cannot be formed by convergent migration with moderate migration rates. 
The final configuration is always a 2:1 resonance, not the observed 3:2 resonance. 

In chapter \ref{ch:threetwo}, we present a successful formation scenario that involves a short phase of rapid inwards type III migration. 
A disc with a mass about five times that of the minimum mass solar nebula is sufficient to drive type III migration. 
Furthermore, a large disc scale height is favourable for the long term sustainability of the resonance. 
These results imply that planets form in discs which are more massive than previously thought.

Finally, the outcome of hydrodynamical simulations are directly compared to the measured radial velocity curve.
We find a new orbital solution, based on the simulation results, that has a lower $\chi^2$ value than the previously reported \textit{best fit}. 
The eccentricities of both planets are considerably smaller.
The results are generic, as in all simulations we perform and that form a 3:2 commensurability, the orbital parameters are almost identical. 
This difference is the first
prediction of the precise orbital configuration of an extra-solar planetary system. 
It can be verified with new radial velocity measurements within the next few years. 

This study opens a new window to quantitative and predictive studies of the dynamics of planetary systems beyond our own solar system.

\subsection{Moonlets in Saturn's rings migrate stochastically}
An analytic estimate of both the random walk in semi-major axis and the mean equilibrium eccentricity of a small moonlet which is embedded in Saturn's rings is presented in chapter \ref{ch:moonlet}.
We show that the most important effects that determine the dynamical evolution of the moonlet are collisions with ring particles. 
If the surface density of the rings is sufficiently large that gravitational wakes can form, then those exert an additional force on the moonlet which becomes dominant eventually.
No matter how the moonlet is excited, it will migrate stochastically, with its semi-major axis changing on average like $\sqrt{t}$ for large $t$. 

We perform realistic three dimensional collisional N-body simulation of up to half a million self-gravitating ring particles and confirm the analytic estimates within a factor of two. 
New, pseudo shear periodic, boundary conditions are used which allow us to use a much smaller shearing patch than previous authors and we therefore get a large performance boost.

The random walk in semi-major axis leads to changes in the longitude that are observable with the Cassini spacecraft.
If confirmed, this can be used to calculate an upper limit on the age of the propellers and eventually the ring system itself.

\subsection{Physical collisions are vital in numerical simulations of planetesimal formation}
In chapter \ref{ch:planetesimals}, we revisit the numerical requirements in simulations of planetesimal formation via gravitational instability.
We find that the inclusion of collisions is indispensable to obtain a consistent model. 
If collisions are not included, gravitational scattering is the only small scale process present. 
It will therefore become very important as soon as clumping occurs.  
Naively, one might think that gravitational close encounters could be treated as if these were physical collisions. 
However, properties such as the cross section are different compared to physical collisions and gravitational scattering can therefore not resemble the real system.

The results show that earlier studies should not be trusted beyond the initial clumping phase. 
Estimates on the minimum number of particles required for convergent results are given. 

Future studies on planetesimal formation will have to address this issue and the method presented here shows a clear way on how to do that.
This will lead to self-consistent results in future simulations.

\section{Future work}
\subsection{Multi-planetary systems}
The results on multi-planetary systems in this thesis, as successful as they might be, have been restricted to only two systems. However, over a third of the discovered planets are confirmed to be in multi-planetary systems.

Having developed all necessary tools in chapters \ref{ch:threetwo} and \ref{ch:randwalk}, it is now an easy task to study many systems in a similar fashion. 
In the following years, I will select up to 15 multi-planetary systems and study their evolution with both N-body as well as two and three dimensional hydrodynamic simulations. This will be the first survey of realistic formation scenarios among multi-planetary systems. 
The results will be more precise and not as degenerate as population synthesis models of single planets. 

New observations suggest that a substantial fraction of hot Jupiters might be in retrograde orbits (Triaud et al. 2010, in preparation). 
There are various ways to form these systems, such as Kozai Migration or planet-planet scattering \citep{FabryckyTremaine2007,Nagasawa2008}.
So far, these methods have not been studied in combination with the traditional ideas of planet-disc interaction. 
I will use Prometheus to study highly inclined systems in the presence of gas discs. 

I also want to continue studying the effects of MRI turbulence on planetary systems. The strength of stochastic forces coming from turbulence is not well constrained yet. For that reason a large interval of possible values of the diffusion coefficient had to be studied in chapter \ref{ch:randwalk}.
Using the MHD module of Prometheus or a different MHD code such as Athena, I will explore the parameter space in a systematic way with global non-ideal MHD simulations. The measurements of statistical properties of MRI turbulence will improve the predictive capability of our analytic random walk model significantly.

One needs at least 16 (maybe 32 or even 64) grid cells per scale-height to simulate the turbulent proto-stellar disc accurately. Assuming one wants to simulate a radial domain ranging from {1~to~4~AU} and 6~scale-heights in the vertical direction, one needs a 96x480x502 grid. This is a challenging task. During my PhD, I have gained knowledge on how to speed up calculations using CPU specific hardware features. If the performance of the code becomes an issue and it turns out that higher resolution is required to simulate the relevant scales in the system, I can implement these changes to speed up the simulations. 

With these new results, I am going to show that a broad selection of systems can be formed in a turbulent disc and prove if there is statistically significant observational evidence of turbulence in the exo-planet distribution or not.

\subsection{Saturn's rings}
The work presented in chapter \ref{ch:moonlet} was limited to a rather small subset of the parameter space. 
I plan to extend this study to a wider range of parameters, especially to higher internal densities and a broader size distribution to reduce granularity.
As soon as observational data has been published, I will investigate the constraints arising from the measured random walk of moonlets. 
If the random walk is indeed as strong as predicted, it is possible to obtain an upper limit on the age of the ring system.

Direct simulations of planetary rings are constrained to a local model. 
Although this is a valid approximation in most circumstances, it has also certain drawbacks. 
For example, simulations presented in chapter \ref{ch:moonlet} capture the random walk in semi-major axis but do not capture the net migration over long timescales \citep{Crida2010}.

I plan to extend direct simulations of Saturn's rings to a \textit{global} model. 
Periodic boundary conditions in the azimuthal and pseudo shear periodic boundary conditions in the radial direction, in a similar manner to those presented in chapter \ref{ch:moonlet} and appendix \ref{app:gravtree}, can be used to extend the shearing box to a wedge that extends radially over several tens or hundres of kilometres. 
We can get rid of Hill's equations and solve the planet's gravitational force directly.

Numerically, this is a challenging task, as the number of particles will be very large. 
Therefore, the particles have to be distributed across processors.
As outlined in some detail in appendix \ref{app:gravtree}, MPI is not ideally suited for parallelising tree structures. 
In this case, however, the fact that one dimension is much larger than the others comes in handy because boxes far away do not have to communicate directly with each other so that their interaction can be pre-calculated.

\subsection{Planetesimal formation}

The results on planetesimal formation presented in this thesis are preliminary, showing that it is indeed possible to simulate the gravitational collapse of a self-gravitating planetesimal population. To do this self consistently, one has to include collisions and gas dynamics. The former part has been completed in chapter \ref{ch:planetesimals}. The latter part has been done by other authors \citep[e.g.][]{Johansen2007}.  A combination of both is the logical next step.  

Many groups are currently working on planetesimal formation in a turbulent disc. I plan to focus on the differences between global simulations and the local shearing box approximation which has been used by most authors. This is essential to understand the importance of effects such as the streaming instability which rely on a large over-density in the solid to gas ratio. In a local simulation with periodic boundary conditions, this ratio is set by the initial conditions and cannot change. However, in a real proto-stellar disc, as well as in global models, the ratio is controlled by large scale structures in the disc such as pressure and temperature gradients.

To gain a  better understanding of these global effects, I will couple the new method that treats particles correctly to a global MHD simulation. The method to solve the non-ideal MHD equations will be the same as for the study of multi-planetary systems. To simulate the self-gravity and collisions in the dust layer, I will re-use the GravTree code, presented in appendix \ref{app:gravtree}, which is ideally suited for this project and already fully developed and tested.

\subsection{Numerical codes}
The primary codes that have been written for this thesis, Prometheus and GravTree, will be made publicly available. 
Both codes will be hosted on github at \url{http://github.com/hannorein/} under the open source license GPL. 
They are easily portable to different architectures as they are written in C and C++, respectively, using only standard libraries such as OpenGL, OpenMP and libpng. 
Compilation is controlled by the GNU autotool suite. 

Providing the source code is, in my opinion, a necessity as numerical codes are part of the initial conditions. 
All results should be easily reproducible. 
Firstly, this helps to make progress in the subject, and secondly, it also increases the chance to find bugs in the code. It is therefore a win-win situation.

\appendixpage
\appendix
\chapter{Orbital elements}\label{app:orbit}
In this appendix, the orbital elements of an elliptical orbit are defined. It is a solution to the two body problem, the simplest non-trivial problem in celestial mechanics. 

We consider a test particle on an orbit around a central object with mass $m$ at the origin. The specific force on the particle is given by 
$\mathbf{F} = -\frac{Gm}{r^3}\mathbf{r},$
where $\mathbf{r}$ is the position of the particle. Thus, the equation of motion is simply
\begin{eqnarray}
\ddot {\mathbf r} &=& -\frac{Gm}{r^3}\mathbf{r}.\label{eq:orbit:motion}
\end{eqnarray}
Taking the vector product of the above equation with $\mathbf r$ evaluates to zero on the right hand side and gives after integrating
\begin{eqnarray}
{\mathbf r} \times \dot {\mathbf r} = \mathbf h,\label{eq:orbit:ang}
\end{eqnarray}
where $\mathbf h$ is a constant of motion.
The absolute value of $\mathbf h$ is the specific angular momentum. 
The position and velocity vectors are in the same plane and allowing us to restrict ourselves to the two dimensional case. 
By switching to polar coordinates $r$ and $\phi$, we can rewrite equation \ref{eq:orbit:ang}, giving $h=r^2\dot\phi$. 
We also rewrite equation \ref{eq:orbit:motion}, giving
\begin{eqnarray}
\ddot r &=& r\dot \phi^2 -\frac{Gm}{r^2}.\label{eq:orbit:motionpolar}
\end{eqnarray}
The solution to equation \ref{eq:orbit:motionpolar} can then be written as
\begin{eqnarray}
r &=& \frac{p}{1+e\cos({\phi-\varpi})} \label{eq:orbit:sol},
\end{eqnarray}
where $e$ is the eccentricity, $\varpi$ is the longitude of pericentre and $p$ the semilatus rectum which can be written as $p={h^2}({Gm}){^{-1}}$. 
We restrict the further discussion to an elliptic orbit for which $0\leq e<1$. 
In that case $p=a(1-e^2)$, where $a$ is the semi major axis. 
We define the true anomaly as $f=\phi-\varpi$. We can further more define an average angular velocity, $n=2\pi/T$, called the mean motion, where $T$ is the orbital period, given by
\begin{eqnarray}
T^2 = \frac{4\pi^2}{Gm}a^3.
\end{eqnarray}
For a circular orbit, the mean motion is simply $v_c/a$, where $v_c=\sqrt{Gm/a}$. Now that we have a mean angular velocity, we can also define a mean angle $M$ (the mean longitude), measured from the pericentre of the planet, via
\begin{eqnarray}
M=n(t-\tau).
\end{eqnarray}
Here, $\tau$ is the time of pericentre passage. $M$ grows linearly in time. 

\begin{sidewaysfigure}[p]
\centering
\scalebox{1.0}{
\begin{picture}(0,0)%
\includegraphics{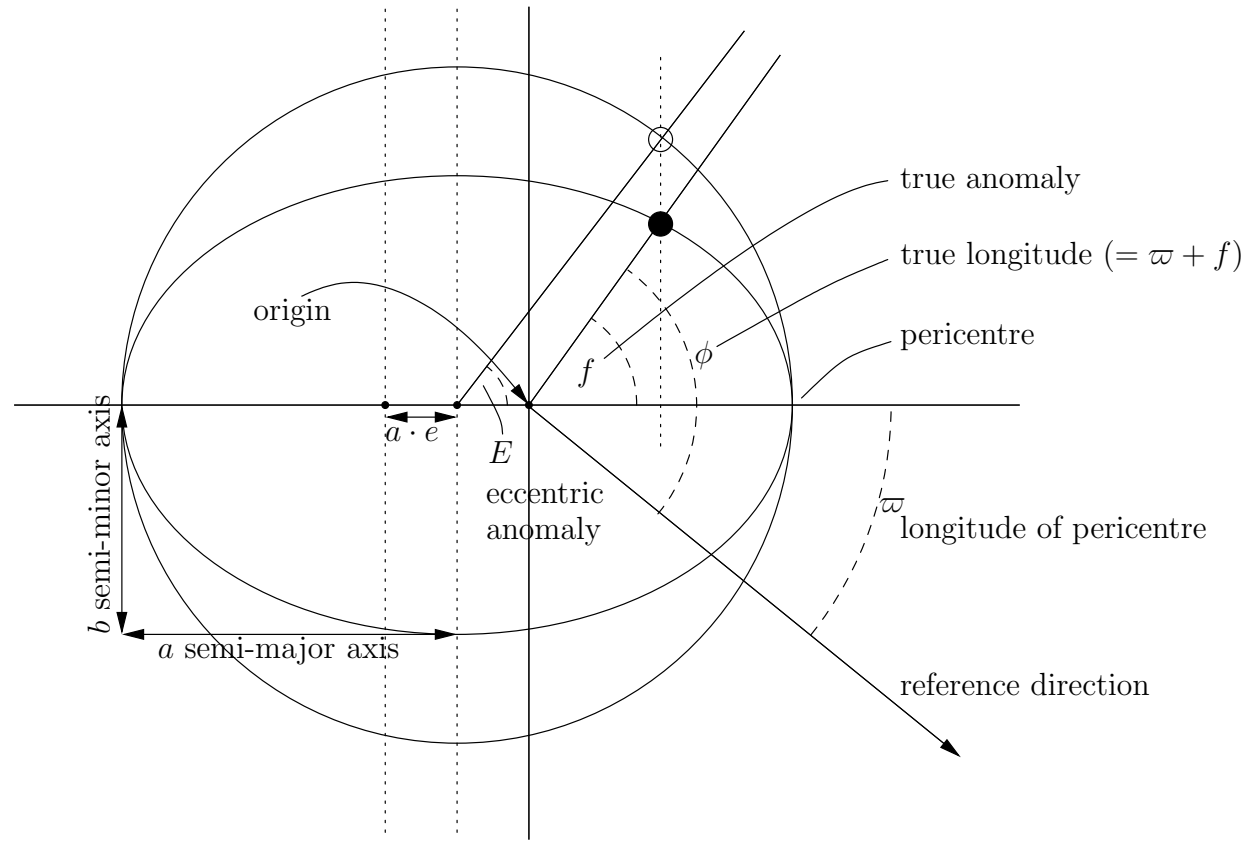}%
\end{picture}%
\setlength{\unitlength}{3947sp}%
\begingroup\makeatletter\ifx\SetFigFont\undefined%
\gdef\SetFigFont#1#2#3#4#5{%
  \reset@font\fontsize{#1}{#2pt}%
  \fontfamily{#3}\fontseries{#4}\fontshape{#5}%
  \selectfont}%
\fi\endgroup%
\begin{picture}(7192,5199)(2764,-5773)
\put(6301,-2911){\makebox(0,0)[lb]{\smash{{\SetFigFont{12}{14.4}{\familydefault}{\mddefault}{\updefault}{\color[rgb]{0,0,0}$f$}%
}}}}
\put(5746,-3421){\makebox(0,0)[lb]{\smash{{\SetFigFont{12}{14.4}{\familydefault}{\mddefault}{\updefault}{\color[rgb]{0,0,0}$E$}%
}}}}
\put(5733,-3671){\makebox(0,0)[lb]{\smash{{\SetFigFont{12}{14.4}{\familydefault}{\mddefault}{\updefault}{\color[rgb]{0,0,0}eccentric}%
}}}}
\put(5733,-3896){\makebox(0,0)[lb]{\smash{{\SetFigFont{12}{14.4}{\familydefault}{\mddefault}{\updefault}{\color[rgb]{0,0,0}anomaly}%
}}}}
\put(5101,-3286){\makebox(0,0)[lb]{\smash{{\SetFigFont{12}{14.4}{\familydefault}{\mddefault}{\updefault}{\color[rgb]{0,0,0}$a\cdot e$}%
}}}}
\put(8326,-2686){\makebox(0,0)[lb]{\smash{{\SetFigFont{12}{14.4}{\familydefault}{\mddefault}{\updefault}{\color[rgb]{0,0,0}pericentre}%
}}}}
\put(8326,-1711){\makebox(0,0)[lb]{\smash{{\SetFigFont{12}{14.4}{\familydefault}{\mddefault}{\updefault}{\color[rgb]{0,0,0}true anomaly}%
}}}}
\put(3676,-4636){\makebox(0,0)[lb]{\smash{{\SetFigFont{12}{14.4}{\familydefault}{\mddefault}{\updefault}{\color[rgb]{0,0,0}$a$ semi-major axis}%
}}}}
\put(3376,-4486){\rotatebox{90.0}{\makebox(0,0)[lb]{\smash{{\SetFigFont{12}{14.4}{\familydefault}{\mddefault}{\updefault}{\color[rgb]{0,0,0}$b$ semi-minor axis}%
}}}}}
\put(8326,-4861){\makebox(0,0)[lb]{\smash{{\SetFigFont{12}{14.4}{\rmdefault}{\mddefault}{\updefault}{\color[rgb]{0,0,0}reference direction }%
}}}}
\put(8326,-3886){\makebox(0,0)[lb]{\smash{{\SetFigFont{12}{14.4}{\familydefault}{\mddefault}{\updefault}{\color[rgb]{0,0,0}longitude of pericentre}%
}}}}
\put(7039,-2797){\makebox(0,0)[lb]{\smash{{\SetFigFont{12}{14.4}{\familydefault}{\mddefault}{\updefault}{\color[rgb]{0,0,0}$\phi$}%
}}}}
\put(4276,-2536){\makebox(0,0)[lb]{\smash{{\SetFigFont{12}{14.4}{\familydefault}{\mddefault}{\updefault}{\color[rgb]{0,0,0}origin}%
}}}}
\put(8329,-2187){\makebox(0,0)[lb]{\smash{{\SetFigFont{12}{14.4}{\familydefault}{\mddefault}{\updefault}{\color[rgb]{0,0,0}true longitude ($=\varpi+f$)}%
}}}}
\put(8196,-3726){\makebox(0,0)[lb]{\smash{{\SetFigFont{12}{14.4}{\familydefault}{\mddefault}{\updefault}{\color[rgb]{0,0,0}$\varpi$}%
}}}}
\end{picture}%
}
\caption{Orbital elements of an elliptical orbit.\label{fig:orbit:def}}
\end{sidewaysfigure}

Finally, we would like to get an equation for $r(t)$ and $\phi(t)$. 
The time dependence of equation \ref{eq:orbit:sol} is hidden in $\phi$. 
It turns out that an explicit form does in general not exist and the solution can only be calculated iteratively. 
We can see this by introducing another angle, the eccentric anomaly $E$ (see figure \ref{fig:orbit:def}). 
It can be shown \citep{solarsystemdynamics} that a relation between the mean longitude and the eccentric anomaly exists such that
\begin{eqnarray}
M=E-e\sin E.
\end{eqnarray}
This equation is transcendental but can be solved iteratively, for example with a Newton-Raphson method.

Most of the quantities defined above are plotted in figure \ref{fig:orbit:def} for illustrative purposes.

\chapter{Dissipative forces}\label{app:leepeale}
Additional terms can be added to the equations of motion in N-body simulations to describe eccentricity and semi-major axis damping.  
These forces could result from various sources, including a gaseous disc, a planetesimal disc or tides.
The derivation follows \cite{LeePeale01} and some expressions are taken from \cite{solarsystemdynamics}. 

\noindent 
We have two additional, non-conservative terms in the equations for both the velocity and acceleration. In a Cartesian coordinate system, these are
\begin{eqnarray}
\frac{dx}{dt}\Bigg|_{\dot a} + \frac{dx}{dt}\Bigg|_{\dot e}
&=& \frac{\partial x}{\partial a}{\dot a} +
    \frac{\partial x}{\partial e}{\dot e} \quad  \quad \text{and}
\label{eq:leepeale:dxdt}\\
& & \nonumber\\
\frac{d\dot x}{dt}\Bigg|_{\dot a} + \frac{d\dot x}{dt}\Bigg|_{\dot e}
&=& \frac{\partial\dot x}{\partial a}{\dot a} +
    \frac{\partial\dot x}{\partial e}{\dot e} .
\label{eq:leepeale:ddotxdt}
\end{eqnarray}
The expressions are analogue for the $y$ and $z$ coordinates.
In order to differentiate with respect to $a$ and $e$, we have to write the cartesian coordinates in terms of orbital elements. The positions can be expressed as
\begin{eqnarray}
x &=& r\cos{\Omega}\cos{(\omega+f)}-r\cos{i}\sin{\Omega}\sin{(\omega+f)},
\nonumber\\
y &=& r\sin{\Omega}\cos{(\omega+f)}+r\cos{i}\cos{\Omega}\sin{(\omega+f)} \quad \quad \text{and}
\label{eq:leepeale:x}\\
z &=& r\sin{i}\sin{(\omega+f)},\nonumber
\end{eqnarray}
which simply follow from the definition of the orbital parameters (see appendix \ref{app:orbit}). By differentiating once, we get expressions for the velocities 
\begin{eqnarray}
{\dot x} &=& \cos{\Omega}[{\dot r}\cos{(\omega+f)}
                          -r{\dot f}\sin{(\omega+f)}] \nonumber\\
&& \quad    -\cos{i}\sin{\Omega}[{\dot r}\sin{(\omega+f)}
                                 +r{\dot f}\cos{(\omega+f)}], \nonumber\\
{\dot y} &=& \sin{\Omega}[{\dot r}\cos{(\omega+f)}
                          -r{\dot f}\sin{(\omega+f)}] \label{eq:leepeale:dotx}\\
&& \quad    +\cos{i}\cos{\Omega}[{\dot r}\sin{(\omega+f)}
                                 +r{\dot f}\cos{(\omega+f)}], \nonumber\\
{\dot z} &=& \sin{i}[{\dot r}\sin{(\omega+f)}
                     +r{\dot f}\cos{(\omega+f)}] \nonumber.
\end{eqnarray}
We also need expressions for the radius $r$, the time derivative of the radius ${\dot r}$, and $r{\dot f}$ in terms of $a$, $e$, and $f$:
\begin{eqnarray}
 r &=& \frac{a(1-e^2)}{1+ e \cos f}\\
\dot r &=& \frac{r \,\dot f \,e \;\sin f}{1+ e \cos f}\\
r \dot f &=& \frac{na}{\sqrt{1-e^2}}  (1+ e \cos f).
\end{eqnarray}
To complete the list of equations, we compute several other derivatives needed to calculate the partial derivatives in equation \ref{eq:leepeale:dxdt} and \ref{eq:leepeale:ddotxdt}
\begin{eqnarray}
\frac{\partial r}{\partial a} &=& \frac{r}{a},\nonumber \\
& & \nonumber \\
\frac{\partial r}{\partial e} &=& \left[-\frac{2er}{1-e^2}-\frac{r^2\cos{f}}
                                        {a(1-e^2)}\right],\nonumber\\
& & \nonumber \\
\frac{\partial\dot r}{\partial a} &=& -\frac{\dot r}{2a},\nonumber\\
& & \nonumber \\
\frac{\partial\dot r}{\partial e} &=& \frac{\dot r}{e(1-e^2)},
                                                 \label{eq:leepeale:drdtae}\\
& & \nonumber \\
\frac{\partial (r\dot f)}{\partial a} &=& -\frac{r\dot f}{2a},\nonumber\\
& & \nonumber \\
\frac{\partial (r\dot f)}{\partial e} &=& \frac{r\dot f(e+\cos{f})}
                                          {(1-e^2)(1+e\cos{f})}.\nonumber
\end{eqnarray}
Putting everything together, we arrive at the following expressions 
\begin{eqnarray}
\frac{dx}{dt}\Bigg|_{\dot a} + \frac{dx}{dt}\Bigg|_{\dot e}
&=& \frac{x}{a}{\dot a}
+\left[\frac{r}{a(1-e^2)}-\frac{1+e^2}{1-e^2}\right]\frac{x}{e}{\dot e}, \label{eq:leepeale:dxdtf}\\
\frac{dy}{dt}\Bigg|_{\dot a} + \frac{dy}{dt}\Bigg|_{\dot e}
&=& \frac{y}{a}{\dot a}
+\left[\frac{r}{a(1-e^2)}-\frac{1+e^2}{1-e^2}\right]\frac{y}{e}{\dot e}, \label{eq:leepeale:dydtf}\\
\frac{dz}{dt}\Bigg|_{\dot a} + \frac{dz}{dt}\Bigg|_{\dot e}
&=& \frac{z}{a}{\dot a}
+\left[\frac{r}{a(1-e^2)}-\frac{1+e^2}{1-e^2}\right]\frac{z}{e}{\dot e}.  \label{eq:leepeale:dzdtf}
\end{eqnarray}
The additional terms for each of $d\dot x/dt$, $d\dot y/dt$, $d\dot z/dt$ are
distinct for variations in $e$:
\begin{eqnarray}
\frac{d\dot x}{dt}\Bigg|_{\dot a} + \frac{d\dot x}{dt}\Bigg|_{\dot e}
&=& -\frac{\dot x}{2a} {\dot a} + \cos{\Omega}\left[
\frac{\partial \dot r}{\partial e}\cos{(\omega+f)}-\frac{\partial (r\dot
f)}{\partial e}\sin{(\omega+f)}\right]{\dot e}\label{eq:leepeale:ddotxdtf}\\
&&\nonumber\\
&&\quad -\cos{i}\sin{\Omega}\left[\frac{\partial\dot r}{\partial e}
\sin{(\omega+f)}+\frac{\partial(r\dot f)}{\partial e}\cos{(\omega+f)}
\right]{\dot e},\nonumber\\
&&\nonumber\\
\frac{d\dot y}{dt}\Bigg|_{\dot a} + \frac{d\dot y}{dt}\Bigg|_{\dot e}
&=& -\frac{\dot y}{2a} {\dot a}+\sin{\Omega}\left[
\frac{\partial \dot r}{\partial e}\cos{(\omega+f)}-\frac{\partial (r\dot
f)}{\partial e}\sin{(\omega+f)}\right]{\dot e}\label{eq:leepeale:ddotydtf}\\
&&\nonumber\\
&&\quad +\cos{i}\cos{\Omega}\left[\frac{\partial \dot r}{\partial e}
\sin{(\omega+f)}+\frac{\partial (r\dot f)}{\partial e}\cos{(\omega+f)}
\right]{\dot e},\nonumber\\  
&&\nonumber\\
\frac{d\dot z}{dt}\Bigg|_{\dot a} + \frac{d\dot z}{dt}\Bigg|_{\dot e}
&=& -\frac{\dot z}{2a} {\dot a}+\sin{i}\left[
\frac{\partial \dot r}{\partial e}\sin{(\omega+f)}+\frac{\partial (r\dot f)}
{\partial e}\cos{(\omega+f)}\right]{\dot e}\label{eq:leepeale:ddotzdtf}. 
\end{eqnarray}
The final equations \ref{eq:leepeale:dxdtf}-\ref{eq:leepeale:ddotzdtf} are then used to calculate the dissipative forces and included in the N-body code (see section \ref{sec:threetwo:nbody}). Note that the orbital elements have to be calculated every sub-timestep. 
Depending on the integrator and the required number of force evaluations, this can be numerically expensive.
\chapter{Response calculation of particles on horseshoe orbits}\label{app:response}
In this appendix, we calculate the amplitude of epicyclic motion that is induced by an eccentric moonlet in a small particle on a horseshoe orbit.

\section{Interaction potential}
Due to some finite eccentricity, the moonlet undergoes a small oscillation about the origin in the local Hill Cartesian coordinate system (see section \ref{moonlet:sec:coordinates}). 
Its Cartesian coordinates then become $(X,Y)$, with these being considered small in magnitude.
The components of the equation of motion for a ring particle with Cartesian coordinates $(x,y)\equiv {\bf r}_1$ are
\begin{eqnarray}
\frac{d^2 x}{dt^2}-2\Omega\frac{d y}{dt} &=&3\Omega^2x - \frac{1}{m_1}\frac{\partial \Psi_{1,2}}{\partial x}\hspace{3mm} {\rm and}\\
\frac{d^2 y}{dt^2}+2\Omega\frac{d x}{dt} &=&-\frac{1}{m_1} \frac{\partial \Psi_{1,2}}{\partial y},\label{equmot}
\end{eqnarray}
where the interaction gravitational potential due to the moonlet is
\begin{eqnarray}
\Psi_{1,2} = -\frac{Gm_1m_2}{(r^2+R^2-2rR\cos(\phi -\Phi))^{1/2}}
\end{eqnarray} 
with the cylindrical coordinates of the particle and moonlet being $(r, \phi)$ and $(R,\Phi)$ respectively.
This may be expanded correct to first order in $R/r$ in the form
\begin{eqnarray}
\Psi_{1,2} = -\frac{Gm_1m_2}{r} -\frac{Gm_1m_2R\cos(\phi -\Phi)}{r^2}.\label{pertpot}
\end{eqnarray}
The moonlet undergoes small amplitude epicyclic oscillations
such that $X={\mathcal E}_2\cos(\Omega t +\epsilon), Y=-2{\mathcal E}_2\sin(\Omega t +\epsilon)$
where $e$ is its small eccentricity and $\epsilon$ is an arbitrary phase.
Then $\Psi_{1,2}$ may be written as
\begin{eqnarray}
\Psi_{1,2} = -\frac{Gm_1m_2}{r} - \Psi'_{1,2},
\end{eqnarray}
where
\begin{eqnarray}
\Psi'_{1,2} = - \frac{Gm_1m_2r\,{\mathcal E}_2(x\cos(\Omega t +\epsilon) -2y\sin(\Omega t +\epsilon))}{r^3}
\label{pertpot1}\end{eqnarray}
gives the part of the lowest order interaction potential associated with 
the eccentricity of the moonlet.

We here view the interaction of a ring particle with the moonlet as involving two components.
The first, due to the first term on the right hand side of equation \ref{pertpot} operates when ${\mathcal E}_2=0$
and results in standard horseshoe orbits for ring particles induced by a moonlet in circular orbit.
The second term, equation \ref{pertpot1}, perturbs this motion when ${\mathcal E}_2$ is small.
We now consider the response of a ring particle undergoing horseshoe motion to this perturbation.
In doing so we make the approximation that the variation of the leading order potential equation \ref{pertpot}
due to the induced ring particle perturbations may be neglected. This is justified by the fact
that the response induces epicyclic oscillations of the particle which are governed
by the dominant central potential.

\section{Response calculation}
Setting $x\rightarrow x + \xi_x, y\rightarrow y + \xi_y,$ where \mbox{\boldmath${\xi}$}
is the small response displacement induced by equation \ref{pertpot1} and linearising equations \ref{equmot},
we obtain the following equations for the components of \mbox{\boldmath${\xi}$}
\begin{eqnarray}
\frac{d^2\xi_x}{dt^2}-2\Omega\frac{d \xi_y}{dt} &=&3\Omega^2\xi_x - \frac{1}{m_1}\frac{\partial \Psi'_{1,2}}{\partial x}\hspace{3mm} {\rm and}\\
\frac{d^2 \xi_y}{dt^2}+2\Omega\frac{d \xi_x}{dt} &=&-\frac{1}{m_1} \frac{\partial \Psi'_{1,2}}{\partial y},\label{equpert}.
\end{eqnarray}
From these we find a single equation for $\xi_x$ in the form
\begin{eqnarray}
\frac{d^2\xi_x}{dt^2}+\Omega^2\xi_x &=& - \frac{1}{m_1}\frac{\partial \Psi'_{1,2}}{\partial x}-
\int \frac{1}{m_1}\frac{\partial \Psi'_{1,2}}{\partial y}dt = F
\label{equpert1}.
\end{eqnarray}
When performing the time integral on the right hand side of the above, as we are concerned with a potentially resonant epicyclic response,
we retain only the oscillating part. 

The solution to equation \ref{equpert1} which is such that \mbox{\boldmath${\xi}$} vanishes in the distant past $(t=-\infty)$ 
when the particle is far from the moonlet may be written as
\begin{eqnarray}
\xi_x=\alpha\cos(\Omega t)+ \beta\sin(\Omega t),
\end{eqnarray}
where
\begin{eqnarray}
\alpha  =  -\int_{-\infty}^t\frac{F\sin(\Omega t)}{\Omega}dt \quad\quad {\rm and} \quad\quad
\beta  =  \int_{-\infty}^t\frac{F\cos(\Omega t)}{\Omega}dt.
\end{eqnarray}
After the particle has had its closest approach to the moonlet and moves to a large distance from it
it will have an epicyclic oscillation with amplitude and phase determined by
\begin{eqnarray}
\alpha_{\infty}  = -\int_{-\infty}^{\infty}\frac{F\sin(\Omega t)}{\Omega}dt \quad \quad {\rm and} \quad\quad
\beta_{\infty}  = \int_{-\infty}^{\infty} \frac{F\cos(\Omega t)}{\Omega}dt.\label{epiamp}
\end{eqnarray}
In evaluating the above we note that, although $F$ vanishes when the particle
is distant from the moonlet at large $|t|,$ it also oscillates with the epicyclic angular frequency $\Omega$
which results in a definite non zero contribution. This is the action of the co-orbital resonance.
It is amplified by the fact that the encounter of the particle with the moonlet in general
occurs on a horseshoe libration time scale which is much longer than $\Omega^{-1}.$
We now consider this unperturbed motion of the ring particles.
\section{Unperturbed horseshoe motion}
The equations governing the unperturbed horseshoe motion are equations \ref{equmot} 
with 
\begin{eqnarray}
\Psi_{1,2} = \Psi^0_{1,2}= -\frac{Gm_1m_2}{r}\label{horsepot}.
\end{eqnarray}
We assume that this motion is such that $x$ varies on a time scale much longer than $\Omega^{-1}$ so that we may approximate
the first of these equations as
$x=-\left[2/(3\Omega)\right](dy/dt).$ Consistent with this we also neglect $x$ in comparison to $y$ in equation \ref{horsepot}.
From the second equation we than find that
\begin{eqnarray}
\frac{d^2 y}{dt^2} &=&\frac{3}{m_1} \frac{\partial \Psi^0_{1,2}}{\partial y},\label{equhorse}
\end{eqnarray}
which has a first integral that may be written
\begin{eqnarray}
\left(\frac{d y}{dt}\right)^2 = -\frac{6Gm_2}{|y|} +\frac{9\Omega^2b^2}{4},\label{HSHOE}
\end{eqnarray}
where as before $b$ is the impact parameter, or the constant value of $|x|$ at large distances from the moonlet.
The value of $y$ for which the horseshoe turns is then given by
$y=y_0=24Gm_2/(9\Omega^2b^2).$
At this point we comment that we are free to choose the origin of time $t=0$
such that it coincides with the closest approach at which $y=y_0.$
Then in the approximation we have adopted, the horseshoe motion 
is such that $y$ is a symmetric function of $t.$
\section{Evaluation of the induced epicyclic amplitude}
Using equations \ref{pertpot1}, \ref{equpert1} and \ref{HSHOE}
we may evaluate the epicyclic amplitudes from equation \ref{epiamp}.
In particular we find 
\begin{eqnarray}
F= - \frac{Gm_2\,{\mathcal E}_2\cos(\Omega t +\epsilon)}{r^3}\left(5 -\frac{(3x^2+12y^2)}{r^2}\right). 
\end{eqnarray}
When evaluating equation \ref{epiamp}, consistent with our assumption that the epicyclic oscillations
are fast, we average over an orbital period assuming that other quantities in the integrands
are fixed, and use equation \ref{HSHOE} to express the integrals with respect to $t$ as integrals with respect to $y.$
We then find $\alpha_{\infty}=-A_{`\infty}\sin\epsilon $ and $\beta_{\infty}=-A_{`\infty}\cos\epsilon$ where
\begin{eqnarray}
A_{\infty}&=& \int^{\infty}_{y_0}\frac{\sqrt{6Gm_2y_0}\;{\mathcal E}_2}{6\Omega r^3\sqrt{1-y_0/y}}
\cdot\left[5 -{\left(\frac{8Gm_2}{\Omega^2}
\left(\frac{1}{y}-\frac{1}{y_0}\right)+12y^2\right)}{r^{-2}}\right]dy,
\end{eqnarray}
with $r^2 = 8Gm_2/(3\Omega^2y_0)(1-y_0/y)+y^2.$
From this we may write
\begin{eqnarray}
|A_{\infty}|={\mathcal E}_{1f} =\frac{Gm_2\,{\mathcal E}_2}{y_0^3\Omega^2}|\mathcal{I}|,\label{inducede}
\end{eqnarray}
where ${\mathcal E}_{1f}$ is the final epicyclic motion of the ring particle and the dimensionless integral $\mathcal{I}$
is given by
\begin{eqnarray}
{\mathcal{I}} &=& \sqrt{\frac{\Omega^2 y^3_0}{6Gm_2}}\int^{\infty}_{y_0}\frac{y_0^2r^{-3}}{\sqrt{1-y_0/y}}
\cdot\left(5 -\frac{\left(\frac{8Gm_2}{\Omega^2}
\left(\frac{1}{y}-\frac{1}{y_0}\right)+12y^2\right)}{r^2}\right)dy.
\end{eqnarray}
We remark that the dimensionless quantity $\eta = 8Gm_2/(3\Omega^2y_0^3)=8(r_H/y_0)^3,$ with $r_H$ being the Hill radius
of the moonlet. It is related to the impact parameter $b$ by $\eta =2^{-6}(b/r_H)^6.$
Thus for an impact parameter amounting to a few Hill radii, in a very approximate sense, $\eta$ is of order unity, $y_0$ is of order $r_H$ and
the induced eccentricity ${\mathcal E}_{1f}$ is of order ${\mathcal E}_2$.

\section{Numerical verification}
We verify the analytic calculation with the numerical code GravTree (see appendix \ref{app:gravtree}). 
A moonlet with Hill radius $r_H=32\,\mathrm{m}$ is setup on an eccentric orbit with ${\mathcal E}_2=130\,\mathrm{m}$ and $\Omega=0.0001314\,\mathrm{s}^{-1}$. 
A ring particle is placed on an initially circular orbit far away from the moonlet with different impact parameters $b$ that all lead to horseshoe orbits. 

\begin{figure}[tb]
\centering
\includegraphics[angle=270,width=\textwidth]{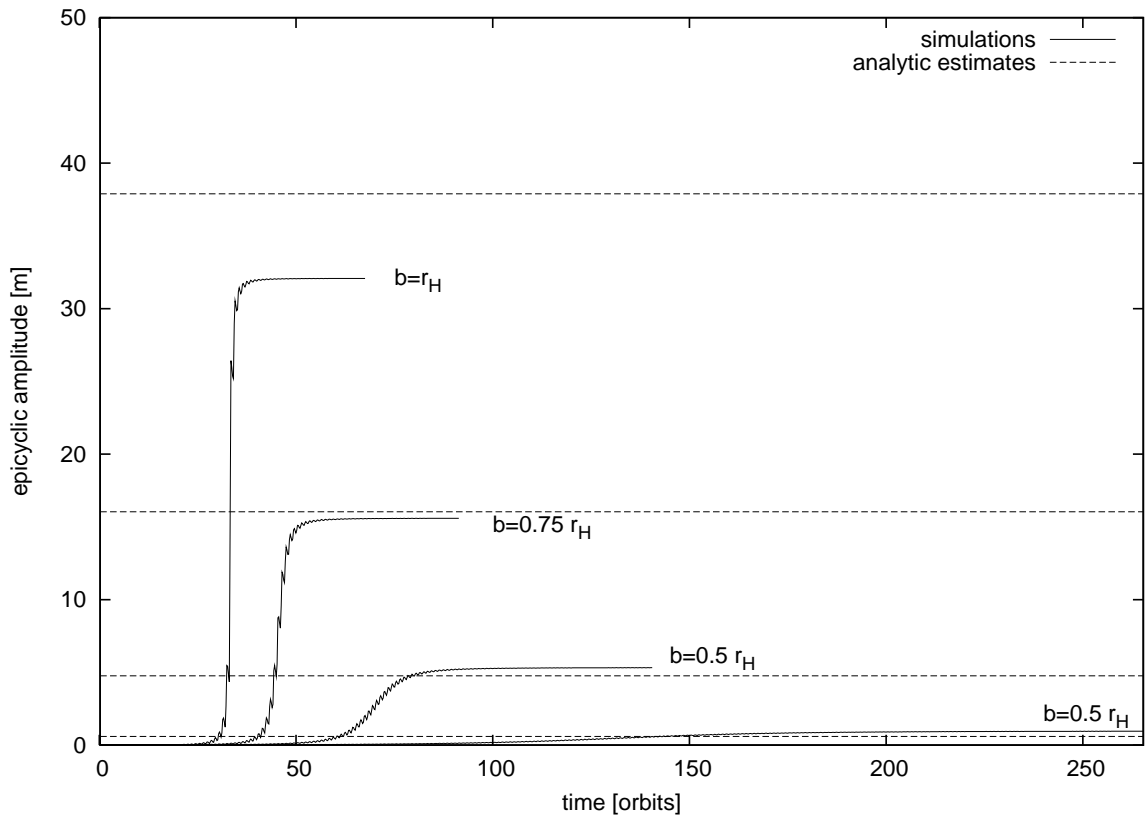}
\caption{Epicyclic amplitude ${\mathcal E}_1$ of a ring particle on a horseshoe orbit as a function of time for different impact parameters $b$. The horizontal lines are the estimate given by equation \ref{inducede}.
\label{fig:responsetest}}
\end{figure}

The amplitude of epicyclic motion of the ring particle, ${\mathcal E}_1$, is plotted in figure \ref{fig:responsetest} for four test cases with $b=0.25r_H$, $b=0.5r_H$, $b=0.75r_H$ and $b=r_H$. 
We also plot the analytic estimates of the final epicyclic motion given by equation \ref{inducede} as straight lines. The agreement is very good, showing some differences at large $b$. Note that an impact parameter greater than $1r_H\sim 2r_H$ leads to impacts so that the departure from the analytic estimate does not influence the circularisation time calculated in chapter \ref{ch:moonlet} significantly.

\chapter{N-body simulations with GravTree}\label{app:gravtree}
This appendix describes the GravTree code that has been written to solve many body systems. The particles might interact gravitationally and through physical collisions. Modules incorporating gas drag and stochastic forces have also been implemented.

First, the relevant equations of motions are presented, followed by a description of the special symplectic integrator used for these. Second, the Barnes Hut tree code is explained. Finally, tests and optimisation strategies that have been implemented are discussed.

\section{Hill's equations}
In many situations it is impossible to simulate an entire accretion disc or planetary ring. For example, in the case of a proto-planetary disc there are about one trillion ($10^{12}$) metre-sized particles. 
Numerically, we can only simulate a few million, at most.

Fortunately, it is usually sufficient to simulate a small patch of the disc that includes most of the physics. In such a patch (also known as a shearing sheet) the shear is linearised and geometric effects are ignored. This leads to a well defined set of equations with shear periodic boundary conditions in the radial and azimuthal directions \citep{Wisdom1988}.  

Let us start from the equations of motion of a test particle around a single star at the origin with mass $M$ such that the gravitational force can be described by
\begin{eqnarray}
\ddot r &=& -\frac{GM}{r^2},\label{eq:app:gravtree:inertial} 
\end{eqnarray}
where $r$ is the distance from the star.
We now move to a non-inertial frame, rotating with angular frequency $\Omega = \left|\mathbf{\Omega}\right|$. The time derivative of a vector quantity $\mathbf v$ in the new frame is given by
\begin{eqnarray}
\dot {\mathbf v}_{\mathrm{r}} &=& \dot {\mathbf v}_{\mathrm{i}} - \mathbf{\Omega} \times {\mathbf v},
\end{eqnarray}
where subscripts i and r denote the inertial and rotating frames, respectively. The acceleration in the rotating frame is thus given by
\begin{eqnarray}
\mathbf{a}_{\mathrm{r}} &=& \mathbf{a}_{\mathrm{i}} - 2 \;\mathbf{\Omega} \times \mathbf{v}_{\mathrm{r}} - \mathbf{\Omega} \times (\mathbf{\Omega} \times \mathbf{r}_{\mathrm{r}}), 
\end{eqnarray}
where ${\mathbf v}_r$ is the velocity and $\mathbf{r}_{\mathrm{r}}$ is the position vector in the rotating frame. Adopting a coordinate system in the rotating frame where $x$ corresponds to the radial direction, $y$ to the azimuthal direction and $z$ to the rotation axis, setting the origin at a distance $a$ away from the star, we can simplify the above equation to get 
\begin{eqnarray}
\ddot x &=& \ddot r_{\mathrm{i}} + 2 \Omega \;\dot y_{\mathrm{r}} + \Omega^2\; (x+a)_{\mathbf{r}}\quad \mathrm{and}\\ 
\ddot y &=&  - 2 \Omega\; \dot x_{\mathrm{r}} + \Omega^2 \;y_{\mathbf{r}}.
\end{eqnarray}
Linearising \Eq \ref{eq:app:gravtree:inertial} around $r=a+x$ gives 
\begin{eqnarray}
\ddot r &=& -\frac{GM}{r^2} =   -\frac{GM}{(a+x)^2} =  -\frac{GM}{a^2} + 2\frac{GM}{a^3} x + \ldots\\
&=& -\Omega^2 a + 2 \Omega^2 x + \ldots
\end{eqnarray}
where we have assumed a Keplerian rotation, $\Omega^2 r^3= GM$. Thus, the linearised equations of motion for an unperturbed particle orbiting in the rotating frame are
\begin{eqnarray}
\ddot x &=&  3 \Omega^2\; x  + 2 \Omega \;\dot y \label{eq:app:gravtree:hills1}\quad \mathrm{and}\\ 
\ddot y &=&  - 2 \Omega \;\dot x , \label{eq:app:gravtree:hills2}
\end{eqnarray}
where we have dropped the subscript r. We can also linearise the gravity in the vertical direction to get 
\begin{eqnarray}
\ddot z &=&  - \Omega^2\; z. \label{eq:app:gravtree:hills3}
\end{eqnarray}
This set of local approximations is known as Hill's equations.
We solve these with a symplectic leap frog, kick drift kick, time-stepping scheme which is discussed in more detail in the next section.

\section{Symplectic integrator}
Symplectic integrators in celestial mechanics have several attractive features, such as conservation of energy up to machine precision. We would therefore like to use such an integrator for solving Hill's equations numerically. A widely used standard symplectic integrator is leap-frog which can be described by a kick, followed by a drift and another kick sub-time-step (KDK). A possible C implementation of one full time-step with length \texttt{dt} is
\begin{verbatim}
void kdk(double* x, double* v, double* a, double dt){
   // Kick
   v[0] += 0.5 * dt * a[0];
   v[1] += 0.5 * dt * a[1];
   v[2] += 0.5 * dt * a[2];

   // Drift
   x[0] += dt * v[0];
   x[1] += dt * v[1];
   x[2] += dt * v[2];

   // Kick
   update_force(x,v,a);
   v[0] += 0.5 * dt * a[0];
   v[1] += 0.5 * dt * a[1];
   v[2] += 0.5 * dt * a[2];
}
\end{verbatim}
Here, \texttt{x}, \texttt{v} and \texttt{a} are pointers to the position, velocity and acceleration vectors, respectively.
This integrator is symplectic if the forces calculated in \texttt{update\_force(x,v,a)} do not depend on the velocities. However, Hill's equations do depend on the velocities (see \Eqs \ref{eq:app:gravtree:hills1} and \ref{eq:app:gravtree:hills2}). Note that the $z$ component (\Eq \ref{eq:app:gravtree:hills3}) does not depend on the velocities and thus can be integrated with a standard method. \cite{Quinn2010} describe a new symplectic integrator for Hill's equations. The changes to the KDK algorithm are minimal and only one additional value per particle has to be saved. A modified version of the above code is
\begin{verbatim}
void kdk_quinn(double* x, double* v, double* a, double* Py, double dt){
   // Kick
   v[0] += -0.5*dt * (OMEGA*OMEGA * r[0] - a[0]);
   Py    = v[1] + 2.0*OMEGA * r[0] + 0.5*dt * a[1];
   v[0] += dt * OMEGA * Py;
   v[1]  = Py - OMEGA * r[0] - OMEGA * (r[0] + dt * v[0]);
   v[2] += 0.5*dt * a[2];

   // Drift
   x[0] += dt * v[0];
   x[1] += dt * v[1];
   x[2] += dt * v[2];

   // Kick
   update_force(x,a);
   v[0] += dt * OMEGA * Py;
   v[0] += -0.5*dt * (OMEGA*OMEGA * r[0] - a[0]);
   v[1]  = Py - 2.0*OMEGA * r[0] + 0.5*dt * a[1];
   v[2] += 0.5*dt * a[2];
}
\end{verbatim}
Here, \texttt{OMEGA} is the orbital frequency and \texttt{Py} is initialised to \texttt{v[1] + 2.0*OMEGA * r[0]}. \texttt{Py} is a conserved quantity for an unperturbed orbit. Note that the drift step is unmodified. This allows for an unchanged collision detection algorithm, usually performed during the drift sub-time-step. The function \texttt{update\_force()} does not include the terms due to Hill's equation anymore, only additional forces (e.g. gravity from other particles, gas drag,~\ldots). 

The symplectic integrator has proven to be very useful for integrations of moonlets in Saturn's rings, where the eccentricity is generally very small $(\sim10^{-8})$ and the system is integrated for hundreds of orbits.

\section{Gravity calculation}
In the $N$-body problem that we are considering here, we have to solve Newton's equations of universal gravitation for a large number of particles $N$. The gravitational force on the $i$-th particle is given by 
\begin{eqnarray}
\bm{F}_\text{grav} = \sum_{j\neq i} G \frac{m_im_j}{\left( r_{ij} +b \right)^2}\mathbf{e}_{ij}, \label{eq:nbody}
\end{eqnarray}
where $b$ is the smoothing length used to avoid divergencies in numerical simulations. In simulations where particles have a finite size and physical collisions are included we can safely set $b = 0$. The unit vector in the direction of the gravitational force between the $i$-th  and $j$-th particle is $\mathbf{e}_{ij}$. 
Calculating the gravity for each particle from each other particle results in $O\left(N^2\right)$ operations. To reduce the number of operations, one can use different approximations to \Eq~\ref{eq:nbody}. Here, we use a Barnes-Hut (BH) tree code \citep{Barnes1986}.
The BH tree reduces the number of calculations to $O\left(N\log N\right)$. 

A tree code in three dimensions works by sub-dividing the original box into eight smaller cells with half the length of the original box. 
This can be done iteratively, by adding one particle to the box at a time. 
The new particle is given to the root box first. 
If the root box has already been sub-divided into smaller boxes, the particle is passed down to the next level. If the particle ends up in a box that has not yet been sub-divided, but already contains a particle, then this box is divided into smaller boxes and the particle that was in this cell is removed. 
The original and the new particles are then re-added. This process is repeated until all particles are inserted.
The depth of the tree in a homogeneous medium is approximately $\log_8 N$. 
The next step is to pre-calculate the total mass and the centre of mass for each cell at each level of the tree. 

\begin{figure}[tb]
\center
\resizebox{0.5\columnwidth}{!}{
\begin{picture}(0,0)%
\includegraphics{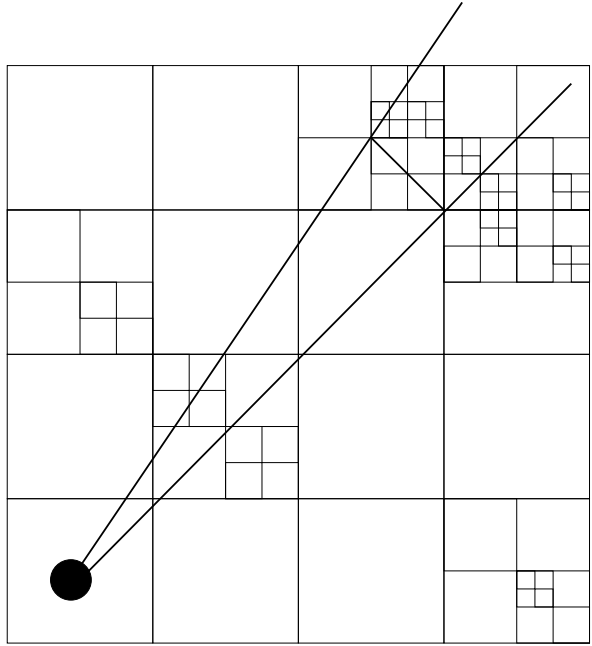}%
\end{picture}%
\setlength{\unitlength}{3947sp}%
\begingroup\makeatletter\ifx\SetFigFont\undefined%
\gdef\SetFigFont#1#2#3#4#5{%
  \reset@font\fontsize{#1}{#2pt}%
  \fontfamily{#3}\fontseries{#4}\fontshape{#5}%
  \selectfont}%
\fi\endgroup%
\begin{picture}(4824,5359)(3739,-7048)
\end{picture}%
}
\caption{The opening angle $\theta$ is the acute angle in the triangle and defined as the ratio of cell width divided by the distance from the particle to the cell. If $\theta \leq 0.5$, the cell is sufficiently large away and any substructure can be ignored.
\label{fig:app:gravtree:openingangle}}
\end{figure}

To get the force acting on a particle, one starts at the top of the tree and descends into the tree as far as necessary to achieve a given accuracy. 
If the current cell is far away from the particle for which the force is calculated, then the detailed density structure within this cell is not important. 
All that matters is the box's monopole moment (total mass and centre of mass)\footnote{This could be easily extended to include higher order moments.}. 
One therefore does not have to descend into this branch any further. 
This criteria can be quantified by an opening-angle $\theta$, which is defined as the ratio of the box width and the distance from the particle to the centre of the box, as shown in figure \ref{fig:app:gravtree:openingangle}.
In all simulations we use  $\theta \leq 0.5$.

\begin{figure}[tbp]
\center
\resizebox{0.5\columnwidth}{!}{
\begin{picture}(0,0)%
\includegraphics{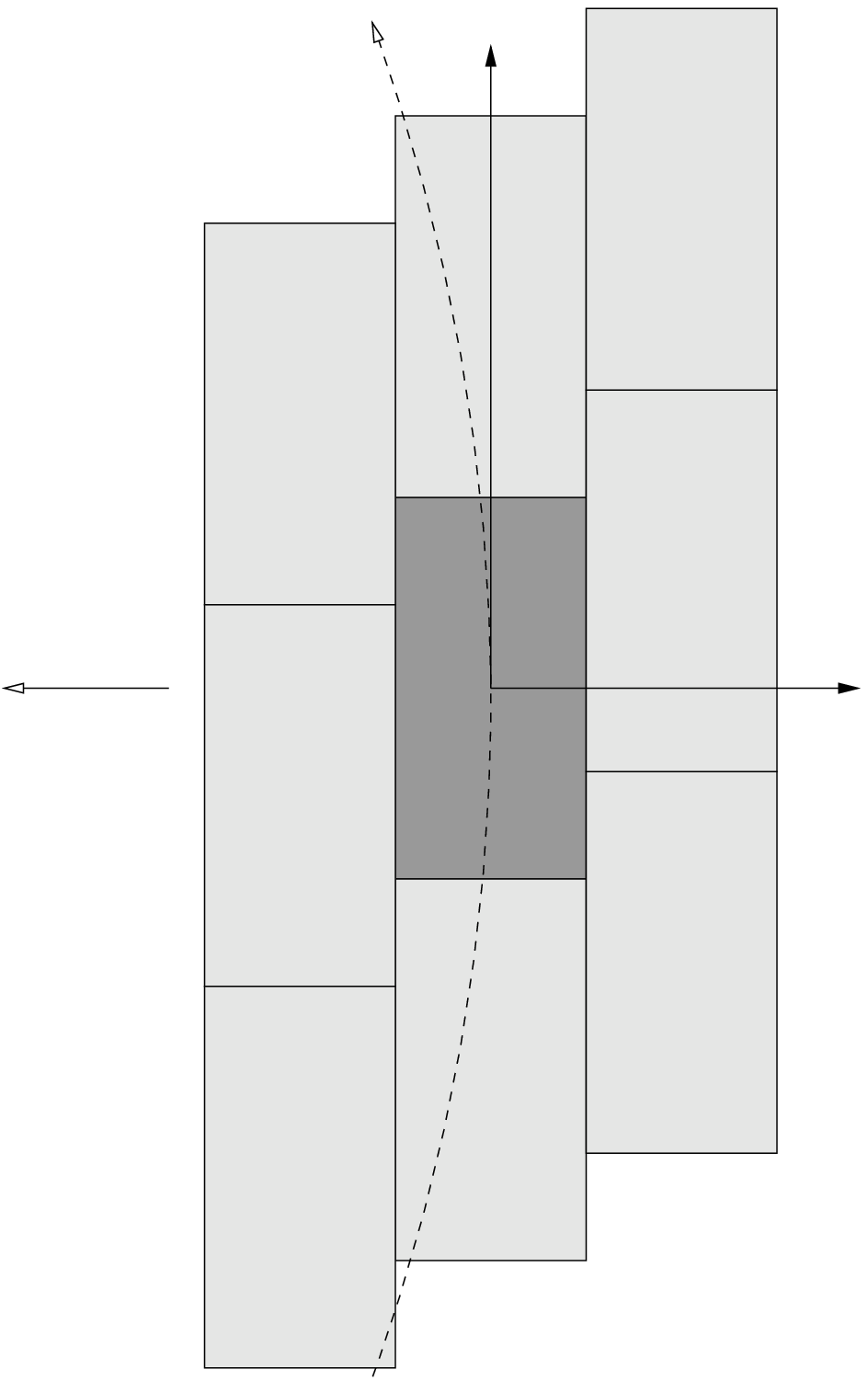}%
\end{picture}%
\setlength{\unitlength}{3276sp}%
\begingroup\makeatletter\ifx\SetFigFont\undefined%
\gdef\SetFigFont#1#2#3#4#5{%
  \reset@font\fontsize{#1}{#2pt}%
  \fontfamily{#3}\fontseries{#4}\fontshape{#5}%
  \selectfont}%
\fi\endgroup%
\begin{picture}(5424,8645)(1039,-7794)
\put(4201,389){\makebox(0,0)[lb]{\smash{{\SetFigFont{10}{12.0}{\familydefault}{\mddefault}{\updefault}{\color[rgb]{0,0,0}y}%
}}}}
\put(6226,-3361){\makebox(0,0)[lb]{\smash{{\SetFigFont{10}{12.0}{\familydefault}{\mddefault}{\updefault}{\color[rgb]{0,0,0}x}%
}}}}
\put(4951,-1411){\makebox(0,0)[lb]{\smash{{\SetFigFont{10}{12.0}{\familydefault}{\mddefault}{\updefault}{\color[rgb]{0,0,0}Ghost boxes}%
}}}}
\put(3976,-3286){\makebox(0,0)[lb]{\smash{{\SetFigFont{10}{12.0}{\familydefault}{\mddefault}{\updefault}{\color[rgb]{0,0,0}Main box}%
}}}}
\put(1276,-3286){\makebox(0,0)[lb]{\smash{{\SetFigFont{10}{12.0}{\familydefault}{\mddefault}{\updefault}{\color[rgb]{0,0,0}to star}%
}}}}
\end{picture}%
}
\caption{Shearing box and ghost boxes. The ghost-boxes move in the $y$ direction, following the mean shear, leading to \textit{shear-periodic} boundary conditions.
Here, the boxes have an aspect ratio of 2, but any other aspect ratio is possible.
\label{fig:app:gravtree:boxes}}
\end{figure}

\section{Boundary conditions}
Our implementation can handle arbitrary aspect ratios of the box, which is especially useful if a small annulus has to be simulated.
We use ghost rings made of ghost boxes in the radial and azimuthal directions as shown in \Fig \ref{fig:app:gravtree:boxes}. The figure shows one ghost ring.
A ghost box is either simply a shifted copy of the main box or a completely separate simulation, depending on the type of simulation (see below).

The gravity on each particle is then calculated by summing over contributions from each (ghost) box. 
This setup approximates a medium of infinite horizontal extent and avoids large force discontinuities at the boundaries.
In general no ghost boxes are used in the vertical direction because the disc is stratified.

\subsection{Shear periodic boundary conditions}
In a standard simulation, we consider the main box together with eight ghost boxes that are all identical copies of the main box shifted according to the mean shear.
On account of this motion ghost boxes are removed when their centres are shifted in azimuth by more than $1.5$ times their length from the centre
of the main box and then reinserted in the same orbit on the opposite side of the main box so that the domain under consideration is prevented from shearing out. 

If a particle in the main box crosses one of its boundaries, it is reinserted on the opposite side so that the total number of particles is conserved.
The gravity acting on a particle in the main box, is calculated by summing over the particles in the main box and all ghost boxes. 

A finite number of ghost rings can only act as an approximation of an infinite medium and the gravitational force will tend to concentrate particles horizontally in the centre of the box. 
In simulations of planetesimals (see chapter \ref{ch:planetesimals}), we use 8 ghost rings, corresponding to 63 ghost boxes. 
The asymmetry of the gravitational force between the centre and the faces of the box is reduced by a factor of 20 compared to using no ghost rings at all. 
Note, however, that a small asymmetry remains in every simulations and leads to higher a concentration of particles in the box centre.
In other simulations, such as \cite{Tanga2004}, this effect can not be seen because the system is initially gravitationally unstable and integrated for only one orbit. 
Some systems that we are interested in are marginal gravitationally unstable and this slight asymmetry might become important after many orbits (see chapter \ref{ch:planetesimals}).

\subsection{Pseudo shear periodic boundary conditions}
If a strong perturber is present in the simulation, such as a moonlet in chapter \ref{ch:moonlet}, we use a different setup that we call \emph{pseudo} shear periodic boundary conditions.
In that case, all ghost, or auxiliary, boxes are identical copies whose centres are shifted according to the background shear as in a normal shearing sheet.
However, the auxiliary boxes are a completely separate simulation, mimicking the background state of the ring system without a perturber.  

If a particle in the main box crosses one of its boundaries, it is discarded.
If a particle in the auxiliary box crosses one of its boundaries, it is reinserted on the other side of this box, according to normal shear periodic boundary conditions.
But in addition it is also copied into the corresponding location in the main box.

Other simulations \citep{LewisStewart2009} use a very long box (about 10 times longer than the typical box size that we use) to ensure that particles are completely randomised between encounters with the moonlet. 
We are in general not interested in the long wavelength response that is created by the moonlet.
Effects that are most important for the moonlet's dynamical evolution are found to happen within a few Hill radii. 
Using the pseudo shear periodic boundary conditions, we ensure that incoming particles are uncorrelated and do not contain prior information about the perturber.

The gravity acting on a particle in the main box, which also contains the moonlet, is calculated by summing over the particles in the main box and all auxiliary boxes. 
The gravity acting on a particle in the auxiliary box is calculated the standard way, 
by using ghost boxes which are identical copies of the auxiliary box.

This setup speeds up our calculations by more than an order of magnitude.

\section{Collision detection}
\begin{figure}[tbp]
\center
\resizebox{0.7\columnwidth}{!}{ 
\begin{picture}(0,0)%
\includegraphics{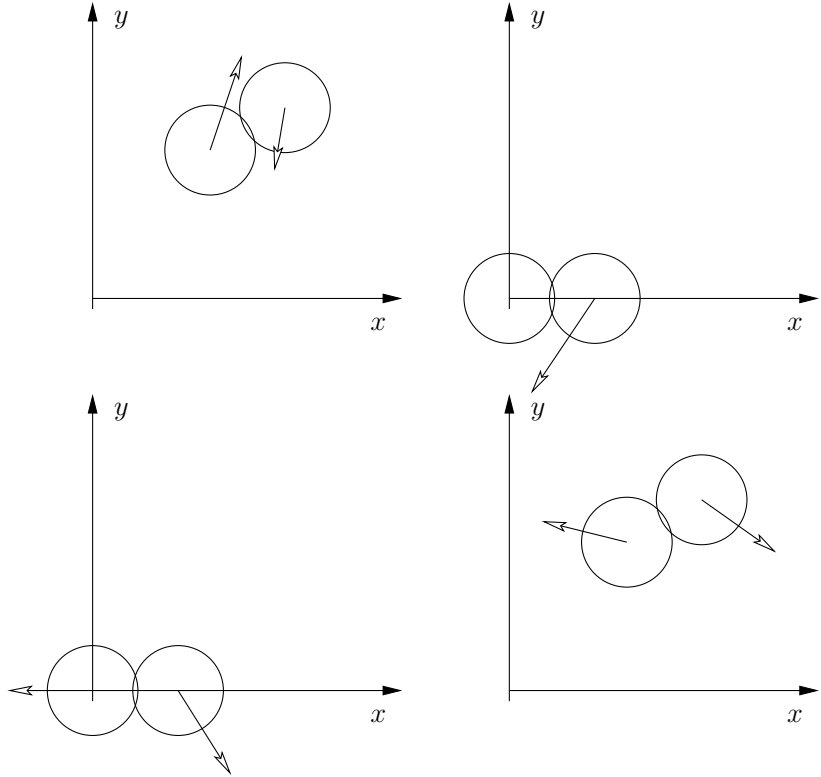}%
\end{picture}%
\setlength{\unitlength}{3947sp}%
\begingroup\makeatletter\ifx\SetFigFont\undefined%
\gdef\SetFigFont#1#2#3#4#5{%
  \reset@font\fontsize{#1}{#2pt}%
  \fontfamily{#3}\fontseries{#4}\fontshape{#5}%
  \selectfont}%
\fi\endgroup%
\begin{picture}(5763,5499)(2239,-8098)
\put(4801,-4936){\makebox(0,0)[lb]{\smash{{\SetFigFont{12}{14.4}{\familydefault}{\mddefault}{\updefault}{\color[rgb]{0,0,0}$x$}%
}}}}
\put(3001,-2761){\makebox(0,0)[lb]{\smash{{\SetFigFont{12}{14.4}{\familydefault}{\mddefault}{\updefault}{\color[rgb]{0,0,0}$y$}%
}}}}
\put(7726,-4936){\makebox(0,0)[lb]{\smash{{\SetFigFont{12}{14.4}{\familydefault}{\mddefault}{\updefault}{\color[rgb]{0,0,0}$x$}%
}}}}
\put(5926,-2761){\makebox(0,0)[lb]{\smash{{\SetFigFont{12}{14.4}{\familydefault}{\mddefault}{\updefault}{\color[rgb]{0,0,0}$y$}%
}}}}
\put(4801,-7711){\makebox(0,0)[lb]{\smash{{\SetFigFont{12}{14.4}{\familydefault}{\mddefault}{\updefault}{\color[rgb]{0,0,0}$x$}%
}}}}
\put(3001,-5536){\makebox(0,0)[lb]{\smash{{\SetFigFont{12}{14.4}{\familydefault}{\mddefault}{\updefault}{\color[rgb]{0,0,0}$y$}%
}}}}
\put(7726,-7711){\makebox(0,0)[lb]{\smash{{\SetFigFont{12}{14.4}{\familydefault}{\mddefault}{\updefault}{\color[rgb]{0,0,0}$x$}%
}}}}
\put(5926,-5536){\makebox(0,0)[lb]{\smash{{\SetFigFont{12}{14.4}{\familydefault}{\mddefault}{\updefault}{\color[rgb]{0,0,0}$y$}%
}}}}
\end{picture}%
}
\caption{Resolving collisions. Top left: An overlap has been detected in the original frame. Top right: Reference frame where one particle is at the origin and at rest. Bottom left: Collision is resolved using momentum and energy conservation. Bottom right: Back in original frame.
\label{fig:app:gravtree:coll}}
\end{figure}
The collision detection algorithm makes use of the already existing tree structure to search for nearest neighbours. 
Again, this reduces the complexity from $O(N^2)$ to $O(N\log N)$, and can easily be done in parallel. 
We search for overlapping particles that are approaching each other.  
The order in which multiple collisions involving the same particle are resolved are not important in our case. 
We can ignore this effect as long as the time-step is sufficiently small. 

This approach might lead to another problem. Consider two particles that are constantly touching;
they might sink into each other over long timescales due to numerical errors and other dissipative effects. 
To avoid this fate, we limit the minimal impact velocity $v_{|,\mathrm{min}}$ to a tiny fraction of the random velocity dispersion.
$v_{|,\mathrm{min}}$ has to be small enough to not affect the global evolution of the system.

Collisions are then resolved using energy and angular momentum conservation and a (velocity dependent) coefficient of restitution.
This is done by moving to a reference frame in which one particle is at rest at the origin and the other particle is on the $x$-axis, as shown in figure \ref{fig:app:gravtree:coll}.

\section{Optimisations}
\subsubsection{Cache}
The tree structure leads to a large number of random memory accesses. 
This can potentially harm the performance of the code if many cache misses occur. 
When a cache miss occurs, the CPU has to request the data from the main memory, which is much slower than a request that can be handled by the cache.
Using a simple heuristic branch prediction, one can speed up the calculation. We do this by pre-caching lower level tree branches before working on the current level. 
In C++, using non-standard \texttt{MMX} extensions, a typical tree walking subroutine looks as follows:
\begin{verbatim}
#include <xmmintrin.h>  
void cell::walk_tree(){
    // Try to cache position and tree structures of next lower level
    for (int i =0; i<8; i++) {
         if (d[i]!=NULL){
              _mm_prefetch((const char*) &(d[i]->pos), _MM_HINT_NTA);
              _mm_prefetch((const char*) d[i]->d,      _MM_HINT_NTA);
         }
    }
    
    // Do calculations here
    // ...

    // Then descent into next level
    for (int i =0; i<8; i++){
         if (d[i]!=NULL){
               d[i]->walk_tree();
         }
    }
}
\end{verbatim}
While calculations on the current level are performed, the CPU can try to cache data for future calculations in idle CPU cycles. 
This trick can give a performance boost of up to 20\%.

\subsubsection{MPI}
Parallelising tree codes on distributed memory systems is hard. This is not a shortcoming of the algorithm itself but rather the implementation of interconnect libraries such as MPI. 
A standard MPI program knows in advance when to send and receive data. This is a reasonable approach for grid based hydrodynamic codes for example, where exactly the same boundary cells have to be transmitted each time-step. 

In a tree code, as described above, the situation is different. It is not possible to predict which cells will have to be opened during the calculation of the gravitational force for one particle. Thus, if the cell structures are distributed over many machines, we have to \textit{pull} the required data instantaneously. 

With the current implementations of the MPI library, this is not possible without large overhead. Thus, we use a brute force method to do calculations on distributed memory machines.

Each machine holds an exact copy of the entire tree. The tree construction and reconstruction is done on a single node and only the computationally expensive calculations are done in parallel. These include gravity calculation and collision detection. 

This kind of parallelisation is limited by the number of particles, as the memory usage of each node scales with $N\log N$. Furthermore, the communication (usually all-to-all and one-to-all) scale also as $N\log N$. 
Nevertheless, it has proven to be very practical up to  $N\sim10^{6}-10^{7}$.

\section{Tests}
\begin{figure}[htb]
\centering
\includegraphics[angle=270,width=0.85\columnwidth]{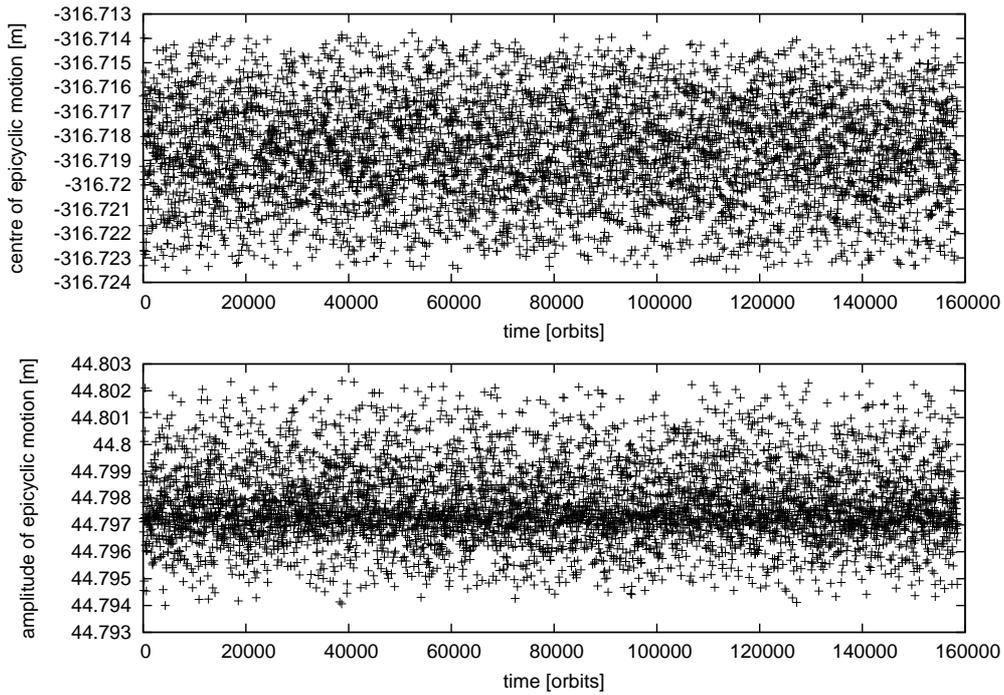}
\caption{Conserved quantities in the shearing box. The top panel shows the centre of epicyclic motion relative to the centre of the box. The bottom panel shows the amplitude of epicyclic motion.
\label{fig:conserv}}
\end{figure}

We begin with testing several conserved quantities. 
In figure \ref{fig:conserv} the amplitude and the centre of epicyclic motion of a test particle in the shearing box are plotted as a function of time. 
One can see that both quantities are conserved accurately without any asymmetry over many dynamical timescales, as expected for the symplectic integrator. 
Furthermore, we made sure that the treatment of collisions is working correctly, especially across ghost cell boundaries. 

\begin{sidewaysfigure}[p]
\begin{flushright}
\setlength{\unitlength}{0.9\columnwidth}
\begin{picture}(1,0.6)
\epsfxsize=0.9\columnwidth
  \epsffile{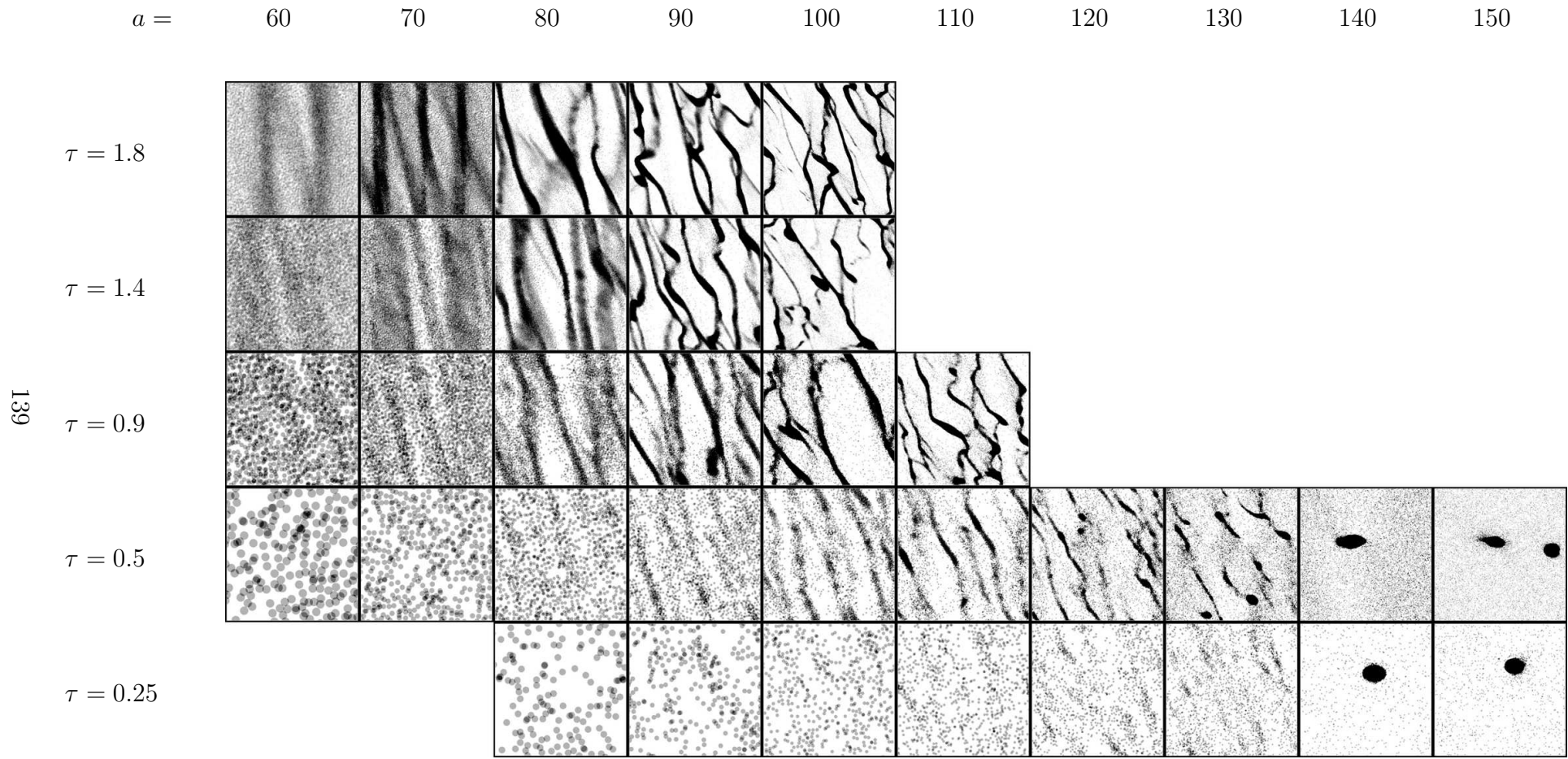} 
  \put(-1.12,0.441){$\tau=1.8$}
  \put(-1.12,0.341){$\tau=1.4$}
  \put(-1.12,0.241){$\tau=0.9$}
  \put(-1.12,0.141){$\tau=0.5$}
  \put(-1.12,0.041){$\tau=0.25$}
  
  \put(-1.07,0.541){$a=$}
  \put(-0.97,0.541){$60$}
  \put(-0.87,0.541){$70$}
  \put(-0.77,0.541){$80$}
  \put(-0.67,0.541){$90$}
  \put(-0.57,0.541){$100$}
  \put(-0.47,0.541){$110$}
  \put(-0.37,0.541){$120$}
  \put(-0.27,0.541){$130$}
  \put(-0.17,0.541){$140$}
  \put(-0.07,0.541){$150$}
\end{picture}
\end{flushright}
\caption{Dependence of self-gravity wakes on the geometrical optical depth $\tau$ and the semi major axis $a$ (in units of 1000km).}
\label{fig:gravtree:salo}
\end{sidewaysfigure}

Next, we present a comparison between GravTree and the results obtain by \cite{Schmidt2009}. 
This is a comprehensive sample of runs including of both self-gravity and collisions. 
In every simulation, the particle density is $\rho=0.9\,\mathrm{g/cm^3}$, and the box width corresponds to four critical Toomre wavelengths. 
The coefficient of restitution is $0.5$.
By changing the optical depth and semi-major axis of the planetary ring, one can sweep through different regimes. 
Snapshots of the density distribution are plotted in figure \ref{fig:gravtree:salo}. 
Those should be compared to figure 14.7 in \cite{Schmidt2009}.

One can clearly see different regimes. On the right hand side of the plot, particles clump into aggregates. On the top right boundary, strong gravitational wakes are visible. 
For $\tau=1.8$, $a=70000\,\mathrm{km}$ and simulations close by, viscous overstability can be observed as strong azimuthal structures.
Overall, the agreement with \cite{Schmidt2009} is very good. 
For large $a$, the simulations performed with GravTree have a trend to form aggregates slightly earlier. 

\begin{figure}[htb]
\centering
\includegraphics[angle=270,width=0.85\columnwidth]{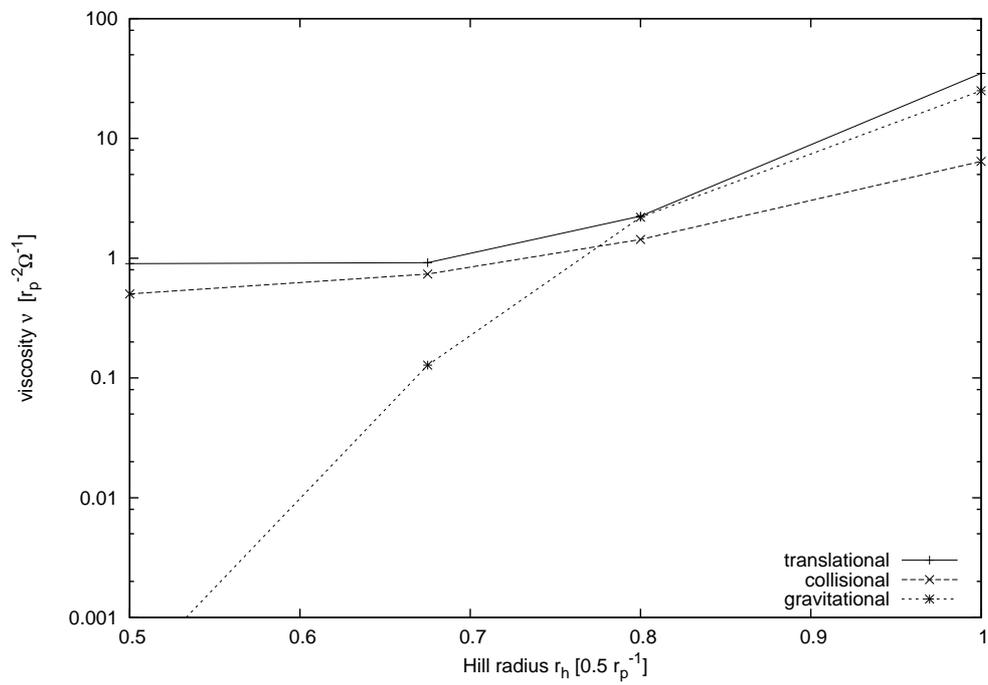} 
\caption{Translational, collisional and gravitational viscosity components as a function of normalised Hill radius of the ring particles.
\label{fig:daisaka}}
\end{figure}

As a further, more quantitative test, we reproduce the results from \cite{Daisaka2001}.
Therefore, the translational, collisional and gravitational components of the ring viscosity are measured, as defined by equation 19 in said paper.
The results are shown in figure \ref{fig:daisaka} and should be compare to the left hand side of figure 7 in \cite{Daisaka2001}. 
The viscosity is in units of the particle radius and the orbital frequency, the Hill radius (we here use the slightly different convention $r_h=\left({2m_p}/({3M})\right)^{1/3} a$) is normalised by the particle radius. 
One can see that for small Hill radii (low density particles), the effects of self-gravity are not important.
Again, very good agreement with the previous study is achieved. 


\section{Visualisations}
\begin{figure}[p]
\centering
\includegraphics[width=0.85\columnwidth]{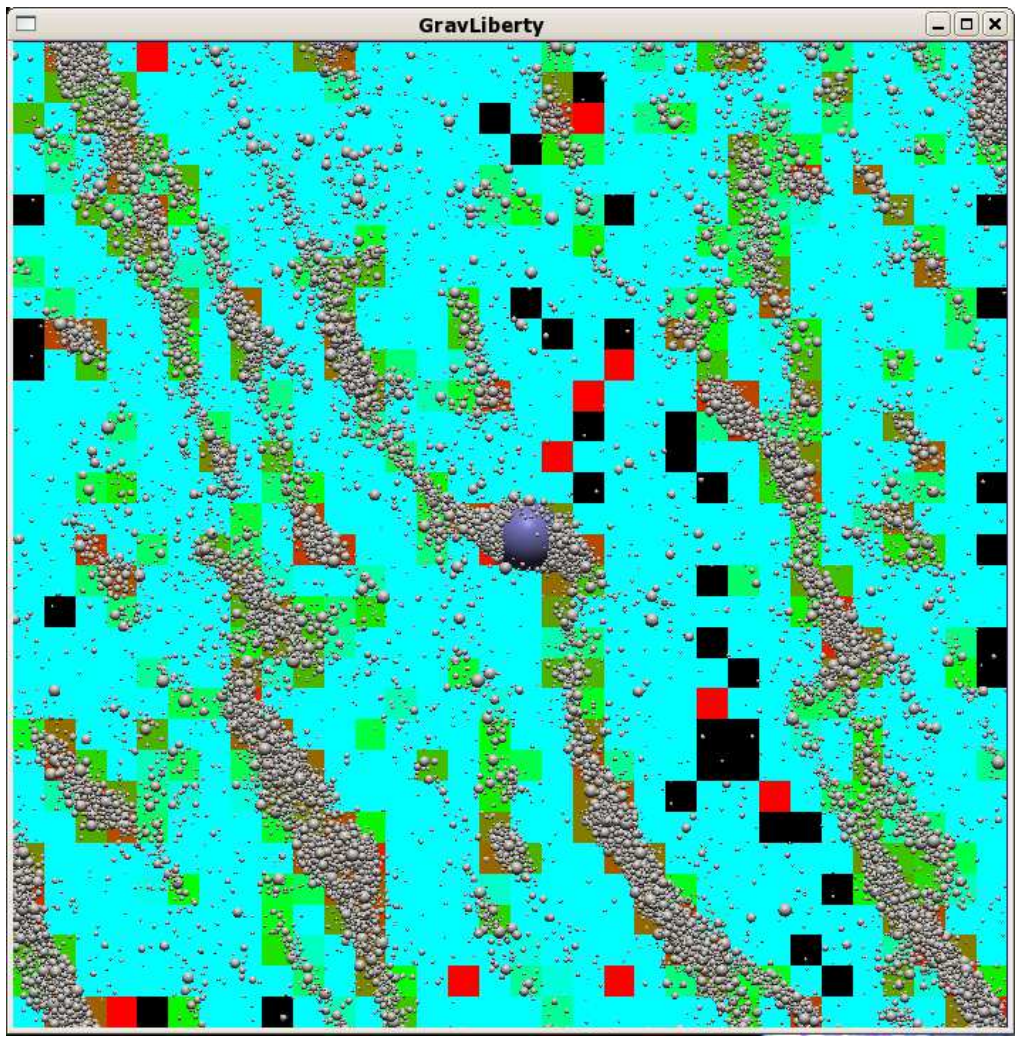} 
\caption{ Screenshot of the OpenGL visualisation module of GravTree, showing a simulation of a moonlet embedded in Saturn's rings. The entire computational domain is shown with an overlay, showing the Toomre $Q$ parameter as coloured squares.
\label{fig:app:gravtree:screen1}}
\end{figure}

\begin{figure}[p]
\centering
\includegraphics[width=0.85\columnwidth]{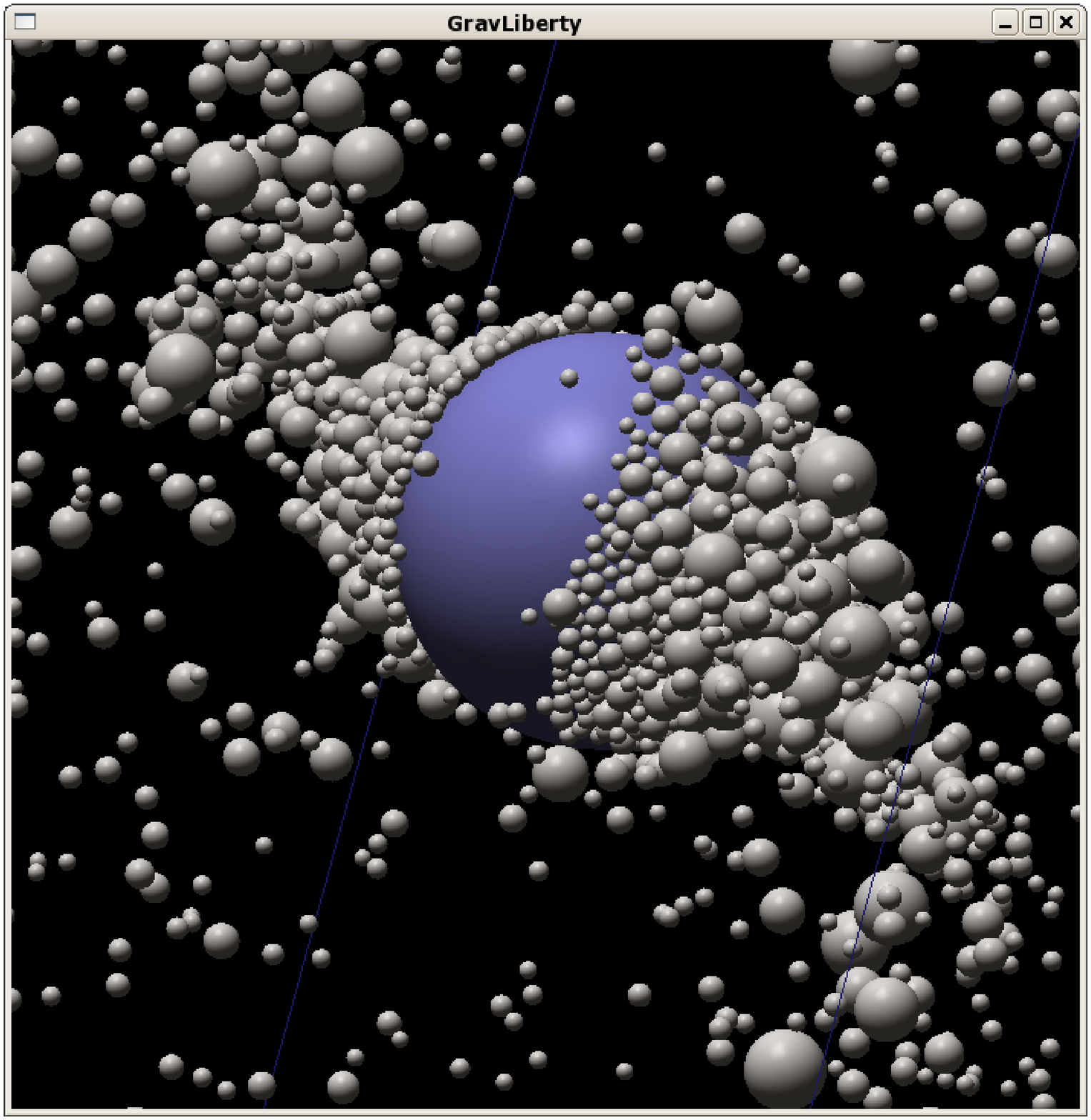} 
\caption{ Screenshot of the OpenGL visualisation module of GravTree, showing a simulation of a moonlet embedded in Saturn's rings. A closeup of the moonlet with particles resting on the moonlet's surface (blue) is shown.
\label{fig:app:gravtree:screen2}}
\end{figure}

The visualisation of large data sets plays an increasingly important role in the scientific process. 
GravTree comes with a visualisation module, based on the OpenGL API. It can be used to visualise simulations in real time while they are running
or in a standalone playback mode. The real time mode has proven to be especially useful during the development process.

Beside the particle distribution itself, average quantities such as the velocity dispersion and the filling factor can be shown as a transparent
overlay. The tree structure can also be plotted. All graphic routines are easily expandable to incorporate any diagnostic that the current 
project requires.
Two screenshots are shown in figures \ref{fig:app:gravtree:screen1} and \ref{fig:app:gravtree:screen2}.

\chapter{Hydrodynamical simluations with Prometheus}\label{app:dpmhd3d}
Simulations of planets embedded in an accretion disc have been performed by a large number of authors in recent years. 
There are two main types of algorithm that can be used to simulate gas dynamics: finite difference methods such as ZEUS \citep{StoneNorman1992}, NIRVANA \citep{ZieglerYorke1997}  and FARGO \citep{Masset00} or Godunov methods like RODEO \citep{PardekooperMellema2006} and ATHENA \citep{Athena2008}. As the name suggests, finite difference codes evaluate partial derivatives with a finite difference approximation. Godunov methods solve (exactly or approximately) a Riemann problem at every cell interface. 

In this appendix, the code Prometheus is presented, which is a finite difference code in spherical coordinates, specialised for planet disc interactions. It is similar to the FARGO code but can also run simulations in three dimensions and has a module for solving the full MHD equations. This appendix can not be a comprehensive discussion of finite difference codes in general and merely lists the main concepts and implementation specific features of Prometheus.

Prometheus is written entirely in C and requires no external libraries. It can be compiled with a visualisation module, in which case it has to be linked to OpenGL, GLUT and libpng. Prometheus will be made publicly available at \url{http://github.com/hannorein/prometheus}.

\section{Navier-Stokes equations}
For many astrophysical processes collisions between gas particles are frequent enough such that a continuum description can be used. Here, the proto-planetary disc is treated as a gas whose motion is described by the Navier-Stokes equations. Although an MHD module has already been written for this code, it is not described here, due to current lack of testing and it has also not been used in scientific runs yet. We therefore restrict this discussion to the classical Navier-Stokes equations without magnetic fields, which we list in the following for both two and three dimensions.

\subsection{Three dimensions}
\begin{figure}
\begin{picture}(0,0)%
\includegraphics{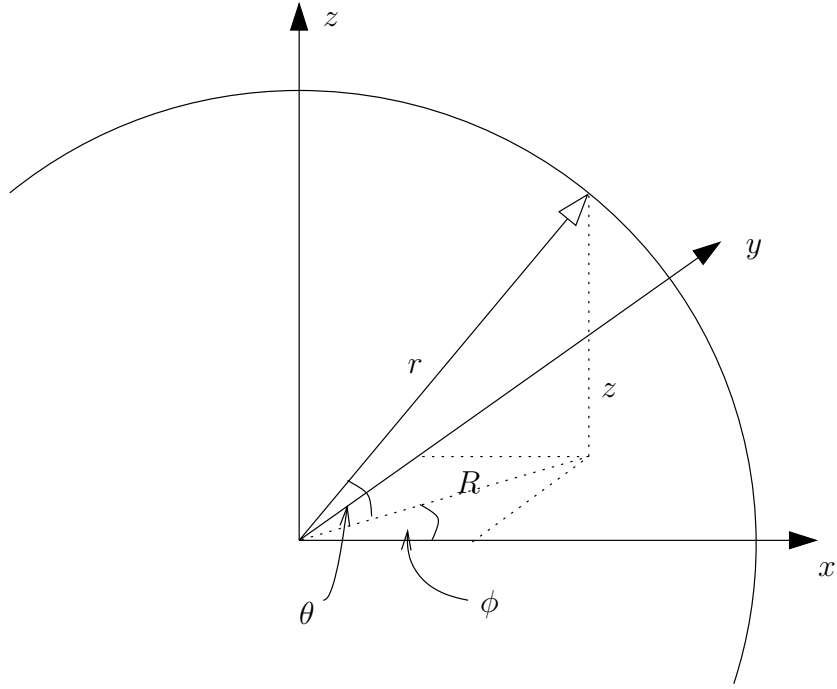}%
\end{picture}%
\setlength{\unitlength}{3947sp}%
\begingroup\makeatletter\ifx\SetFigFont\undefined%
\gdef\SetFigFont#1#2#3#4#5{%
  \reset@font\fontsize{#1}{#2pt}%
  \fontfamily{#3}\fontseries{#4}\fontshape{#5}%
  \selectfont}%
\fi\endgroup%
\begin{picture}(5309,4295)(4193,-6069)
\put(6151,-1936){\makebox(0,0)[lb]{\smash{{\SetFigFont{12}{14.4}{\familydefault}{\mddefault}{\updefault}{\color[rgb]{0,0,0}$z$}%
}}}}
\put(9226,-5386){\makebox(0,0)[lb]{\smash{{\SetFigFont{12}{14.4}{\familydefault}{\mddefault}{\updefault}{\color[rgb]{0,0,0}$x$}%
}}}}
\put(8776,-3361){\makebox(0,0)[lb]{\smash{{\SetFigFont{12}{14.4}{\familydefault}{\mddefault}{\updefault}{\color[rgb]{0,0,0}$y$}%
}}}}
\put(7126,-5611){\makebox(0,0)[lb]{\smash{{\SetFigFont{12}{14.4}{\familydefault}{\mddefault}{\updefault}{\color[rgb]{0,0,0}$\phi$}%
}}}}
\put(6001,-5686){\makebox(0,0)[lb]{\smash{{\SetFigFont{12}{14.4}{\familydefault}{\mddefault}{\updefault}{\color[rgb]{0,0,0}$\theta$}%
}}}}
\put(6676,-4111){\makebox(0,0)[lb]{\smash{{\SetFigFont{12}{14.4}{\familydefault}{\mddefault}{\updefault}{\color[rgb]{0,0,0}$r$}%
}}}}
\put(6976,-4861){\makebox(0,0)[lb]{\smash{{\SetFigFont{12}{14.4}{\familydefault}{\mddefault}{\updefault}{\color[rgb]{0,0,0}$R$}%
}}}}
\put(7876,-4261){\makebox(0,0)[lb]{\smash{{\SetFigFont{12}{14.4}{\familydefault}{\mddefault}{\updefault}{\color[rgb]{0,0,0}$z$}%
}}}}
\end{picture}%
\centering
\caption{Spherical coordinate system used in Prometheus \label{fig:app:dpmhd3d:spherical}}
\end{figure}

In spherical coordinates as defined in \Fig~\ref{fig:app:dpmhd3d:spherical}, the inviscid Navier-Stokes equations in conservative form are
\begin{align}
&\frac{\partial \rho}{\partial t } &&+ \nabla \cdot \left(\rho \mathbf u \right) &&=0 &&\label{eq:app:dpmhd3d:navier1}\\
&\frac{\partial \rho u_r}{\partial t } &&+ \nabla \cdot \left(\rho u_r \mathbf u \right) &&=
     \rho\;\frac{u_\theta^2 + u_\phi^2}{r}  &&- \frac{\partial p}{\partial r}&&+ \rho a_r  \label{eq:app:dpmhd3d:navier2}\\
&\frac{\partial \rho r u_\theta}{\partial t } &&+ \nabla \cdot \left(\rho r u_\theta \mathbf u \right) &&=
    - \rho \; \tan\theta\;\; u_\phi^2 &&-\frac{\partial p}{\partial\theta} &&+ \rho a_\theta\;r \label{eq:app:dpmhd3d:navier3}\\
&\frac{\partial \rho r u_\phi\cos\theta}{\partial t } &&+ \nabla \cdot \left(\rho r u_\phi \cos\theta \; \mathbf u \right) &&=
   && - \frac{\partial p}{\partial \phi}  &&+ \rho  a_\phi \;r \cos \theta\label{eq:app:dpmhd3d:navier4}.
\end{align}
Here, $\rho$ is the density, $\mathbf{u}=(u_r,u_\phi,u_\theta)$ the velocity, $p$ the pressure and $\mathbf{a}=(a_r, a_\phi,a_\theta)$ the acceleration due to the star, planets or other forces (e.g. from a rotating coordinate system). The terms on the left hand side describe advection, the terms on the right hand side are source terms. 
Equations~\ref{eq:app:dpmhd3d:navier1}-\ref{eq:app:dpmhd3d:navier4} are the continuity equation, the radial momentum equation, the meridional momentum equation and the angular momentum equation, respectively. The system of equations is closed with an equation of state, determining the pressure $p$. All simulations presented in this thesis use an isothermal equation of state $p=c_s^2\rho$, where the sound speed $c_s$ might vary with radius.

\subsection{Two dimensions}\label{sec:dpmhd:twodim}
One can simplify the equations in the previous section by averaging over the $\theta$ component, or $z$ for cylindrical coordinates, to obtain a two dimensional system of equations. 
The vertically averaged surface density is given as the integral over the density
\begin{eqnarray}
\Sigma(r,\phi) =\int \rho(r,\phi,z)dz.
\end{eqnarray}
Thus, in two dimensional polar coordinates the Navier-Stokes equations can be written as
\begin{align}
&\frac{\partial \Sigma}{\partial t } &&+ \nabla \cdot \left(\Sigma \mathbf u \right) &&=0 &&\label{eq:app:dpmhd3d:2dnavier1}\\
&\frac{\partial \Sigma u_r}{\partial t } &&+ \nabla \cdot \left(\Sigma u_r \mathbf u \right) &&=
     \Sigma\frac{u_\phi^2}{r}  &&- \frac{\partial p}{\partial r}&&+ \Sigma a_r  \label{eq:app:dpmhd3d:2dnavier2}\\
&\frac{\partial \Sigma r u_\phi}{\partial t } &&+ \nabla \cdot \left(\Sigma r u_\phi  \; \mathbf u \right) &&=
   && - \frac{\partial p}{\partial \phi}  &&+ \Sigma  a_\phi \;r \label{eq:app:dpmhd3d:2dnavier3}.
\end{align}
Here, $\mathbf{u}$ is the velocity two-vector $(u_r,u_\phi)$, sometime also written as $(u,v)$. Equation \ref{eq:app:dpmhd3d:2dnavier1} is the two dimensional continuity equation, \Eq \ref{eq:app:dpmhd3d:2dnavier2} is the radial momentum equation and \Eq \ref{eq:app:dpmhd3d:2dnavier3} is the angular momentum equation.

In general the two dimensional system is a good approximation in planet disc simulations. In two dimensions, a smoothed planet potential is used to simulate the vertical structure of the circum-planetary disc (see equation \ref{eq:nbody}). The smoothing length $b$ is approximately half the local disc scale height.

\subsection{Viscosity}
Strong physical viscosity is often used to model the effective viscosity resulting from turbulence (see section \ref{sec:introduction:mri}). This can be achieved by an additional source term in \Eqs \ref{eq:app:dpmhd3d:navier1}-\ref{eq:app:dpmhd3d:navier4} and \ref{eq:app:dpmhd3d:2dnavier2}-\ref{eq:app:dpmhd3d:2dnavier3}.
Here, we restrict ourselves to the two dimensional case in which the additional terms to \Eqs \ref{eq:app:dpmhd3d:2dnavier2} and \ref{eq:app:dpmhd3d:2dnavier3} are given by \citep[see e.g.][]{Masset2002}
\begin{eqnarray}
f_r &=& \frac 1r \frac{\partial r \tau_{rr}}{\partial r} + \frac 1 r \frac{\partial \tau_{r\phi}}{\partial \phi}-\frac{\tau_{\phi\phi}}r\\
f_\phi &=& \frac 1r \frac{\partial r \tau_{\phi r}}{\partial r} + \frac 1 r \frac{\partial \tau_{\phi\phi}}{\partial \phi}+\frac{\tau_{r\phi}}r.
\end{eqnarray}
The components of the symmetric viscous stress tensor $\tau$ are given by
\begin{eqnarray}
\tau_{rr} 	&=&	2\, \Sigma \nu\, D_{rr} 		- \frac 23\, \Sigma \nu\; \nabla \mathbf{u},\\
\tau_{\phi\phi}	&=&	2\, \Sigma \nu\, D_{\phi\phi} 	- \frac 23\, \Sigma \nu\; \nabla \mathbf{u}\quad\text{and}\\
\tau_{r\phi}	=\tau_{\phi r}	&=&	2\, \Sigma \nu\, D_{r\phi},
\end{eqnarray}
where we have used 
\begin{eqnarray}
D_{rr} 		&=&	\frac{\partial u_r}{\partial r}\\
D_{\phi\phi} 	&=&	\frac1r\, \frac{\partial u_\phi}{\partial \phi}+\frac{u_r}r\\
D_{r\phi} 	&=&	\frac12 \left(r \frac{\partial \left(u_\phi/r\right)}{\partial r} + \frac 1r \frac{\partial u_r}{\partial\phi}\right).
\end{eqnarray}
Here, $\nu$ is the shear viscosity and the bulk viscosity has been assumed to be zero. $\nu$ can be expressed in terms of the $\alpha$ parameter \citep{shakurasyunyaev73} via $\alpha =\nu/(c_s^2/ \Omega)$, with $c_s$ being the local sound speed and $2\pi/\Omega$ being the orbital period.

\section{Discretisation}
\begin{figure}[tb]
\begin{picture}(0,0)%
\includegraphics{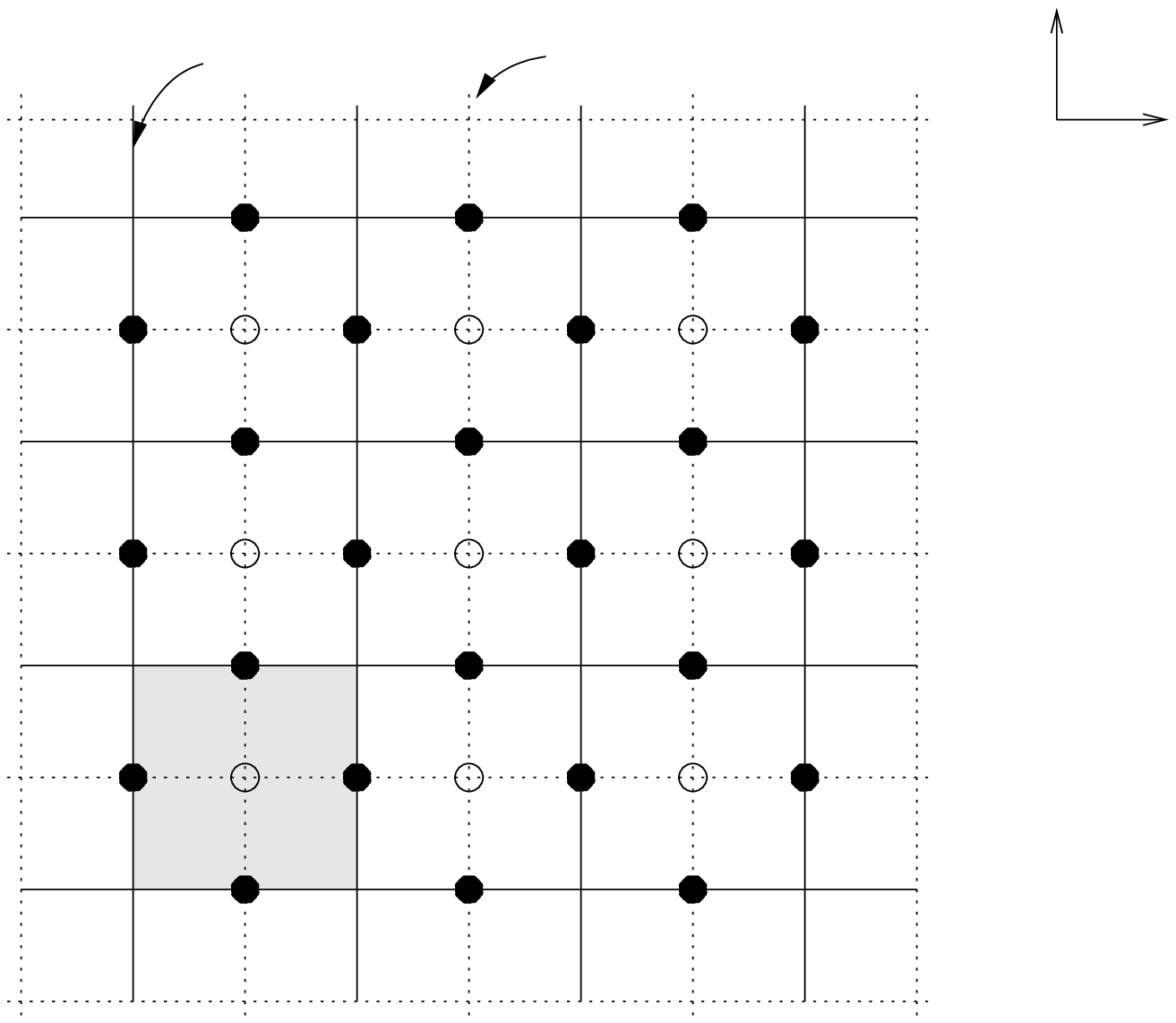}%
\end{picture}%
\setlength{\unitlength}{3947sp}%
\begingroup\makeatletter\ifx\SetFigFont\undefined%
\gdef\SetFigFont#1#2#3#4#5{%
  \reset@font\fontsize{#1}{#2pt}%
  \fontfamily{#3}\fontseries{#4}\fontshape{#5}%
  \selectfont}%
\fi\endgroup%
\begin{picture}(6846,5827)(1576,-6176)
\put(5101,-5911){\makebox(0,0)[lb]{\smash{{\SetFigFont{12}{14.4}{\familydefault}{\mddefault}{\updefault}{\color[rgb]{0,0,0}$i+2$}%
}}}}
\put(6301,-5911){\makebox(0,0)[lb]{\smash{{\SetFigFont{12}{14.4}{\familydefault}{\mddefault}{\updefault}{\color[rgb]{0,0,0}$i+3$}%
}}}}
\put(3901,-5911){\makebox(0,0)[lb]{\smash{{\SetFigFont{12}{14.4}{\familydefault}{\mddefault}{\updefault}{\color[rgb]{0,0,0}$i+1$}%
}}}}
\put(2776,-5911){\makebox(0,0)[lb]{\smash{{\SetFigFont{12}{14.4}{\familydefault}{\mddefault}{\updefault}{\color[rgb]{0,0,0}$i$}%
}}}}
\put(5701,-6136){\makebox(0,0)[lb]{\smash{{\SetFigFont{12}{14.4}{\familydefault}{\mddefault}{\updefault}{\color[rgb]{0,0,0}$i+2$}%
}}}}
\put(6901,-6136){\makebox(0,0)[lb]{\smash{{\SetFigFont{12}{14.4}{\familydefault}{\mddefault}{\updefault}{\color[rgb]{0,0,0}$i+3$}%
}}}}
\put(4501,-6136){\makebox(0,0)[lb]{\smash{{\SetFigFont{12}{14.4}{\familydefault}{\mddefault}{\updefault}{\color[rgb]{0,0,0}$i+1$}%
}}}}
\put(3376,-6136){\makebox(0,0)[lb]{\smash{{\SetFigFont{12}{14.4}{\familydefault}{\mddefault}{\updefault}{\color[rgb]{0,0,0}$i$}%
}}}}
\put(8176,-1186){\makebox(0,0)[lb]{\smash{{\SetFigFont{12}{14.4}{\familydefault}{\mddefault}{\updefault}{\color[rgb]{0,0,0}$r$}%
}}}}
\put(7576,-586){\makebox(0,0)[lb]{\smash{{\SetFigFont{12}{14.4}{\familydefault}{\mddefault}{\updefault}{\color[rgb]{0,0,0}$\phi$}%
}}}}
\put(5176,-661){\makebox(0,0)[lb]{\smash{{\SetFigFont{12}{14.4}{\familydefault}{\mddefault}{\updefault}{\color[rgb]{0,0,0}staggered grid}%
}}}}
\put(3301,-661){\makebox(0,0)[lb]{\smash{{\SetFigFont{12}{14.4}{\familydefault}{\mddefault}{\updefault}{\color[rgb]{0,0,0}grid}%
}}}}
\put(1576,-3361){\makebox(0,0)[lb]{\smash{{\SetFigFont{12}{14.4}{\familydefault}{\mddefault}{\updefault}{\color[rgb]{0,0,0}$j+1$}%
}}}}
\put(1576,-2161){\makebox(0,0)[lb]{\smash{{\SetFigFont{12}{14.4}{\familydefault}{\mddefault}{\updefault}{\color[rgb]{0,0,0}$j+2$}%
}}}}
\put(1576,-961){\makebox(0,0)[lb]{\smash{{\SetFigFont{12}{14.4}{\familydefault}{\mddefault}{\updefault}{\color[rgb]{0,0,0}$j+3$}%
}}}}
\put(1726,-3961){\makebox(0,0)[lb]{\smash{{\SetFigFont{12}{14.4}{\familydefault}{\mddefault}{\updefault}{\color[rgb]{0,0,0}$j+1$}%
}}}}
\put(1726,-2761){\makebox(0,0)[lb]{\smash{{\SetFigFont{12}{14.4}{\familydefault}{\mddefault}{\updefault}{\color[rgb]{0,0,0}$j+2$}%
}}}}
\put(1726,-1561){\makebox(0,0)[lb]{\smash{{\SetFigFont{12}{14.4}{\familydefault}{\mddefault}{\updefault}{\color[rgb]{0,0,0}$j+3$}%
}}}}
\put(3526,-4336){\makebox(0,0)[lb]{\smash{{\SetFigFont{12}{14.4}{\familydefault}{\mddefault}{\updefault}{\color[rgb]{0,0,0}$\rho_{i,j}$}%
}}}}
\put(3526,-5011){\makebox(0,0)[lb]{\smash{{\SetFigFont{12}{14.4}{\familydefault}{\mddefault}{\updefault}{\color[rgb]{0,0,0}$v_{i,j}$}%
}}}}
\put(3526,-3811){\makebox(0,0)[lb]{\smash{{\SetFigFont{12}{14.4}{\familydefault}{\mddefault}{\updefault}{\color[rgb]{0,0,0}$v_{i,j+1}$}%
}}}}
\put(3526,-2611){\makebox(0,0)[lb]{\smash{{\SetFigFont{12}{14.4}{\familydefault}{\mddefault}{\updefault}{\color[rgb]{0,0,0}$v_{i,j+2}$}%
}}}}
\put(4126,-4711){\makebox(0,0)[lb]{\smash{{\SetFigFont{12}{14.4}{\familydefault}{\mddefault}{\updefault}{\color[rgb]{0,0,0}$u_{i+1,j}$}%
}}}}
\put(2926,-4711){\makebox(0,0)[lb]{\smash{{\SetFigFont{12}{14.4}{\familydefault}{\mddefault}{\updefault}{\color[rgb]{0,0,0}$u_{i,j}$}%
}}}}
\put(2926,-3511){\makebox(0,0)[lb]{\smash{{\SetFigFont{12}{14.4}{\familydefault}{\mddefault}{\updefault}{\color[rgb]{0,0,0}$u_{i,j+1}$}%
}}}}
\put(5326,-4711){\makebox(0,0)[lb]{\smash{{\SetFigFont{12}{14.4}{\familydefault}{\mddefault}{\updefault}{\color[rgb]{0,0,0}$u_{i+2,j}$}%
}}}}
\put(3526,-3136){\makebox(0,0)[lb]{\smash{{\SetFigFont{12}{14.4}{\familydefault}{\mddefault}{\updefault}{\color[rgb]{0,0,0}$\rho_{i,j+1}$}%
}}}}
\put(4726,-3136){\makebox(0,0)[lb]{\smash{{\SetFigFont{12}{14.4}{\familydefault}{\mddefault}{\updefault}{\color[rgb]{0,0,0}$\rho_{i+1,j+1}$}%
}}}}
\put(5926,-3136){\makebox(0,0)[lb]{\smash{{\SetFigFont{12}{14.4}{\familydefault}{\mddefault}{\updefault}{\color[rgb]{0,0,0}$\rho_{i+2,j+1}$}%
}}}}
\put(4726,-4336){\makebox(0,0)[lb]{\smash{{\SetFigFont{12}{14.4}{\familydefault}{\mddefault}{\updefault}{\color[rgb]{0,0,0}$\rho_{i+1,j}$}%
}}}}
\put(5926,-4336){\makebox(0,0)[lb]{\smash{{\SetFigFont{12}{14.4}{\familydefault}{\mddefault}{\updefault}{\color[rgb]{0,0,0}$\rho_{i+2,j}$}%
}}}}
\put(2026,-5161){\makebox(0,0)[lb]{\smash{{\SetFigFont{12}{14.4}{\familydefault}{\mddefault}{\updefault}{\color[rgb]{0,0,0}$j$}%
}}}}
\put(1876,-4561){\makebox(0,0)[lb]{\smash{{\SetFigFont{12}{14.4}{\familydefault}{\mddefault}{\updefault}{\color[rgb]{0,0,0}$j$}%
}}}}
\put(4726,-3811){\makebox(0,0)[lb]{\smash{{\SetFigFont{12}{14.4}{\familydefault}{\mddefault}{\updefault}{\color[rgb]{0,0,0}$v_{i+1,j+1}$}%
}}}}
\put(4726,-5011){\makebox(0,0)[lb]{\smash{{\SetFigFont{12}{14.4}{\familydefault}{\mddefault}{\updefault}{\color[rgb]{0,0,0}$v_{i+1,j}$}%
}}}}
\end{picture}%
\centering
\caption{Discretisation and staggered grid in two dimensions, showing the location of the cell and face centred variables.\label{fig:dpmhd3d:discrete}}
\end{figure}

To solve \Eqs \ref{eq:app:dpmhd3d:navier1}-\ref{eq:app:dpmhd3d:navier4} or \ref{eq:app:dpmhd3d:2dnavier1}-\ref{eq:app:dpmhd3d:2dnavier3} numerically, we discretise them on a grid. A staggered grid, where the velocities are face-centred and the densities are cell-centred has been used, as illustrated in the two dimensional case in \Fig \ref{fig:dpmhd3d:discrete}.  The discretisation is identical to that used in the ZEUS code. A staggered grid has the advantage that spacial derivatives are automatically second order accurate, but does also add some complexity, especially in three dimensions.

In the radial direction, the boundary conditions can be chosen to be either reflecting or damping. In the azimuthal direction, the boundary conditions are periodic. It is also possible to simulate only a wedge of the disc (without any planets). In three dimensions, periodic boundary conditions are used in the $\theta$ direction.

\subsection{Operator splitting}
\begin{figure}
\resizebox{\columnwidth}{!}{
\begin{picture}(0,0)%
\includegraphics{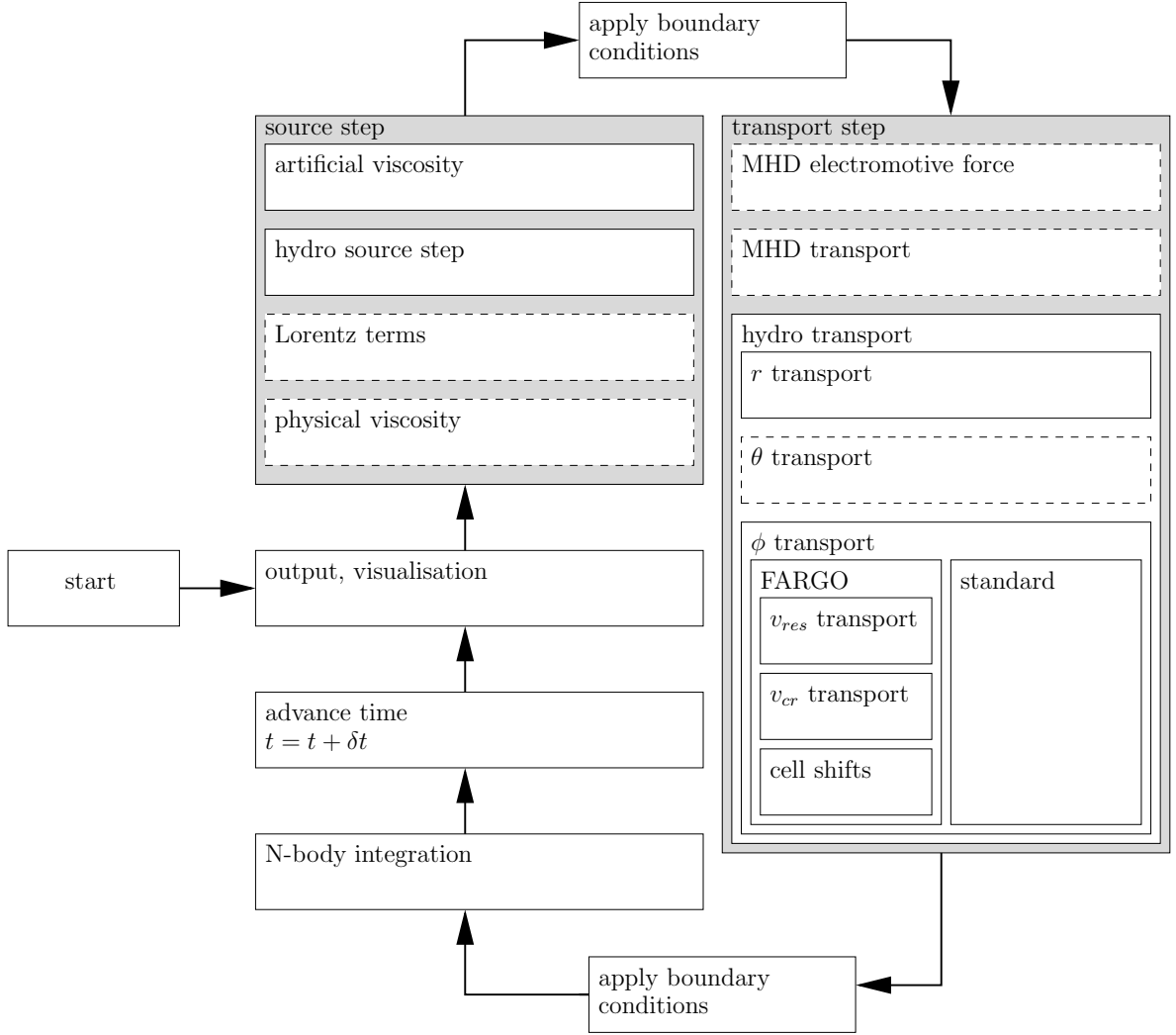}%
\end{picture}%
\setlength{\unitlength}{3947sp}%
\begingroup\makeatletter\ifx\SetFigFont\undefined%
\gdef\SetFigFont#1#2#3#4#5{%
  \reset@font\fontsize{#1}{#2pt}%
  \fontfamily{#3}\fontseries{#4}\fontshape{#5}%
  \selectfont}%
\fi\endgroup%
\begin{picture}(9174,8199)(-2111,-11473)
\put(3601,-4336){\makebox(0,0)[lb]{\smash{{\SetFigFont{12}{14.4}{\familydefault}{\mddefault}{\updefault}{\color[rgb]{0,0,0}transport step}%
}}}}
\put(3676,-4636){\makebox(0,0)[lb]{\smash{{\SetFigFont{12}{14.4}{\familydefault}{\mddefault}{\updefault}{\color[rgb]{0,0,0}MHD electromotive force}%
}}}}
\put(3676,-5311){\makebox(0,0)[lb]{\smash{{\SetFigFont{12}{14.4}{\familydefault}{\mddefault}{\updefault}{\color[rgb]{0,0,0}MHD transport}%
}}}}
\put(3676,-5986){\makebox(0,0)[lb]{\smash{{\SetFigFont{12}{14.4}{\familydefault}{\mddefault}{\updefault}{\color[rgb]{0,0,0}hydro transport}%
}}}}
\put(3751,-6286){\makebox(0,0)[lb]{\smash{{\SetFigFont{12}{14.4}{\familydefault}{\mddefault}{\updefault}{\color[rgb]{0,0,0}$r$ transport}%
}}}}
\put(3751,-6961){\makebox(0,0)[lb]{\smash{{\SetFigFont{12}{14.4}{\familydefault}{\mddefault}{\updefault}{\color[rgb]{0,0,0}$\theta$ transport}%
}}}}
\put(3751,-7636){\makebox(0,0)[lb]{\smash{{\SetFigFont{12}{14.4}{\familydefault}{\mddefault}{\updefault}{\color[rgb]{0,0,0}$\phi$ transport}%
}}}}
\put(3826,-7936){\makebox(0,0)[lb]{\smash{{\SetFigFont{12}{14.4}{\familydefault}{\mddefault}{\updefault}{\color[rgb]{0,0,0}FARGO}%
}}}}
\put(5401,-7936){\makebox(0,0)[lb]{\smash{{\SetFigFont{12}{14.4}{\familydefault}{\mddefault}{\updefault}{\color[rgb]{0,0,0}standard}%
}}}}
\put(3901,-8836){\makebox(0,0)[lb]{\smash{{\SetFigFont{12}{14.4}{\familydefault}{\mddefault}{\updefault}{\color[rgb]{0,0,0}$v_{cr}$ transport}%
}}}}
\put(3901,-8236){\makebox(0,0)[lb]{\smash{{\SetFigFont{12}{14.4}{\familydefault}{\mddefault}{\updefault}{\color[rgb]{0,0,0}$v_{res}$ transport}%
}}}}
\put(3901,-9436){\makebox(0,0)[lb]{\smash{{\SetFigFont{12}{14.4}{\familydefault}{\mddefault}{\updefault}{\color[rgb]{0,0,0}cell shifts}%
}}}}
\put(-74,-4336){\makebox(0,0)[lb]{\smash{{\SetFigFont{12}{14.4}{\familydefault}{\mddefault}{\updefault}{\color[rgb]{0,0,0}source step}%
}}}}
\put(  1,-4636){\makebox(0,0)[lb]{\smash{{\SetFigFont{12}{14.4}{\familydefault}{\mddefault}{\updefault}{\color[rgb]{0,0,0}artificial viscosity}%
}}}}
\put(  1,-6661){\makebox(0,0)[lb]{\smash{{\SetFigFont{12}{14.4}{\familydefault}{\mddefault}{\updefault}{\color[rgb]{0,0,0}physical viscosity}%
}}}}
\put(  1,-5986){\makebox(0,0)[lb]{\smash{{\SetFigFont{12}{14.4}{\familydefault}{\mddefault}{\updefault}{\color[rgb]{0,0,0}Lorentz terms}%
}}}}
\put(-74,-7861){\makebox(0,0)[lb]{\smash{{\SetFigFont{12}{14.4}{\familydefault}{\mddefault}{\updefault}{\color[rgb]{0,0,0}output, visualisation}%
}}}}
\put(-74,-10111){\makebox(0,0)[lb]{\smash{{\SetFigFont{12}{14.4}{\familydefault}{\mddefault}{\updefault}{\color[rgb]{0,0,0}N-body integration}%
}}}}
\put(-1649,-7936){\makebox(0,0)[lb]{\smash{{\SetFigFont{12}{14.4}{\familydefault}{\mddefault}{\updefault}{\color[rgb]{0,0,0}start}%
}}}}
\put(  1,-5311){\makebox(0,0)[lb]{\smash{{\SetFigFont{12}{14.4}{\familydefault}{\mddefault}{\updefault}{\color[rgb]{0,0,0}hydro source step}%
}}}}
\put(2476,-3511){\makebox(0,0)[lb]{\smash{{\SetFigFont{12}{14.4}{\familydefault}{\mddefault}{\updefault}{\color[rgb]{0,0,0}apply boundary}%
}}}}
\put(2476,-3736){\makebox(0,0)[lb]{\smash{{\SetFigFont{12}{14.4}{\familydefault}{\mddefault}{\updefault}{\color[rgb]{0,0,0}conditions}%
}}}}
\put(2551,-11086){\makebox(0,0)[lb]{\smash{{\SetFigFont{12}{14.4}{\familydefault}{\mddefault}{\updefault}{\color[rgb]{0,0,0}apply boundary}%
}}}}
\put(2551,-11311){\makebox(0,0)[lb]{\smash{{\SetFigFont{12}{14.4}{\familydefault}{\mddefault}{\updefault}{\color[rgb]{0,0,0}conditions}%
}}}}
\put(-74,-9211){\makebox(0,0)[lb]{\smash{{\SetFigFont{12}{14.4}{\familydefault}{\mddefault}{\updefault}{\color[rgb]{0,0,0}$t=t+\delta t$}%
}}}}
\put(-74,-8986){\makebox(0,0)[lb]{\smash{{\SetFigFont{12}{14.4}{\familydefault}{\mddefault}{\updefault}{\color[rgb]{0,0,0}advance time}%
}}}}
\end{picture}%
}
\centering
\caption{A full time-step performed by Prometheus\label{fig:dpmhd:timestep}. Dashed boxes represent optional modules.}
\end{figure}

Prometheus makes use of the operator splitting technique, where the equations are solved in small chunks.
To do that, a time-step is divided into several sub-time-steps. Figure \ref{fig:dpmhd:timestep} illustrates the main loop in \Prom, including the additional function calls for the MHD module, which are not discussed any further here. 

First, the source terms (right hand side of \Eqs \ref{eq:app:dpmhd3d:navier1}-\ref{eq:app:dpmhd3d:navier4} and \ref{eq:app:dpmhd3d:2dnavier1}-\ref{eq:app:dpmhd3d:2dnavier3}) are added.
Then, the advection terms (left hand side of \Eqs \ref{eq:app:dpmhd3d:navier1}-\ref{eq:app:dpmhd3d:navier4} and \ref{eq:app:dpmhd3d:2dnavier1}-\ref{eq:app:dpmhd3d:2dnavier3}) are solved using the half updated values after the source step, ignoring all terms on the right hand side. 
We further use dimensional splitting, solving the advection in the radial, azimuthal and meridional directions consecutively. 

This method is only a simplified approximation to the real solution, but has proven to be more accurate than a single time-step. 
It is (approximately) second order accurate in time.
Furthermore, it has the advantage that new physics can be easily incorporated as additional sub-time-steps. In this code we make use of this when an MHD module is added.
However, operator splitting can also lead to spurious results in special circumstances, as shown in section \ref{sec:dpmhd:dispersionerror}.

\subsection{Constrained transport}
The transport step is solved using the integral representation of \Eqs \ref{eq:app:dpmhd3d:navier1}-\ref{eq:app:dpmhd3d:navier4} and \ref{eq:app:dpmhd3d:2dnavier1}-\ref{eq:app:dpmhd3d:2dnavier3} to ensure conservation of mass and (angular-)momentum up to machine precision. We therefore need to have the fluxes at cell boundaries.  There are different ways of estimating the cell faced, time centred flux at each interface, whose discussion goes beyond the scope of this section. In general, higher order methods are less diffusive but computationally more expensive. Both the first order donor cell method and the second order van Leer method have been implemented. 

Additionally, a concept called constrained transport is used, where the momentum flux calculations are constrained, such that they are consistent with the mass flux. 
This leads to improved conservation properties. It is physically motivated, as all conserved quantities are advected by the fluid.

\subsection{Splitting the angular momentum equation}
\cite{Masset00} proposes a method, called FARGO (Fast Advection in Rotating Gaseous Objects), to speed up calculations in which there is a large mean background flow, as it is the case in a sheared accretion disc. 
The idea is to rewrite the advection terms, the second column in equations \ref{eq:app:dpmhd3d:navier1}-\ref{eq:app:dpmhd3d:navier4} and \ref{eq:app:dpmhd3d:2dnavier1}-\ref{eq:app:dpmhd3d:2dnavier3}, by replacing the angular velocity $u_\phi$ by the sum  $\bar{u}_\phi+\hat{u}_\phi+  u'_\phi$. We choose $\bar{u}_\phi$ to be the mean angular velocity of the annulus, rounded such that $\bar{u}_\phi \;\delta t$ is an integer multiple of the grid spacing $\delta \phi$. $\hat{u}_\phi$ is the remainder of the rounding operation and thus less or equal than $0.5\,\delta \phi/\delta t$. Both $\bar{u}_\phi$ and $\hat{u}_\phi$ are independent of $\phi$ and defined for an annulus with constant $r$. The residual velocity is given by $u'_\phi$. Thus, the contribution from the $\phi$ component in a generic advection term of the form $\nabla \cdot \left(a\, \mathbf u \right)$ can be written as
\begin{eqnarray}
\frac{\partial}{\partial \phi}  \left(a\, (\bar{u}_\phi+\hat{u}_\phi+u'_\phi) \right)
&=&\underbrace{\bar{u}_\phi \frac{\partial}{\partial \phi}  \left(a \right)}_{\text{(a)}}}
+\underbrace{\hat{u}_\phi \frac{\partial}{\partial \phi}  \left(a \right)}_{\text{(b)}}
+\underbrace{\frac{\partial}{\partial \phi}  \left(a\,u'_\phi \right),}_{\text{(c)}
\end{eqnarray}
where $a$ can be any quantity. The term (c) is just like the old advection term, but with a different velocity, as if we were in a rotating frame. The terms (a) and (b) can be interpreted as a simple shift. Because we've chosen $\bar{u}_\phi$ such that $\bar{u}_\phi/\delta t$ is an integer multiple of the grid spacing $\delta \phi$, the term (a) corresponds to shifting the whole ring of cells, which is numerically very inexpensive. Because $|u'_\phi|\ll|u_\phi|$ and $|\hat{u}_\phi|\ll|u_\phi|$ the time-step constraint arising from the CFL condition allows us now to use much larger time-steps and the terms (b) and (c) can be solved using a standard advection algorithm. The splitting of the advection term gives an efficiency boost of a factor $\sim 10$, in a typical simulation.

\section{Tests}
\subsection{Shock-tube problems}\label{sec:strongshocktube}
\begin{figure}
\centering
\includegraphics[angle=270,width=1.0\columnwidth]{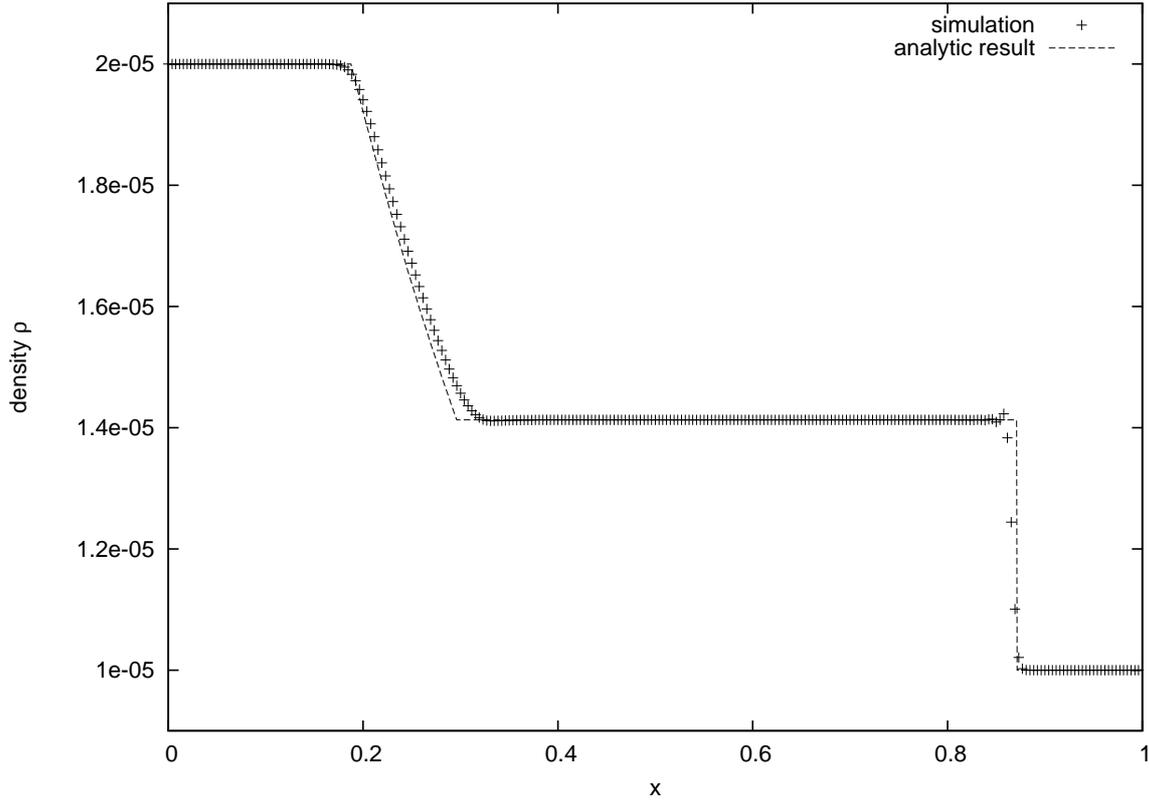}
\caption{Density profile for a strong shock-tube problem at time $t=10$ with Mach number $v/c=2.9$.  \label{fig:strongshocktube}}
\end{figure}
Shock-tube problems are a standard test in numerical codes \citep[see e.g.][]{StoneNorman1992}. Even though these test are in general not very thorough and easy to pass, we present a one dimensional example in Cartesian geometry, mainly for comparison with a low Mach number shock in section \ref{sec:dpmhd:dispersionerror}.

The Mach number of the shock front is $v/c_s = 2.9$. An equidistantly spaced grid with 256 grid points, a CFL condition of 0.5 and a sound speed of $c_s=0.05$ is used. The density profile after $t=10$ together with the analytic solution is plotted in \Fig \ref{fig:strongshocktube}. One can see that the artificial viscosity smears out the discontinuity over several grid cells.
Furthermore, the rarefaction wave results in some undershooting.  
These effects are expected and present in any finite difference code.

\subsection{Dispersion error}\label{sec:dpmhd:dispersionerror}

Finite difference codes are known to have a numerical dispersion error because they do not, contrary to Godunov methods, solve the equations along wave characteristics.
The fluid equations are solved using operator splitting. Each part, namely the source and transport step, are evaluated successively. Propagating sound waves involve both operators. The advection of a passive scalar, contrariwise, requires only one operator.

 In most situations, this does not create any problems, as the error associated with the operator splitting is small and mostly affects the highest frequency modes which appear on the grid scale. However, if there is a discontinuity in the flow, large frequencies are present that can lead to a large dispersion error. For supersonic shocks, the artificial viscosity takes care of this, by smearing out discontinuities over a few grid cells, as shown in section \ref{sec:strongshocktube}. For low amplitude, \textit{subsonic} shocks this might not be the case. The shock behaves like a travelling wave package, where the package includes a discontinuity. 

Unfortunately, we are exactly in this situation for a low mass planet being embedded in an accretion disc. The wake of the planet contains a discontinuity and its amplitude depends on the planet's mass. For sufficiently small masses, the discontinuity becomes subsonic and we expect noticeable numerical dispersion to occur. 
Since the wake is stationary in a frame co-rotating with the planet, the result will also be a stationary pattern. 

In the following, we will first illustrate the problem with a shock tube test in a Cartesian box. We will then show that the same pattern appears in simulations of low mass planets and outline possible solutions such as physical viscosity and Godunov methods.

\subsubsection{Low Mach number shock-tube tests}\label{sec:shocktube}
\begin{figure}
\centering
\includegraphics[angle=270,width=1.0\columnwidth]{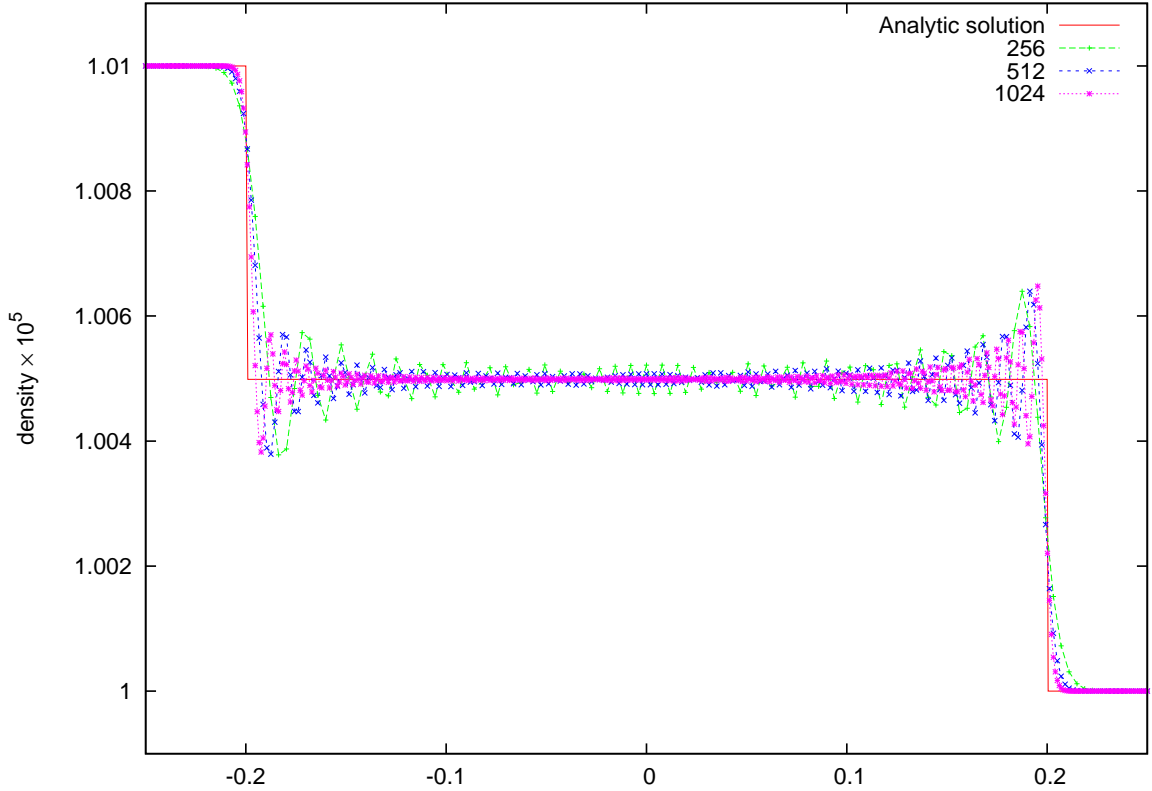}
\caption{Density profiles for different grid resolutions in the isothermal ($c_s=0.05$) shock-tube problem at time $t=4$. The initial density perturbation is 1\%.  \label{fig:shock}}
\end{figure}
We set up a simple test problem that shows the problematic behaviour. We work in a Cartesian grid with resolutions 256x256, 512x512 and 1024x1024 and use an isothermal equation of state with sound speed $c_s = 0.05$ in dimensionless units.
The box has periodic boundary conditions and we set up a shock-tube problem by increasing the density in the left part of the box by 1\%. The analytic solution after $t=4$ is plotted as a solid line in figure \ref{fig:shock}. Due to the low amplitude of the initial perturbation, the waves are basically sound packages travelling at the speed of sound, i.e. the rarefraction wave is very steep. Large ringing can be observed in the numerical solutions, which are also plotted in figure \ref{fig:shock}. This is because the numerical dispersion error is largest for the smallest wavelength. As expected for a Gibbs-like phenomenon, the over and under-shoots do not die out with increasing resolution. However, they get more concentrated near the discontinuity. 

Note that in all simulations, the Neumann and Richtmyer artificial viscosity is switched on. The artificial viscosity does not kill the overshoots as it is quadratic in the velocities. This is reasonable approach in strong shocks \citep[][section \ref{sec:strongshocktube}]{StoneNorman1992}. However, the velocity perturbations are very small compared to the sound speed in this case. 


\subsubsection{Low mass planets}
\begin{figure}[p]
\centering
\includegraphics[angle=270,width=1.0\columnwidth]{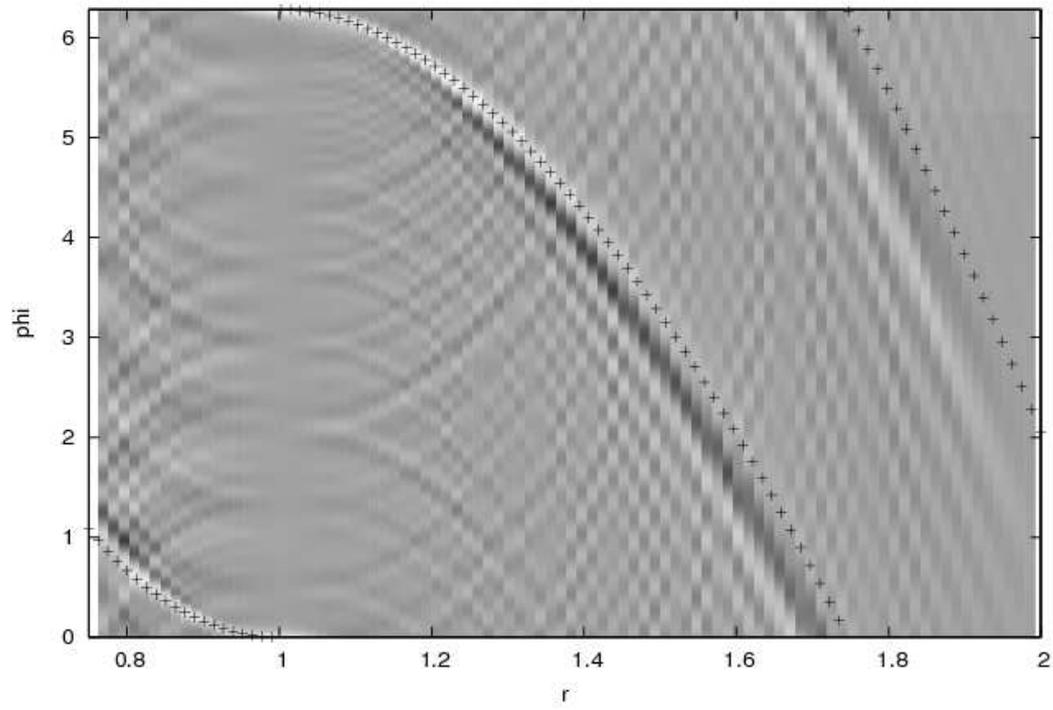}
\includegraphics[angle=270,width=1.0\columnwidth]{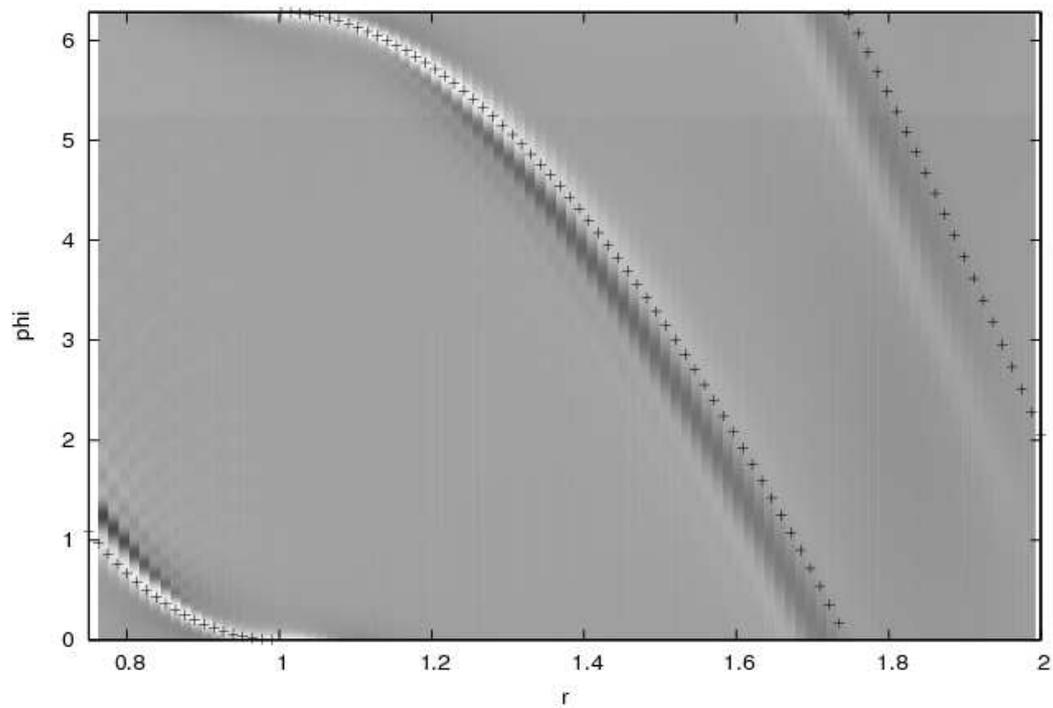}
\caption{Surface density after 10 orbits in simulations with a resolution of 128x384. The simulation on the top uses no physical viscosity, whereas the simulation on the bottom uses $\nu=2\cdot10^{-5}$. \label{fig:shock128}}
\end{figure}

A similar phenomenon can be observed in simulations of planet disc interactions. All integrations were performed with Prometheus. The results have been confirmed with a variety of codes (FARGO, NIRVANA and RH2D) and do not depend on the implementation. 

The phenomenon is stronger for low mass planets, as the shocks that appear in the wake are weaker and eventually become subsonic. Therefore, a planet with a very low mass of $m_p=1.26\cdot10^{-6}M_\odot=0.42M_\mathrm{earth}$ is used in these simulations. The cylindrical grid is spaced equidistantly in the radial direction and has a resolution of 128x384. We plot the surface density contours of one simulation with and without physical viscosity in figure \ref{fig:shock128}. We over-plot the analytic position of the wake \citep{OgilvieLubow2002}.

One can clearly see that there are over-shootings occurring, similar to those discussed above. Including a small amount of physical viscosity smears out the phenomenon. Increasing the resolution also helps to reduce the strength of the over-shootings. 
Numerical diffusion becomes more important if more grid cells, and therefore also a smaller time-step, are used. 

For reasonable resolutions ($N_\phi>512$), the torque on the planet is not significantly affected by this effect for two reasons. Firstly, the over-shootings appear a few scale heights away from the planet. Secondly, the gravitational torque from the over and under-shootings cancel and give almost no net effect. However, note that this is not necessarily the case for multi planetary systems or situations in which the saturation of co-rotation torques plays an important role \citep{PaardekooperPapaloizou2009}. In those cases, finite difference codes should only be used with great care and possibly with explicit physical viscosity. 

The phenomenon described here does not occur in Godunov codes such as RODEO (S.-J. Paardekooper, private communication). This is expected, as those schemes do solve the hydrodynamical equations along characteristics and therefore do not have a dispersion error.

\subsection{Planet torques}
\begin{figure}[htbp]
\centering
\includegraphics[angle=270,width=1.0\columnwidth]{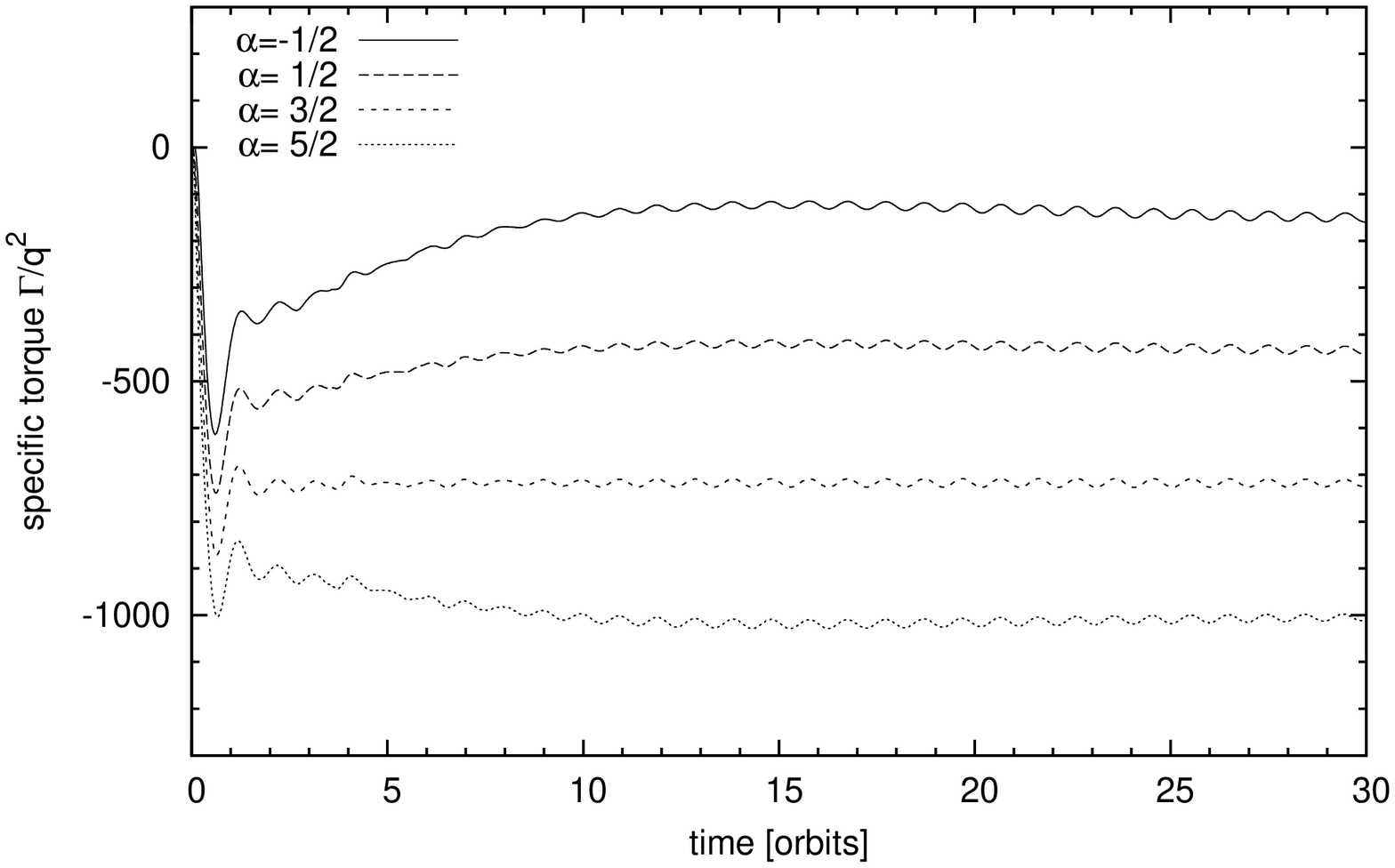}
\caption{Specific torque felt by a planet with mass $q=1.26\cdot10^{-5}$ in an inviscid disc with varying surface density slope $\alpha$. The results are in good agreement with \cite{PaardekooperPapaloizou2009}. \label{fig:dpmhd:torque1}}
\end{figure}
As a further test, we calculate the torques for low mass planets in various disc models. A planet mass of $q=1.26\cdot10^{-5}$ and a smoothing length of $b=0.6$ is used, so that the simulations can be easily compared to those performed by \cite{PaardekooperPapaloizou2009}. We first test the two dimensional version of Prometheus without physical viscosity. The specific torque $\Gamma=\left|\mathbf{f}\times\mathbf{r}\right|$ felt by the planet is plotted in \Fig \ref{fig:dpmhd:torque1}, normalised by the square of the planet mass ratio time the disc surface density for four different surface density profiles with slope $\alpha$. It can be seen by comparing \Fig \ref{fig:dpmhd:torque1} to figure 7 in \cite{PaardekooperPapaloizou2009} that the measured torques (and therefore the migration rates) are in very good agreement.  
In particular the case $\alpha=3/2$ is in accord with linear theory (since it does not vary on a libration timescale). 
The long term trend visible towards the end of the simulation is due to different boundary conditions used (damping).

\subsection{Planet disc interaction in three dimensions}
\begin{figure}[httb]
\centering
\includegraphics[angle=270,width=1.0\columnwidth]{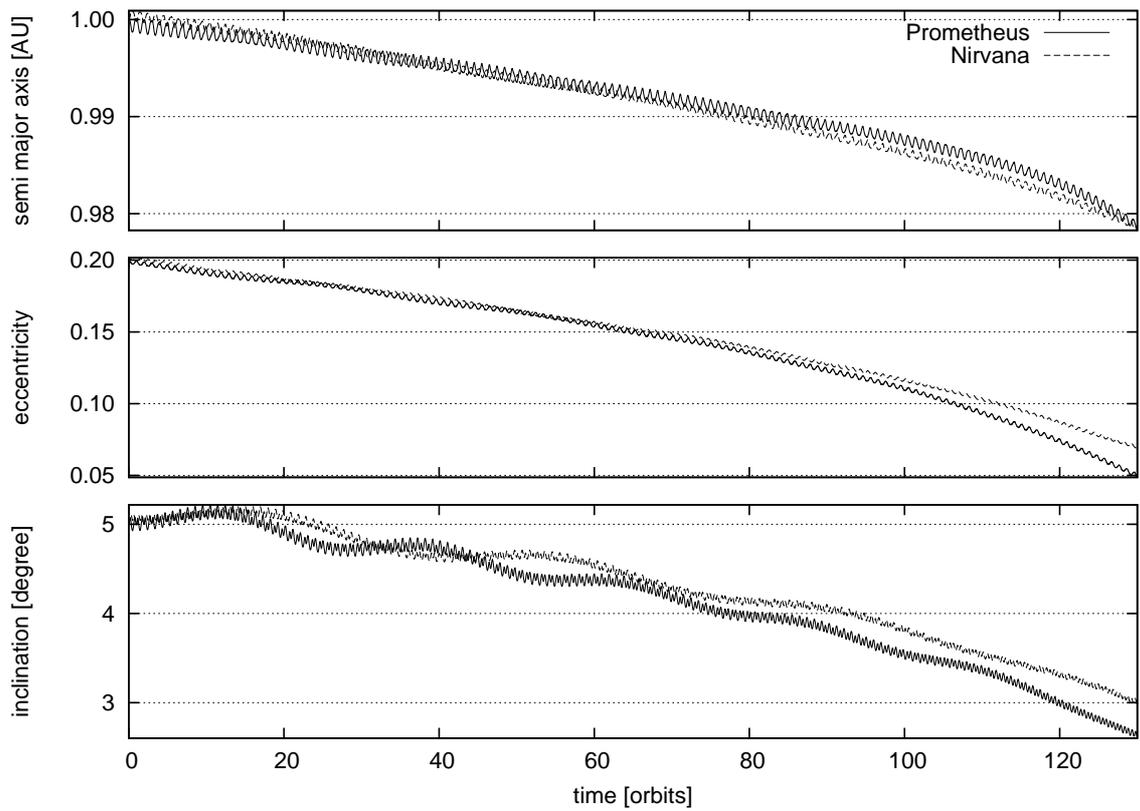}
\caption{Orbital parameters of a planet embedded in a three dimensional disc. The solid curves have been obtained with Prometheus, the dashed curves have been obtained with Nirvana by \cite{KleyBitschKlahr2009}. \label{fig:dpmhd:test3d}}
\end{figure}
We now go on and test a three dimensional simulation of planet disc interaction. Therefore, the results from Prometheus are compared to those of \cite{KleyBitschKlahr2009}. 

A 20 Earth mass planet is set up on an eccentric and inclined orbit ($e=0.2,\,i=5\degree$) and embedded in a stratified, three dimensional disc. The disc mass in the computational domain ($r=0.4\ldots2.5$) is $0.01\text{ M}_\odot$ and the aspect ratio of the disc is 0.05. The grid resolution used is $128\times384\times34$. We use no viscosity and a planet potential which is smoothed over a fraction (0.43) of a scale height. Note that this is a slightly different setup, compared to \cite{KleyBitschKlahr2009}.

Both, the results from Prometheus and those of \cite{KleyBitschKlahr2009} are plotted in \Fig \ref{fig:dpmhd:test3d}. They are in good agreement and show only some small variation at large times.

\clearpage
\section{Visualisations}
\begin{figure}[tb]
\centering
\includegraphics[width=0.8\columnwidth]{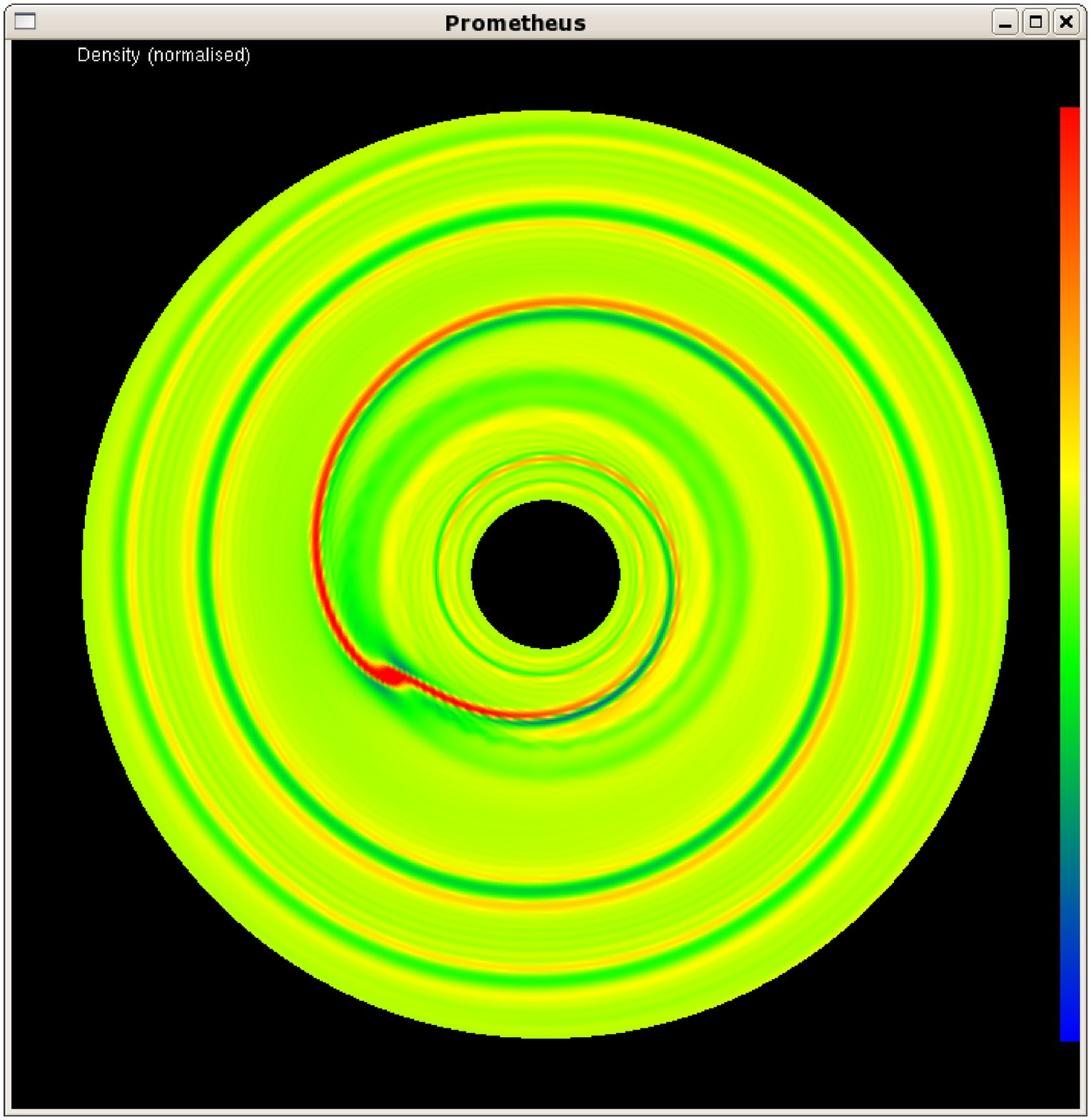} 
\caption{Screenshot of the OpenGL visualisation module of Prometheus, showing the surface density in a two dimensional simulation with an embedded planet.
\label{fig:dpmhd:screen1}}
\end{figure}
\begin{figure}[tb]
\centering
\includegraphics[width=0.8\columnwidth]{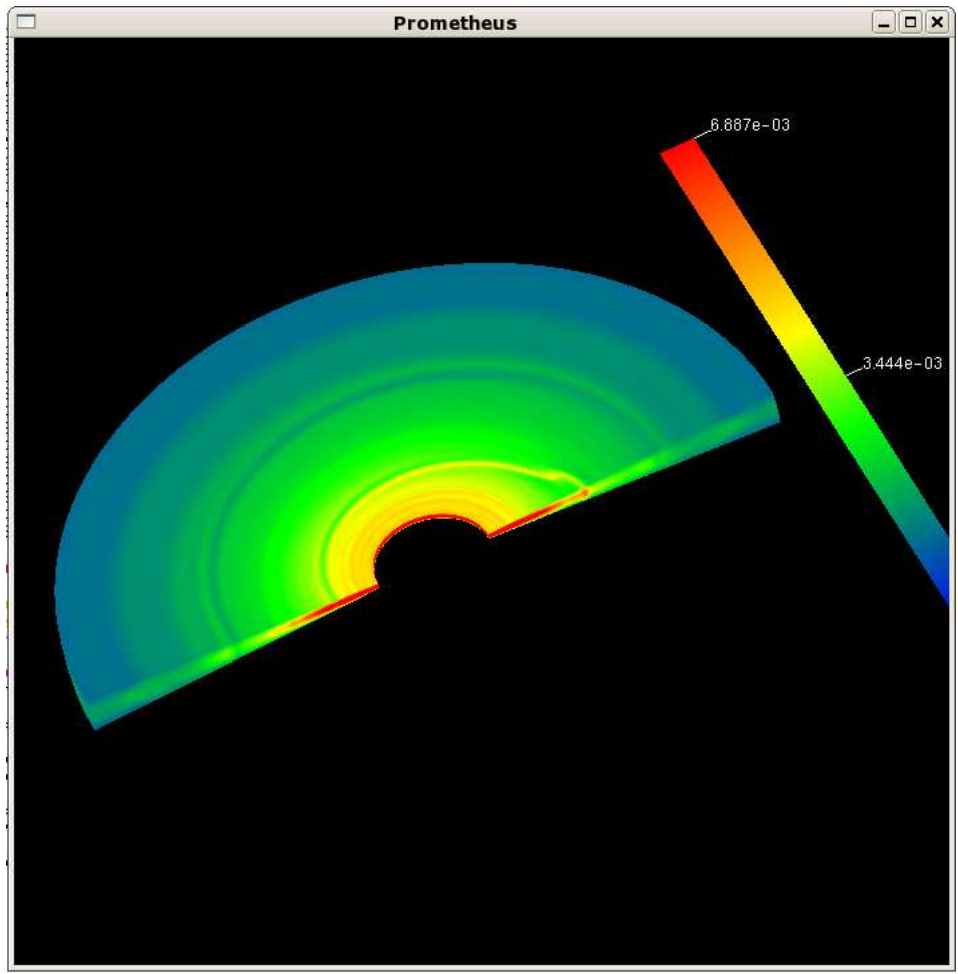} 
\caption{Screenshot of the OpenGL visualisation module of Prometheus, showing the density in a three dimensional simulation of a stratified proto-planetary disc with an embedded planet. 
\label{fig:dpmhd:screen2}}
\end{figure}
Similar to the visualisation module presented in appendix \ref{app:gravtree}, Prometheus comes with 
a visualisation module, based on the OpenGL API. It can be used to visualise simulations in real time while they are running
or in a standalone playback mode. 
Standard quantities, such as density, velocity, divergence of velocity and vorticity can be plotted in both, two and three dimensional simulations in either a cylindrical/spherical or Cartesian representation.
Two screenshots are shown in figures \ref{fig:dpmhd:screen1} and \ref{fig:dpmhd:screen2}.

The visualisation module is also easily adaptable to allow for direct user interaction with the simulation. For example, this can be used to add perturbations or additional planets to the simulation and watch the simulation change instantaneously.
This provides an astonishing new way to easily understand physical phenomena. 

Furthermore, an iPhone version of Prometheus is available on the Apple AppStore at \url{http://itunes.apple.com/us/app/hydro/id341643251?mt=8}.

\backmatter
\bibliographystyle{aa}
\bibliography{../bibliography/full}
\newpage \thispagestyle{empty}\mbox{}

\end{document}